\documentclass[twocolumn,floats,pra,eqsecnum,showpacs]{revtex4-1}

\usepackage{amsmath}
\usepackage{amsfonts}
\usepackage{amssymb}

\usepackage{tikz}

\newcommand {\be}{\begin{equation}}
\newcommand {\ee} {\end{equation}}
\newcommand {\bea}{\begin{eqnarray}}
\newcommand {\eea} {\end{eqnarray}}
\newcommand{\non}{\nonumber}

\newcommand{\bx}{{\bf x}}
\newcommand{\bbZ}{{\mathbb{Z}}}
\newcommand{\bbN}{{\mathbb{N}}}
\newcommand{\bbE}{{\mathbb{E}}}
\newcommand{\bbR}{{\mathbb{R}}}

\begin{document}


\title{Structural analysis of Gibbs states and metastates in short-range classical spin glasses:\\indecomposable
metastates, dynamically-frozen states, and metasymmetry}
\author{N. Read}
\affiliation{Department of Physics, Yale University, P.O. Box 208120, New Haven, CT 06520-8120\\
Department of Applied Physics, Yale University, P.O. Box 208284, New Haven, CT 06520-8284}
\date{July 28, 2024}

\begin{abstract}
We consider short-range classical spin glasses, or other disordered systems, consisting of Ising spins. For a low-temperature 
Gibbs state in infinite size in such a system, for given random bonds, it is controversial whether its decomposition into pure states 
will be trivial or non-trivial. 
We undertake a general study of the overall structure of this problem, based on metastates, which are essential to prove the existence of a 
thermodynamic limit. A metastate is a probability distribution on Gibbs states, for given disorder, that satisfies certain covariance properties. 
First, we prove that any metastate can be decomposed as a mixture of 
indecomposable metastates, and that all Gibbs states drawn from an indecomposable metastate are alike macroscopically. 
Next, we consider stochastic stability of a metastate under random perturbations of the disorder, and prove that any metastate is
stochastically stable. Using related methods and older results, we prove that if the pure-state decomposition of any Gibbs states drawn from 
an indecomposable metastate is countably infinite, then the weights in the decomposition follow a Poisson-Dirichlet distribution 
with a fixed value of the single parameter describing such distributions, and also that if the overlap takes a finite number of values,
then the pure states are organized as an ultrametric space, and the overlaps are non-negative. Dynamically-frozen states 
play a role in the analysis of Gibbs states drawn from a metastate, either as states
or as parts of states. Using a mapping into real Hilbert space, we prove further results about Gibbs states, and classify them into six types. 
Any indecomposable metastate has a compact symmetry group, though it may be trivial; we call this a metasymmetry. 
Metastate-average states are studied, and can be related to states arising dynamically at long times after a quench from 
high temperature, under some conditions. Many features that are permitted by general results are already present in replica 
symmetry breaking (RSB). Our results are for cases both with and without spin-flip 
symmetry of the Hamiltonian and, technically, we use mixed $p$-spin--interaction models.
\end{abstract}


\maketitle

\section{Introduction}
\label{intro}

The spin-glass (SG) problem is a very old one in the field of disordered systems. In the standard form of SG model, due to 
Edwards and Anderson (EA) \cite{ea}, classical spins on a lattice interact with their neighbors via bonds that are taken 
to be independent random variables. Then equilibrium thermodynamic properties or correlation functions should be 
calculated for given bonds, before an average over the disorder is performed. The EA work was followed by the formulation
of an infinite-range model of Ising spins [the Sherrington-Kirkpatrick (SK) model \cite{sk}], for which mean-field theory should 
be exact. The correct (and highly unusual) mean-field theory for this model was found by Parisi \cite{par79}, and its formulation 
is known as replica symmetry breaking (RSB). Subsequently, it was shown that RSB describes a countable infinity of ``ordered states'',
with a particular distribution of the relative weights of each \cite{par83,mpv_book}. The correctness of (at least some of) the key RSB 
results for the SK model has been proved in rigorous work \cite{guerra03,talagrand_book,panchenko_book}. 

For short-range systems, say of Ising spins, on a lattice in $d$-dimensional space the situation is less clear and remains controversial. A natural 
expectation would be that, as with conventional mean-field theories, the RSB theory describes at least the form of the ordered phases of the 
system, involving many ordered or ``pure'' states (though critical exponents might not be correct in low dimensions, and the SG phase 
presumably disappears in very low dimension); we explain the meaning of the term ``pure state'' in a moment.
A leading alternative scenario for a SG phase is known as the scaling-droplet (SD) picture \cite{bm1,macm,fh}. In this, the structure 
in the SG phase is simpler than in RSB, in that in zero magnetic field there are only two pure states, which are related by spin-flip 
symmetry of the Hamiltonian; complex behavior is still possible within this scenario \cite{fh}. The question of whether
one of these two, or some other, picture describes the SG phase has remained unresolved, and is a major motivation for work
on this problem, both theoretical and experimental. In this paper, these questions are addressed in a particular sense: we consider what 
structures, including those for pure states, are allowed for the probability distributions on the states from a general point of view valid 
for all equilibrium disordered short-range classical spin systems, but focused on Ising spins. In order to explain the underlying logic 
of the whole and give an outline of this, with additional background, we need first to introduce some definitions, which we do here informally, 
and more formally later.

\subsection{Gibbs states, pure states, and metastates}

The notion of there being more than one pure state is only properly defined in an infinite size system, and this requires some explanation. 
In a finite-size system, we have the well-known Gibbs statements about thermal equilibrium: for a Hamiltonian,
say
\be
H=-{\sum_{\{i,j\}}}'J_{ij}s_is_j
\label{inteq:eaham}
\ee
(where $s_i=\pm 1$ are Ising spins, and $J_{ij}$ are the bonds, which are random; the sum is over distinct pairs of nearest-neighbor
sites $i$, $j$), the probability distribution on spin configurations is
\be
p(s)=e^{-H/T}/Z,
\label{inteq:gibbs}
\ee
where $0<T<\infty$ is temperature and $Z=\sum_s e^{-H/T}$ is the partition function [the sum in $Z$ is over all configurations
of the spins $s\equiv (s_i)_i$]. In an infinite system, these do not make sense because the sums would be infinite. The solution
to this is also fairly well known (see e.g.\ \cite{georgii_book,simon_book}). The prescription is to require that the distribution on the spins
in any finite subregion, conditioned on any given spin configuration on the complementary region, be described by a similar Gibbsian formula,
in which the spins outside the region play the role of a boundary condition.
These conditional distributions on finite subregions fit together to form a ``state'' (probability distribution) on all the spins (it is 
characterized completely by its correlation functions for any finite set of spins); such states are
called Dobrushin-Lanford-Ruelle (DLR) states. 

In an infinite-size system, a DLR state may not be unique for given $H$ and $T$
(in a finite-size system, one is led back to the usual Gibbs distribution as above, which is determined uniquely by $H$ and $T$). 
But a general such state $\Gamma$ can always be decomposed into a mixture of so-called pure (or extremal) states $\Gamma_\alpha$, 
which are themselves DLR states, in a unique fashion. For correlation functions, this takes the form
\be
\langle\cdots\rangle_\Gamma=\sum_\alpha w_\Gamma(\alpha)\langle\cdots\rangle_\alpha.
\ee
Here $\cdots$ stands for the product of any finite number of spins, $\langle\cdots\rangle_\Gamma$ or $\langle\cdots\rangle_\alpha$
stand for ``thermal''  expectation in $\Gamma$ or $\Gamma_\alpha$, respectively, and $w_\Gamma(\alpha)\geq 0$ are a set of probabilities or 
``weights'', $\sum_\alpha w_\Gamma(\alpha)=1$; for simplicity, we have written this for a mixture that is a sum rather than an integral. 
A pure state is by definition a DLR state that cannot be expressed as a mixture of DLR states other than itself. Conversely, 
if more than one pure state exists for the given $H$ and $T$, then there is a continuum of possible DLR states, given by such mixtures. 
We usually refer to a DLR state simply as a Gibbs state, and we almost always mean one in infinite size.
 
If there are many possible Gibbs states, then there is the question of which one is physically relevant. One way to settle this would
be to begin with finite size, and then take some sort of limit to arrive at a DLR state. (Indeed, real systems are finite, and a natural way 
to attempt to approximate large but finite size is by passing to the limit.) In systems that are not SGs this is usually not particularly
problematic, though there is always a question of the boundary conditions to be used. But for a SG, considered with bonds
drawn from some distribution (say, in which the bonds $J_{ij}$ have mean zero), and not averaged, it is not at all clear that, 
for a given sample of bonds, the limit of the finite-size Gibbs states even exists. Instead \cite{aw,ns92}, there could be the phenomenon 
termed ``chaotic size dependence'' (CSD) by Newman and Stein (NS) \cite{ns92}. 

To overcome this, Aizenman and Wehr (AW) \cite{aw} and NS introduced what NS called ``metastates'' \cite{ns96b}; 
see Refs.\ \cite{ns_rev,nrs_rev} for reviews.  
A metastate $\kappa_J$ is a probability distribution on Gibbs states (in infinite size) for given disorder (bonds) $J\equiv (J_{ij})_{ij}$ 
in the Hamiltonian. 
AW and NS both constructed metastates as a limit from finite-size systems, but in different ways. Both of these types of metastates contain 
information about what occurs in equilibrium in a large finite-size system. (We do not need the details of the constructions at the moment.)
The possibility of chaotic size dependence---that is, the lack of a proof of its absence for SG systems at low temperature and weak 
magnetic field---means that metastates, or some alternative construction, are {\em necessary} (not optional), in order to discuss the 
infinite-size problem; we know of no alternative construction. If one simply assumes that the thermodynamic limit
of the Gibbs state is unproblematic, one is effectively assuming that the metastate is trivial, a case that we analyze in depth in this paper; 
the possible structures of the Gibbs states are then constrained much more than we can prove to be the case in nontrivial metastates.  In fact, 
there are SG models (but not highly realistic ones) in which the presence of a nontrivial metastate can be shown \cite{ns94,wf}. In relation 
to the scenarios introduced earlier, in the SD picture the metastate is implicitly trivial, while certain results found in RSB can occur in the 
short-range case only if the metastate is nontrivial \cite{ns96b}, and indeed RSB directly predicts that the AW metastate is nontrivial in the 
SG phase \cite{read14}.

Metastates have, or can have, various covariance properties; these will play a key role in this paper. If the finite-size systems are defined
using periodic boundary conditions, and the distribution on bonds is translation invariant, then either of the AW and NS metastates obtained in 
the limit is translation covariant, meaning that
$\kappa_J$ is unchanged if the bonds as well as the states (i.e.\ the spins) are translated. This does not need much discussion, but the 
other type of covariance may seem more subtle at first sight. In finite size, if a single bond, say $J_{i_0j_0}$, is changed by 
$\Delta J_{i_0j_0}$, then the Gibbs distribution, eq.\ (\ref{inteq:gibbs}) above, can be replaced by that for the modified bonds. Thus the 
(unique, in the present case) Gibbs state changes covariantly, and in a particular way, with a change in the bonds. This property survives 
in the limit in the AW and NS constructions of a metastate: If, starting with $\kappa_J$ for given $J$, a single bond (or, by extension, 
any finite number of bonds) is changed by some amount, then there is a corresponding change (similar to that 
which occurs in finite size) in any Gibbs state, and the probability distribution (metastate) on the modified Gibbs states is simply that of 
the unmodified ones, moved to the modified Gibbs states (a formal definition is in the following section) \cite{aw}. This is referred to 
as covariance of the metastate under local transformations (i.e.\ of bonds). 

The two constructions possess these covariance properties, but following NS and co-authors \cite{answ14} we will {\em define} 
a metastate abstractly in the 
infinite-size system, without reference to a particular construction, as a probability distribution on states for given bonds $J$,
such that the states are Gibbs states (for those bonds), and the distribution possesses the local covariance property. It is sometimes 
useful to allow non--translation-covariant metastates, but usually we will require the metastate to be translation covariant as well. 
These classes of metastate are the first main objects analyzed in this paper. It should be kept in mind that this abstract definition 
may not capture some properties that occur when, for example, the AW or NS constructions of a metastate are used, in 
other words properties that may even be generic when taking the limit using finite systems. But even in this broad setting 
many constraints on the possible behavior of the disordered spin systems arise.

\subsection{Indecomposable metastates}

In this paper, the first main result is that there is a notion of an {\em indecomposable} metastate, and that any metastate can be decomposed
into indecomposable parts. The idea of such a decomposition of a metastate into parts is that a metastate may be a mixture of other
``simpler'' ones (always having the same covariance properties), for example $\kappa_J=\lambda\kappa_J^{(1)}
+(1-\lambda)\kappa_J^{(2)}$, where $0<\lambda<1$. In order to respect covariance, this decomposition must be valid for any bonds $J$, 
with $\lambda$ independent of $J$. Such decompositions can be extended (similarly to the decomposition of a Gibbs state in terms of pure 
states, as above) to sums with more terms or even to integrals involving a distribution on metastates. An indecomposable metastate is then 
one that cannot be so decomposed in any such way, other than trivially. 

To obtain a theory of this, we first reformulate local covariance. We introduce a system of probability kernels such that the joint distribution 
$\kappa^\dagger$ on bonds $J$ and Gibbs states $\Gamma$ (the conditional distribution of $\kappa^\dagger$ given $J$ is just $\kappa_J$) 
is invariant under convolution with the kernels; this requirement builds in both the probability distribution on the bonds
and the local covariance of the metastate. (The use of these kernels is analogous to the way that the DLR conditions can be formalized
\cite{georgii_book}.) Translation covariance can also be included, if desired. With this technology, we 
can prove that any metastate, translation covariant or not, can be decomposed into indecomposable ones. 

The indecomposable metastates have attractive properties:
the Gibbs states drawn from an indecomposable metastate all ``look alike'' macroscopically, meaning that they are all identical when viewed
only in terms of properties that are invariant under local transformations (and also are translation invariant in the case with translation 
covariance). Distinct indecomposable metastates for the same $J$ put all their probability on disjoint sets of Gibbs states $\Gamma$;
this again is analogous to the case of distinct pure Gibbs states, which put all their probability on disjoint sets of spin configurations 
(see Ref.\ \cite{georgii_book}, Chapter 7). It seems likely that indecomposable metastates will be crucial for future developments in this topic. 
Some of the models in Refs.\ \cite{ns94,wf} already exhibit indecomposable metastates.

We will give some immediate applications; these typically require translation covariance of the metastate. For example, in recent work 
\cite{nrs23} (to be referred to as NRS23), it was proved (under some conditions)
that similar macroscopic properties,
but for pure states, are the same for all pure states in the decomposition of each Gibbs state drawn from a metastate. One of these
is the self-overlap. The simplest type of overlap of two pure states, say $\alpha$, $\alpha'$, is similar to the definition of EA \cite{ea},
\be
q_{\alpha\alpha'}={\rm Av}\, \langle s_i\rangle_\alpha\langle s_i\rangle_{\alpha'},
\label{inteq:overl}
\ee
where $\rm Av$ stands for average over translations of site $i$; the latter average is done first for a finite subvolume of the infinite system, 
and then the limit as the subvolume tends to the entire space is taken (see Sec.\ \ref{models}; this definition, from Ref.\ \cite{ns96b}, 
is not subject to the criticisms made in Ref.\ \cite{hf}). If $\alpha'=\alpha$, we obtain the self-overlap  
(or EA order parameter). The self-overlap is 
translation and locally invariant, and from Ref.\ \cite{nrs23} it is the same for all pure states for given $\Gamma$, but not necessarily the
same for all $\Gamma$. (In this introduction, most statements hold with probability one in the relevant distribution, but we omit
that language, and, as here, simply say ``all'' or ``every''; later we will be much more precise.) But now for an indecomposable 
metastate $\kappa_J$, the self-overlap must be the same for all pure states drawn from every Gibbs state drawn from $\kappa_J$. 

Another application concerns the pure-state structure of Gibbs states. The weight $w_\Gamma$ is in general a probability distribution
on pure states, so it may consist of some $\delta$-functions (``atoms''), each meaning that some single pure state has nonzero probability 
on its own,
or it may be a continuum (``atomless''), in which individuals have zero probability, but (uncountable) sets of pure states have nonzero 
probability, or it may be a mixture of both. NS proved \cite{ns09} (to be referred to as NS09) that the number of atoms can be $1$ 
(and only if the pure-state decomposition is trivial) or a countable infinity, but not any number in-between, or it can be zero in the 
presence of an atomless part. For an indecomposable metastate, the same one of these alternatives must occur for every Gibbs state. 
In fact, we will later prove that 
a nontrivial mixture having both atoms and an atomless part in the decomposition of a Gibbs state is also ruled out in any metastate. So 
for an indecomposable metastate, the pure-state decomposition of each Gibbs state is the same one of just three distinct ``characters'': 
its pure-state decomposition is either trivial, or a countable infinity of atoms (only), or atomless.

[Strictly speaking, in these last results we need to pay attention to some symmetry considerations. The Ising EA Hamiltonian as in eq.\
(\ref{inteq:eaham}) has spin-flip symmetry, under which every spin is reversed, $s_i\to -s_i$. In this case each pure state either is invariant
or else has a flip-reversed partner; we consider only metastates that are covariant under the symmetry (e.g.\ Ref.\ \cite{nrs23}), and then 
the statement above should be modified to include the effect of symmetry. In our discussion here and in most of the paper, we consider for 
simplicity the case of a Hamiltonian without spin-flip symmetry, for example by adding $-\sum_i h_i s_i$, with $h_i$ a random field 
for every site $i$, to eq.\ (\ref{inteq:eaham}), to which the preceding statements apply directly. Modifications to allow for spin-flip symmetry  
can be made easily, as we will explain later.]

\subsection{Stochastic stability and dynamically-frozen states}

To make further progress, and to prove some of the results mentioned already, we make use of a version of ``stochastic stability'';
for this circle of ideas, see Refs.\ \cite{guerra,ac,gg}. In our version, similar to Ref.\ \cite{ad}, we consider adding additional short-range
random bonds to the Hamiltonian (in infinite system size), 
independent of those already present, and nonzero only within a finite subregion, and then taking a limit as the region tends to the whole 
system while the strength of the added disorder tends to zero in a certain way. In the limit, this has the effect of changing a given Gibbs 
state $\Gamma$, but only by changing the weight $w_\Gamma$ of the pure states, leaving each pure state, and the bonds $J$, unchanged. 
The change in weight amounts to first multiplying it by $e^{g(\alpha)}$, where $g(\alpha)$ is a mean-zero Gaussian random variable that 
depends on the pure state $\alpha$, and has covariance (here we use the term ``covariance'' with its standard meaning in probability theory, 
which is distinct from the sense used above; we hope this will not cause confusion!)
\be
\bbE\, g(\alpha) g(\alpha')=q_{\alpha\alpha'},
\ee
then finally normalizing the modified weight. Thus $g(\alpha)$ is in effect a random change in the relative free energy of the pure 
states (up to a factor of $-1/T$). By replacing $g$ here by $\lambda g$, we obtain a stochastic process similar to a diffusion process,
under which the weight evolves with ``time'' $t=\lambda^2/2$. Using the result of NRS23 \cite{nrs23}, we prove that 
any metastate is invariant under this evolution; this is another main result of this paper. This invariance is what we call stochastic stability 
\cite{ac} of the metastate; it says that the metastate is stable under random perturbations of the Hamiltonian. We emphasize that
this holds at given disorder $J$, and for the full metastate, and not only for its restriction to, for example, 
translation-invariant quantities (cf.\ Ref.\ \cite{ad,ans15}). Using only the latter weaker version, we prove an extension of the 
NRS23 ``zero-one law'' \cite{nrs23};
the extension was used in some results mentioned already. We also obtain sets of identities of the form of some of those in the work 
of Aizenman and Contucci (AC) \cite{ac,cg_book} (part of this was already mentioned in Ref.\ \cite{nrs23}) and, in addition, for an 
indecomposable metastate, others of the form of some of those in the work of Ghirlanda and Guerra (GG) \cite{gg} (the latter have been 
especially useful in work on the SK model \cite{talagrand_book,panchenko_book}).

Stochastic stability can be very effective in constraining the pure-state structure of Gibbs states drawn from a metastate. 
For example, using the same approach \cite{ad,ans15} as in the proof of the extension of the zero-one law, we prove using ideas and results
from Ref.\ \cite{aa} that, first, if the Gibbs states drawn from an indecomposable metastate have countably-infinite pure state 
decomposition, then the distribution of their weights in that decomposition follow a so-called Poisson-Dirichlet (PD) distribution; 
the latter form a one-parameter family, and the parameter takes a fixed value for that metastate. Second, if in addition the overlap takes 
only a finite number $k+1$ of values, then for every Gibbs state drawn from the indecomposable metastate, the overlaps of pairs 
of pure states define an ultrametric structure on the set of pure states, the distribution of weights for the pure states is then a Ruelle 
cascade that involves the overlap values, and the overlap values are non-negative.
These results are just what is found in RSB, where the PD distribution is related to the relative free energies of the pure states
being distributed as an exponentially increasing Poisson process, ultrametricity is a widely-known property of RSB, and the case of the 
overlap taking only $k+1$ non-negative values is referred to as $k$-step RSB, or $k$-RSB \cite{mpv_book}. The results then suggest that 
non-trivial Gibbs states are always described with ultrametric structure and Ruelle cascades. (Rigorous results of this form but
for the SK model can be found in Ref.\ \cite{panchenko_book}.)

Another particular case occurs whenever a metastate has an atom at some Gibbs state; this includes the case of a trivial metastate. 
From stochastic stability, it follows that the Gibbs state must be invariant under the evolution described by random $g$. We prove 
that this can occur in only two ways: either the Gibbs state is trivial (its decomposition is a single pure state), or else the overlap  
$q_{\alpha\alpha'}$ must take the same value for (almost) every pair of pure states (and differ from the self-overlap of each; 
this is possible only when $w_\Gamma$ is atomless, and this refines the conclusion in Ref.\ \cite{ns07}, to be referred to as NS07). 
The latter case gives rise to the notion of what we will call a ``dynamically-frozen (DF) state'' (the precise definition 
we use is a little more involved than we describe here; see Sec.\ \ref{subsec:ginv}). 

We use this term because this DF behavior arose in a mean-field solution for infinite-range Potts and 
$p$-spin models found in Ref.\ \cite{kirk} (the solution is a degenerate limit of RSB), and was identified with a corresponding phase 
found in a dynamical treatment 
below a temperature at which the dynamics froze (i.e.\ ceased being ergodic), and so the phase was said to be dynamically frozen.
(Any Gibbs state whose decomposition involves more than one pure state has broken ergodicity of the dynamics, but
for this particular case the phase transition from a trivial Gibbs state was invisible in thermodynamics in the infinite-range models \cite{kirk},
hence the name.) For short-range systems, it was argued \cite{ktw,bb} that the phase is destroyed, but that argument appears to tacitly 
assume that the outcome of the argument must be a single pure state, and ignored the possibility that (in our terms) a DF state with 
subextensive complexity of pure states survives \cite{hr,read22}. We know of no reason to rule out such a DF state.

\subsection{Hilbert-space distribution and Gibbs states}

The overlaps, as in eq.\ (\ref{inteq:overl}), resemble inner products of vectors in an inner product space, and indeed, if we consider 
overlaps among a collection of pure states, we obtain a (Gram) matrix that can easily be seen to be positive semidefinite (i.e.\ 
the self-overlap of a linear combination of pure states is non-negative). This might suggest somehow mapping
the pure states in a given Gibbs state $\Gamma$ to vectors $\bf v$ in a Hilbert space $\cal H$, and so mapping the probability 
distribution $w_\Gamma$ to a distribution on ${\bf v}\in{\cal H}$. This can be accomplished by the Dovbysh-Sudakov (DS) distribution 
\cite{ds} (it was used for SGs in Ref.\ \cite{aa}). 
DS actually obtained such a distribution in a different, more general, setting, so we give a self-contained proof in the setting 
of a distribution on an affine probability space equipped with a positive-semidefinite bilinear form on pairs of points, as just discussed for 
pure states. In our version of the DS distribution, we have an essentially unique map into a separable real Hilbert space $\cal H$, 
and then we can also find the conditional distribution of the original distribution for a given vector $\bf v$ in $\cal H$. Quite generally, 
the map has the property that, expressed in terms of our situation, it preserves the overlap (taking it to the inner product) 
for almost every pair of pure states. Further, the conditional distribution $w_{\Gamma{\bf v}}$ for any given $\bf v$ has the property 
that it is either trivial (a $\delta$-function on a single pure state) or else describes a DF state; in the latter case, the overlap
of (almost) any pair of pure states is equal to the Hilbert-space norm square, $||{\bf v}||^2$, of $\bf v$. 

Combined with earlier results such as the zero-one laws, we then show that, for given $\Gamma$ (drawn from $\kappa_J$),
the norm $||{\bf v}||$ of $\bf v$ is constant for every pure state drawn from $w_\Gamma$, and for an indecomposable $\kappa_J$,
constant for all $\Gamma$ as well (a similar result that $||{\bf v}||$ is constant was found in the SK model, based on the GG identities 
\cite{panchenko_book}). Further, the DS distribution obeys the same trichotomy of characters as the pure states did already, that is, 
either it is trivial, or it consists of a countable infinity of atoms, or it is atomless. We can then classify each Gibbs state drawn from 
$\kappa_J$ according to the form of its DS distribution; we call these types I, II, and III, respectively.
For each of these types, we can then further subclassify the Gibbs states according to whether the conditional distribution 
$w_{\Gamma{\bf v}}$ is an atom or atomless (DF); we call these types a and b, respectively. Then we have a total of six types, and this 
gives a refinement of the earlier classification into three characters, in which the third (i.e.\ the atomless) case has now been broken 
into four distinct types. The earlier example concerning an atom in the metastate is a special case, in which the result was that the 
Gibbs state must be type I (either a or b); hence for type II or III Gibbs states, a nontrivial, not purely atomic, metastate is necessary. 
For an indecomposable metastate, all Gibbs states drawn from it are of the same type, for all $J$. These are more of the main conclusions 
of this work.

Because the overlaps describe the covariance of $g$ in stochastic stability, and are preserved in the DS distribution, stochastic stability 
works together naturally with the DS distribution \cite{aa,ad}, but cannot distinguish type a from type b states. Proofs of ultrametricity 
(as in RSB \cite{par83,mpv_book}) of the overlap structure in the SK model \cite{aa,panchenko_book} made use of the DS distribution 
as well as of the stochastic stability or GG identities. Our earlier results can now be extended to show that, for type II Gibbs states, 
1) the weights of the atoms in the DS distribution follow the PD distribution, and 2) if the overlap, or inner product, between the
vectors at the locations of the atoms in the DS distribution takes only a finite number $k+1$ of values, then there is an ultrametric structure,
the distribution of the weights of the atoms is a Ruelle cascade, and the overlaps are non-negative. These again are significant conclusions.

\subsection{Metasymmetry and relative weak mixing}

Finally, we return to the structure of a metastate. First we show, using methods similar to those earlier, that 
for two pure states, each drawn from a Gibbs state drawn independently from a metastate $\kappa_J$, their overlap does not depend
on the two pure states, though it may depend on the two Gibbs states. Without this result, it would not be obvious that
overlaps of Gibbs states are locally invariant in general, because the weights (when nontrivial) are not. But this result implies that
the overlaps of Gibbs states are locally invariant, which is needed in what follows. Then we consider the DS distribution for Gibbs states 
drawn from a metastate $\kappa_J$. If $\kappa_J$ is indecomposable, then all Gibbs states drawn from it ``look alike'' macroscopically, 
and this must apply to the statistics of their overlaps with other Gibbs states. It then follows that if the DS distribution is not trivial
then it must have a high degree of symmetry: its support must be a homogeneous space (a subset of a sphere in $\cal H$), and 
so can be expressed as a quotient $G/H$ of a group $G$ by a subgroup $H$, where the group $G$ acts by orthogonal linear maps 
of $\cal H$ and is compact; the DS distribution must be invariant under $G$. Thus if $G$ is not trivial there is a ``hidden'' symmetry of 
the metastate, which one might call a ``metasymmetry'', a symmetry of the 
metastate, rather than of the Gibbs states. The conditional metastates for given ${\bf v}\in{\cal H}$, denoted $\kappa_{J{\bf v}}$,
again have the property that each either is trivial or else has the property (similar to a DF state) that any two Gibbs states drawn from 
it have the same overlap. In either case, the overlap of two pure states drawn from $w_\Gamma$ for each of two Gibbs states drawn 
from $\kappa_{J{\bf v}}$ must be a constant. 

If the group $G$, or the DS distribution of $\kappa_J$, is trivial, then the indecomposable metastate is the same as its conditional, 
and so has one of the two latter forms. If so, then the pairwise overlaps of two pure states drawn as described 
at the end of the last paragraph are all equal. This is striking because it is a property that holds within RSB, according to a 
main result from Ref.\ \cite{read14}; in RSB, in terms of Parisi's function $q(x)$, the value of that overlap is $q(0)$ \cite{read14}
(it also holds for a trivial metastate, including the SD case). We call this property ``relative weak mixing'', for reasons given later. 
Thus if relative weak mixing does not hold, then there must be a nontrivial metasymmetry. This is another of the main results of 
this work. Some of the examples of indecomposable metastates resemble the case with nontrivial $G$ \cite{wf}. 

These last results can also be described in terms of the metastate-average state (MAS) $\rho_J$ of the metastate which, as its name implies,
is the average under $\kappa_J$ of the Gibbs states $\Gamma$, and is itself a Gibbs state (for the given $J$). Its pure state decomposition
is denoted $\mu_J$, where $\mu_J=\int \kappa_J(d\Gamma) w_\Gamma$. The DS distribution of $\mu_J$ is the same as that of $\kappa_J$,
so has the same metasymmetry. The conditional distributions $\mu_{J{\bf v}}$ either are trivial or else describe DF states. We then show that
for two MASs obtained from two indecomposable metastates, either their $\mu_J$s are the same or they put their probability on disjoint sets of 
pure states; thus the MASs themselves are either identical or disjoint in the same way.

\subsubsection{Maturation MAS}

As an extension of this result, we consider so-called maturation metastates \cite{wf} and the corresponding maturation MAS (MMAS). 
These arise
when considering an instantaneous quench of an infinite-size system to a temperature in the SG phase (this has been much studied, but see 
especially Ref.\ \cite{ns99}). The state that evolves from the initial state, which involves an average over both dynamics and the initial spin 
configuration (for given bonds $J$) is a state that, at asymptotically long times, resembles a MAS. We show that, assuming the equilibrium 
metastate is of the relative weak mixing type, then under mild assumptions the MMAS is identical to the equilibrium MAS. Hence these 
MMASs are either trivial or DF states. This conclusion is in agreement with earlier suggestions and with 
numerical work \cite{ns99,wf,mhy,wy,jry}. 

\subsection{Summary of structures}

In case readers feel overwhelmed by the number of probability distributions involved, we attempt here to summarize them in a logical sequence
of conditional distributions, which we could call ``top down''. First, we have the joint distribution on bonds $J$ and Gibbs states $\Gamma$. 
This consists of, at the top, the marginal distribution, the disorder distribution on bonds only. Conditionally, for given bonds, we then have the 
metastate, a distribution on states for given bonds. The metastate may be decomposable; then it can be analyzed as a distribution on 
indecomposable metastates. Conditioning on a choice of indecomposable metastate, it is again a distribution on Gibbs states. It can be analyzed 
as the DS distribution on vectors in Hilbert space; this is where metasymmetry may appear. For a given vector, we have a 
conditional metastate which is, yet again, a distribution on Gibbs states, and is either trivial or atomless. Drawing a Gibbs state from 
the conditional metastate, its DS distribution is one of types I, II, or III (for the trivial case, it must be type I, while for the atomless case 
it can be any type; nontrivial ultrametricity may appear only for types II and III, and this is where stochastic stability plays a nontrivial role). 
For any of these, conditioning on (another) vector, we obtain a conditional Gibbs state, which is either trivial (a single pure state) 
or a DF state, so type a or b; type b gives a possible additional level of ultrametricity. In the latter case, the DF state involves an atomless 
distribution on pure states. For the final level, each pure state is a distribution on spin configurations, and there is no lower level than that. 
For the MAS, we replace the conditional metastate by a conditional MAS, which takes us directly to a distribution on pure states, 
and then finally to the distribution of a given pure state on spin configurations.

It should be noted that all our results are logically of the ``universal'' type, saying that any object satisfying some conditions
has certain properties, thus ruling out others as disallowed. We do not prove any results of the ``existential'' type, which would say that some 
such objects actually exist (however, we do prove that metastates as we define them exist). Hence it is still possible that some possibilities 
allowed here can be ruled out, generically if not universally, perhaps even with arguments similar to those used here. (We comment on some 
candidates for this later.) A particular case, which we suspect will be done in the future, would be to rule out nontrivial Gibbs states 
(drawn from the metastate) other than ones with ultrametric structure in the pure-state decomposition. 

It is noteworthy that there were many points in this work at which RSB might have been 
ruled out, but as it turned out it always survived. The structure described above 
resembles RSB at many points, except for metasymmetry (if it is nontrivial). In fact, we conjecture that 
RSB (including its degenerate limits) and unbroken replica symmetry (RS) together describe the possible structures of all indecomposable 
metastates with the relative weak mixing property, including the possible forms of the Gibbs states drawn from such a metatate; 
RS corresponds to high temperature and also to the SD picture. More generally, RSB and RS 
may together describe all possible conditional metastates (defined above), similarly.

\subsection{Organization of the paper}

In the paper, the topics are treated in essentially the same sequence as in this Introduction, though the organization differs somewhat.
In Section \ref{models}, we explain the background to the work, concerning Gibbs states, metastates, overlaps, and so forth,
together with basic results about these. Section \ref{zeroone} is a very short section, which describes a recent result \cite{nrs23}
and an extension to be proved in this paper. Section \ref{indecomp} describes the decomposition of a metastate, and indecomposability 
and its consequences, with some examples. In Sec.\ \ref{sec:stst}, we prove results about stochastic stability and the AC and GG
identities in the setting of metastates, and discuss some first consequences. We formulate stochastic stability as a Markov process 
we term $\Sigma$ evolution, prove the invariance of any metastate under $\Sigma$ evolution, and also formulate a related random process 
called $g$ evolution that applies to individual Gibbs states. We also prove the extended result from Sec.\ \ref{zeroone}. The results
(for a countable infinity of pure states) on the PD distribution, and ultrametricity when overlaps take only $k+1$ values, are in 
Subsec.\ \ref{subsec:pdultr}. In Sec.\ \ref{trivmet}, 
we discuss results for Gibbs states, including what happens if the metastate is trivial, or more generally contains any atoms, in which case
a DF state can appear. Then we formulate the DS distribution, and prove the statements for our formulation directly. With that result, 
we then extend some earlier results and classify Gibbs states drawn from a metastate into six types. In Sec.\ \ref{sec:mas}, we discuss 
the notion of relative weak mixing, and prove results that include the metasymmetry of any indecomposable metastate. We also discuss 
the relation with the MAS and its properties, and finally the maturation MAS is related to the equilibrium MAS. Section \ref{sec:disc} is a 
final discussion. Two Appendices discuss additional material and results, and an additional example, respectively.

\section{Models, definitions, and properties}
\label{models}

We begin the detailed discussion with the general set-up and some background that is needed before the Theorems
can be stated and proved in later sections. Well-informed readers may be able to skim this section for notation and 
definitions, referring back to it as needed, however the discussion of local transformations and local covariance is crucial for what 
follows, and should be carefully noted, as should the treatment of symmetry described at the end of the Section. We also devote
Subsec.\ \ref{sec:rsb} to a description of what RSB implies for short-range systems.

\subsection{Disorder, Gibbs states, and pure states}

Let $i$ enumerate the sites of a $d$-dimensional lattice $\bbZ^d$; site $i$ is at position $\bx_i\in\bbZ^d$ (we also write $i\in\bbZ^d$). 
For Ising spins, a spin configuration $s$ is a function on $\bbZ^d$ taking values $s_i=\pm 1$ for all $i$; then $s=(s_i)_{i \in \bbZ^d}$ 
is the indexed set of $s_i$ for all $i$. The most general Hamiltonian we can consider is a function of $s$ that takes the form
\be 
H(s)=-\sum_{X\in{\cal X}} J_Xs_X,
\ee
where $\cal X$ is the set of all nonempty finite subsets $X\subset\bbZ^d$, and $s_X=\prod_{i\in X}s_i$ [we sometimes
write $p=|X|$ for the size of (number of sites in) $X$]; $J_\emptyset$ is omitted, or set to zero, because that term
is a constant that cancels in correlation functions. [The original Edwards-Anderson (EA) model
\cite{ea} at zero magnetic field has terms with $p=2$ only, and only for $i$, $j$ that are nearest neighbors.] We will write 
$J=(J_X)_{X\in{\cal X}}$ for the indexed set of all ``bonds'' $J_X$. The sum here is ill-defined for the system on $\bbZ^d$ as we 
have defined it; we deal with that below, and at present the infinite sum is merely a convenient formal way to present the Hamiltonian.
For a finite system consisting of sites in $\Lambda\subset \bbZ^d$, we restrict $s$ to $s|_\Lambda=(s_i)_{i\in\Lambda}$ 
and the sum to  $X\subseteq \Lambda$ (or write $X\in {\cal X}(\Lambda)$, the collection of nonempty finite subsets of $\Lambda$);
such a Hamiltonian will be denoted $H_\Lambda$ or $H_\Lambda(s|_\Lambda)$.
When we wish to use periodic boundary conditions on a hypercube $\Lambda$ we can make suitable modifications to this $H_\Lambda$. 

The bonds $J_X$ are random variables, and the joint distribution of $J$ is (informally) $\nu(J)$; we will write 
$\bbE$ for expectation under $\nu$ (the term ``random variable'' means we assume that $\bbE\, J_X$ and $\bbE\, |J_X|$ exist and are finite, 
for every $X\in{\cal X}$; see Refs.\ \cite{chung_book,breiman_book,royden_book}). $\nu$ is usually invariant under translations of $\bbZ^d$, 
though when restricted to a finite-size 
sample the translation invariance might be broken by the boundary conditions. In the spirit of the EA model, models with all $J_X$s
independent, so that $\nu(J)$ is the product over $X$ of the distributions $\nu_X(J_X)$ for each $X$, will be used throughout the paper,
and termed ``mixed $p$-spin models''. (We will also make some references to other models in which the $J_X$s are not independent,
but each is a function of members of a countable collection of independent random variables, in a translation-covariant way.
If we give up translation invariance, then we may also consider the sites to be vertices of an infinite graph in place of $\bbZ^d$.) 
One could assume that the bonds are Gaussian random variables, and possibly that they are centered (i.e., the mean is zero), 
but most of our results are much more general than that. 

An important condition which we will impose on $\nu$ throughout the paper is what we call the ``short-range'' condition,
\be
\sum_{X:i\in X}\mathbb{E}|J_X|<\infty
\label{eq:shortrange}
\ee
for all sites $i$. This is particularly important in case $\bbE |J_X|<\infty$ for each $X$, and there are 
infinitely many nonzero terms in the sum. It does not require that the $J_X$s be independent, or translation invariance 
of $\nu$, though clearly when the latter holds, the sum is independent of $i$. We explain the reason for the condition
and for its name in a moment. Note that the condition implies that $\sum_{X:i\in X}|J_X|<\infty$ for all $i$ 
($J$ is ``absolutely summable''), $\nu$-almost surely.
We will also say that $\nu$ has the ``n.i.p.\ property'', or simply ``is n.i.p.'', if, for each $X$, 
the support of $\nu_X$ (note that the support of a probability distribution is the smallest closed set that has probability one) 
has {\em no isolated points}; 
this means that, for $\nu_X$-almost every $J_X$ and any $\varepsilon>0$, the set of $J_X'\neq J_X$ such that $|J_X'-J_X|<\varepsilon$ has
nonzero probability. [A closed subset of a topological space that has no isolated points, or equivalently is dense in itself 
(i.e.\ all its points are limit points), is sometimes called a ``perfect'' set \cite{kechris_book}.] A distribution on a space is said to be
continuous, or atomless, if any single point of the space has probability zero (thus, there are no $\delta$-functions in the distribution). 
A continuous distribution $\nu_X$ on $J_X$ has the n.i.p.\ property, but such behavior is not required in most of the proofs; the weaker 
n.i.p.\ requirement is often sufficient \cite{nrs23}, and will be used later. 

Some later results require that there be an independent Gaussian piece
in $J_X$ for some $X$ and all its translations. This may be viewed as a, possibly small, deformation of the model, but has to be introduced
from the beginning of the constructions. The definition is that
\be
J_X=J_X^{(1)}+J_X^{(2)},
\ee
so that there may be two random variables for the same given $X$. $J_X^{(1)}$ can have any distribution (including a trivial $\delta$-function),
but $J_X^{(2)}$ is assumed to be a centered Gaussian random variable of nonzero variance. We might use this deformation for all $X$,
and in the mixed $p$-spin models all resulting $J_X^{(i)}$ ($X\in{\cal X}$, $i=1$, $2$) are independent, and condition (\ref{eq:shortrange})
is again in force. The presence of a Gaussian piece for all $X$ implies that the distribution is n.i.p., and that ${\rm Var}\,J_X>0$, for all $X$
(the latter condition will be used later, as in Ref.\ \cite{nrs23}).

Although only Ising spins will be considered explicitly (except in one class of examples), many results extend {\it mutatis 
mutandi} to other cases of classical spins, such as Potts spins and $m$-component unit vector spins [called $O(m)$ 
models], provided the analogs of $|s_X|$ are bounded above by $1$. The changes required are mostly notational, 
and were detailed in another paper recently \cite{read22}. 

In general, we use the term ``state'' for a probability distribution $\Gamma(s)$ on spin configurations $s$;
${\cal P}(S)$ is the space of all states $\Gamma$ on the space $S$ of all 
spin configurations. We will denote expectation of a function $f(s)$ of $s$ in a state $\Gamma(s)$ by $\langle f(s)
\rangle_\Gamma$ (with a minor abuse of notation). An important fact is that a state is uniquely determined by the values
of expectations of the form $\langle s_X\rangle$ as $X$ runs through $\cal X$. Strictly speaking, states $\Gamma$ and 
disorder distributions $\nu$ 
are both probability measures, and cannot generally be viewed simply as functions (probabilities, or probability densities) 
$\Gamma(s)$ or $\nu(J)$ as we have written them informally so far. The function notation is legitimate for states 
$\Gamma(s|_\Lambda)$ of Ising spins in a finite region $\Lambda$, and for $\nu_X(J_X)$ when $\nu_X$ is absolutely 
continuous with respect to Borel-Lebesgue (BL) measure on $\mathbb{R}$, but not in general for $\Gamma(s)$ or $\nu(J)$ in 
an infinite system. In more formal settings, we will use notation like $\Gamma(A)$ or $\nu(A)$ for the probability of 
a measurable set $A$ of $s$ or $J$, respectively, and expectations will be written likewise, for example as 
$\bbE\cdots=\int \nu(dJ)\cdots$, using notation (in this example) $\nu(dJ)$ for the measure of an ``infinitesimal'' measurable 
set of bonds (centered at some given $J$), whose precise meaning is part of the definition of (Lebesgue) integrals
\cite{chung_book,breiman_book,royden_book}.
[Thus, if e.g.\ $\nu_X$ can be expressed using a density $\nu_X(J_X)$ relative to BL measure on $J_X$, 
this means $\nu_X(dJ_X)=\nu_X(J_X)dJ_X$.]
Also, for conditional probabilities, we will use notation like $\Gamma(A\mid s|_\Lambda)$ for the conditional probability
of a set $A$ when (in this example) $s|_\Lambda$ is given, even though more formally this should be (and sometimes will be) expressed as 
conditioning on the $\sigma$-algebra generated by (in this case) $s|_\Lambda$, or generally on some sub--$\sigma$-algebra of the 
$\sigma$-algebra generated by $s$ \cite{chung_book,breiman_book}; the latter produces a measurable function, which then can be 
evaluated at some given $s|_\Lambda$.

In a finite-size system, a Gibbs state for a given Hamiltonian $H_\Lambda$ on a set $\Lambda$ of sites is defined by
\be
\Gamma(s|_\Lambda)=e^{-\beta H_\Lambda(s|_\Lambda)}/\sum_{s|_\Lambda} e^{-\beta H_\Lambda(s|_\Lambda)},
\label{finsizgibbs}
\ee
where $\beta=1/T$ and $T$ is temperature, $0\leq T\leq \infty$ (the $T=0$ case can be defined as a limit $T\to0$).
In an infinite-size system, this formula cannot be used directly, so instead a Gibbs state $\Gamma$ for the Hamiltonian $H$ 
is defined for $0\leq T\leq \infty$ as a state satisfying the Dobrushin-Lanford-Ruelle (DLR) conditions \cite{georgii_book}, which state
that the conditional probability distribution for the spin configuration $s|_\Lambda$ at sites in any finite subset $\Lambda$ conditioned on the  
remaining spins $s|_{\Lambda^c}$ in the complement $\Lambda^c$ of 
$\Lambda$ is, for $\Gamma$-almost every $s$,
\be
\Gamma(s|_\Lambda\mid s|_{\Lambda^c})=\gamma_{J\Lambda}(s|_\Lambda\mid s|_{\Lambda^c}),
\label{eq:specdef}
\ee
where $\gamma_J=(\gamma_{J\Lambda})_{\Lambda\in{\cal X}}$ is an indexed set
of fixed functions $S_{\Lambda^c}\to{\cal P}(S_\Lambda)$ [called ``probability kernels''; $S_\Lambda$ ($S_{\Lambda^c}$) 
is the space of all $s|_\Lambda$ ($s|_{\Lambda^c}$)], defined for all $s$ by
\be
\gamma_{J\Lambda}(s|_\Lambda\mid s|_{\Lambda^c})=e^{-\beta H_\Lambda'(s)}/\sum_{s_\Lambda} 
e^{-\beta H'_\Lambda(s)},
\ee
and 
\be
H_\Lambda'(s)=-\sum_{X\in{\cal X}:X\cap\Lambda\neq\emptyset}J_Xs_X,
\ee 
that is, $H_\Lambda'$ consists only of the terms in $H$ that involve spins in $\Lambda$. The point here is that the 
so-called specification $\gamma_J$ is fixed (specified) in advance, and defined for literally all $s$. In infinite size the solutions
$\Gamma$ to the DLR conditions for $\gamma_J$ (for given $J$) may not be unique, so that phase transitions
at which the number of solutions changes (perhaps breaking a symmetry) could occur. The space of all Gibbs states 
for the given $\gamma_J$ is denoted ${\cal G}(\gamma_J)$, or simply ${\cal G}_J$. (Additional technical details regarding the topology of, 
and measure theory and integration over, the relevant spaces is provided in Appendix \ref{app:back}; again see also Refs.\ 
\cite{chung_book,breiman_book,royden_book}.) With these definitions, the short-range condition (\ref{eq:shortrange}) 
ensures that $H_\Lambda'$
is absolutely convergent for all $s$ and $\Lambda$, $\nu$-almost surely (see also Ref.\ \cite{read22}), so that the definition 
of $\gamma_J$ makes sense for $\nu$-almost every $J$ (even when the $J_X$ are not all independent). While the term 
``short range'' might be taken to mean that the range of the 
bonds (i.e., the diameter of $X$ for nonzero $J_X$) is bounded, which would ensure that the condition holds provided only 
that the first absolute moment of $J_X$ is finite for all $X$, the more general usage captures the fact that models 
satisfying this condition appear to behave mathematically just like those with bounded-range (or ``strictly short-range'' 
\cite{read22}) bonds. At the same time, it will be useful to consider models with independent $J_X$ and ${\rm Var}\,J_X>0$ 
for all $X$ when we prove some later results, as in NRS23 \cite{nrs23}.
Further, most results in this paper should extend to the case of so-called finite-range bonds as defined in Ref.\ \cite{read22}, where it 
was established that the metastate constructions extend to such cases, provided the definitions of a DLR state or of a 
specification are extended somewhat. For simplicity, we do not discuss those cases in this paper. 

The DLR condition can be re-expressed as the invariance of the Gibbs state 
under the set of kernels $\gamma_{J\Lambda}$ \cite{georgii_book,simon_book} (we include this here because we use 
parallel constructions in the following section). For this we need a broader definition: 
$\gamma_{J\Lambda}(A\mid s)$ is defined in terms of the previous expression as a function $S\to{\cal P}(S)$ given by
(see Georgii \cite{georgii_book}, Chs. 1 and 2)
\be
\gamma_{J\Lambda}(A\mid s)= \sum_{s|_\Lambda}\gamma_{J\Lambda}(s|_\Lambda \mid s|_{\Lambda^c}) {\bf 1}_A(s),
\ee
which is a probability measure on sets $A$ for each $s$. (Here the indicator function ${\bf 1}_A(x)$ of a set $A$ is the function 
that takes the value $1$ if $x\in A$, $0$ otherwise.) Clearly $\gamma_{J\Lambda}(A\mid s)$ is independent of $s|_\Lambda$,
and so is measurable with respect to the $\sigma$-algebra generated by $s|_{\Lambda^c}$. 
Then the previous DLR condition becomes more broadly 
$\Gamma(A\mid s|_{\Lambda^c})=\gamma_{J\Lambda}(A\mid s)$ for all $A$, all $\Lambda$, and $\Gamma$-almost 
every $s$. To reformulate this, we use $\gamma_J$ as a set of probability kernels; a probability kernel $\pi$ from one space 
(say $S$) into distributions on another, possibly different one [here, into ${\cal P}(S)$] is a probability distribution $\pi$ on $S$ 
defined for all $s$, and for each measurable set $A$, $\pi(A\mid s)$ is a measurable function of $s$; in general, it might be 
measurable with respect to a sub-$\sigma$-algebra of the full one. A probability kernel $\pi$ from a space 
$\Omega$ into probability distributions ${\cal P}(\Omega)$ on that same space acts on a probability measure $\mu$ 
by convolution: for $A$ a measurable set of $\omega\in\Omega$, we define
\be
\mu(A) \mapsto (\mu\cdot \pi) (A)\equiv\int \mu(d\omega) \pi(A |\omega).
\label{eq:dotnot1}
\ee 
(Georgii \cite{georgii_book} writes $\mu\cdot \pi$ as $\mu \pi$, but that would conflict with other notation we use.)
Also, a probability kernel $\pi$ is said to be ``proper'' if 
\be
\pi(B\mid\omega)={\bf 1}_B(\omega)
\ee
for all measurable sets $B$ in the sub-$\sigma$-algebra with respect to which $\pi(A|\omega)$ is a measurable function of $\omega$
(see Ref.\ \cite{georgii_book}, p.\ 14); for us, this holds for $\gamma_{J\Lambda}$ for each $\Lambda$ when $B$ is $S_\Lambda$ times
a set of $s|_{\Lambda^c}$, by the remark above. Then by the definition of conditional probabilities 
\cite{chung_book,breiman_book}, or by remark (1.20) in Georgii \cite{georgii_book}, the DLR condition is equivalent to the 
invariance property
\be
\Gamma(A)=(\Gamma\cdot\gamma_{J\Lambda})(A)
\label{dlreqs}
\ee
for all measurable sets $A$ of $s$, and for all $\Lambda$. 
As $\gamma_J$ is fixed, these equations are linear 
(more properly, affine) in the state $\Gamma$, and so if two states $\Gamma_1$, $\Gamma_2$ satisfy these conditions,
the state $\lambda\Gamma_1+(1-\lambda)\Gamma_2$ does too, for any $\lambda\in[0,1]$. This shows that the
space of Gibbs states ${\cal G}(\gamma_J)$ is convex; it is also closed and compact.  

For a fixed $J$ or corresponding specification $\gamma_J$, the extremal or ``pure'' Gibbs states in ${\cal G}(\gamma_J)$ 
are those that cannot be expressed as a (generalized) convex combination of other Gibbs states in ${\cal G}(\gamma_J)$; 
they form a subset ${\rm ex}\,{\cal G}(\gamma_J)\subseteq{\cal G}(\gamma_J)$, also called the ``extreme boundary'' 
of ${\cal G}(\gamma_J)$. Each Gibbs state $\Gamma$, say, for the given $J$, can be expressed {\em uniquely} as a convex 
combination of pure Gibbs states in ${\rm ex}\,{\cal G}(\gamma_J)$ \cite{georgii_book,phelps_book}. In general, this may 
require an integral with some measure, as
\be
\Gamma=\int w_\Gamma(d\Psi)\Psi,
\ee
where $w_{\Gamma}$ is a probability distribution on ${\cal P}(S)$  
with $w_\Gamma({\rm ex}\,{\cal G}(\gamma_J))=1$, so a state drawn from $w_\Gamma$ is $w_\Gamma$-almost 
surely pure. We refer to this form as the pure-state decomposition of
a Gibbs state $\Gamma$; $w_\Gamma$, which depends on both $\gamma_J$ and the chosen Gibbs state $\Gamma$, 
is called the weight of the decomposition. Here and elsewhere we use $\Psi\in{\cal P}(S)$ as a variable 
for the space in which the pure states are found; hence in applications, $\Psi$ will be a pure state $w_\Gamma$-almost surely.
We briefly discuss the fact that $w_\Gamma$ is jointly measurable in $(J,\Gamma)$ in Appendix \ref{app:furth}, which will be used later,
sometimes tacitly.

The Hamiltonian may possess some ``internal'' symmetry, by which we mean a global symmetry that acts on the spins
on all the sites at once, and leaves the sites unchanged. For Ising spins, this means the spin-flip symmetry 
$\theta_\pm$ that acts as $\theta_\pm:s_i\to-s_i$ for all $i$. This is a symmetry of the Hamiltonian 
(or of the specification) provided $J_X$ is nonzero only when $p=|X|$ is even. This symmetry might be broken spontaneously 
at low temperature, leading to the appearance of distinct pure states that are mapped to one another by $\theta_\pm$. 
When symmetry is present, we will consider only Gibbs states that are flip invariant; at low temperature, examples of these 
would be the equal-weight superposition of two pure states that are connected by spin flip. Flip-invariant Gibbs states 
arise if one constructs states (perhaps as a limit from finite size) using a procedure that maintains the 
spin-flip symmetry of the finite-size systems. The flip-invariant Gibbs states form a subset 
${\cal G}_{\theta_\pm}(\gamma_J)$ whose members can be decomposed into extremal flip-invariant states; 
an extremal flip-invariant state is the symmetry average of a single pure state (which may or may not itself 
be flip-invariant). In general, we define a Gibbs state to be {\em trivial} if either it is pure, in the case that 
the Hamiltonian is not spin-flip invariant, or it is extremal in ${\cal G}_{\theta_\pm}(\gamma_J)$, in the 
case that it is flip-invariant. We return to how we will handle internal symmetry at the end of this Section.

\subsection{Covariance properties}

A central role in the argument will be played by certain covariance properties emphasized by AW and NS \cite{aw,ns98}. 
There are two classes of these. The first is translation covariance. 
If we define $\theta_{\bx}:{\bx}_i \to\bx_i+\bx$ for any $\bx \in \bbZ^d$, with a corresponding action 
on $i$ and $X$ then, if $\theta_\bx s$ means configuration $s$ shifted by $\bx$, it is represented by the vector 
with $i$-component $(\theta_\bx s)_i=s_{\theta_\bx^{-1}i}$, and similarly $(\theta_\bx s)_X=
s_{\theta_\bx^{-1}X}$ and $(\theta_\bx J)_X=J_{\theta^{-1}_\bx X}$. We can then define the translation 
$\theta_\bx \Gamma$ of any state $\Gamma$ by $\theta_\bx\Gamma(s)=\Gamma(\theta_\bx^{-1} s)$. 
It will be useful when discussing covariance to make explicit the $J$ for which $\Gamma$ is a DLR state,
by denoting it by $\Gamma_J$.
If Gibbs states are given for $J$ and for $\theta_\bx J$ then they are related translation covariantly if 
they look the same when one not only translates all the bonds $J_X$, but also translates the region in 
which one looks at the correlations of spins. That is, they are related translation covariantly if
\be
\Gamma_{\theta_\bx J}(s)=\theta_\bx \Gamma_J( s).
\ee 

The second, local, covariance property corresponds to the effect of changing the bonds from $J\to J+\Delta J=J'$,
so $H\to H+\Delta H$, where, when all $J_X$ are independent, we require $\Delta J$ to be nonzero 
for only a finite number of $X$s (hence the term ``local''), and otherwise unrestricted. We will also express 
the local transformation by writing $\theta_{\Delta J} J=J+\Delta J$. For any state $\Gamma$, we can define
$\theta_{\Delta J} \Gamma$ by
\be
\langle\cdots\rangle_{\theta_{\Delta J}\Gamma}=\frac{\langle\cdots e^{-\beta\Delta H}\rangle_{\Gamma}}
{\langle e^{-\beta\Delta H}\rangle_{\Gamma}}.
\label{loctransgibbs}
\ee
Here, for any state $\Gamma$, $\langle\cdots\rangle_{\Gamma}$ is the (``thermal'') expectation in $\Gamma$ 
of a bounded continuous function of the spins. In this paper, in general we consider this only for $0< T <\infty$, because at $T=0$
changing a $J_X$ might reverse an infinite set of spins \cite{ns22}, something we wish to avoid; however, we will consider $T=0$
in some examples in which there is no such difficulty. The case $T=\infty$ is not interesting: all Gibbs states are the uniform 
distribution on $s$. We might also write the transformation for the probabilities (with 
$\sum_s\Gamma_J(s)=1$) as
\be
\theta_{\Delta J}\Gamma(s)=\frac{\Gamma(s)e^{-\beta\Delta H}}{\langle e^{-\beta\Delta H}\rangle_\Gamma},
\label{eq:covgam}
\ee
as it gives simple formulas; this is correct for the marginal distribution on any finite set of spins, but not strictly
correct for the infinite set of all of them, when $\Gamma$ is really a probability distribution, not a countable set of
probabilities. (In general, our discussion of local covariance is for all $J_X$ independent. If the $J_X$s are not all independent, then
each reference to $\Delta J_X$ for some $X$ should be replaced by a change in one of the independent variables instead,
and we require that only a finite number of them be changed, which induces a change $\Delta J$. In this case it will be necessary 
to require that $\Delta H$ be finite, and some locality of the dependence of $J$ on the independent variables will be needed, 
which we will not attempt to specify precisely; when we consider examples, locality will be clear. After this, we will not comment
on these cases further until Sec.\ \ref{sec:examples}.) 
Finally, we say that Gibbs states $\Gamma_J$, $\Gamma_{\theta_{\Delta J} J}$ are related 
covariantly under this transformation if 
\be
\theta_{\Delta J}\Gamma_J=\Gamma_{\theta_{\Delta J}J}.
\label{loccovgibbs}
\ee 
[For later purposes, it may be useful to know 
that both transformations $\theta_\bx$ and $\theta_{\Delta J}$ as we defined them, acting on general $\Gamma$, 
are {\em continuous} maps of ${\cal P}(S)$ into itself, for all $\bx$ and allowed $\Delta J$.]

Both covariance properties can easily be seen to hold for all $J$, $\Delta J$, and $\bx$ for the Gibbs state 
$\Gamma_J$ obtained from a Hamiltonian $H$ [determined for all $J$ by the formula (\ref{finsizgibbs})] 
in a finite size system, assuming periodic boundary conditions to ensure translation covariance. In infinite size, 
there may be many Gibbs states for a given $J$, and the notation $\Gamma_J$ here is not meant to suggest that we specify 
a choice of Gibbs state for each $J$. In general, it may not be obvious how a particular Gibbs state $\Gamma_J$ 
at some $J$ is related to another $\Gamma_{J'}$ at a different value $J'$. Hence the use below of the covariance 
properties must be carefully justified.

In a general infinite-size Gibbs state, the local covariance, eq.\ (\ref{loccovgibbs}), always holds within 
a family of Gibbs states 
when they are obtained by starting from a Gibbs state for a particular $J$, and then using local transformations
to produce a corresponding Gibbs state for other $J'=J+\Delta J$, because it agrees with the change in the
specification under a change $\Delta J$. (We should point out that the transformations obey $\theta_{\Delta J}
\theta_{\Delta J'}=\theta_{\Delta J+\Delta J'}$, and so commute, and so this procedure for obtaining a family 
is well defined.) In general the transformation formula, eq.\ (\ref{loctransgibbs}), may be meaningless if we 
change $J_X$ for infinitely many $X$ simultaneously. But the change in $\Gamma_J$ will be well-defined 
(the sums converge) if the $l^1$-norm
\be
||\Delta J||_1=\sum_X|\Delta J_X|
\ee
of $\Delta J$ is finite, because $|\Delta H|\leq ||\Delta J||_1$ and so it is finite also.
(This is still a local change, because convergence of the sum implies that $\Delta J_X$ must go to zero 
at infinity, sufficiently fast.) For our purposes it will be sufficient to continue to consider only $\theta_{\Delta J}$ 
in which $\Delta J_X\neq 0$ for only a finite number of $X$, and often only one $X$. We will abuse notation
by writing $\theta_{\Delta J_X}$ for the latter case. For each $X$, let $\Theta_X$ denote the Abelian group
of all $\theta_{\Delta J_X}$ for that $X$.

The preceding considerations lead us to define the Abelian group of local transformations $\Theta_0$ 
to be the subset of the Cartesian (or direct) product $\prod_X \Theta_X$ of the groups $\Theta_X$ (with the obvious group 
operation) that consists of all elements $\theta_{\Delta J}$ in the product such that, for all except a finite number of $X$, the 
transformation $\theta_{\Delta J_X}$ is the identity; in other words, $\Delta J_X=0$ except for a finite number of $X$. 
This group, denoted by $\Theta_0=\coprod_X \Theta_X$, differs from the direct product $\prod_X \Theta_X$ of 
groups when there is an infinite number of nontrivial $\Theta_X$; it is a natural algebraic construction in the category 
of Abelian groups, called the coproduct or direct sum of the $\Theta_X$s. 
Let $\Theta$ be the group generated by $\Theta_0$ and the translations. $\Theta_0$ and $\Theta$ act on both the set of $J$ and 
the set of states $\Gamma$ as described already. We use the symbol $\theta$ for generic elements of $\Theta_0$ or $\Theta$,
and $\Phi$ to stand for either $\Theta_0$ or $\Theta$.

The form of the local covariance is of course unchanged if $\Gamma_J$ is a pure Gibbs state; 
if a symmetry is present we consider only $\Delta H$ that preserves the symmetry. In addition, pure
states remain pure under such a change; this is because they can also be characterized by the decay of
a general class of correlations \cite{georgii_book}, while the perturbation is local. 
If we now compare the local transformation for a Gibbs state and for its decomposition into
pure (or trivial) Gibbs states, then we find that in the decomposition of $\theta_{\Delta J}\Gamma_J$ 
the weight (at $\Psi$) is
\be
w_{J+\Delta J,\theta_{\Delta J}\Gamma_J}(d\Psi)
=\frac{\langle 
e^{-\beta\Delta H}\rangle_{\theta_{\Delta J}^{-1}\Psi}}{\langle e^{-\beta\Delta H}\rangle_{\Gamma_J}}
w_{J\Gamma_J}(d[\theta_{\Delta J}^{-1}\Psi]).
\label{loctransweight}
\ee
[Here we used notation for the pure-state decomposition that makes $J$ explicit:
\be
\langle\cdots\rangle_{\Gamma_J}=\int w_{J\Gamma_J}(d\Psi)\,\langle \cdots
\rangle_{\Psi}.]
\ee
If the covariance relation, eq.\ (\ref{loccovgibbs}), holds, then we can say that $(\theta_{\Delta J} 
w_{J\Gamma_J})(d\Psi)
= w_{J+\Delta J,\theta_{\Delta J}\Gamma_J}(d\Psi)=w_{J+\Delta J,\Gamma_{J+\Delta J}}(d\Psi)$.
We do not call this a covariance relation. For translations, if the Gibbs state and the 
pure states are translation covariant, then the weight $w_{J\Gamma_J}(d\Psi)$ is translation covariant: 
\be
w_{\theta_\bx J \theta_\bx \Gamma_ J}(d\Psi)=w_{J \Gamma_J}(d[\theta_\bx^{-1}\Psi]).
\ee 
Translation invariance of the distribution of $J$ is expressed similarly, as $\nu(dJ)=\nu(d[\theta_\bx^{-1}J])$,
which holds for all $\bx$. $\nu$ is never invariant under local transformations.

We point out here that in the local transformation behavior of the weight of pure states, the fact that the states are pure
was never used. Hence the exact same behavior also applies to {\em any} decomposition of the Gibbs state as a weighted 
average of parts, using a partition of the pure states into disjoint subsets, where each part of the Gibbs state is a partial Gibbs state, 
constructed as the (normalized) weighted average of the pure states in one part of the partition. 

\subsection{Metastates and their covariance}

Now we turn to metastates; we only outline the construction, as most details will not be important (see
Refs.\ \cite{aw} and \cite{ns97} or \cite{newman_book} for full discussion). A metastate $\kappa_J$ 
is a probability distribution on states $\Gamma$, and is required to be a measurable function of $J$ for $\nu$-almost 
all values of the disorder $J$, and such that $\Gamma$ drawn from $\kappa_J$ is $\kappa_J$-almost surely a Gibbs state for $\gamma_J$
(the precise definition of a metastate that we use is stated below). Constructions of a metastate 
involve first considering the pairs $(J,\Gamma_J)$ consisting of a disorder configuration and the corresponding finite-size Gibbs state,
and a probability distribution on these. The infinite-size limit can then be taken as a weak* limit of distributions 
\cite{chung_book,breiman_book,billingsley_book2},
giving a distribution $\kappa^\dagger$ on pairs $(J,\Gamma)$ where $\Gamma$ is a state on the infinite 
system (the existence of the limit may involve taking a subsequence of sizes, but is guaranteed 
by compactness arguments). The marginal distribution of $\kappa^\dagger$ for $J$ is simply $\nu$, and 
then the conditional probability distribution \cite{chung_book,breiman_book}
$\kappa_J$ on $\Gamma$ for given $J$ can be defined; we will write this as 
\be
\kappa^\dagger=\nu\kappa_J
\ee 
[we will also do the same for other conditional distributions; informally, as if the distributions are represented
by densities, this means $\kappa^\dagger(J,\Gamma)=\nu(J)\kappa_J(\Gamma)$ for all $(J,\Gamma)$, while more formally
and more generally it means $\kappa^\dagger(dJ,d\Gamma)=\nu(dJ)\kappa_J(d\Gamma)$ at all $(J,\Gamma)$].
It can be proved \cite{aw,ns97,newman_book} that, in this construction, $\kappa_J$ is supported on Gibbs states 
$\Gamma\in{\cal G}(\gamma_J)$ $\nu$-almost surely, as we already required in terms of $\kappa^\dagger$. 
This also confirms that Gibbs states exist for $\nu$-almost every $\gamma_J$.
There are two constructions of such metastates in the literature: in the earlier one due to AW \cite{aw} (which is the one sketched here), 
$\kappa_J$ can be viewed as induced from the dependence of the state in each fixed finite size on the disorder 
in a distant outer region, while the disorder in the inner region is fixed, and then the limits are taken. 
In the other, due to NS \cite{ns96b}, an average over a range of finite sizes takes the place of that 
over the disorder in the outer region. Thus the resulting two types of metastate contain 
distinct information about the behavior in finite size, even after the limit has been taken. 
The two constructions work at $T\geq 0$, but as stated above, we do not consider $T=0$ except in certain examples.

If periodic boundary conditions are used in the metastate constructions, then the resulting metastates enjoy two
classes of covariance properties \cite{aw}. One is covariance under translations, 
\be
\kappa_{\theta_\bx J}(d\Gamma)=\kappa_J(d[\theta_\bx^{-1}\Gamma])
\ee
[or alternatively $\kappa^\dagger(dJ,d\Gamma)=\kappa^\dagger(d[\theta_\bx^{-1}J],
d[\theta_\bx^{-1}\Gamma])$, translation invariance of $\kappa^\dagger$]. The other is
covariance under local transformations,
\be
\kappa_{J+\Delta J}(d\Gamma)=\kappa_J
(d[\theta_{\Delta J}^{-1}\Gamma]).
\label{eq:loccov}
\ee
These hold for $\nu$-almost all $J$, all $\bx$, and all $\Delta J$ such that $||\Delta J||_1<\infty$.
These properties (like those of the Gibbs states) hold in finite size in the early stage (sketched above) 
of each construction, and consequently are inherited in the limit. In the particular case in which the metastate
is trivial, that is, supported on a single Gibbs state for $\nu$-almost all $J$, then that Gibbs state is 
covariant, obeying the covariance relations for $\nu$-almost all $J$, all $\bx$, and all $\Delta J$ such 
that $||\Delta J||_1<\infty$. (One can also make use of other boundary conditions in the construction, 
and then metastates without translation covariance can be obtained \cite{ns96b}.) We will use notation like 
$\mathbb{P}_{\kappa^\dagger}$ for probability, and $\bbE_{\kappa^\dagger}$ for
expectation, under the probability distribution identified in the subscript (here $\kappa^\dagger$),
and similar for those under $\kappa_J$. When we say a property holds $\kappa^\dagger$-almost surely, 
or -almost everywhere (and so forth) we mean that it holds except for a set of measure zero with respect 
to the identified distribution on the variables in question. 

We define a metastate in general without reference to a construction from finite size \cite{answ14}. The definition
is that $\kappa_J$ is a metastate if it is supported on Gibbs states [i.e.\ $\kappa_J({\cal G}(\gamma_J))=1$]
and it is covariant under all local transformations $\theta\in\Theta_0$, both properties for $\nu$-almost every $J$. 
We frequently use translation-covariant metastates, that is, covariant under all $\theta\in\Theta$, but not exclusively.
The existence of metastates under this definition follows from the AW and NS constructions.

The metastate constructions avoid such things as boundary conditions that depend on $J$, and thus are
physically reasonable; they resemble conditions that one could imagine implementing in an experiment
on a finite-size system. Admittedly periodic boundary conditions are not easy to implement experimentally, 
but there is a long tradition of their use in theoretical work in order to model homogeneity. In fact, as remarked
by NS \cite{ns01}, given a joint distribution $\kappa^\dagger$ that is not translation invariant {\it a priori}, 
we can obtain one that is translation invariant by averaging over translations (we thank D.L. Stein for raising this point;
the average involves a weak* limit, and at least a subsequence limit will exist, again by a relative compactness argument). 
Because we always use a disorder distribution $\nu$ that is translation invariant, the marginal distribution $\nu$ is 
unaffected by averaging, and we then obtain a translation-covariant $\kappa_J$ for $\nu$-almost every $J$. A 
consequence of this is that, when we consider a property [i.e.\ a measurable function of $(J,\Gamma)$ that does not depend 
directly on $\kappa^\dagger$] that is translation invariant, such as a property of the weights $w_\Gamma$, then even if 
the metastate happens not to be translation covariant, the probability distribution of that property induced from $\kappa^\dagger$ is 
unchanged by translation averaging, and so our results for such properties will hold even for such metastates. 

We are not aware 
of constructions of a thermodynamic limit for Gibbs states in the low-temperature region of a SG other 
than those using, or equivalent to using, a metastate. Of course, if there are many pure states for each 
$\gamma_J$, then many other Gibbs states can be constructed from them as convex combinations of 
more than one pure state. These Gibbs states may be unrelated for different $J$, or could be constructed 
so that the weight does not transform as above under local transformations (and the Gibbs state is not 
covariant), unlike the case in the metastate construction. We regard only the metastate 
constructions as sufficiently physical (and we mostly use only the translation-covariant version). It should be noted 
that for {\em any} construction that is claimed to produce a single Gibbs state as a thermodynamic limit for almost all $J$, and 
in which that Gibbs state possesses the natural covariance properties discussed above, the result corresponds to the case 
of a trivial metastate (even if the construction makes no overt reference to metastates). Hence various results in the present paper
apply to any such construction.

We also define the metastate average state (MAS), or barycenter, of $\kappa_J$ \cite{aw} as 
\be
\rho=\bbE_{\kappa_J}\Gamma.
\ee 
As an average of Gibbs states for given $J$, $\rho$ (sometimes written as $\rho_J$) is itself a Gibbs state. We define 
$\mu_J=\bbE_{\kappa_J} w_\Gamma$ 
to be the pure state decomposition of $\rho$, and $\mu^\dagger=\nu \mu_J$ to be the joint distribution on pairs $(J,\Psi)$ 
of bonds and pure states. 

\subsection{Overlaps and pseudometrics}
\label{subsec:overl}

We will also make use of overlaps, a concept that goes back to Ref.\ \cite{ea}, and of related (pseudo-)metrics;
we emphasize that we use ``window'' overlaps, calculated first in a finite window of the infinite system, followed by a limit.
For any $X\in{\cal X}$, and for any two Ising spin configurations $s$, $s'$, we can define the (normalized) overlap
in $\Lambda$,
\be
\widehat{q}_{X\Lambda} (s,s')=\frac{1}{|\Lambda|}\sum_{\bx:\bx\in\Lambda}
s_{\theta_\bx X} s_{\theta_\bx X}',
\ee
where $\Lambda$ is a finite region of $\bbZ^d$, such as a hypercube of side $W$ centered at the origin, which we denote by 
$\Lambda_W$. Then 
\be
1-\widehat{q}_{X\Lambda} (s,s')
\ee
is a generalization of a Hamming (or $l^1$-) distance between configurations $s$, $s'$, defined for the region $\Lambda$. 
For example, if $X=\{i\}$ consists of a single site $i$, then this is precisely the normalized Hamming distance 
between $s|_{i+\Lambda}$, $s'|_{i+\Lambda}$ (where $i+\Lambda$ means the set of $i'$ such that $\bx_{i'}=\bx_i+\bx$ for some 
$\bx\in\Lambda$), taking values between $0$ and $2$, and is a genuine metric on spin configurations (that is, 
it is symmetric, non-negative, equals zero if $s|_{i+\Lambda}=s'|_{i+\Lambda}$, obeys the triangle inequality, and is zero 
only if $s_i=s'_i$ for all $i\in\Lambda$). [For other $X\subseteq\Lambda$, 
the first four properties of a metric still hold exactly. But even then, the last property may not hold;
consider for example $X$ a fixed nearest-neighbor pair, then for $s|_\Lambda$ and its spin flip, $\widehat{q}_{X\Lambda}=1$.] 
We can also consider the limit $\Lambda\to\infty$, which is always taken in such a way that $\Lambda$ eventually contains all of $\bbZ^d$
(for example, choose $\Lambda=\Lambda_W$ and let $W\to\infty$). 
The limit as $\Lambda\to\infty$, when it exists, is in general only a pseudometric, even for $X=\{i\}$, meaning that it has the 
properties of a metric except that it might be zero even for configurations that are not identical. The $\Lambda\to\infty$ limit
of $\widehat{q}_{X\Lambda}(s,s')$, when it exists, is invariant under a translation of the chosen $X$, and so will be denoted 
$\widehat{q}_{[X]}(s,s')$, because it depends only on the translation equivalence class $[X]$, not on $X$. We may write $[{\cal X}]$
for the set of all (distinct) $[X]$, for $X\in{\cal X}$.

We define similar overlaps $q_{X\Lambda}(\Gamma,\Gamma')$ (useful also for non-Ising spins) using thermal averages in two states 
$\Gamma$, $\Gamma'$ (which at the moment do not have to be Gibbs states):
\be
q_{[X]\Lambda} (\Gamma,\Gamma')=\frac{1}{|\Lambda|}\sum_{\bx:
\bx\in\Lambda}\langle s_{\theta_\bx X}\rangle_\Gamma \langle s_{\theta_\bx X}\rangle_{\Gamma'},
\ee
and then define 
\be
q_{[X]} (\Gamma,\Gamma')=\lim_{\Lambda\to\infty}q_{[X]\Lambda} (\Gamma,\Gamma')
\ee 
whenever the limit exists (this definition was advocated for short-range systems in Ref.\ \cite{ns96b} and, as we mentioned,
it evades the issues in Ref.\ \cite{hf}). 
[In addition, for later use we will also define the hybrid between these cases, the overlap $\widehat{q}_{[X]}(s,\Gamma')$ (or the same 
with the arguments switched), using sums of $s_{\theta_\bx X}  \langle s_{\theta_\bx X}\rangle_{\Gamma'}$ likewise.]
In general the limit might depend on the limiting procedure used; in relation to the ergodic theorem, which may apply to such 
an average, it is useful to assume that $\Lambda$ is a hypercube $\Lambda_W$ of side $W$ centered at the origin, and then 
$\Lambda\to\infty$ means $W\to\infty$. This type of translation average will occur often enough that it is 
worthwhile to define a notation for it: for a function $f$ of, say, a single state, define
\be
{\rm Av}\,f=\lim_{\Lambda\to\infty}\frac{1}{|\Lambda|}\sum_{\bx:
\bx\in\Lambda}\theta_\bx f
\ee
(if the limit exists).
The construction used to obtain the Hamming distance may not give a pseudometric here, but we can instead
define a pseudometric $d_{[X]}$ as a limit of $l^2$ metrics by
\be
d_{[X]}(\Gamma,\Gamma')^2={\rm Av}\, (\langle s_{X}\rangle_\Gamma - \langle s_{X}
\rangle_{\Gamma'})^2
\label{psemet}
\ee
(note the square on the left-hand side), and (when the limit exists) we define $d_{[X]}(\Gamma,\Gamma')$ to be the non-negative square root; 
it clearly obeys the triangle inequality. We note the evident fact that, for each pair $\Gamma$, $\Gamma'$, if the quantities exist,
then they obey $d_{[X]}^2(\Gamma,\Gamma')=q_{[X]}(\Gamma,\Gamma)+q_{[X]}(\Gamma',\Gamma')
-2q_{[X]}(\Gamma,\Gamma')$. This definition gives a family of pseudometrics, indexed by $[X]$, whenever the limit exists, 
and similarly for the overlaps (which form the family of all so-called bond overlaps). (For $|X|=1$, we will write $[\{i\}]$ as 
$[1]$.) The existence and translation 
invariance of the $\Lambda\to\infty$ limit is guaranteed (almost surely) by the ergodic theorem, if the states $\Gamma$, $\Gamma'$ 
are drawn from translation-invariant joint distributions \cite{ns97,newman_book,ad}. In particular, this is the case for Gibbs states 
and for pure states when the 
distributions are $\kappa^\dagger$ or $\kappa^\dagger w_\Gamma$, respectively, when a translation-invariant $\kappa^\dagger$ 
is used. When $\Gamma$, $\Gamma'$ are pure states $\Psi_1$, $\Psi_2$, we will also write $q_{[X]}(1,2)$ and $d_{[X]}(1,2)$ 
for the overlap and pseudometric. We repeat (from the Introduction) that this definition of an overlap, generalized from Ref.\ \cite{ns96b}, 
is not subject to the criticisms made in Ref.\ \cite{hf}.

In addition, we can define total overlaps and pseudometrics, as follows. Suppose $(a_{[X]})_{[X]}$ is
an indexed set of numbers such that $a_{[X]}>0$ for all $[X]$ and $\sum_{[X]} a_{[X]}<\infty$. Then we can define a total overlap 
$q_{\rm tot}=\sum_{[X]}a_{[X]}q_{[X]}$ and a total pseudometric as the non-negative square root of 
\be
d_{\rm tot}^2=\sum_{[X]}a_{[X]}d_{[X]}^2;
\ee 
$d_{\rm tot}$ is indeed a pseudometric (one can also replace all $d^2$s 
with $d$s on both sides of the definition of $d_{\rm tot}$ to obtain another total pseudometric, which we denote $\widetilde{d}_{\rm tot}$). 
It is clear that $d_{\rm tot}=0$ if and only if $d_{[X]}=0$ for all $[X]$, for any 
choice of $(a_{[X]})_{[X]}$ that satisfies the conditions (and similarly for $\widetilde{d}_{\rm tot}$). These total quantities
will appear in the following and later, and may be useful in applications. Many statements that we will give which involve $q_{[X]}$ for given 
$[X]$ also hold {\it mutatis mutandi} for $q_{\rm tot}$ in place of such $q_{[X]}$; we will not state them all explicitly. 

We point out that, in our setting, the overlaps $\widehat{q}_{[X]}$, $q_{[X]}$ and the (pseudo-)metrics $d_{[X]}$ are measurable functions 
on pairs of spin configurations or pure states (or of a single pure state for a self-overlap). This is because they were defined above 
as limits of functions for finite $W$ that are measurable, and the limit exists almost surely, so the limit is measurable and defined almost 
everywhere. 

Next we point out a simple relation of $\widehat{q}$ with $q$. Suppose the pair $(s^{(1)},s^{(2)})$ 
is drawn from $\Psi_1\times \Psi_2$ for {\em pure} states $\Psi_1$, $\Psi_2$, which are drawn from $w_{\Gamma_1}\times w_{\Gamma_2}$, 
and the pair $(\Gamma_1,\Gamma_2)$ is drawn from $\kappa_J\times\kappa_J$ and finally $J$ was drawn from $\nu$ (the same result 
will hold if $\Gamma_1=\Gamma_2$ is drawn from a single copy of $\kappa_J$ instead), then the distribution on $(s^{(1)},s^{(2)})$ 
is translation invariant, and so the limit $\widehat{q}_{[X]}(s^{(1)},s^{(2)})$
exists $\nu(\kappa_J\times\kappa_J)(w_{\Gamma_1}\times w_{\Gamma_2})(\Psi_1\times\Psi_2)$-almost surely, by the ergodic theorem, 
and so does its thermal expectation $q_{[X]}(\Psi_1,\Psi_2)$ under $\Psi_1\times\Psi_2$. 
Now we note the characterization of pure states in terms of the decay of connected correlations
\cite{georgii_book}, which in particular implies that for any pure state $\Psi$ and any $X$, $X'$,
\be
\langle s_X s_{\theta_\bx X'}\rangle_{\Psi} - \langle 
s_X\rangle_{\Psi} \langle s_{\theta_\bx X'}\rangle_{\Psi} \to 0
\ee
as $|\bx|\to\infty$; this is called clustering of correlations in a pure state.
If we take the expectation under $\Psi_1\times\Psi_2$ of 
$[\widehat{q}_{[X]\Lambda_W}(s^{(1)},s^{(2)}) -q_{[X]\Lambda_W}(\Psi_1,\Psi_2)]^2$ 
for finite $W$ and any pair of pure states $\Psi_1$, $\Psi_2$ then, by the clustering property of both pure states, it tends to zero as 
$W\to\infty$. The limit of the difference $\widehat{q}_{[X]\Lambda_W}(s^{(1)},s^{(2)}) -q_{[X]\Lambda_W}(\Psi_1,\Psi_2)$
exists for $\Psi_1\times\Psi_2$-almost every $(s^{(1)},s^{(2)})$, and for $\nu(\kappa_Jw_{\Gamma_1}\times \kappa_Jw_{\Gamma_2})$ 
almost every $(J,(\Gamma_1,\Psi_1),(\Gamma_2,\Psi_2))$, so by the clustering property it is equal to zero almost surely. That is, 
\be
\widehat{q}_{[X]}(s^{(1)},s^{(2)})=q_{[X]}(\Psi_1,\Psi_2),
\label{qhatq}
\ee 
$\Psi_1\times \Psi_2$-almost surely, for $\nu(\kappa_J\times\kappa_J)(w_{\Gamma_1}\times w_{\Gamma_2})$-almost every 
$(J,(\Gamma_1,\Gamma_2),(\Psi_1,\Psi_2))$ (and similarly when $\Gamma_1=\Gamma_2$). 

It is natural to ask whether a given pseudometric distance $d_{[X]}$, say, is in fact a metric, so that if it is zero then the two 
states are identical, at least almost surely for pairs of states (which could be, but do not have to be, Gibbs or pure states). 
In the translation-invariant setting
if, for some $X$, $d_{[X]}=0$ then $\langle s_{X'}\rangle$ almost surely takes the same value in both of the two states for 
all $X'\in[X]$ (i.e., all translates of $X$), as a consequence of the existence of the limit (this will be a corollary to the following proof), 
but it is not clear if this implies that the states are equal. We can prove a weaker statement: if $d_{[X]}=0$ for all $X$, 
then the two states are almost surely the same. It is worth recording this formally as a Lemma.
\newline 
{\bf Lemma 1}: Consider a translation-invariant joint distribution $\kappa^\dagger$ on $(J,\Gamma)$ and, for given $(J,\Gamma)$
($\Gamma$ a Gibbs state $\kappa^\dagger$-almost surely), the pure-state decomposition $w_\Gamma$ on states $\Psi$ 
and the set $B_{J\Gamma}$ of pairs of states $\Psi_1$, $\Psi_2$ such that (i) $d_{[X]}(\Psi_1,\Psi_2)=0$ for all $X\in{\cal X}$ 
and (ii) $\Psi_1\neq\Psi_2$. Then for $\kappa^\dagger$-almost every $J$, $\Gamma$, the $w_\Gamma\times w_\Gamma$-probability 
of the set $B_{J\Gamma}$ is zero.
\newline
{\bf Proof}: For given $(J,\Gamma)$, consider the set $A_{J\Gamma}$
of pairs $(\Psi_1,\Psi_2)$ such that $d_{[X]}(\Psi_1,\Psi_2)=0$ for all $[X]$. 
Then by taking an expectation with the indicator for that set we have, for any $X$,
\bea
0&=&{\bbE}_{\kappa^\dagger(w\times w)}{\bf 1}_{A_{J\Gamma}}(\Psi_1,\Psi_2)d_{[X]}(\Psi_1,\Psi_2)^2\\
&=&\lim_{W\to\infty}\!\!W^{-d}{\bbE}_{\kappa^\dagger(w\times w)}{\bf 1}_{A_{J\Gamma}}
{\sum_{X'\subset\Lambda_W}}'(\langle s_{X'}\rangle_{\Psi_1} - \langle s_{X'}
\rangle_{\Psi_2})^2\non\\
&=&{\bbE}_{\kappa^\dagger(w\times w)}{\bf 1}_{A_{J\Gamma}}(\langle s_{X}\rangle_{\Psi_1} - \langle s_{X}
\rangle_{\Psi_2})^2,\non
\eea
where the sum in the middle line is over $X'$ such that $X'\in [X]$, and we used the bounded convergence theorem 
\cite{chung_book,breiman_book,royden_book} to move 
the limit outside the expectation, and then used translation invariance.
Hence, for each $X$, 
$\langle s_{X}\rangle_{\Psi_1} = \langle s_{X}\rangle_{\Psi_2}$ for every pair $(\Psi_1,\Psi_2)\in A_{J\Gamma}$
(except for a set of $w_\Gamma\times w_\Gamma$ probability zero), and for $\kappa^\dagger$-almost every $(J,\Gamma)$. 
Because the set of $X$ is countable, we can then say that, for $\kappa^\dagger$-almost every $(J,\Gamma)$, 
$\langle s_{X}\rangle_{\Psi_1} = \langle s_{X}\rangle_{\Psi_2}$ for {\em all} $X$, 
for every pair $(\Psi_1,\Psi_2)\in A_{J\Gamma}$ except for a set with $w_{\Gamma}\times w_{\Gamma}$ probability zero. 
That implies that, for $\kappa^\dagger$-almost every $(J,\Gamma)$, $\Psi_1=\Psi_2$ for every pair $(\Psi_1,\Psi_2)\in A_{J\Gamma}$,
except for a set of pairs with $w_{\Gamma}\times w_{\Gamma}$ probability zero, which is the desired result $\Box$ 

Similar statements and proofs hold for the pseudometric distances between a pair $\Gamma_1$, $\Gamma_2$ 
of Gibbs states drawn from $\kappa_J\times \kappa_J$ for $\nu$-almost every $J$, and for the pseudometric
distances between pure states $\Psi_1$, $\Psi_2$ drawn from $\mu_J\times \mu_J$ for $\nu$-almost every $J$. 
In fact, the proof of Lemma 1 establishes a general principle: for a translation-invariant distribution on pairs of 
states $(\Gamma_1,\Gamma_2)$ (where the states can be arbitrary, and the distribution need not be symmetric, 
nor have any covariance property other than translation invariance of the distribution), and again for the distances 
$d_{[X]}(\Gamma_1,\Gamma_2)$ (as defined above) which are well defined and translation invariant for almost every pair 
$(\Gamma_1,\Gamma_2)$, the probability of the set of pairs such that $d_{[X]}=0$ for all $[X]$ and $\Gamma_1\neq\Gamma_2$ 
is zero. Thus the family of all pseudometrics ``separates points'' in our applications in practice. These results will be used frequently 
later, and the general principle will be referred to as a ``version of Lemma 1''. 

The result can be restated in terms of $d_{\rm tot}$ (or similarly $\widetilde{d}_{\rm tot}$), by the equivalence $d_{\rm tot}=0$ 
if and only if $d_{[X]}=0$ for all $[X]$. Then any version of Lemma 1 states that $d_{\rm tot}$ (or $\widetilde{d}_{\rm tot}$) in fact 
behaves as a metric in the situation as stated in that 
version, that is, it is positive definite for almost every pair of states. We note that the standard (i.e.\ weak*) topology on the space of 
states on the infinite-size spin system can always be metrized
(see e.g.\ \cite{georgii_book,billingsley_book2,aliborder_book}), but that such a metric is typically 
not translation invariant, whereas $d_{\rm tot}$ and $\widetilde{d}_{\rm tot}$ are translation invariant when the states concerned are drawn 
from a translation-invariant distribution, as above. Each of these metrics induces a topology on the states that is essentially Hausdorff, and 
differs from the weak* one.

It is also simple to establish that, if the two states used to construct an overlap $q_{[X]}$ are pure states 
$\Psi_1$, $\Psi_2$ admitted by $\gamma_J$, then under a local transformation $\theta\in\Theta_0$ the overlap 
is unchanged (assuming the overlap itself exists as a limit). This is connected with the clustering properties of pure 
states. It is easy to show by elementary analysis that, because of the average over translations
over the region $\Lambda$, which tends to infinity, the overlap is unchanged by a local transformation of one 
or both pure states. The same holds for a pseudometric $d_{[X]}(\Psi_1,\Psi_2)$.

Using an overlap and the pure-state decomposition $w_{\Gamma}$ of a Gibbs state 
$\Gamma$, we define the probability distribution for overlaps of type $[X]$, for given $J$ and 
$\Gamma$:
\be
P_{J\Gamma [X]}(q)=\int w_{\Gamma}(d\Psi_1)w_{\Gamma}(d\Psi_2)\delta\left(q
-q_{[X]}(\Psi_1,\Psi_2)\right)
\ee
(or similarly for a total overlap $q_{\rm tot}$).
The metastate and $\kappa^\dagger$ expectations of $P_{J\Gamma [X]}(q)$ are denoted 
$P_{J [X]}(q)=\bbE_{\kappa_J}P_{J\Gamma [X]}(q)$ and $P_{[X]}(q)=\bbE_{\kappa^\dagger}P_{J\Gamma [X]}(q)$, respectively
(with translation covariance of the metastate and ergodicity of $\nu$, in fact $P_{[X]}(q)=P_{J [X]}(q)$; one says it self-averages).
We can also define the overlap distribution in the MAS $\rho$, called $P_{J\rho [X]}(q)$, in which $w_\Gamma$ is replaced by $\mu_J$,
which again self-averages under the same conditions. Later we will make use of all of these distributions in describing our results. 
We also point out that, by eq.\ (\ref{qhatq}), the overlap $q_{[X]}$ of pure states can be replaced in the definition of the overlap distributions
with $\widehat{q}_{[X]}$, with the average under $w_\Gamma$ or $\mu_J$ replaced by $\Gamma$ or $\rho_J$ (respectively), 
the corresponding distributions on spin configurations $s$. This is the usual approach in numerical work and in rigorous studies
of infinite-range models, though there the overlap may be defined as the average over all sites in a finite system, rather than using the limit 
of a finite window $\Lambda$ in an infinite system as we did here.

\subsection{RSB in short-range spin glasses}
\label{sec:rsb}

We sometimes hear people say that ``we don't know what RSB means for short-range spin glasses''. 
We certainly do know what it means and so, because it will also be useful later as some additional background, we will outline what RSB, 
which of course is not a rigorous approach, means in the short-range models considered in this paper, 
in terms of the rigorous concepts defined so far in this Section. Much of this was 
done already in the 1980s \cite{par83,mpv_book}, but requires some additional care with definitions for the short-range case, and also must 
account for the metastate. These refinements were done in Ref.\ \cite{read14}, but some points were not spelled out there (see also Ref.\ 
\cite{nrs_rev}). As in that 
reference, the starting point for RSB theory in the short-range case is not the SK model, but the replicated field theory for the short-range 
(Ising) EA model with only 
$p=2$-site interaction terms, for a derivation of which see Ref.\ \cite{bm79}. [There the field theory derived is the Landau-Ginzburg 
(LG) theory, but one may also choose to stop the derivation at an earlier stage, thus obtaining a functional at each site in the lattice field 
theory (the functional is the same as for the SK model), rather than the LG form. We will 
assume either form.]  The fluctuating field in the theory is a matrix $Q_{ab}(\bx)$ at each point $\bx$ in space, where $a$, $b=1$, \ldots, 
$n$, and the $n\to0$ limit must be taken. The theory gives access to expectations of correlations of the spins in one or more ``real'' replicas 
(copies of the system), where expectation means with respect to both thermal average and disorder average over all bonds. On the lattice, 
$Q_{ab}$ has the meaning $Q_{ab}(\bx_i)\sim s_i^{(a)}s_i^{(b)}$ (when appropriately normalized), where $s^{(a)}_i$ is again the Ising spin 
at site $i$ in the $a$th replica. We will write expectation in the replicated field theory at $n=0$ as $\langle\langle\cdots\rangle\rangle$. 

The system can sensibly be taken to be infinite at some stage in the derivation, and in that limit
mean-field theory can be used as a zeroth-order approximation of the analysis. (Thus the approach assumes that taking the limit
 for the disorder average of correlations is unproblematic.) That is, $Q_{ab}$ is taken independent of $\bx$, and the LG 
free energy per site must be maximized in the $n\to0$ limit (the maximization, rather than usual minimization, is a consequence of the 
replicas). We will not enter here into the well-known hierarchical structure of Parisi's ansatz for the mean-field $Q_{ab}$ matrix at $n=0$;
see Refs.\ \cite{par79,mpv_book}. It leads to an order parameter that is a non-decreasing function $q(x)$ of $x$ in the closed interval 
$[0,1]$. In terms of rigorous concepts, the assumption of 
the limit means that disorder expectation in finite size becomes $\bbE_{\kappa^\dagger}$, the expectation under both the disorder distribution 
$\nu$ and a metastate $\kappa_J$.  

To obtain the meaning of $q(x)$, we take an overlap $\widehat{q}_{X\Lambda}(s,s')$ for some choice of $X$ and for a subregion 
$\Lambda$ of the system that is already infinite in size, and look at the $\bbE_{\kappa^\dagger}$ expectation of the thermal average 
over $s=s^{(1)}$, $s'=s^{(2)}$ drawn from two copies $1$, $2$, of the same Gibbs state $\Gamma$ drawn from the metastate $\kappa_J$. 
For $\Lambda=\Lambda_W$ a hypercube, the expectation gives in the $W\to\infty$ limit
\be
\bbE_{\kappa^\dagger(w_\Gamma\times w_\Gamma)}q_{[X]}(\Psi_1,\Psi_2),
\ee
and if the first moment of $\widehat{q}$ is replaced by the $m$th, $q_{[X]}$ is replaced by $q_{[X]}^m$. Here the quantities involved 
were defined in the preceding subsection, and for $m>1$ the argument that translation averages of $s_i^{(a)}$ can be replaced by those of 
$\langle s_i^{(a)}\rangle_{\Psi_a}$ because of clustering in pure states is exactly what was used by Parisi in Ref.\ \cite{par83} (though for 
the SK model that he discussed, the meaning of a pure state or of clustering is much less clear; also, the use of the ergodic theorem 
to ensure existence of the $W\to\infty$ limit \cite{ad}, as discussed above, 
which involves translation invariance of $\kappa^\dagger$, was 
overlooked in Ref.\ \cite{read14}). On the field-theory side, the ansatz $s_i^{(a)}s_i^{(b)}\sim Q_{ab}(\bx_i)$, together with 
the fact that correlations of $Q_{ab}(\bx)-\langle\langle Q_{ab}\rangle\rangle$ tend to zero as the separation of points $\bx$ tends to infinity,
implies as in Ref.\ \cite{par83} (using the RSB structure) that the same moments are given by, for $[X]=[1]$,
\be
\int_0^1 dx\, q_{[1]}(x)^m.
\ee
The probability distribution for $\widehat{q}_{[1]}\in [-1,1]$ [or for $q_{[1]}(\Psi_1,\Psi_2)$] can be reconstructed from its moments,
with the result $P_{[1]}(q)=(dq_{[1]}/dx)^{-1}$ 
\cite{par83}. (Here the inverse function $x(q_{[1]})$ can defined to have a jump if $q_{[1]}(x)$ is constant over an open interval of $x$, and 
to be constant if $q_{[1]}(x)$ has a jump.)

We define $x_1$ to be the infimum of the set of values of $x<1$ 
such that $q_{[1]}(x)$ is constant on the open interval $(x,1)$, or $x_1=1$ if the latter set is empty. We note that $q_{[1]}(x)$ may not 
be (left-) continuous at $x=1$, but that is a separate question. If $x_1<1$, then $q_{[1]}(x)$ has a ``plateau'', that is, it is constant, 
for $x$ in either $(x_1,1)$ or $(x_1,1]$, and there is a $\delta$-function in $P_{[1]}(q)$ 
at the corresponding $q$ value. Such a $\delta$-function requires that there also be a $\delta$-function at the same location in 
$P_{J\Gamma[1]}(q)$ with positive $\kappa^\dagger$ probability. That in turn implies that there is positive $w_\Gamma\times w_\Gamma$ 
probability for that 
overlap value, for the set of $(J,\Gamma)$ for which the former $\delta$-function appears. In the cases studied in the early work on 
the SK model, $x_1<1$, and $q_{[1]}(x)$ is continuous at $x=1$. Then the maximum overlap should be the self-overlap of a pure state, 
and $1-x_1=\bbE_{\kappa^\dagger w_\Gamma}w_\Gamma(\Psi)$, that is, the $\kappa^\dagger$ expectation of the sum of the squares 
of the $w_\Gamma$ weight of each atom in the decomposition of $\Gamma$. In the Refs.\ \cite{par79,par83,mpv_book}, 
$w_\Gamma$ is treated as purely atomic, 
and then it follows that the self-overlap of every pure state drawn from $w_\Gamma$ is the same, and also the same for almost all $\Gamma$.

From here, the remainder of the interpretation of RSB theory (other than aspects touching directly on the metastate) can be obtained by 
similar means, following the arguments of Ref.\ \cite{mpv_book} and references therein. [A caveat is that in the SK model, higher overlaps like 
$q_{[X]}$, $|X|>1$, are simply powers of $q_{[1]}$, but in short-range systems presumably $q_{[X]}\neq q_{[1]}^{|X|}$, so the following 
applies to $q=q_{[1]}$ only.] These conclusions include: (i) ultrametricity, the fact that overlaps (at least $q_{[1]}$) obey 
\be
q(1,2)\leq \max(q(1,3),q(2,3))
\ee
for almost any three pure states $1$, $2$, $3$ drawn from almost any given $\Gamma$; (ii) so-called non-self-averaging of 
$P_{J\Gamma[1]}(q)$, which we now understand to be a consequence of a non-trivial metastate \cite{ns96b,read14}; and (iii) again 
assuming the case in which $w_\Gamma$ is purely atomic, the weights of the pure states follow the Ruelle cascade, for 
$\kappa^\dagger$-almost every $(J,\Gamma)$ \cite{mpv_book}. Finally, regarding the metastate, in RSB theory it is non-trivial 
and the overlap of 
$\mu_J\times \mu_J$-almost every pair of pure states drawn independently from the MAS $\rho$ takes the value $q(0)$, $\nu$-almost 
surely \cite{read14}. That is, $P_{J\rho[1]}(q)$ is almost surely a $\delta$-function at $q=q(0)$. We note that $q(x)$ may not be (right-) 
continuous at $x=0$.

While these results do not answer every possible question that might be raised about the RSB scenario, they surely invalidate the claim with 
which we began the Subsection.

\subsection{Symmetry aspects}

Finally, we return to questions involving global internal symmetry. In order to avoid tedious repetition and parallel constructions with and without 
the Ising spin-flip symmetry, we will adopt the following strategy. We explicitly discuss only the case without spin-flip symmetry. In this case,
for many purposes we must consider Hamiltonians with nonzero random $J_X$ for all $X\in{\cal X}$ (possibly, ${\rm Var}\,J_X>0$ 
for all $X$), and local transformations $\theta_{\Delta J_X}$, overlaps $q_{[X]}$, and pseudometrics $d_{[X]}$ for all $X\in{\cal X}$ also 
(then the total overlap $q_{\rm tot}$ and pseudometric $d_{\rm tot}$ involve a sum over all $X$ also). But the results still apply {\it mutatis 
mutandis} for models with symmetry, as follows.
In addition to a flip-invariant Hamiltonian, we assume a flip-covariant metastate. This arises from finite-size constructions provided the 
models in finite size give flip-invariant finite-size Gibbs states, due to boundary conditions that respect the symmetry, and if a subsequence 
limit is required, it too preserves the symmetry. In general, a flip-covariant metastate is one that puts full probability on flip-invariant Gibbs 
states. As we mentioned earlier, such Gibbs states can be decomposed uniquely as mixtures of flip-invariant trivial Gibbs states, each of which is 
either a flip-invariant pure state, or the symmetry average of two pure states, each the spin flip of the other. All statements we prove 
continue to hold in the presence of spin-flip symmetry if (including in later definitions) every instance of the term ``pure'' Gibbs state 
is replaced by ``trivial'' Gibbs state, the weight $w_\Gamma$ is viewed as the weight on trivial Gibbs states, and if whenever $X$ or $[X]$ 
is mentioned, a restriction to $|X|$ even is imposed, including in total overlaps and pseudometrics, $q_{\rm tot}$ and $d_{\rm tot}$. 
In this way, the symmetry is respected, and the results extend to this case. 

\section{Zero-One Laws}
\label{zeroone}

Before embarking on the main agenda of this paper, we first discuss and extend some recent work, for later use. 
In the recent work Ref.\ \cite{nrs23}, to be referred to as NRS23, the authors proved a basic principle for the Gibbs states drawn 
from a translation-covariant metastate (fully stated below). The statement involves the notions of invariant sets, or invariant functions 
of, pairs $(J,\Psi)$, and first we recall their definitions; these are used throughout the paper. 

First, consider bonds only; here we assume the model is a mixed $p$-spin model,
so the $J_X$ are independent. We can define the sub-$\sigma$-algebra (of the $\sigma$-algebra of all Borel measurable sets of $J$) 
of sets of $J$ that are invariant under $\Phi$ (recall that $\Phi=\Theta_0$ or $\Theta$). Thus $A$ is $\Theta_0$ invariant if $\theta A=A$ for all 
$\theta\in\Theta_0$ (i.e.\ under any local transformation), and $\Theta$ invariant if the same holds for all $\theta\in\Theta$ (i.e.\ under 
translations as well); denote the respective $\sigma$-algebras by ${\cal I}_\Phi$, or by ${\cal I}_{\Theta_0}={\cal I}_0$, ${\cal I}_{\Theta}
={\cal I}$, and note that ${\cal I}\subseteq{\cal I}_0$. It is clear that ${\cal I}_0$ consists of measurable sets that are independent of any 
finite set of $J_X$, that is they are in the intersection over all $Y$ of the collections of sets that include a factor $\mathbb{R}$ for each 
finite set $X$, for $X\in Y$, times some (measurable) set of $J|_{Y^c}$. This $\sigma$-algebra ${\cal I}_0$ on a product space is usually 
called the $\sigma$-algebra of remote (or tail) events, and denoted $\cal T$ \cite{chung_book,breiman_book,georgii_book}. The sets 
in ${\cal I}$ are the translation-invariant members of ${\cal I}_0$. 

The $\sigma$-algebras ${\cal I}_{1\Phi}$ (or ${\cal I}_{10}$ and 
${\cal I}_1$) are modeled on these: they are the $\sigma$-algebras of sets $A$ of $(J,\Psi)$ [or $(J,\Gamma)$; the symbol 
for the state makes no difference to the $\sigma$-algebras] such that $\theta A=A$ [where $\theta A=\{(\theta J,\theta \Psi):
(J,\Psi)\in A\}$] for all $\theta\in\Phi$. Once again, clearly ${\cal I}_1\subseteq{\cal I}_{10}$. We can define $\Phi$-invariant 
functions of $(J,\Psi)$ (or also of $J$ only) similarly: they are defined as
measurable functions $f$ that are invariant, so $f(\theta J,\theta\Gamma)=f(J,\Gamma)$ for all $\theta\in\Phi$, 
or equivalently, that are ${\cal I}_{1\Phi}$ (or ${\cal I}_{10}$-, ${\cal I}_1$-) measurable. We note that the group structure of
$\Theta_0$ will not be used, though for given $J$, $\Gamma$, $\Psi$, the possibility of choosing a $\theta_{\Delta J}$ will be; 
otherwise what are important are the invariant $\sigma$-algebras defined here. We also extend this definition to 
$\sigma$-algebras ${\cal I}_{n\Phi}$($n=1$, $2$, \ldots) that consist of sets of bonds and $n$ states (which may each be 
denoted $\Gamma$ or $\Psi$) that are invariant under the simultaneous action of any $\theta\in\Phi$ on the bonds and all $n$ states, 
and invariant (or ${\cal I}_{n\Phi}$-measurable) functions of $n+1$ arguments, similarly to the $n=1$ cases just mentioned.
For sets $A$ of $(J,\Gamma)$, we write $A_J$ for $\{\Gamma:(J,\Gamma)\in A\}$, the {\em section} of $A$ at given $J$.
For sets of bonds and $n>1$ states, we can define sections such as $A_{J\Gamma}$ similarly, for a choice of $J$ and the 
first state $\Gamma$, and so on. We will apply this notation especially to sets $A$ in ${\cal I}_{n\Phi}$. 

The results proved in NRS23 \cite{nrs23} are contained in the following.
\newline
{\bf Proposition 1}: Consider a short-range mixed $p$-spin model with translation invariant n.i.p.\ $\nu$ such that ${\rm Var}\,J_X>0$ for all 
$X\in{\cal X}$, and a translation-covariant metastate $\kappa_J$. Then (i) for $\kappa^\dagger$-almost 
every pair $(J,\Gamma)$, and any set $A$ of $(J,\Psi)$ in ${\cal I}_1$, the distribution (weight) $w_\Gamma$ is trivial on the set 
$A_J$ of $\Psi$, that is, $w_\Gamma(A_J)=0$ or $1$.
Equivalently, (ii) for any invariant (i.e.\ ${\cal I}_1$-measurable) function $O(J,\Psi)$ and for $\kappa^\dagger$-almost 
every pair $(J,\Gamma)$, $O(J,\Psi)$ takes the same value for $w_\Gamma$-almost every $\Psi$. $\Box$
\newline
In NRS23, the form (i) was referred to as the zero-one law, while the equivalent form (ii) was termed ``single-replica equivalence'';
note that in (ii) the constant value of $O$ can depend on $\Gamma$, though $O$ is defined without reference to $\Gamma$. 
The invariant function $O(J,\Psi)$ can be viewed as a ``macroscopic'' observable property of a pure state $\Psi$ for given $J$, $\Gamma$,
so Proposition 1 states that the pure states drawn from a given $\Gamma$, itself drawn for the metastate $\kappa_J$ for given $J$, 
all look alike macroscopically, though the value of the observable could still depend on $\Gamma$. 

In using Proposition 1, one must always be careful to check that a set or function is invariant, and also independent of $\Gamma$. 
It is natural to ask if the result in Proposition 1 can be extended to sets and functions that depend on $\Gamma$ as well as on $(J,\Psi)$.
The following ``strong'' zero-one law will be proved later, in Sec.\ \ref{subsubsec:prop2} (without circularity).
\newline
{\bf Proposition 2}: Consider the same hypotheses as in Proposition 1. Then (i) for $\kappa^\dagger$-almost 
every pair $(J,\Gamma)$, and any set $A$ of $(J,\Gamma,\Psi)$ in ${\cal I}_2$, the distribution (weight) $w_\Gamma$ is trivial on the set 
$A_{J\Gamma}$ of $\Psi$. 
Equivalently, (ii) for any invariant (i.e.\ ${\cal I}_2$-measurable) function 
$O(J,\Gamma,\Psi)$ and for $\kappa^\dagger$-almost every pair $(J,\Gamma)$, 
$O(J,\Gamma,\Psi)$ takes the same value for $w_\Gamma$-almost every $\Psi$. 
\newline 
This stronger result will be used in the Sections to follow.

\section{Analysis of metastates: indecomposability}
\label{indecomp}

In this section, we introduce for any disordered classical spin system with disorder distribution $\nu$ a semigroup of transformations, 
consisting of a family of probability kernels. If a joint distribution $\kappa^\dagger$ on $(J,\Gamma)$ is invariant under all members of this 
family, then its marginal distribution on $J$ must be $\nu$, and its conditional distribution $\kappa_J$ must be covariant; further, 
this invariance is compatible with $\kappa_J$ being supported on Gibbs states only, so $\kappa_J$ can be a metastate.  
We then consider $\kappa^\dagger$s that are extremal for this semigroup, or the corresponding metastates $\kappa_J$, which are termed 
indecomposable. Any metastate can be uniquely decomposed as a $J$-independent mixture of indecomposable metastates. 
The Gibbs states $\Gamma$ drawn from an indecomposable metastate all ``look alike'' in that any invariant (or ``macroscopic'') observable
function of $(J,\Gamma)$ takes the same value for almost every $\Gamma$. We discuss some 
examples, and show the utility of these concepts by deriving a number of results about metastates 
and Gibbs states in short-range disordered classical spin systems.

The first two Subsections in this Section are somewhat technical, but only the conclusions, not the machinery behind them, 
will be used later in the paper.

\subsection{Invariance of $\kappa^\dagger$}

First we point out that if $\kappa_J^{(1)}$, $\kappa_J^{(2)}$ are two metastates based on $J$ with the same distribution
$\nu$, then a convex combination
\be
\kappa_J=\lambda\kappa_J^{(1)}+(1-\lambda)\kappa_J^{(2)}
\ee
($0\leq\lambda\leq 1$) is also a metastate, if $\lambda$ is a constant, that is, independent of $(J,\Gamma)$. 
(Recall that metastates are defined in this paper as probability measures that are supported
on Gibbs states and are covariant under local transformations.) The same
is true for more general mixtures (convex combinations), and there are corresponding forms for the joint distribution 
$\kappa^\dagger$. The most general form would be an integral over a set of $\kappa^\dagger$s, using in place of 
$(\lambda,1-\lambda)$ a probability measure that is invariant under local transformations. Thus metastates 
form a covariant convex set of probability distributions, and $\kappa^\dagger$s belong to a convex set of distributions.

This observation suggests that it may be useful to consider the extreme points of the convex set (if it has any), that is, 
$\kappa^\dagger$s that 
cannot be expressed as mixtures of other $\kappa^\dagger$s, and 
a decomposition of a general metastate or $\kappa^\dagger$ into an integral over the set of extreme points
(also known as the extreme boundary). The covariance properties are already described using certain transformations that form
a group $\Phi$. Translation invariance can be imposed on $\nu$ and $\kappa^\dagger$ if desired, but for local transformations 
as usually considered $\nu$ is not invariant. It would be useful to consider transformations under which $\nu$ and $\kappa^\dagger$ are
always invariant; then methods and results of ergodic theory could be used. So our first goal will be to redefine 
the transformations to obtain a family of transformations under which $\nu$ and $\kappa^\dagger$ are invariant,
and which retains the essential features of local transformations (translations will act in the usual way). 

The construction, and many of the results to follow, work for any probability distribution $\nu=\prod_X \nu_X$; 
here $\prod_X$ means product over all the $X$ such that there is
a term containing $J_X$ in the Hamiltonian. (It does not use the detailed way in which
the independent random variables $J_X$ enter the Hamiltonian, and so the construction and the general theorems that follow also work
for the more general models mentioned in Sec.\ \ref{models}, in which there is a countable infinity of independent random variables, 
on which each $J_X$ depends, with only minor changes in some places. It also continues to work in the cases in which, for each $X$, 
$J_X=J_X^{(1)}+J_X^{(2)}$ with $J_X^{(2)}$ centered Gaussian and independent of $J_X^{(1)}$; here there is a separate $\nu_X^{(i)}$ 
for both of $i=1$, $2$, and the generalization to this case should be obvious. We will not emphasize these extensions further.) Each $\nu_X$ 
is the unique stationary distribution
of some continuous- or discrete-time stationary Markov process (a semigroup, i.e.\ the elements may not all have an inverse) on 
$\mathbb{R}$. For example, for $\nu_X$ Gaussian, we 
can use the continuous-time Ornstein-Uhlenbeck process. For general $\nu_X$, we can always construct a discrete-time process by defining a 
probability kernel (see Sec.\ \ref{models} or Ref.\ \cite{georgii_book})
which we denote by $\pi_X$, as follows. First, suppose we have only the random variable $J_X$ to consider.
Then $\pi_X$ acts on a distribution $\mu$ on $J_X$ as, for $A$ a measurable set of $J'_X$,
\be
\mu(A) \mapsto (\mu\cdot \pi_X) (A)=\int \mu(dJ_X) \pi_X(A \mid J_X)
\ee 
(throughout, we will use the symbol $\pi_X$ for both the operation and the corresponding probability kernel).
We take $\pi_X(\cdot \mid J_X)$ to be equal to the probability distribution $\nu_X(\cdot)$ itself, independent of the given value of $J_X$. 
It is clear that, for this choice of probability kernel, this maps any $\mu$ to $\nu_X$, so $\nu_X$ is the unique stationary distribution
(applying $\pi_X$ more than once gives the same result as applying it once, $\pi_X\cdot \pi_X=\pi_X$; i.e.\ it is idempotent, similar to a 
projection operator). Note that $\pi_X$ is not the identity map ${\rm id}$, even when $\nu_X$ is a single atom. $\pi_X$ and the identity 
generate (in fact, are the only elements of) a semigroup $\Pi_X=\{{\rm id},\pi_X\}$ of transformations acting on the space of 
probability distributions on $\mathbb{R}$. 

Now we extend the definition, letting $\pi_X=\pi_X(\cdot\mid J)$ act on the space of 
all $J_{X'}$, not only all $J_X$ for the selected $X$, by defining its kernel to be that above times a $\delta$-function 
$\delta_{J_{X'}}$ for each $X'\neq X$. For any $X$, $X'$, 
$\pi_X$ and $\pi_{X'}$ clearly commute. Then we form the coproduct $\Pi_0=\coprod_X \Pi_X$ over $X$ of the semigroups 
$\Pi_X=\{{\rm id},\pi_X\}$ (in the 
category of Abelian semigroups, or more properly of Abelian ``monoids''), the elements of which are the identity transformation 
at each $X$, except for at most a finite number of $X$. $\Pi_0$ has $\nu$ as its unique stationary (i.e.\ invariant) 
distribution. For translation invariance, we include the translations in the semigroup as well (they too can be expressed as 
probability kernels \cite{georgii_book}), to arrive at a semigroup $\Pi$ which, acting on probability distributions on $J$, has the 
translation-invariant $\nu$ as its unique stationary or invariant distribution.

For any change $J_X\to J_X'$, we know how $\Gamma$ should transform by local transformation, so we can immediately extend the probability
kernel for a local transformation to obtain a probability kernel, which will be denoted by $\pi_X$ also, acting on probability distributions 
on the space of $(J,\Gamma)$. This kernel can be written explicitly as [for $A$ a set of $(J',\Gamma')$]
\bea
\pi_X(A\mid (J,\Gamma))&=&\int \nu_X(dJ_X')\prod_{X':X'\neq X}\delta_{J_{X'}}(dJ_{X'}')\non\\
&&{}\times\delta_{\theta_{J_X'-J_X}\Gamma}(d\Gamma'){\bf 1}_{A}(J',\Gamma').\quad
\eea
[Here the $\delta$-functions are viewed as measures,
and the integrals are over the space of all pairs $(J',\Gamma')$.]
$\pi_X$ has the effect of spreading the initial point $(J,\Gamma)$ along its orbit under $\theta_{J_X'-J_X}$, distributed according
to $\nu_X(dJ_X')$ times $\delta$-functions. 

From the definition, we can see that, if $(J,\Gamma)$ is replaced by a transformed pair 
$(\theta J,\theta\Gamma)$ for any $\theta=\theta_{\Delta J_X}\in\Theta_X$, then $\pi_X(A\mid(J,\Gamma))$ is invariant: 
\be
\pi_X(A\mid (\theta_{\Delta J_X}J,\theta_{\Delta J_X}\Gamma)))=\pi_X(A\mid (J,\Gamma)).
\ee 
Consequently, for any measurable $A$,
$\pi_X(A\mid (J,\Gamma))$ is a function of $(J,\Gamma)$ that is measurable with respect to the $\sigma$-algebra ${\cal I}_{1X}$ 
(a sub-$\sigma$-algebra of the $\sigma$-algebra of all pairs) that consists of Borel sets invariant under all elements of 
$\Theta_X$. So now $\pi_X(\cdot\mid\cdot)$ can be viewed as a probability kernel \cite{georgii_book} from ${\cal I}_{1X}$ to the full 
$\sigma$-algebra of Borel sets of pairs $(J,\Gamma)$, and further $\pi_X$ is proper (see Sec.\ \ref{models}, or Ref.\ \cite{georgii_book}, 
p.\ 14). ${\cal I}_{1X}$ can also be characterized by the statement:
\be
\pi_X(A\mid (J,\Gamma))={\bf 1}_A(J,\Gamma)
\ee
for all $(J,\Gamma)$ if and only if $A\in {\cal I}_{1X}$ [similar to Ref.\ \cite{georgii_book}, Remark (7.6) 1)], which can serve as 
an alternative definition. Then it is clear that a non-empty set in ${\cal I}_{1X}$ must consist of pairs $(J,\Gamma)$ that form complete orbits 
under the action of $\Theta_X$; if we forget the states $\Gamma$, then the sets of bonds are independent of $J_X$. 
(Similar statements hold for the action of $\pi_X$ on distributions on $J$ only.)

${\rm id}$ and $\pi_X$ again form a semigroup, denoted $\Pi_X$, of mappings of probability distributions on pairs $(J,\Gamma)$.
and again $\pi_X$ and $\pi_{X'}$ commute. Then we define the coproduct, again denoted $\Pi_0=\coprod_X\Pi_X$. [It is the same
abstract semigroup, but now it acts on pairs $(J,\Gamma)$ instead of on $J$ only.]
By including all translations, we obtain a semigroup 
that we denote by $\Pi$; $\Pi_0\subset\Pi$ is a normal sub-semigroup of $\Pi$. The sub-$\sigma$-algebras of invariant sets 
for the actions of $\Pi_0$ and $\Pi$ (i.e.\ invariant under all elements of $\Pi_0$ or $\Pi$), the definitions of which should be obvious, 
are ${\cal I}_{1\Pi_0}=\cap_{X\in{\cal X}}{\cal I}_{1X}$ and ${\cal I}_{1\Pi}$ respectively. It follows from the preceding discussion 
that, in terms of the $\sigma$-algebras defined in Sec.\ \ref{models}, we have
${\cal I}_{1\Pi_0}={\cal I}_{10}$, ${\cal I}_{1\Pi}={\cal I}_{1}$ [again, see Ref.\ \cite{georgii_book}, Remark (7.6) 1)];
thus these do not depend on the choice of $\nu$. $\Pi_0$ is Abelian, so the product of $\pi_X$ and $\pi_{X'}$
can be written as $\pi_{X,X'}$. More generally, we can construct $\pi_Y$ for any finite subset 
$Y\subset {\cal X}$, including $Y=\emptyset$, as the product $\pi_Y=\prod_{X:X\in Y}\pi_X$, or the identity if $Y=\emptyset$, 
and let $\cal Y$ be the set of all such $Y$; again, $\cal Y$ is countable. Then we have $\Pi_0=\{\pi_Y:Y\in{\cal Y}\}$. Later it will be 
useful to consider $\pi_Y$ for $Y={\cal X}(\Lambda)$ for finite regions $\Lambda$; then ${\cal X}(\Lambda)\cup\{\emptyset\}$ 
is the power set of $\Lambda$. From now on, we let $\Upsilon$ stand for either $\Pi_0$ or $\Pi$, and it will be convenient
to take the correspondence with $\Phi=\Theta_0$ or $\Theta$ as understood.

The usefulness of all this hinges on the fact that, not only is $\nu$ $\Upsilon$-invariant under the action on probability distributions
on $J$, but also we have the following (which the reader may have anticipated).
\newline
{\bf Lemma 2}: In the above construction, $\kappa_J$ is covariant under all $\theta\in\Phi$, $\nu$-almost surely, 
if and only if
\be
\kappa^\dagger\cdot\pi=\kappa^\dagger
\ee 
for all elements $\pi\in \Upsilon$, that is $\kappa^\dagger$ is $\Upsilon$-invariant.
\newline
{\bf Proof} (outline): As the translation part of the statement is clear, we focus on local transformations. To consider a 
single $X$, it is sufficient
to condition on $J_X$ only, so we set $\kappa^\dagger=\nu_X\kappa_{J_X}$. Then a one-line calculation shows that
$(\kappa^\dagger\cdot\pi_X)(A)=\kappa^\dagger(A)$ for all measurable sets $A$ of $(J,\Gamma)$ if and only if 
(to simplify writing, we suppress the dependence on the spectators, $J_{X'}$ for all $X'\neq X$)
\be
\int \nu_X(dJ'_X)\kappa_{\theta_{J'_X-J_X}J_X}(\theta_{J'_X-J_X}d\Gamma)=\kappa_{J_X}(d\Gamma)
\ee
as measures on $(J_{X'})_{X':X'\neq X}$ and $\Gamma$, for $\nu_X$-almost every $J_X$; the integral is over $J_X'$, not $\Gamma$. 
Then if $(J_X,(J_{X'})_{X':X'\neq X},\Gamma)$ is transformed by the action of $\theta_{J_X''-J_X}$ 
for any $J_X''$, the integral on the left is unchanged, because writing $\Delta J_X=J_X''-J_X$, 
$\theta_{J'_X-J_X-\Delta J_X}(J_X+\Delta J_X) = \theta_{J'_X-J_X}
J_X=J_X'$ and $\theta_{J'_X-J_X-\Delta J_X}\theta_{J_X''- J_X}d\Gamma = \theta_{J'_X-J_X}d\Gamma$. Hence the 
right-hand side is invariant under the same action of $\theta_{J_X''- J_X}$, which is equivalent to covariance; the result 
extends immediately to all elements of $\Theta_0$. The converse is trivial. This proves the Lemma. $\Box$
\newline
Thus $\Pi_0$-invariance of $\kappa^\dagger$ encodes both the distribution $\nu$ and the covariance of $\kappa_J$ under 
local transformations. Use of the kernels $\pi_X$ or $\pi_Y$ 
enables us to obtain all essential information from local transformations with only a countable infinity of these objects, indexed 
by $X$ or $Y$.

An additional fact is illuminating and will motivate later results. First, we extend the definition of ${\cal I}_{1X}$: 
for each nonempty finite $Y\subset {\cal X}$, there is a $\sigma$-algebra ${\cal I}_{1Y}$ of the sets that are invariant under all $\pi_X$ 
for $X\in Y$, and notice that
for $Y\subseteq Y'$, ${\cal I}_{1Y'}\subseteq  {\cal I}_{1Y}$. Then $\pi_Y$ is a proper probability kernel from ${\cal I}_{1Y}$
to the full $\sigma$-algebra, and we can state the following. Recall that conditional probability, in full generality, is conditioned
on a sub-$\sigma$-algebra, and is a function that is measurable with respect to the latter; for ${\cal I}_{1Y}$, that means independent
of $J_X$ for $X\in Y$ (where $\Gamma$ changes covariantly with $J_X$). 
\newline
{\bf Lemma 3}: For each nonempty $Y$, $\kappa^\dagger\cdot \pi_Y=\kappa^\dagger$ if and only if the conditional probability conditioned
on ${\cal I}_{1Y}$ obeys (for any measurable $A$)
\be
\kappa^\dagger(A\mid{\cal I}_{1Y})(\cdot)=\pi_Y(A\mid\cdot)
\label{kdagker}
\ee
$\kappa^\dagger$-almost surely [the $\cdot$ stands for any choice of $(J,\Gamma)$]. 
\newline
{\bf Proof}: essentially just from the definitions [see Georgii \cite{georgii_book}, Remark (1.20)]. $\Box$
\newline
So $\pi_Y$ is a version \cite{chung_book,breiman_book} of the conditional probability; the Lemma is closely analogous to the 
two equivalent DLR 
characterizations of a Gibbs state $\Gamma$ in terms of a specification $\gamma=(\gamma_\Lambda)_\Lambda$. 
There $\gamma_\Lambda$ is a function of given $s|_{\Lambda^c}$, taking ``values'' that are probability distributions on $s$, 
and the two equivalent statements describe it as the conditional probability conditioned on  $s|_{\Lambda^c}$ ($\Gamma$-almost 
surely), as we explained in Sec.\ \ref{models}.

In addition to invariant sets, we can define invariant observable properties of $(J,\Gamma)$, and in fact we did so, in terms of $\Phi$, 
in Sec.\ \ref{models}. Alternatively, given a semigroup $\Pi_0$ as above, we can define the transformed function 
(see Georgii \cite{georgii_book}, p.\ 14), in terms of $\omega=(J,\Gamma)$, 
\be
f(\omega)\mapsto \pi_X\cdot f(\omega) =\int \pi_X(d\omega'\mid \omega)f(\omega'),
\label{eq:dotnot2}
\ee
and then $\Pi_0$ invariance would be $\pi_X\cdot f=f$ as functions for all $X$. The two definitions are equivalent by a somewhat 
similar argument as in Lemma 2. If $f(\omega)={\bf 1}_A(\omega)$, then this includes the definition of invariant sets in 
${\cal I}_{10}$ as a special case ($A$ is invariant if and only if its indicator is invariant). Again, it makes no difference whether
we consider sets of $(J,\Gamma)$ or $(J,\Psi)$ here.

\subsection{Extremality and indecomposability}

From here on, we will refer to a semigroup $\Upsilon=\Pi_0$ or $\Pi$ for some choice of $\nu$. For generality, we will include 
in this subsection cases that do not possess translation invariance. While covariance of $\kappa_J$ under local transformations is a general 
property of all metastates, and may be viewed as part of the definition, metastate constructions for which $\kappa_J$ is not translation 
covariant do occur, for example, constructions that use free or fixed (but, say random, independent of $J$) boundary conditions on finite 
sizes, and these may be of interest. Theorems 1 and $1'$ and Proposition 1 below, and the further results in Appendix \ref{app:furth}, 
apply to these also. Later we will specialize to translation invariant cases, for which much more can be proved. We note that, given a 
$\Pi_0$-invariant joint distribution, a $\Pi$-invariant joint distribution can be obtained by translation averaging \cite{ns01}. 
We will not explore the consequences of this, but have added a note at the end of the paper. 

As pointed out already, the joint distributions $\kappa^\dagger$ that are invariant under $\Upsilon$ form a convex set. 
For given $\Upsilon$, we now inquire about the extremal $\kappa^\dagger$s (if they exist), that is those $\kappa^\dagger$s 
that cannot be expressed as a mixture of other ($\Upsilon$-invariant) $\kappa^\dagger$s. We will also say that a metastate $\kappa_J$ 
is indecomposable if $\kappa^\dagger$ is extremal. (When we speak of ``a'' metastate $\kappa_J$, we mean a
distribution that is random because it depends on $J$, with distribution $\nu$ for $J$.
Also, at the moment, it is immaterial whether or not we impose the condition that $\kappa_J$ is supported on Gibbs states. Because the
Gibbs, i.e.\ DLR, conditions are $\Theta$-covariant, we can impose that condition on $\kappa^\dagger$ afterwards without affecting 
the result.) We will relate extremal $\kappa^\dagger$s to those $\kappa^\dagger$ that are trivial on the $\sigma$-algebra ${\cal I}_{10}$
of invariant sets. 
First, enlarge the $\sigma$-algebra ${\cal I}_{10}$ by defining the sub-$\sigma$-algebra of measurable sets that are $\Pi_0$-invariant 
modulo $\kappa^\dagger$-null sets:
\be
{\cal I}_{1\Pi_0}(\kappa^\dagger)=\{A: \pi(A|\cdot)=1_A\hbox{ $\kappa^\dagger$-almost surely, }\forall \pi\in \Pi_0\}.
\ee
Then we say $\kappa^\dagger$ is trivial on ${\cal I}_{1\Pi_0}(\kappa^\dagger)$ if, for any $A\in{\cal I}_{1\Pi_0}(\kappa^\dagger)$, 
$\kappa^\dagger(A)=0$ or $1$. [This corresponds to what is called {\em ergodicity} in many references. 
In this definition we follow Phelps \cite{phelps_book}. Many authors would instead define ergodicity as triviality on the 
sub-$\sigma$-algebras of {\em strictly} invariant sets ${\cal I}_{1\Pi_0}={\cal I}_{10}$ that we already 
discussed. In some situations, though not all, the two definitions are equivalent \cite{phelps_book,georgii_book}; 
we return to this point later. We will not use the term ergodic in either way for $\Upsilon$, but reserve it for the case of invariance under a 
group, such as the translation group, even though in other areas, such as Markov chain theory, the term ergodic is applied to probability 
kernels.] By a similar argument as for the strictly invariant sets, we can see that in fact
\bea
{\cal I}_{1\Pi_0}(\kappa^\dagger)&=&{\cal I}_{10}(\kappa^\dagger)\\
&\equiv&\{A: \kappa^\dagger(A\bigtriangleup \theta A)=0\;\forall \theta\in\Theta_0\},\label{eq:I10}
\eea 
the sets that are $\Theta_0$ invariant modulo $\kappa^\dagger$-null sets.
(Here $\bigtriangleup$ is symmetric difference, and $\theta A=\{(\theta J,\theta \Gamma): (J,\Gamma)\in A\}$ as usual.)
Here the equivalent definition as ${\cal I}_{10}(\kappa^\dagger)$ makes no reference 
to $\Pi_0$, though it does depend on $\kappa^\dagger$. 

A very general result of ergodic theory now implies that a $\Pi_0$-invariant distribution $\kappa^\dagger$ is extremal if and only 
if it is trivial on the sets in ${\cal I}_{10}(\kappa^\dagger)$; we omit the straightforward proof (see Ref.\ \cite{phelps_book}, Section 12, 
or Ref.\ \cite{georgii_book}, Chapter 7, which works in the broader setting of probability kernels we require). For the corresponding 
metastate $\kappa_J$, first notice that, if $A_J=\{\Gamma:(J,\Gamma)\in A\}$, then for any $\theta\in\Theta_0$,
\bea
\theta A_J\equiv \theta (A_J)&=&\{\theta \Gamma:(J,\Gamma)\in A\},\\ 
A_{\theta J}&=&\{\Gamma:(\theta J,\Gamma)\in A\},
\eea
and
\bea
(\theta A)_J&=&\{\theta \Gamma:(\theta^{-1}J,\Gamma)\in A\}\\
&=&\theta A_{\theta^{-1} J}.
\eea
Then for $A\in{\cal I}_{10}(\kappa^\dagger)$, $A_J$ is, $\nu$-almost surely, a covariant set, modulo $\kappa_J$-null sets, 
in the sense that, for every $\theta\in \Theta_0$, $\kappa_J(A_J\bigtriangleup \theta A_{\theta^{-1}J})=0$ for $\nu$-almost every $J$,
which follows from the invariance of $A$ by conditioning on $J$. Hence, for such $A$, $\kappa_J(A_J)$ is $\Pi_0$ invariant for 
$\nu$-almost every $J$, and as $\nu$ is trivial on ${\cal I}_{0}(\nu)$ [defined similarly to ${\cal I}_{10}(\kappa^\dagger)$], 
it is $\nu$-almost surely 
constant and equal to its $\nu$ expectation $\kappa^\dagger(A)$, and when $\kappa_J$ is indecomposable this is $0$ or $1$. 
Thus we have established the following.
\newline
{\bf Theorem 1}: With respect to the $\Pi_0$ action, a metastate $\kappa_J$ 
is indecomposable if and only if it is $\nu$-almost surely trivial on the covariant sets $A_J$ for $A\in{\cal I}_{10}
(\kappa^\dagger)$. 
\newline
It is from this statement (and related ones to follow) that we will soon obtain a number of 
results about disordered spin systems.

We can repeat all of the preceding definitions, Theorem 1, and its proof, for $\Pi$ in place of $\Pi_0$, by dropping the subscript $0$ 
throughout (we refer to that theorem as Theorem 1 also). When discussing extremality or indecomposability we should specify 
whether we mean $\Pi_0$ or $\Pi$. We will do so by stipulating that $\kappa^\dagger$ be $\Upsilon$ invariant, for $\Upsilon$ 
one or other of $\Pi_0$ or $\Pi$,
using the unmodified terms extremality or indecomposability in the corresponding sense, even though $\Pi_0$ invariance of 
$\kappa^\dagger$ is always in force. Note that we have 
${\cal I}_1\subseteq{\cal I}_{10}$, and similarly for ${\cal I}_1(\kappa^\dagger)$, ${\cal I}_{10}(\kappa^\dagger)$,
and so for a translation-invariant $\kappa^\dagger$, triviality on ${\cal I}_{10}(\kappa^\dagger)$ implies the same for 
${\cal I}_{1}(\kappa^\dagger)$. On the other hand, a $\Pi$-invariant $\kappa^\dagger$ that is extremal might not be 
extremal when viewed as only $\Pi_0$ invariant, because the $\sigma$-algebra ${\cal I}_{10}$ may be larger. Hence, for generality,
we do need both notions. 

As an aside, we emphasize that the result that a $\Upsilon$-invariant distribution is extremal if and only if it is trivial
on ${\cal I}_{1\Upsilon}(\kappa^\dagger)$ also applies to $\nu$, on which we defined the action of $\Upsilon$ before 
that on $\kappa^\dagger$. 
Thus $\nu$, which is extremal because it is the unique invariant distribution under $\Upsilon$, is trivial
with respect to ${\cal I}_{\Upsilon}(\nu)$. For $\Upsilon=\Pi_0$, this gives a proof \cite{georgii_book} of Kolmogorov's zero-one law, 
which says that a product distribution such as $\nu$ is trivial on the $\sigma$-algebra ${\cal I}_0(\nu)$ of sets of $J$ that are $\nu$-almost 
surely independent of any finite number of $J_X$ \cite{chung_book,breiman_book}. 
(Actually, the zero-one law states this for the $\sigma$-algebra ${\cal I}_0$ 
of {\em strictly} invariant sets; we return to the relation of the two shortly. ${\cal I}_0$ is also referred to as the tail $\sigma$-algebra 
$\cal T$ in this context.) Likewise, for translations, ergodicity of a distribution if and only if it is extremal, and ergodicity of a 
translation-invariant product, are basic results of ergodic theory (and the strictly translation-invariant sets form a sub-$\sigma$-algebra 
$\cal I$ of the tail $\sigma$-algebra \cite{georgii_book}). Thus Theorem 1 might be viewed as an extension to $\kappa^\dagger$ 
of both of these classic results for $\nu$.

Next we reformulate Theorem 1 in terms of $\kappa^\dagger$-almost surely $\Upsilon$-invariant observables $O(J,\Gamma)$, which we can 
also define as ${\cal I}_{1\Upsilon}(\kappa^\dagger)$ measurable functions (the need for the almost-sure part of the definition will disappear 
shortly). Then by a standard measure-theoretic argument (cf.\ e.g.\ Ref.\ \cite{nrs23}), Theorem 1 
is equivalent to the following (Theorems 1 and 1$'$ are analogous to the two parts of Proposition 1 above).
\newline
{\bf Theorem} ${\bf 1}'$: Consider an $\Upsilon$-invariant $\kappa^\dagger$. $\kappa_J$ is indecomposable if and only if every
$\kappa^\dagger$-almost surely $\Upsilon$-invariant observable $O(J,\Gamma)$ is $\kappa^\dagger$-almost surely constant.
\newline
Thus such an observable takes the same value for $\kappa_J$-almost every $\Gamma$, for $\nu$-almost every $J$. 
This says that states $\Gamma$ drawn from an indecomposable $\Upsilon$-invariant $\kappa_J$ ``look alike'' in terms of any macroscopic
(i.e.\ $\Phi$- or $\Upsilon$-invariant) observable property, for $\nu$-almost any given $J$. 

Having now characterized, at least to some extent, indecomposable metastates, it would be very useful if we could establish 
a few more facts. The first would be to describe the relation between ${\cal I}_{1\Upsilon}$ and ${\cal I}_{1\Upsilon}
(\kappa^\dagger)$ for $\kappa^\dagger$ invariant under $\Upsilon=\Pi$ or $\Pi_0$. The second would be 
to relate extremality to decay of connected correlation functions. The third would be to show that any metastate, 
or any $\kappa^\dagger$, 
can be decomposed as a mixture of $\Upsilon$-indecomposable (respectively, extremal) ones. All three of these can in fact be done,
using fairly standard methods described in Georgii \cite{georgii_book}, Chapters 7 and 14. As these results require longer proofs, 
and will not be used much in the main text, we postpone the proof sketches to Appendix \ref{app:furth}.

For the first fact, the result is that ${\cal I}_{1\Upsilon}(\kappa^\dagger)$ is the $\kappa^\dagger$-completion of 
${\cal I}_{1\Upsilon}$, in the sense that, for any set $A\in{\cal I}_{1\Upsilon}(\kappa^\dagger)$, there is a set $B\in{\cal I}_{1\Upsilon}$
such that $\kappa^\dagger(A\bigtriangleup B)=0$.
This implies that triviality of $\kappa^\dagger$ on one $\sigma$-algebra implies it on the other (so the different definitions 
analogous to those of ergodicity become equivalent). 
We now state without proof some consequences that follow easily (cf.\ Ref.\ \cite{georgii_book}, Theorem 7.7): 
\newline
1) Theorem 1 becomes: a $\kappa_J$ is indecomposable if and only if it is trivial on covariant sets $A_J$ for 
$A\in{\cal I}_{1\Upsilon}$, for $\nu$-almost every $J$; 
\newline
2) any $\kappa^\dagger$ is determined among the $\Upsilon$-invariant distributions 
by its restriction to ${\cal I}_{1\Upsilon}$;  
\newline
3) Distinct extremal joint distributions $\kappa^{\dagger(1)}$, $\kappa^{\dagger(2)}$ are mutually singular on 
${\cal I}_{1\Upsilon}$, that is, there is a set $A\in {\cal I}_{1\Upsilon}$ such that $\kappa^{\dagger(1)}(A)=0$, 
$\kappa^{\dagger(2)}(A^c)=0$. As ${\cal I}_{1\Upsilon}$ is a sub-$\sigma$-algebra of the full Borel $\sigma$-algebra
of sets of $(J,\Gamma)$, this implies that $\kappa^{\dagger(1)}$, $\kappa^{\dagger(2)}$ are mutually singular on that also,
and that $\kappa^{(1)}_J$, $\kappa^{(2)}_J$ are mutually singular, $\nu$-almost surely. 
\newline
Result 2) says that an extremal $\kappa^\dagger$ is essentially determined uniquely by its support, that is, a set of $(J,\Gamma)$ 
in ${\cal I}_{1\Upsilon}={\cal I}_{1\Phi}$. More precisely, this can be stated as in Appendix \ref{app:furth} (during the proof of existence
of a decomposition into extremal joint distributions): there is a probability kernel $\pi_\Upsilon$ such that an extremal $\Upsilon$-invariant 
$\kappa^\dagger$ obeys $\kappa^\dagger(\cdot)=\pi_\Upsilon(\cdot\mid (J,\Gamma))$ as distributions, 
for $\kappa^\dagger$-almost every $(J,\Gamma)$. Here $\pi_\Upsilon$ is a probability kernel from $(J,\Gamma)$ 
to probability distributions, and is $\Phi$ invariant in $(J,\Gamma)$. It can be viewed as a version of the 
conditional distribution of any $\Upsilon$-invariant $\kappa^\dagger$, conditioned on ${\cal I}_{1\Upsilon}$,
and so puts probability $1$ on the ``smallest'' invariant set in ${\cal I}_{1\Upsilon}$ that contains $(J,\Gamma)$
(see  Ref.\ \cite{einsward_book}, Theorem 5.14, for this and a discussion of the subtleties involved; a related discussion
appears in Ref.\ \cite{georgii_book}, Theorem 7.12).
We caution that result 3) asserts only that states drawn from the respective indecomposable metastates are 
[$\nu(\kappa_J^{(1)}\times\kappa_J^{(2)})$-almost surely] distinct but, when the states are Gibbs states, that does not necessarily mean 
that their respective pure-state decompositions have disjoint support: in principle, they could even be supported on the same set of pure states, 
though necessarily as distinct distributions. But in Sec.\ \ref{subsec:ovsing} below, we show that in fact the pure-state decompositions are  
(almost surely) either identical or disjoint.

Second, we have now seen that there are close analogies between the present situation and that for Gibbs and pure states.
[As Gibbs states are characterized as invariant distributions under a set of probability kernels $\gamma_\Lambda$ (making up the 
specification), this is more than just an analogy; it is a remarkable degree of similarity of structure.] 
A third characterization of pure states, after extremality and triviality on a $\sigma$-algebra of invariant sets, is in terms
of decay of connected correlations. This too can be done for extremal $\kappa^\dagger$, in a fashion similar to Georgii \cite{georgii_book}, 
Theorem 7.9 (and see his remark just before 7.13): a distribution $\kappa^\dagger$ is trivial on ${\cal I}_{10}$ if and only if
for any Borel set $A$, 
\be
\lim_{n\to\infty}\sup_{B\in{\cal I}_{1{\cal X}(\Lambda_n)}}|\kappa^\dagger (A\cap B)-\kappa^\dagger(A)\kappa^\dagger(B)|=0,
\ee
where $(\Lambda_n)_n$ is a cofinal sequence of finite regions, so $\Lambda_n$ eventually includes all sites $i$ as $n\to\infty$.
(This result will not be used in the remainder of this paper.) For example, consider an extremal $\kappa^\dagger$, 
a function of a state $\Gamma$, such as an expectation $\langle s_X\rangle_\Gamma$ for fixed $X$, and a function of some bonds, 
say ${\bf 1}_{\{J_{X'}\in[a,b]\}}$ for $X'\subset \Lambda_n^c$ and real $a$, $b$ ($a<b$).  Then
\bea
&&\left|\bbE_{\kappa^\dagger} (\langle s_X\rangle_\Gamma {\bf 1}_{\{J_{X'}\in[a,b]\}})-\bbE_{\kappa^\dagger}
(\langle s_X\rangle_\Gamma )
\bbE_{\kappa^\dagger}({\bf 1}_{\{J_{X'}\in[a,b]\}})\right|\non\\
&&{}\qquad\qquad\to 0
\eea
as $n\to\infty$, no matter how $X'$, $a$, $b$ depend on $n$. 

For the third desirable fact mentioned above, the existence of such a decomposition is a more delicate issue than the characterization 
of extreme distributions as trivial on a $\sigma$-algebra. Nonetheless, in Appendix \ref{app:furth} we sketch a proof that, just as a Gibbs state 
can be uniquely decomposed as a mixture of extremal (or pure) states, so for a metastate we have the following.
\newline
{\bf Proposition 3}: an $\Upsilon$-invariant joint distribution $\kappa^\dagger$ (or corresponding metastate $\kappa_J$) can be uniquely 
decomposed as a invariant mixture of extremal distributions (respectively, a mixture of indecomposable $\kappa_J$s). 
Thus, for joint distributions,
\be
\kappa^\dagger=\int \lambda_{\kappa^\dagger}(d\eta)\kappa_\eta^\dagger,
\ee
where $\eta$ parametrizes extremal joint distributions $\kappa_\eta^\dagger$;
the latter have no dependence on the given $\kappa^\dagger$. The probability distribution $\lambda_{\kappa^\dagger}$ 
is independent of $(J,\Gamma)$ but depends on the given $\kappa^\dagger$; it puts probability $1$ on the set of extremal elements 
$\kappa^\dagger_\eta$. 
(Conditioning $\kappa$s on both sides on $J$ gives the 
respective decomposition for $\kappa_J$.) In particular, extremal joint distributions $\kappa_\eta^\dagger$ and indecomposable metastates 
$\kappa_{J\eta}$ do exist. 
\newline
[In the following, we will have no occasion to use $\lambda_{\kappa^\dagger}$ explicitly. We can denote the (convex) space of 
$\Upsilon$-invariant 
joint distributions $\kappa^\dagger$ by ${\cal K}^\dagger(\Upsilon)$, and its extreme boundary by 
${\rm ex}\,{\cal K}^\dagger(\Upsilon)$, by analogy with the space of Gibbs states ${\cal G}(\gamma_J)$ for the specification 
$\gamma_J$, and its subspace ${\rm ex}\,{\cal G}(\gamma_J)$ of pure states.]

It then remains to describe the properties and general classification of indecomposable metastates. The general form of the latter 
problem is analogous to one from ergodic theory in the case of group actions, in which there is a group $G$ acting on the space, 
say of pairs $(J,\Gamma)$, and in that case it looks as follows. The space can be mapped to the space of bonds $J$ only, and that 
map commutes with the action of $G$. The distribution $\nu$ on $J$ is ergodic with respect to the $G$ action, and we are interested 
in ergodic distributions $\kappa^\dagger$ on pairs $(J,\Gamma)$ that map to the ergodic $\nu$ when we forget the $\Gamma$ variable. 
This system is an extension $\kappa^\dagger$ of $\nu$ (as systems with $G$ actions) \cite{einsward_book}, and we wish to classify 
the ergodic extensions $\kappa^\dagger$ of the ergodic $\nu$. Our problem has exactly the same form, once we replace $G$ invariance 
by $\Upsilon$ invariance, and ergodicity by triviality on the $\sigma$-algebra of invariant sets.

Here we will present only one basic result about decomposability of metastates (we return to a fuller analysis of the structure 
of indecomposable metastates, which extends that given here, in Sec.\ \ref{sec:mas} below). For given $J$, a metastate can be expressed 
as a mixture of a number (at most 
countable) of atoms, and of an atomless metastate. {\it A priori} it may not be clear whether the weights of each of these parts in the 
mixture are $\nu$-almost surely constant functions of $J$. We refer to the nonzero terms of such a mixture as the ``parts'' of the 
metastate. We have the following, which is a general result, presumably well-known in ergodic theory (and the alternate version
using translation invariance, mentioned after the following proof, is largely contained in Ref.\ \cite{answ14}, Sec.\ 7). It goes through 
whether or not the metastate is translation covariant.
\newline
{\bf Proposition 4}: Consider an $\Upsilon$-invariant $\kappa^\dagger$. If the corresponding metastate $\kappa_J$ consists of one or more 
atoms and 
an atomless part, or contains atoms of different weights, then it is decomposable. 
\newline
{\bf Proof}: For given $J$, we can express $\kappa_J$ (a distribution on $\Gamma$) as 
\be
\kappa_J=\sum_{\alpha}\kappa_{J\alpha}\delta_{\Gamma_\alpha} + \int_D \kappa_J(d\Gamma') \delta_{\Gamma'},
\ee
where here $\alpha$ belongs to a countable set of cardinality at least one, $\Gamma_\alpha$ are Gibbs states, 
$\kappa_{J\alpha}=\int_{\{\Gamma_\alpha\}} \kappa_J(d\Gamma)$ are larger than zero for all $\alpha$, and the domain $D$ is the 
complement, in the set of all $\Gamma$, of the set of all atoms $\Gamma_\alpha$. Rank order the metastate weights 
$\kappa_{J\alpha}$ in strictly decreasing order of the distinct values, with a multiplicity $m_\alpha$ to describe any ties 
or degeneracy, meaning some $\kappa_{J\alpha}$s could be equal. This defines sets of atoms for all $J$. By covariance of the
metastate under local transformations (elements of $\Theta_0$), each distinct value and each multiplicity is independent of
any finite set of $J_X$, and so is $\nu$-almost surely constant by Kolmogorov's zero-one law. Hence the metastate weight $\int_D 
\kappa_J(d\Gamma')=1-\sum_\alpha\kappa_{J\alpha}$ of the atomless part is also almost surely constant. This describes the metastate
as a $\nu$-almost surely $J$-independent mixture of finite sets of atoms, with weight $m_\alpha \kappa_{J\alpha}$ for each set
(where members of the same set have equal weights $1/m_\alpha$ within each finite set),
and an atomless part, proving that it is decomposable. $\Box$
\newline
We remark that, in the $\Pi$- (translation-) invariant case, we could give a similar proof using only translation invariance 
and translation ergodicity of $\nu$, in place of the zero-one law for the product form.

Note that, if there is a set of $m_\alpha>1$ weights equal to $\kappa_{J\alpha}$ for some $\alpha$, it is not clear that 
each such corresponding atom constitutes an indecomposable metastate; it may be only that the metastate decomposes as a mixture of
the sum (with weights $1/m_\alpha$) of those atoms plus a complementary part. This is because 
it is not clear {\it a priori} whether one of the set of $\alpha$s, or a subset of fewer than $m_\alpha$ atoms, can be 
selected for each $J$ in a measurable and covariant way for $\nu$-almost every $J$ to decompose the set further. 
Similarly, it is not clear at present whether the atomless part (if any) can be decomposed further; we present further analysis in Sec.\ 
\ref{sec:mas} below. 

In any case, we have established that an indecomposable metastate is only allowed either to consist of a finite number 
of atoms of equal weight, or to be atomless. We consider some examples of indecomposable nontrivial metastates below. 

It is worth pointing out that, if there is a unique $\Upsilon$-invariant $\kappa^\dagger$, so ${\cal K}^\dagger(\Upsilon)$ is a single point, 
then it is necessarily extremal. On the other hand, in the constructions of metastates as ($J$-independent) subsequence limits of 
finite-size approximations, there is the possibility that the limit is not unique, giving rise to non-unique metastates. A unique metastate
would certainly imply a unique limit, but the converse is not clear: uniqueness of such a limit may not
imply uniqueness or indecomposability of the metastate.

\subsection{Examples}
\label{sec:examples}

There are examples of systems with nontrivial metastates that are decomposable and others that are indecomposable. 
In this subsection, some of the results are obtained only heuristically, at a theoretical physics level of rigor.

A decomposable example arises for the random-field ferromagnet (RFFM), which was 
the context in which a metastate first appeared \cite{aw}. In this case, with constant ferromagnetic short-range bonds and independent 
single-site magnetic fields, each of which has a distribution that is symmetric under change of sign of the field, for $d>2$ and low $T$
there is a metastate that (by symmetry) puts equal weight on each of two pure Gibbs states. The two pure states have nonzero 
magnetization of equal magnitude and opposite signs. As emphasized in Ref.\ \cite{nrs23}, at least in a version of the model in
which the variance of $J_X$ is positive for all $X$, each Gibbs state drawn from $\kappa_J$ must be one of the pure states,
not a mixture of both. Alternatively, for the original form of the model, it is not difficult to see that there must be chaotic size 
dependence, and that the disorder selects one or other ordered state in each system size \cite{ns92}, giving rise to the metastate 
as described. It now follows from Theorem $1'$ that this metastate is decomposable as a mixture of the two trivial ones, 
each with support on a single pure Gibbs state, because the magnetization distinguishes the Gibbs states. The RFFM is
also an example of a case with a unique limit in the metastate construction, and that limit is this decomposable metastate.

A very different situation is found in a simple but instructive ``toy''  model, the ground states of a one-dimensional chain of sites 
$i=1$, $2$, \ldots, 
$L$, with nearest neighbor bonds $J_{i,i+1}= \pm 1$ (with probability $1/2$ for either sign, independently for all bonds), and a fixed spin 
boundary condition at $i=L$, $s_L=\pm 1$ with probability $1/2$ for each, independent of the bonds. This fits in our general class 
of models, except that the lattice is $\subseteq \mathbb{N}$ not $\mathbb{Z}$, and it lacks translation invariance; we consider $\Pi_0$-, 
not $\Pi$-, invariance. For $L$ finite, the ground state can be determined from 
the given spin $s_L$ by using the rule $s_is_{i+1}=J_{i,i+1}$ for $i=1$, \ldots, $L-1$. It is clear that, if we consider the probability 
distribution for the spins in a fixed finite region (contained in $[1,L]$), there is equal probability for each configuration, and this remains
true for any fixed finite region as $L\to\infty$. As the probability distribution in the infinite system is determined by its marginals
for all finite sets of spins, the distribution is simply the product of uniform distributions on each spin.
This determines $\kappa^\dagger$, because the bonds $J$ can be recovered by the rule above, leaving a single bit (or Ising spin)
of information for each $J$. Thus the (AW or NS) metastate as $L\to\infty$ is nontrivial, with equal probability for a ground state and its 
global spin flip for given $J$, so it belongs to the case of finitely-many atoms of equal metastate weight (the atoms being the two 
ground states). The model has no frustration whatsoever, the ground states transform correctly under local transformations,
and in this model there is no difficulty in defining $\Pi_0$ on ground states. 
We can prove that this $\kappa^\dagger$ is extremal among $\Pi_0$-invariant $\kappa^\dagger$s (the metastate $\kappa_J$ is 
indecomposable), as follows. We need consider only the $\sigma$-algebra of spin configurations, which is the product $\sigma$-algebra 
over $i\in\mathbb{N}$ of the $\sigma$-algebra for each $s_i$; the latter is just the power set of $\{\pm 1\}$, and as we have seen
the probability distribution $\kappa^\dagger$ is the product of independent random spins. Due to the lack of frustration, a change 
in a finite number of bonds corresponds one-to-one with a change in a finite number of spins. So the $\sigma$-algebra 
${\cal I}_{1\Theta_0}$  (or ${\cal I}_{1\Pi_0}$) is the tail $\sigma$-algebra of the spins. By Kolmogorov's zero-one law, 
$\kappa^\dagger$ is trivial on the tail $\sigma$-algebra, so there are no measurable $\Theta_0$-invariant subsets
with probability not equal to $0$ or $1$, and $\kappa^\dagger$ is extremal. 

A generalization of the preceding ``toy'' model is obtained if we replace the chain of length $L$ by a tree ${\cal T}_L$ of $L$ 
sites, in which one vertex (i.e.\ site) is the root, at which the spin is fixed at random. We consider a sequence $({\cal T}_L)_L$ of such 
trees in which, for any $L$, $L'$ with $L<L'$, ${\cal T}_L$ is a subtree of ${\cal T}_{L'}$ such that the root of ${\cal T}_L$ is the unique 
vertex connecting ${\cal T}_L$ to its complement in ${\cal T}_{L'}$, and the root of ${\cal T}_{L'}$ does not lie in ${\cal T}_L$. 
Then we can take the limit through the sequence as $L\to\infty$, the root moves off to infinity, and we obtain a similar metastate, 
An almost identical argument proves that it is indecomposable. 

The most general versions of these unfrustrated models are 
Mattis-type models, as follows: the sites $i$, $j$, \ldots, lie on a finite graph; when $i$ and $j$ are adjacent, the bonds are 
$J_{ij}=\xi_i\xi_j$, where $\xi_i=\pm 1$ is a independent uniformly-distributed random sign for each site; and at one site only, 
say $i=0$, we impose a random fixed-spin boundary condition $s_0=\pm 1$. This can even be done for a lattice 
with periodic boundary conditions, with $s_0$ on the boundary. Then in the limit as the graph becomes infinite, with a finite number 
of bonds that involve any given site, and site $i=0$ moving off to infinity, an indecomposable ground-state metastate is obtained.
Heuristically at least, we can extend this case to nonzero temperature, say for graphs that are subsets $\Lambda$ of $\bbZ^d$, 
which form a sequence that tends to the full $\bbZ^d$. For $d>1$ these models possess a low-temperature ordered phase, 
because the uniform Ising ferromagnet does. Then with the random single fixed spin condition, we obtain a similar indecomposable 
metastate as before, but now at $T>0$. 

We can further extend this to EA spin glasses, if we assume that the SD picture holds. In that case, there are supposed 
to be just two pure states at low temperature and zero magnetic field. With spin-flip symmetry, the metastate is then trivial, consisting 
of a single atom at a Gibbs state that is itself trivial, being the equal weight mixture of the two pure states. If we instead impose a 
single fixed-spin condition at a site that moves off to infinity in the infinite-size limit, we again obtain an indecomposable metastate 
(with equal weight for each of two atoms, each a pure state) at $T>0$. In the SD scenario, we would also expect the same if a random 
fixed-spin boundary condition is applied over the full boundary of the finite regions before taking the limit.

In the strongly-disordered model of NS \cite{ns94,jr}, for $d>6$ there is a countable infinity of infinite clusters (trees) forming a minimum 
spanning forest (MSF); the MSF is determined by the ordering of the magnitudes of the bonds, while the signs of the bonds within 
each tree are independent of each other and of the magnitudes and equally likely to be plus or minus (bonds between trees can be 
dropped). In this model, for a given MSF, at $T=0$ the sign of a spin depends on the signs of the bonds on the (unique) path to 
infinity along the tree on which the spin is located. In the version of this model with fixed-spin boundary conditions in each finite size,
each tree in the MSF in infinite size behaves like the infinite tree described above.  Consequently, for that strongly-disordered model, 
if we view $J$ as the set of signs of the bonds only, for a given MSF the metastate is indecomposable, and also atomless. 
If instead the MSF geometry is regarded as part of the randomness $J$, we do not know whether the metastate is decomposable
(indeed, it is unclear if the general Theorems extend to this case). 

Some other examples, all somewhat related to the RFFM, present different behavior. A first set of examples is the slab models of 
White and Fisher (WF) \cite{wf}. In these, the $d$-dimensional lattice is partitioned into $d'$-dimensional ``slabs''  
or layers $\mathbb{Z}^{d'}$, where $2<d'<d$, stacked in parallel. All bonds in $J$ are independent nearest-neighbor pair 
interactions. Those within a slab are ferromagnetic and constant, while those between neighboring slabs are weak and are each 
symmetrically distributed with mean zero. At low $T$, each slab orders ferromagnetically, but the relative signs of the 
magnetizations of adjacent slabs vary chaotically with system size, similar to the magnetization in the RFFM. If the system 
has size $\ell$ in $d-d'$ directions perpendicular to the slabs, while the sizes of the slabs are equal and go to infinity, then 
the metastate will be a uniform distribution on $2^{\ell^{d-d'}-1}$ Gibbs states, each of which is an equal admixture of two 
spin-flip related pure states (due to a global symmetry of this model; we assume periodic boundary conditions in the $d'$ directions
in the finite systems). We can also consider $\ell\to\infty$, either simultaneously 
with or subsequent to the first limit. In either case, the Gibbs states can be labeled by the ordered set of signs of magnetizations 
of the slabs, modulo the effect of global spin flip, and these are covariant under changes in any finite set of the random bonds. 
For decomposability, we should be careful to state which semigroup we consider. For $\ell$ finite or infinite, we can always consider
$\Pi'$, which we define (in this subsection only) to be $\Pi_0$ together with translations in $\mathbb{Z}^{d'}$ only, so 
$\Pi_0\subset \Pi'\subset \Pi$; the metastate will be $\Pi'$-covariant, thanks to the periodic boundary conditions in the 
$d'$ directions in finite size. Then for $\ell>1$ we have described a way to decompose the metastate, independent of $J$, so the 
metastate is decomposable into $\Pi'$-covariant trivial metastates (for $\ell=1$ it is trivial). For $\ell\to\infty$, we can consider $\Pi$ 
also, using periodic boundary conditions in all $d$ directions, so this metastate is $\Pi$-covariant. This metastate is {\em in}decomposable 
in terms of $\Pi$-covariant metastates; the ordered set of magnetizations of all the slabs (modulo global spin flip)  is not a 
$\mathbb{Z}^d$-invariant observable, so cannot be used to distinguish the Gibbs states. All these results are in agreement with 
Theorems 1 and $1'$ for $\Pi$, and with similar Theorems for $\Pi'$.

A similar description of the metastate goes through if the slabs are replaced by infinite fractal clusters that jointly 
span all the lattice sites, again with ferromagnetic bonds within each cluster and weak random bonds between distinct clusters. 
Under some conditions \cite{wf}, a similar description of the ordered states holds. For a given set (of cardinality 
larger than one) of clusters, again the metastate is decomposable (here for $\Theta_0$-covariance). (The condition
here on the bonds within a cluster is more restrictive than that in similar models considered by WF \cite{wf}, who imposed 
only that the bonds be unfrustrated within each cluster, which could give indecomposable behavior similar to the Mattis-type 
and strongly-disordered models above, though now with spin-flip invariant Gibbs states. The free--boundary-condition
version of the strongly-disordered model for a given MSF behaves in this way also.)

Finally, yet another set of models considered by WF involve classical XY, not Ising, spins with short-range constant ferromagnetic 
bonds and independently random, isotropically-distributed, easy-axis anisotropy of infinite strength at each site. In the point of view 
put forward by WF \cite{wf}, we find that these models have an indecomposable metastate, but trivial Gibbs states. As the analysis is 
somewhat lengthy, but will be referred to again later, it is placed in Appendix \ref{app:infanis}. 

Thus we have exhibited examples of both decomposable and indecomposable nontrivial metastates, and for either case
the metastate could be either a finite number of atoms or atomless. There is
an interesting contrast between the strongly-disordered model and the ferromagnetic-within-each-cluster version
of the WF fractal cluster models, even though in both cases the bonds within each cluster are unfrustrated.

\subsection{Properties of indecomposable metastates}
\label{subsec:prop}

Now we turn to properties of $\Gamma$s drawn from an indecomposable metastate $\kappa_J$ for given $J$ (and fixed $\nu$). 
The overarching principle in the discussion is that, under the conditions of Theorem 1, such Gibbs states are macroscopically 
indistinguishable, or ``look alike'', where by ``macroscopic'' we refer to properties that are invariant under local transformations 
and possibly translations (i.e.\ under $\Phi$). Several examples and applications will be presented. From Subsection \ref{subsec:overl} on, 
all results use translation covariance of the indecomposable metastate; that property is needed in the underlying proofs of properties of 
the Gibbs states. The fact that $w_\Gamma$ is jointly measurable in $(J,\Gamma)$ (see App.\ \ref{app:furth}) will be used 
in some applications here, and also in later sections, without comment.

\subsubsection{Ultrametricity and overlap equivalence}
\label{subsec:ultra}

We begin here with some comparatively ``soft'' results about indecomposable metastates,
by considering properties of a Gibbs state that can be defined in terms of overlap distributions. Here we will consider
ultrametricity of a pseudometric, and equivalence of different overlaps or pseudometrics. The results are not deep, but further illustrate 
the use of Theorem 1 or 1$'$.

Ultra(-pseudo-)metricity of a pseudometric, say $d_{[X]}$, is the statement that for a given Gibbs state $\Gamma$ with pure-state 
decomposition $w_\Gamma$, for $w_\Gamma\times w_\Gamma\times w_\Gamma$-almost any three pure states $\Psi_1$, $\Psi_2$, 
$\Psi_3$, the pairwise $[X]$-distances $d_{[X]}(\Psi_i,\Psi_j)\equiv d_{[X]}(i,j)$ ($i$, $j=1$, $2$, $3$) obey a strengthened form of the 
triangle inequality,
\be
d_{[X]}(1,2)\leq\max (d_{[X]}(1,3),d_{[X]}(2,3))
\label{eq:ultramet}
\ee 
(and cylic permutations). (It implies that any triangle is either isosceles or equilateral.) Clearly this is equivalent to the same with $d_{[X]}$ 
replaced by $d_{[X]}^2$, and if
the self-overlaps $q_{[X]}(\Psi,\Psi)$ are the same for $w_\Gamma$-almost every $\Psi$, then this is equivalent to 
$q_{[X]}(1,2)\geq\min (q_{[X]}(1,3),q_{[X]}(2,3))$. Under the conditions in Proposition 1 or NRS23, equality of the two self-overlaps 
for a pair of pure states within a Gibbs state drawn from $\kappa^\dagger$ does hold for $w_\Gamma\times w_\Gamma$-almost every pair, 
$\kappa^\dagger$-almost surely. In any case, for an indecomposable $\kappa^\dagger$, ultrapseudometricity of $d_{[X]}$ 
can be formulated in terms of the probability distribution for the pairwise and self- overlaps of three pure states and then, 
under only the assumptions of this subsection, it follows by similar arguments that either it holds for $\kappa^\dagger$-almost 
every $(J,\Gamma)$, or else for $\kappa^\dagger$-almost every $(J,\Gamma)$ it fails to hold (note that
if $w_\Gamma$ is a single atom, ultrapseudometricity holds trivially). 

We can obtain similar ($\kappa^\dagger$-almost surely) ``always-or-never'' behavior for another property sometimes discussed 
for short-range spin glasses, called {\it overlap equivalence} \cite{pr-t}. We will take this to be the assertion that for two 
overlaps $q_{[X]}$, $q_{[X']}$ defined on pairs of pure states, one is a monotonically-increasing function of the other. (As we will not 
make much use of this concept, we will not attempt to formulate it more precisely, for example to deal with whether monotonicity must be 
strict, or what is required if one or other overlap is not a continuous function of Parisi's $x$ variable.) We can similarly define a related 
equivalence for pseudometrics $d_{[X]}$, $d_{[X']}$. Then for either of these, for a given pair $[X]$, $[X']$, either it holds for 
$\kappa^\dagger$-almost every $(J,\Gamma)$, or else for $\kappa^\dagger$-almost every $(J,\Gamma)$ it fails to hold.
Overlap equivalence, which is closely associated with ultrametricity, may play a role in later considerations on the structure of Gibbs states 
drawn from a metastate

\subsubsection{Behavior of overlap distributions}
\label{subsubsec:overl}

From this point on, all further results in the main text involve use of translation invariance, so of $\Pi$, not $\Pi_0$,
and of ${\cal I}_1$, not ${\cal I}_{10}$; this assumption will be made without further comment. In addition, we will use
consequences of the NRS23 result and its extension Proposition 2, and for that we will need the other stronger assumptions about $\nu$ and 
the Hamiltonian used there. 

First, we obtain a simple consequence involving relatives of the probability distribution $P_{J\Gamma[X]}(q)$ and its metastate average 
$P_{J[X]}(q)$, for any $[X]$, which were introduced in Sec.\ \ref{subsec:overl}. We will obtain a result by using the stronger 
zero-one law, Proposition 2,
together with Theorem 1. Drawing $J$ from $\nu$, then $\Gamma$ from $\kappa_J$, then $\Psi$ from $w_\Gamma$, we
consider the $w_\Gamma$ probability that $\Psi'$ drawn from $w_\Gamma$ has $q_{[X]}$ overlap with $\Psi$
is a specified range $B$ (a Borel set of real numbers), that is
\be
w_\Gamma(\{\Psi':q_{[X]}(\Psi,\Psi')\in B\})\equiv  \bbE_{w_\Gamma}{\bf 1}_B(q_{[X]}(\Psi,\Psi')).
\ee
The weight $w_\Gamma$ changes non-covariantly under a local transformation, however, the effect is multiplicative
and cannot change zero to nonzero or {\it vice versa}; the transformed $w_\Gamma$ 
is absolutely continuous with respect to that before (if we compare the measures at corresponding pure states). 
[Recall that for two measures $\mu_1$, $\mu_2$, $\mu_1$ is absolutely continuous with respect to $\mu_2$, written $\mu_1\ll\mu_2$,
if for any measurable set $A$ such that $\mu_2(A)=0$, we have also $\mu_1(A)=0$ \cite{royden_book}.]
Hence the above probability either remains zero or remains nonzero. If we define
\be
A=\{(J,\Gamma,\Psi):\bbE_{w_\Gamma}{\bf 1}_B(q_{[X]}(\Psi,\Psi'))>0\},
\ee
then $A$ is invariant: $A\in{\cal I}_{2}\subseteq{\cal I}_{1}(\kappa^\dagger)$, where $\Upsilon=\Pi$ 
because we will require translation invariance
(of course, the same is true for $A^c$, the set on which the probability of $q_{[X]}(\Psi,\Psi')\in B$ is zero). We now assume the hypotheses
of Proposition 2 and of Theorem 1, in particular an indecomposable translation-covariant metastate $\kappa_J$.
For $\kappa^\dagger$-almost every $(J,\Gamma)$, by Proposition 2 $w_\Gamma(A_{J\Gamma})=0$ or $1$,
and hence is covariant under elements of $\Theta$. Then by Theorems 1 or 1$'$, $w_\Gamma(A_{J\Gamma})=0$ or $1$, the same 
for $\kappa^\dagger$-almost every $(J,\Gamma)$. We have established the following.
\newline
{\bf Corollary 1} (to Theorem 1 and Proposition 2): Consider an indecomposable translation-covariant $\kappa_J$ and assume
the hypotheses of Proposition 1 and Theorem 1. Then for any $[X]$, either $\bbE_{w_\Gamma}{\bf 1}_B(q_{[X]}(\Psi,\Psi'))>0$ for 
$\kappa^\dagger w_\Gamma$-almost every $(J,\Gamma,\Psi)$, or it is $0$ for $\kappa^\dagger w_\Gamma$-almost every $(J,\Gamma,\Psi)$. 
Clearly, the first alternative will occur if and only if $P_{J[X]}(B)>0$, while the second will occur if and only if $P_{J[X]}(B)=0$. 
\newline
(It is interesting that corresponding statements were obtained by Panchenko for the infinite-range SK model, by making use of 
the GG identities; see Lemma 2.7 in Ref.\ \cite{panchenko_book}.) Further, Corollary 1 implies the weaker statement that either 
$P_{J\Gamma[X]}(B)>0$ [that is, positive $w_\Gamma\times w_\Gamma$ probability for $q_{[X]}(\Psi_1,\Psi_2)\in B$], 
$\kappa^\dagger$-almost surely, in the first case, or $P_{J\Gamma[X]}(B)=0$, $\kappa^\dagger$-almost surely, in the second. 
These weaker results can also be obtained by using Theorem 1 alone (for these, $\Upsilon=\Pi$ or $\Pi_0$: translation invariance 
of $\kappa^\dagger$ is not required). 

Even the weaker versions of the statements have a number of consequences, all for an indecomposable metastate. 
One is that the support of $P_{J\Gamma[X]}(q)$ is almost surely independent of $(J,\Gamma)$, even though the probability
distribution itself may not be. In particular, the supremum and infimum of the support, which we may call $\sup q_{[X]}$
and $\inf q_{[X]}$, are constant, $\kappa^\dagger$-almost surely. More generally, as the support of $P_{J\Gamma[X]}(q)$ 
is a closed set (in fact, contained in $[-1,1]$), then its complement in $\mathbb{R}$ is an open set, and so a countable union 
of disjoint open intervals that is the same union for $\kappa^\dagger$-almost every $(J,\Gamma)$, and the same as for $P_{J[X]}(q)$. 
Thus if $w_\Gamma(\Psi)$ is purely atomic, and hence $P_{J\Gamma[X]}(q)$ also consists solely of atoms 
($\delta$-functions in $q_{[X]}$), then the positions of those atoms are dense in the support of $P_{J[X]}(q)$, $\kappa^\dagger$-almost 
surely.

Another consequence arises if $B$ contains a single real number, say $B=\{b\}$, for $b\in [-1,1]$. 
Positive $w_\Gamma\times w_\Gamma$ probability for that set means $P_{J\Gamma[X]}(q)$ has a $\delta$-function at $b$. 
Then a $\delta$-function at $b$ in $P_{J\Gamma[X]}(q)$ is either present for $\kappa^\dagger$-almost every $(J,\Gamma)$ (it is a ``fixed'' 
$\delta$-function), or its presence has $\kappa^\dagger$-probability zero. (As the $\kappa_J$-expectation $P_{J[X]}(q)$ 
is independent of $J$ $\nu$-almost surely, these statements in fact hold for $\kappa_J$ in place of $\kappa^\dagger$, and given $J$.) 
In the latter case, a $\delta$-function in $P_{J\Gamma[X]}(q)$ at $b$ for the 
given $(J,\Gamma)$ is not ruled out; the result means only that for different $\Gamma$, the $\delta$ function changes to another form or 
location. In that case we can consider $B=[a,b]$, an interval, instead. If it overlaps the support then we have seen that its probability 
is $\kappa^\dagger$-almost surely nonzero, and any $\delta$-functions in $P_{J\Gamma[X]}(q)$ for $q\in B$ must be ``roving'' 
$\delta$-functions, the positions of which depend on $\Gamma$ (though they do not change under local transformations).

Conversely, we can examine $P_{J[X]}(q)$ and obtain its decomposition into atoms 
($\delta$-functions) and the complementary atomless (or continuous) part. [Note, incidentally, that a continuous distribution need not be
entirely absolutely continuous with respect to BL measure on $\mathbb{R}$; it can be (Lebesgue-) decomposed into two 
continuous parts, one absolutely continuous, the other singular, with respect to BL measure \cite{royden_book}.] Recall that 
$P_{J[X]}(q)$ is $\nu$-almost surely independent of $J$. Then we have the following weaker, though more general, version of Corollary 1.
\newline
{\bf Corollary 1$'$} (to Theorem 1): Consider $\Upsilon=\Pi$ or $\Pi_0$. For an indecomposable $\kappa_J$ and any $[X]$, 
consider $P_{J\Gamma[X]}(q)$
and its $\kappa_J$-expectation $P_{J[X]}(q)$. There is an atom in $P_{J[X]}(q)$ if and only if there is one in $P_{J\Gamma[X]}(q)$ 
at the same location (a ``fixed'' atom) for $\kappa^\dagger$-almost every $(J,\Gamma)$. The support of $P_{J\Gamma[X]}(q)$ is, 
$\kappa^\dagger$-almost surely, the same as that of $P_{J[X]}(q)$.
\newline
{\bf Proof}: The first statement follows because there must be nonzero $\kappa^\dagger$-probability of a
$\delta$-function occurring in $P_{J\Gamma[X]}(q)$ at the same location, and hence by the preceding remarks
it occurs in $\kappa^\dagger$-almost every $P_{J\Gamma[X]}(q)$ (it is ``fixed''). The atomless part of $P_{J[X]}(q)$ arises 
from $P_{J\Gamma[X]}(q)$ in which any $\delta$-functions, other than fixed ones, occur at that particular location with 
$\kappa^\dagger$-probability zero 
(an absolutely continuous part of $P_{J\Gamma[X]}(q)$ is not ruled out). The final statement was already discussed. $\Box$

Thus the expected distribution, $P_{J[X]}(q)$ (or even $P_{[X]}(q)$), determines many properties of $P_{J\Gamma[X]}(q)$.
These results resemble well-known results in RSB theory. In the case of full RSB, there is generally a plateau
in $q_{[X]}(x)$ (in the usual RSB mean-field theory of infinite-range models, $X$ is a single site)
on an interval $[x_1,1]$, which corresponds to a $\delta$-function at $\sup q_{[X]}$, and $\sup q_{[X]}$ is usually interpreted
as the self-overlap of any pure state (we discuss this point in Subsection \ref{subsec:gibbs}). The result here implies that
the $\delta$-function is present in $P_{J\Gamma[X]}(q)$ for $\kappa^\dagger$-almost every $(J,\Gamma)$. In some
cases, such as when a magnetic field is present, there is also a plateau at $[0,x_0]$, which means a $\delta$-function
at $\inf q_{[X]}$ for $\kappa^\dagger$-almost every $(J,\Gamma)$ as well. RSB solutions with $k+1$ $\delta$-functions, 
$k=1$, $2$, \ldots, in $P_{J[X]}(q)$, and no continuous part, are known as $k$-step or $k$-RSB solutions, and are also consistent 
with the analysis here for an indecomposable metastate. (A $k$-RSB form and a $k\to\infty$ limit are the basic method for 
constructing full RSB solutions.) If a $k$-RSB solution holds, the positions of the $\delta$-functions in $P_{J\Gamma[X]}(q)$ 
are independent of $\Gamma$, though their weights fluctuate with $\Gamma$. Our analysis implies that, for an indecomposable metastate
in which $P_{J[X]}(q)$ consists of $k+1$ $\delta$-functions, the stated behavior of $P_{J\Gamma[X]}(q)$ follows from general principles.

In these comments we used only the weaker form of the statements above. However, in view of the stronger forms, similar comments
apply if we consider a given $(J,\Gamma,\Psi)$ and look at the distribution of overlaps of other $\Psi'$ with $\Psi$. For example, if there
is a $\delta$-function in $P_{J[X]}(q)$, there will be one in that distribution also. That is, there will be nonzero probability of $\Psi'$ with
exactly that value of $q_{[X]}$ overlap with $\Psi$, for $\kappa^\dagger w_\Gamma$-almost every $(J,\Gamma,\Psi)$.

As a final remark, we point out that we only discussed the distribution of pairwise overlaps, and its relative with one pure state given,
but this was done only for the sake of simplicity. We can give a completely parallel discussion for the joint distribution
of any set of overlaps involving $n\geq 2$ pure states, possibly with one of the pure states given. This shows that the combination
of Proposition 2 and Theorem 1 implies a strong homogeneity property of the geometry of pure states occurring in a given Gibbs state, 
in the sense of overlaps: almost every pure state is alike statistically, up to fluctuations of the probabilities of overlaps, 
when the overlaps are nonzero. 

\subsubsection{Character of a Gibbs state}
\label{subsec:char}

Consider the pure-state decomposition $w_\Gamma$ of a Gibbs state $\Gamma$ drawn from a metastate $\kappa_J$, 
and its cardinality, and also that of the atoms in that decomposition. The number of atoms in $w_\Gamma$ must be countable, 
either finite or infinite, while a nonzero atomless part is necessarily a continuum, so uncountable. Then the total cardinality of the set of 
pure states in the support of $w_\Gamma$ might be $1$, a finite number larger than $1$, countably infinite, or uncountable.  NS09 
\cite{ns09} proved a theorem about 
the number of atoms (a hypothesis of equality of self-overlaps used there follows from the zero-one law of NRS23); 
a slightly extended version of their result is that the pure-state decomposition of $\kappa_J$-almost every Gibbs state $\Gamma$ 
has one of the following alternative forms: it consists of either (i) a single atom ($\Gamma$ is trivial), (ii) a countable infinity of atoms, 
(iii) an atomless distribution, or (iv) both a countable infinity of atoms and an atomless part. That is, a finite number of atoms,  
with or without a continuous part, was ruled out, except for the case of a trivial Gibbs state. [The proof of their result already 
appears to require the use of Lemma 1 in Sec.\ \ref{models}.]

This result can be strengthened using the extended zero-one law, Proposition 2 (see Sec.\ \ref{zeroone}): for given $(J,\Gamma)$, 
under a local transformation the atoms remain atoms of the transformed Gibbs state, though their weights change. 
Hence, for each $(J,\Gamma)$, the set of all pure states that are atoms is a [$(J,\Gamma)$-dependent] covariant set, which is 
the $(J,\Gamma)$ section of a set in ${\cal I}_2$.  In case (iv), the total weight of the set of atoms would lie strictly between $0$ and $1$. 
Then by Proposition 2, possibility (iv) is ruled out. We will refer to (i)--(iii) as the three possible ``characters'' of the Gibbs states. 

Note that the statement so far does not say this is the same for all Gibbs states drawn from $\kappa_J$, however, we can now obtain 
a stronger result. 
\newline
{\bf Corollary 2} (to NS09, Proposition 2, and Theorem 1): Under the same hypotheses as Proposition 1, consider
an extremal $\Pi$-invariant $\kappa^\dagger$, $(J,\Gamma)$ drawn from $\kappa^\dagger$, 
and the pure-state decomposition of $\Gamma$. Then one of the three characters (i)--(iii) listed above for $(J,\Gamma)$ has
$\kappa^\dagger$-probability one.
\newline
{\bf Proof}: Again, the cardinalities of the set of pure states and the set of atoms of $w_\Gamma$ 
are invariant under local transformations, and so sets of $(J,\Gamma)$ on which they take particular values are covariant. 
Then for indecomposable $\kappa_J$, the cardinalities are the same for $\kappa_J$ almost every $\Gamma$. Using the discussion before
the statement of the Corollary, the result follows. $\Box$

As we pointed out, this was not clear for a metastate in the general case. However, if $\kappa_J$ is 
trivial for $\nu$-almost every $J$, then by a simple use of translation-ergodicity of $\nu$, again applied to
the cardinalities of the set of pure states and that of atoms in $w_\Gamma$, we find that those cardinalities must be the same for
almost every $J$. Using the result of NS09, we find that the Gibbs state must have one of the characters (i)--(iv) above, and the same 
for almost every $J$. However, in this case (ii) and (iv) can both be ruled out using further results of NS, discussed in Sec.\  
\ref{sec:moregibbs} below.

\subsubsection{Single-replica equivalence}

Now we return to the reformulation, Theorem $1'$, of Theorem 1.
At first sight, it may appear that there are not many examples of invariant observables. One definite example is the 
free energy density (see Ref.\ \cite{nrs23}); some others are obtained from the preceding subsections. Next we state
a widely-applicable principle, based on the extension Proposition 2 of the theorem of NRS23 (Proposition 1 above), in its single-replica 
equivalence form. Proposition 2 (ii) concerns an invariant observable property $O(J,\Gamma,\Psi)$ of the bonds, Gibbs states,
and pure states in the decomposition of $\Gamma$, and says that for $\kappa^\dagger$-almost every $(J,\Gamma)$, $O(J,\Gamma,\Psi)$
is the same for $w_\Gamma$-almost every $\Psi$. We note that the observable may not be defined, or may not be invariant, for all 
$(J,\Gamma,\Psi)$, but only for $\kappa^\dagger w_\Gamma$-almost every triple, in which case it must be defined and invariant for 
$\kappa_J w_\Gamma$-almost every $(\Gamma,\Psi)$, $\nu$-almost surely; this is frequently easier to establish. An 
indicator function $O$ can also be viewed as expressing, for given $J$, a relation between $\Gamma$ and $\Psi$, which is either true or false. 
A property of $O$ of $\Psi$ for given $(J,\Gamma)$ that is $w_\Gamma$-almost surely constant can be viewed as a property of $\Gamma$
for given $J$. Then together with Theorem $1'$, for an indecomposable metastate this now gives single-replica equivalence for $\mu_J 
w_\Gamma$-almost every $(\Gamma,\Psi)$, for $\nu$-almost every $J$. The formal statement is the following.
\newline
{\bf Corollary 3} (of Theorem $1'$ and Proposition 2): Under the same hypotheses as Proposition 1, consider an extremal $\Pi$-invariant 
$\kappa^\dagger$. Then an invariant observable property $O(J,\Gamma,\Psi)$ takes the same value for 
$\kappa^\dagger w_\Gamma$-almost every $(J,\Gamma,\Psi)$. 
\newline
Thus the pure states drawn from $w_\Gamma$ for a $\Gamma$ drawn from an indecomposable $\kappa_J$ are all 
macroscopically indistinguishable, $\nu\kappa_J w_\Gamma$-almost surely. 
Particular examples include the magnetization 
in a pure state, and the self-overlaps of pure states, for which in a short-range model the existence of the limit in the 
translation average, and the invariance properties, follow from the pointwise ergodic theorem for translations 
\cite{nrs23,breiman_book}. The result is in exact agreement with what is found in RSB for the self-overlaps in the SG phase: the plateau 
in $q(x)$ for $x$ approaching $1$ means that $P(q)$ has a $\delta$-function at $q(1)$. If the pure-state decomposition of each Gibbs state
is not atomless, $q(1)$ can be interpreted as $q_{\rm EA}$, 
the self-overlap of the pure states, and the $\delta$-function means that the values have no dispersion due to the fluctuations of 
$\Gamma$ in the nontrivial metastate.
Corollary 3 does not necessarily apply to, e.g., the self-overlap of a Gibbs state $\Gamma$; Theorem $1'$ might apply, 
but the self-overlap of a Gibbs state that is a nontrivial mixture is not necessarily invariant under local 
transformations, due to the presence of the weights $w_\Gamma$. 

We also obtain from Corollary 3, or even from a weaker version that rests only on Proposition 1, the following.
\newline
{\bf Corollary 3}$'$ (of Theorem 1$'$ and Proposition 1): Under the same hypotheses as in Corollary 3, $\mu^\dagger$ is trivial on 
${\cal I}_1(\mu^\dagger)$, the sets of $(J,\Psi)$ that are invariant modulo null sets. 
\newline
[Because we do not define the semigroup $\Pi$ acting on the pure states, the $\sigma$-algebra ${\cal I}_1(\mu^\dagger)$
must here be viewed as defined in terms of the group $\Theta$, as in eq.\ (\ref{eq:I10}), with $\kappa^\dagger$ 
replaced by $\mu^\dagger$, and $\Theta_0$ by $\Theta$.]

\section{Analysis of metastates:\\ stochastic stability} 
\label{sec:stst}
Further results, some but not all of which involve indecomposability of the metastate, merit a separate Section.
We discuss stochastic stability of any metastate, the related GG identities and their applications, and the proof of Proposition 2.
The $\Sigma$- and $g$-evolutions described here play a role at several later points in the paper. Note that translation covariance
of any metastate we use (i.e.\ $\Pi$-invariance of the corresponding $\kappa^\dagger$) is assumed throughout this Section and 
those to follow, but will not be mentioned again.

\subsection{Stochastic stability, GG-type identities, $1$RSB, and Poisson-Dirichlet distribution}
\label{subsec:stst}

Another type of invariant observable is the class of translation averages of thermal expectations of the terms in the Hamiltonian,
$-J_X\langle s_X\rangle_\Psi$ for all $X$, also known as parts of the internal energy (it will be convenient in the discussion to drop 
the overall minus sign shown here). For $\Psi$ a state drawn from $w_\Gamma$ ($w_\Gamma$-almost surely a pure state), it was pointed 
out in NRS23 that the translation average is invariant and, directly from the zero-one law (Proposition 1 above), is the same for 
$w_\Gamma$-almost all $\Psi$ in the decomposition of given $\Gamma$ (drawn from $\kappa_J$), so 
${\rm Av}\,J_X\langle s_X\rangle_{\Psi}=\int w_\Gamma(d\Psi){\rm Av}\,J_X\langle s_X\rangle_{\Psi}={\rm Av}\,J_X\langle 
s_X\rangle_\Gamma$, $\nu\kappa_J w_\Gamma$-almost surely. [Here the proof of the final equality, that for a short-range model and 
$\nu\kappa_J$-almost surely, the $w_\Gamma$-expectation can be exchanged with the $W\to\infty$ limit involved in the 
${\rm Av}$ operation uses a general Lebesgue dominated convergence theorem (see Ref.\ \cite{royden_book}, p.\ 270) and the 
pointwise ergodic theorem, but we skip the details here and in the similar results to follow in this paragraph.] For $\kappa_J$ 
indecomposable, it now follows 
by Corollary 3 (and a similar argument as that preceding this one) that the value is the same for $\kappa_J$-almost every $\Gamma$ 
as well, ${\rm Av}\,J_X\langle s_X\rangle_\Gamma={\bbE}_{\kappa_J}{\rm Av}\,J_X\langle s_X\rangle_\Gamma
={\rm Av}\,J_X{\bbE}_{\kappa_J}\langle s_X\rangle_\Gamma$, $\nu\kappa_J$ almost surely. In addition,  
${\rm Av}\,J_X s_X = {\rm Av}\,J_X\langle s_X\rangle_\Psi$ for $\nu\kappa_Jw_\Gamma\Psi$-almost every $(J,\Gamma,\Psi,s)$,
by an argument similar to that for $\widehat{q}_{[X]}$ in Sec.\ \ref{models}. Finally, the metastate expectation is a 
function only of $J$, and $\nu$ is ergodic (under translations), so we have on general grounds that
${\rm Av}\,J_X{\bbE}_{\kappa_J}\langle s_X\rangle_\Gamma={\bbE}\,{\rm Av}\,J_X{\bbE}_{\kappa_J}\langle s_X\rangle_\Gamma
={\rm Av}\,{\bbE}\,[J_X{\bbE}_{\kappa_J}\langle s_X\rangle_\Gamma]$. Each of these remarks about translation averages $\rm Av$
also holds if each $J_X$ is replaced by its $\nu$ expectation, $\bbE\, J_X$ (the final one becoming trivial in this case), or by any other constant. 
Hence we can also replace each $J_X$ by $J_X-\bbE\, J_X$ in each case. 

These four almost-sure statements can now be turned into $L^p$ forms (we hope no confusion will arise from the brief use of the conventional
symbol $p$ here, which has no relation to the $p$ in mixed $p$-spin models). First notice that
\bea
\lefteqn{J_X s_X-{\bbE}\left[J_X{\bbE}_{\kappa_J}\langle 
s_X\rangle_\Gamma\right]=}\quad\quad&&\non\\
&=&\left(J_X s_X-J_X\langle s_X\rangle_\Psi\right)\non\\
&&{}+\left(J_X\langle s_X\rangle_\Psi-J_X\langle s_X\rangle_\Gamma\right)\non\\
&&{}+\left(J_X\langle s_X\rangle_\Gamma -  J_X{\bbE}_{\kappa_J}\langle s_X\rangle_\Gamma\right)\non\\
&&{}+\left(J_X{\bbE}_{\kappa_J} \langle s_X\rangle_\Gamma-{\bbE}\left[J_X{\bbE}_{\kappa_J}\langle s_X\rangle_\Gamma\right]
\right).\quad
\label{ggmart}
\eea
[The four terms $J_X s_X$, $J_X\langle s_X\rangle_\Psi$, $J_X\langle s_X\rangle_\Gamma$, and 
${\bbE}_{\kappa_J} J_X\langle s_X\rangle_\Gamma$ form a martingale, and the equation is a martingale decomposition 
of the function of $(J,\Gamma,\Psi,s)$ on the left, under the distribution $\nu\kappa_J w_\Gamma\Psi$.] This guides 
the arrangement of the expressions. Then for each of the four differences above, adding and subtracting the ${\rm Av}$ of either term 
in the difference (those $\rm Av$s are equal, as we explained) and using the above and the $L^p$ ergodic theorem for translations,
where $1\leq p<\infty$ 
(which can be obtained as a consequence of the almost-sure ergodic theorem \cite{breiman_book,einsward_book,walters_book}), we obtain the 
following, 
in which (ii) was already in NRS23 for $p=1$, and (iii) is a corollary of Corollary 3 of Theorem 1 [so only (iii) depends on the metastate being 
indecomposable]:
\newline
{\bf Theorem 2}: Under the same hypotheses as in Proposition 1, the following are true for given $[X]$ when $J_X$ is in 
$L^p(\nu_X)$ ($1\leq p<\infty$) for all $X\in[X]$, assuming in (iii) that the metastate $\kappa_J$ is indecomposable: 
\newline
(i)
\bea
\lefteqn{\lim_{W\to\infty}{\bbE}_{\kappa^\dagger}
\bbE_{w_\Gamma}\left\langle\left|\frac{1}{W^d}\sum_{{\bf x}'
\in \Lambda_W}\big(J_{\tau_{{\bf x}'}X} s_{\tau_{{\bf x}'}X}\right.\right.}\quad\;\;\;&&\non\\
&&\qquad\qquad\qquad\;\;\left.\left.\vphantom{\frac{1}{W^d}\sum_{{\bf x}'\in \Lambda_W}}
{}- J_{\tau_{{\bf x}'}X} \langle s_{\tau_{{\bf x}'}X}\rangle_\Psi\big)\right|^p\right\rangle_\Psi=0;\qquad
\eea
(ii)
\bea
\lefteqn{\lim_{W\to\infty}{\bbE}_{\kappa^\dagger}
\bbE_{w_\Gamma}\left|\frac{1}{W^d}\sum_{{\bf x}'
\in \Lambda_W}\big(J_{\tau_{{\bf x}'}X} \langle s_{\tau_{{\bf x}'}X}\rangle_\Psi\right.}\quad\quad\;\;\;&&\non\\
&&\qquad\qquad\qquad\;\;\left.\vphantom{\frac{1}{W^d}\sum_{{\bf x}'\in \Lambda_W}}
{}- J_{\tau_{{\bf x}'}X} \langle s_{\tau_{{\bf x}'}X}\rangle_\Gamma\big)\right|^p=0;\qquad
\eea
(iii)
\bea
\lefteqn{\lim_{W\to\infty}{\bbE}_{\kappa^\dagger}\left|\frac{1}{W^d}\sum_{{\bf x}'
\in \Lambda_W}\big(J_{\tau_{{\bf x}'}X} \langle s_{\tau_{{\bf x}'}X}\rangle_\Gamma\right.}\quad\;\;\;&&\non\\
&&\qquad\qquad\qquad\;\;\left.\vphantom{\frac{1}{W^d}\sum_{{\bf x}'\in \Lambda_W}}
{}-J_{\tau_{{\bf x}'}X}{\bbE}_{\kappa_J}  \langle s_{\tau_{{\bf x}'}X}\rangle_\Gamma\big)\right|^p=0;\qquad
\eea
(iv)
\bea
\lefteqn{\lim_{W\to\infty}{\bbE}\,\left|\frac{1}{W^d}\sum_{{\bf x}'
\in \Lambda_W}\big( J_{\tau_{{\bf x}'}X} {\bbE}_{\kappa_J}\langle s_{\tau_{{\bf x}'}X}\rangle_\Gamma
\right.}\;\;&&\non\\
&&\qquad\qquad\qquad\;\;\left.\vphantom{\frac{1}{W^d}\sum_{{\bf x}'\in \Lambda_W}}
{}- {\bbE}\left[J_{\tau_{{\bf x}'}X} {\bbE}_{\kappa_J}\langle s_{\tau_{{\bf x}'}X}\rangle_\Gamma\right]\big)\right|^p=0.\qquad
\eea
Each statement also holds, under the same conditions, if every term $J _{\tau_{{\bf x}'}X}$ is replaced by 
$J_{\tau_{{\bf x}'}X}-\bbE\, J_{\tau_{{\bf x}'}X}$.
\newline
(The additional condition on $J_X$ is unnecessary for $p=1$, as $L^p(\nu)\subseteq L^1(\nu)$, and we assumed $J_X\in L^1(\nu)$ for all 
$X$.)

The four parts of Theorem 2 describe respectively thermal fluctuations within a pure state, thermal fluctuations from one pure state 
to another in a given Gibbs state, metastate fluctuations from one Gibbs state to another, and fluctuations due to disorder,
all in an $L^p$ average sense.
Taking $p$th roots and using the triangle inequality for the $L^p$ norm, we also obtain 
under the same hypotheses, including an indecomposable metastate, (v)
\bea
\lefteqn{\lim_{W\to\infty}{\bbE}_{\kappa^\dagger}\left\langle\left|\frac{1}{W^d}\sum_{{\bf x}'
\in \Lambda_W}\big(J_{\tau_{{\bf x}'}X} s_{\tau_{{\bf x}'}X}\right.\right.}\;\;\;&&\\
&&\qquad\qquad\qquad\;\;\left.\left.\vphantom{\frac{1}{W^d}\sum_{{\bf x}'\in \Lambda_W}}
{}- {\bbE}\left[J_{\tau_{{\bf x}'}X}{\bbE}_{\kappa_J} \langle s_{\tau_{{\bf x}'}X}\rangle_\Gamma\right]\big)
\right|^p\right\rangle_\Gamma=0,\qquad\non
\eea
and again also with $J_X-\bbE\, J_X$ in place of $J_X$. We point out that the proofs \cite{breiman_book,walters_book} of the 
$L^p$ ergodic theorem still hold if only partial (i.e.\ conditional) expectation is used, and then hold for almost every value of the variable 
conditioned on. Thus we also have versions of (i) for $\Psi$ expectation, for $w_\Gamma$-almost every $\Psi$, $\kappa_J$-almost every 
$\Gamma$ for given $J$, and $\nu$-almost every $J$, and also 
for $w_\Gamma\Psi$ expectation, for $\kappa_J$-almost every $\Gamma$ for given $J$, and $\nu$-almost every $J$, and for
$\kappa_J w_\Gamma\Psi$ expectation for $\nu$-almost every $J$, and also of (ii) for $w_\Gamma$ expectation for $\kappa_J$-almost 
every $\Gamma$ and $\nu$-almost every $J$, and so on. We will use such versions below.

Taking $p=1$, (i)--(iv) and their sum (v) can each be used to obtain sets of identities similar to those in Refs.\ \cite{ac,gg}, where 
they were obtained for infinite range models. Starting from (i) plus (ii), or (v), the AC, or stochastic stability, 
identities \cite{ac} and the GG identities \cite{gg} are recovered, respectively (we do similar derivations below), 
if we view the ${\bbE}_{\kappa^\dagger}$ expectation as disorder average, and replace expectation under $w_\Gamma\Psi$ 
with expectation under $\Gamma$ itself. The GG identities result from a two-term martingale, involving the vanishing of thermal 
and disorder fluctuations, similar to our three-term martingale above, while a third set arises as the differences of the corresponding 
AC and GG identities \cite{cg_book}. 

We now outline the derivation of some sets of identities for the short-range systems. Each of (ii)--(iv) yields a set of identities, 
while (i) gives no additional information; we will not explore all of the identities here. We begin with the stochastic-stability (AC) identities, 
extended to include the metastate. For this case, only (i), (ii) are used [(i) could be omitted, but including it simplifies some intermediate 
expressions], and indecomposability of the metastate is not required; the NRS23 single-replica equivalence result, along with more basic 
observations, is sufficient, as mentioned already there \cite{nrs23}. These identities will be useful in the following Subsection 
and Section. To obtain as much information as possible about both the metastate and the Gibbs states, we will first study functions of 
correlations of spins in Gibbs states (later we will consider also other functions involving the states), which may be drawn independently from 
the metastate $\kappa_J$, and also of bonds. We use one of the models in Sec.\ \ref{models} in which $J_X=J_X^{(1)}+J_X^{(2)}$, 
where $\{J_X^{(1)},J_X^{(2)}:X\in{\cal X}\}$ are all independent, and all $J_X^{(2)}$ are centered Gaussians with nonzero variance 
for all $X$; then we can express $\nu=\nu^{(1)}\times\nu^{(2)}$, by independence, and so $\bbE=\bbE_{(1)}\bbE_{(2)}$,
and we sometimes write $\bbE_{(2)}$ for $\bbE_{\nu^{(2)}}$. [It may well be possible to dispense with the $J^{(2)}$ pieces, and prove 
similar identities using only n.i.p.\ distributions $\nu_X$ for each $X$, provided ${\rm Var}\,J_X<\infty$ (and possibly further conditions 
on $\nu_X$), without using integration by parts, but using instead methods similar to those leading to the central limit theorem; for this, 
the versions of the above with $J_X-\bbE\,J_X$ should be used. While this will not be done explicitly in this paper, its outline may become 
clearer, at least in special cases, later in this Section.] To obtain functions of correlation functions, here we will use in fact monomials of 
correlation functions, meaning a product of correlations, each of which is an expectation of $s_{X'}$ 
for some $X'\in{\cal X}$; $X'$ in distinct expectations can be unrelated, and the Gibbs states in which the correlations are evaluated may 
in some cases be copies of the same Gibbs state (drawn from $\kappa_J$), while others are drawn independently. Thus we draw a finite 
number of Gibbs states $\Gamma_r$, labeled $r=1$, $2$, \ldots independently from $\kappa_J$ for given $J$, and use a monomial 
function $f$ that is 
a product over $r$ of $n_r$ of $s_{X'}$s (each $X'\in{\cal X}$) of spins distributed in $\Gamma_r$, times a product of $J^{(2)}_{X''}$
for a finite set of $X''$, times any integrable function $f^{(1)}(J^{(1)})$ of $J^{(1)}$. Let $l_r=1$, \ldots, $n_r$ for each $r$ such that 
$n_r>0$. Let $n=\sum_r n_r$, and suppose that $n_r>0$ up to some $r$, and $n_r=0$ thereafter (so $n>0$). 

To simplify notation, for given $(n_r)_r$ we form the product over $r$ of $n_r$ copies of $\Gamma_r$ for each $r$ on $n$ copies 
of spin space $S$, with the components labeled 
\be
s^{(r,l_r)}=\left(s_i^{(r,l_r)}\right)_{i\in\bbZ^d},
\ee 
and write the product distribution for brevity as $\times \Gamma$. As each $\Gamma_r$ is $\kappa_J$-almost surely a Gibbs state,
we can also use $\Gamma_r=\bbE_{w_{\Gamma_r}}\Psi$ for each $r$, and write $\Psi_{r,l_r}$ for the integration variable
in the $l_r$th $w_{\Gamma_r}$; then we abbreviate the product of $\Psi_{rl_r}$ as $\times \Psi$, and similarly $\times w$ for 
$w_{\Gamma_1}\times w_{\Gamma_1}\times\cdots\times w_{\Gamma_{(2)}}\times\cdots$, with $n_r$ factors of each 
$w_{\Gamma_r}$. Then $\langle\cdots\rangle_{\times\Gamma}=\bbE_{\times w}\langle\cdots\rangle_{\times\Psi}$. Also we write 
$\kappa_J^\infty$ for $\kappa_J\times \kappa_J \times \cdots$, the product here indexed by $r=1$, $2$, \ldots. Finally, we consider 
functions $f$ that have the form 
\be
f=\left(\prod_{r,l_r}s_{X_{r,l_r}}^{(r,l_r)}\right)f^{(1)}(J^{(1)},J^{(2)})\prod_{\tilde{r}}J^{(2)}_{X_{\tilde{r}}}
\ee 
for a finite indexed set $(X_{r,l_r})_{r,l_r}$ and a finite indexed set $(X_{\tilde{r}})_{\tilde{r}}$, $X_{\tilde{r}}\in{\cal X}$. 
$f$ will be assumed to be a square-integrable function of $J^{(2)}$, so it is in $L^{(2)}(\nu^{(2)})$, for $\nu^{(1)}$ almost every $J^{(1)}$.
In particular, this is true if $f^{(1)}$ is bounded and continuous.

Now consider, for any $X\in{\cal X}$,
\bea
\lefteqn{\left|\bbE_{\nu^{(2)}\kappa_J^\infty}\left[ J^{(2)}_X\langle f s_X^{(1,1)}\rangle_{\times \Gamma}-J^{(2)}_X\langle 
f\rangle_{\times\Gamma}\langle s_X^{(1,1)}\rangle_{\times\Gamma}\right]\right|}\qquad&&\non\\
&\leq&\left(\bbE_{(2)} {f^{(1)}}^2\prod_{\tilde{r}}|J^{(2)}_{X_{\tilde{r}}}|^2\right)^{1/2}\non\\
&&{} \times \left(\bbE_{\nu^{(2)}\kappa_J}\left\langle\left|J^{(2)}_X 
s_X-J^{(2)}_X\langle s_X\rangle_\Gamma\right|^2\right\rangle_\Gamma\right)^{1/2},\qquad
\label{csapp}
\eea
by the Cauchy-Schwarz inequality, and on the right-hand side we reverted to the simpler notation as there is only 
a single $\Gamma$ involved. Similarly, we obtain an inequality with a translation average ${\rm Av}$ over $X$ inside 
the absolute value signs 
on the left and in the second factor on the right. Then the first factor on the right-hand side is, by assumption, finite for 
$\nu^{(1)}$ -almost every $J^{(1)}$, so the right-hand side is zero by statements (i), (ii) in Theorem 2 for $p=2$, 
$\nu^{(1)}$ -almost surely. 
Alternatively, if the set of $\tilde{r}$ is empty and $f^{(1)}=1$, we can use $p=1$ 
and the Holder inequality in place of $p=2$ and Cauchy-Schwarz, and drop the first factor on the right (since $|f|$ is bounded above by $1$). 

To obtain generalized stochastic-stability identities, we return to the left-hand side of the preceding inequality, without the translation average.
We can evaluate it by the well-known technique
of Gaussian integration by parts on $J^{(2)}$. That is, if $g$ is a differentiable function of $J_X^{(2)}$ and possibly also a function 
of other variables held fixed, and $J_X^{(2)}$ is a centered Gaussian, then
\be
\bbE_{(2)} J_X^{(2)}g(J_X^{(2)})={\rm Var}\,J_X^{(2)} \bbE_{(2)}\frac{\partial g(J_X^{(2)})}{\partial J_X^{(2)}}
\ee
Further, the derivative of each Gibbs state is
\be
\frac{\partial \Gamma(ds)}{\partial J_X^{(2)}}=\beta (s_X-\langle s_X\rangle_\Gamma)\Gamma(ds),
\ee
that is, a factor $s_X-\langle s_X\rangle_\Gamma$ inside each $\Gamma$ expectation.
Using this, after some algebra we obtain, using local covariance of the metastate \cite{aw}, and similarly to Refs.\ \cite{ac,gg,cg_book} 
[note there appears to be a mistake in Ref.\ \cite{cg_book} in eqs.\ (5.34), (5.39)]
\bea
\lefteqn{(\beta{\rm Var}\,J_X^{(2)})^{-1}\bbE_{\nu^{(2)}\kappa_J^\infty}\left[ J_X^{(2)}\langle f s_X^{(1,1)}\rangle_{\times \Gamma}
\right.}&&\non\\
&&\qquad\qquad\qquad\qquad{}\left.-J_X^{(2)}\langle 
f\rangle_{\times\Gamma}\langle s_X^{(1,1)}\rangle_{\times\Gamma}\right]\label{ibp}\\
&=&\bbE_{\nu^{(2)}\kappa_J^\infty}\!\!\!\left[{\sum}'\vphantom{\sum_r}\left(\langle f s_X^{(r,l_r)} s_X^{(1,1)}
\rangle_{\times \Gamma}
-\langle f s_X^{(r,l_r)}\rangle_{\times \Gamma}\langle s_X^{(1,1)}\rangle_{\times \Gamma}\right)\right.\non\\
&&\!\!{}\left.-\sum_r(n_r+\delta_{r,1})\left(\langle f s_X^{(1,1)}\rangle_{\times \Gamma}-\langle f\rangle_{\times \Gamma}\langle 
s_X^{(1,1)}
\rangle_{\times \Gamma}\right)\langle s_X^{(r,1)}\rangle_{\times \Gamma}\right],\non
\eea
where $\sum'$ is sum over all $(r,l_r)$ except $(1,1)$. On the right-hand side we omitted terms arising from the $J_{X_{\tilde{r}}}^{(2)}$s
in $f$, because they drop out after the translation average over $X$, performed below (such factors in $f$ are usually omitted in similar 
derivations in the literature). We want the same to be true for the terms arising from the possible dependence of $f^{(1)}$ on $J^{(2)}$,
which also do not appear in the literature. This will be valid if $f^{(1)}$ is a monomial in a finite number of $J_{X'}^{(2)}$, similarly
to the case of the $J_{X_{\tilde{r}}}^{(2)}$s. Again, these span $L^2(\nu^{(2)})$. Hence initially we can use only the monomials, 
and in the end we will obtain a result for $L^2(\nu^{(2)})$; it will be convenient to state the final result just for its subspace, 
the space of all bounded, continuous functions. 

We comment here that further identities, which might be expected to be even more complicated, can be obtained if, for each copy of 
each Gibbs state ($\Gamma$, say), we also allow for the function to depend on a finite number of pure states (let $\Psi$ be one such 
copy) drawn from $w_\Gamma$, that is, expectations of $s_{X'}$ in $\Psi$. As the $\Psi$ also vary with $J_X^{(2)}$, integration by 
parts produces $s_X-\langle s_X\rangle_\Psi$ inside each $\Psi$ expectation. But this time, in the translation average over $X$, when the 
$\widehat{q}_{[X]}(s,s')$ overlap is taken out of the $\Psi$ (and another) expectation, each $s_X$ becomes $\langle s_X\rangle_\Psi$
(by the same remarks from Sec.\ \ref{models} as before), and so these terms cancel. Thus in fact these more general identities are no more 
complicated.

Next we introduce translation average $\rm Av$ over $X$ in both sides of eq.\ (\ref{ibp}), which gives rise on the right-hand side 
to factors $\widehat{q}_{[X]}$ (defined in Sec.\ \ref{models}), the overlap of 
configuration $s^{(r,l_r)}$ with $s^{(1,1)}$. We can use the simple lemma proved in Sec.\ \ref{models}, which says that 
$\widehat{q}_{[X]}(s^{(1)},s^{(2)})=q_{[X]}(\Psi_1,\Psi_2)$, $\Psi_1\times \Psi_2$-almost surely, for $\nu(\kappa_J\times\kappa_J)
(w_{\Gamma_1}\times w_{\Gamma_2})$-almost every $(J,(\Gamma_1,\Gamma_2),(\Psi_1,\Psi_2))$ (and similarly when 
$\Gamma_1=\Gamma_2$). It follows from this that, even if $f$ has a more general dependence on $s^{(r,l_r)}$ than specified here, 
the factors $\widehat{q}_{[X]}$ appearing on the right-hand side (after translation average) can be replaced by 
$q_{[X]}(\Psi_{rl_r},\Psi_{11})$ if we use the pure-state decomposition of the Gibbs states. (It is at this stage that it 
becomes apparent that we could have replaced $s_X^{(1,1)}$ at the beginning with $\langle s_X^{(1,1)}\rangle_{\Psi_{11}}$ 
without affecting the result, so that use of (i) of Theorem 2 could have been omitted, at the cost of making some earlier 
expressions more complicated. It is also at this stage that the terms arising
from the $J_{X_{\tilde{r}}}^{(2)}$ in $f$ and from $J^{(2)}$ in $f^{(1)}$ go to zero, as promised.) Finally, the left-hand side 
goes to zero when translation 
averaged, as we showed just now. Hence we have obtained a set of identities, valid for $\nu^{(1)}$-almost every $J^{(1)}$:
\bea
\lefteqn{\bbE_{\nu^{(2)}(\kappa_Jw_\Gamma)^\infty}\!\!\!\left[{\sum}'
\vphantom{\sum_r}
\langle f\rangle_{\times \Psi} \,\left(\vphantom{\Gamma_1)}
q_{[X]}(\Psi_{rl_r} ,\Psi_{11})
-q_{[X]}(\Psi_{rl_r}, \Gamma_1)\right)\right.}&&\non\\
&&{}\left.-\sum_r(n_r+\delta_{r,1})\langle f\rangle_{\times \Psi}\,\left(q_{[X]}(\Gamma_r, \Psi_{11})
-q_{[X]}(\Gamma_r ,\Gamma_1)\right)\right]\non\\
&&\qquad\qquad{}=0.
\eea
The factors $\prod_{\tilde{r}}J_{X_{\tilde{r}}}$ appear only in $f$, not in the other factors under the $E_{\nu^{(2)}}$ expectation.
As the identities hold for all choices of those factors, which form a dense set of functions of $J^{(2)}$ (an orthonormal basis, after Gram-
Schmidt orthogonalization) in the (separable) Hilbert space $L^2(\nu^{(2)})$, 
this implies that the similar expression with $\bbE_{(2)}$ removed and $f$ replaced by the function $\widetilde{f}$ in which the 
$J^{(2)}_{X_{\tilde{r}}}$ are removed is orthogonal to all basis vectors in $L^2(\nu^{(2)})$. Hence that expression has $L^2(\nu^{(2)})$ 
norm zero, and so is zero for $\nu^{(2)}$-almost every $J^{(2)}$ as well as for $\nu^{(1)}$-almost every $J^{(1)}$. 
To summarize explicitly, we have proved, in the same class of models, for all $[X]$ and for functions $f$ (i.e.\ what was just called 
$\widetilde{f}$, but we now drop the tilde) that we can take to be bounded 
continuous functions of the spins in any finite number $n$ of $\Gamma$s (drawn from $\kappa_J$ independently) and of the bonds 
$J$, the extended stochastic stability identities [as a Corollary to Theorem 2 (i), (ii)]  
\bea
\lefteqn{\bbE_{(\kappa_Jw_\Gamma)^\infty}\!\!\!\left[{\sum}'
\vphantom{\sum_r}
\langle f\rangle_{\times \Psi}\left(\vphantom{\Gamma_1)} \,
q_{[X]}(\Psi_{rl_r} ,\Psi_{11})
-q_{[X]}(\Psi_{rl_r}, \Gamma_1)\right)\right.}&&\non\\
&&{}\left.-\sum_r(n_r+\delta_{r,1})\langle f\rangle_{\times \Psi}\,\left(q_{[X]}(\Gamma_r, \Psi_{11})
-q_{[X]}(\Gamma_r ,\Gamma_1)\right)\right]\non\\
&&\qquad\qquad{}=0,
\eea
for $\nu$-almost every $J$. Identities similar to these will play a significant role in later developments.
Results in the literature \cite{ac,cg_book} are for the $\nu$ expectation, not $\nu$-almost surely, and correspond to the special case 
$n_r=0$ for $r>1$, so $n=n_1$, that is, a single $\kappa_J$ average and with $f$ independent of $J$.
[In some instances of the latter \cite{ac}, the choice $(1,1)$ is replaced by $(1,l)$, with a uniform average over $l$. Some versions 
include an average over an additional parameter that may represent a range of temperatures; that is {\em not} required here.]
The result is unchanged in form if further translation averages are applied to $X_{rl_r}$ in some fashion, so that, for example, $f$ 
becomes a product of overlaps $\widehat{q}_{[X']}$ of pairs of spin configurations $s^{(r,l_r)}$ drawn from $\Psi_{rl_r}$ for various choices 
of pairs of $(r,l_r)$ (or of more general overlaps \cite{nrs23}), and of $X'$. 

Next we turn to the similar derivation of some extended GG identities, in which indecomposability of the metastate is assumed. 
The notation is the same as before, and the $\Gamma$ dependence of $f$ can be
a product of correlation functions or of overlaps of spins drawn from $\Psi_{rl_r}$. We use inequality (\ref{csapp}), but now with 
$\bbE_{(2)}\bbE_{\kappa_J}$ applied to the second term, as in (iv) after Theorem 2. A nearly identical derivation (similar to 
Refs.\ \cite{gg,cg_book}) leads this time, for an indecomposable $\Theta$-covariant metastate, to
\bea
\lefteqn{\bbE_{(\kappa_Jw_\Gamma)^\infty}\!\!\!\left[{\sum}'
\vphantom{\sum_r}
\langle f\rangle_{\times \Psi}\vphantom{\Gamma_1)} \,
q_{[X]}(\Psi_{rl_r} ,\Psi_{11})
\right.}&&\non\\
&&{}\left.-\sum_r n_r\langle f\rangle_{\times \Psi} q_{[X]}(\Psi_{11}, \Gamma_r)+\langle f\rangle_{\times \Gamma}
\bbE_{\kappa_J}q_{[X]}(\Gamma',\Gamma')\right]\non\\
&&\qquad{}=0.
\eea
Here, in the final term, we used self averaging of $\bbE_{\kappa_J} q_{[X]}(\Gamma',\Gamma')=\bbE_{\nu^{(2)}\kappa_J} 
q_{[X]}(\Gamma',\Gamma')$ ($\Gamma'$ is the integration variable for the expectation under ${\kappa_J}$), which holds for the 
metastate expectation of any translation invariant function of $\Gamma$, by ergodicity of $\nu$. In the case $n=n_1$ (a single $\Gamma$), 
these identities simplify to (note we can drop $r$ from the notation)
\bea
\lefteqn{\bbE_{\kappa_Jw_\Gamma^\infty}\!\!\left[\sum_{l=2}^n
\vphantom{\sum_r}
\langle f\rangle_{\times \Psi}\vphantom{\Gamma_1)} \,
q_{[X]}(\Psi_{l} ,\Psi_{1})
+\langle f\rangle_{\times \Gamma}
\bbE_{\kappa_J}q_{[X]}(\Gamma',\Gamma')
\right]}\qquad\qquad\qquad&&\non\\
&=&n\bbE_{\kappa_Jw_\Gamma^\infty}\langle f\rangle_{\times \Psi} q_{[X]}(\Psi_{1}, \Gamma)\qquad\qquad\qquad
\eea
This corresponds term by term with the basic original cases of GG identities \cite{gg}, though here the expectation over $J$ has 
been removed, while that under $\kappa_J$ remains. In the literature, the identities are extended
further to obtain ones in which each $[X]$ overlap $q_{[X]}(\Psi_1,\Psi_2)$ of pure states (when all Gibbs states are fully decomposed 
into pure states) is replaced by $q_{[X]}(\Psi_1,\Psi_2)^k$ for $k=2$, $3$, \ldots; these do not seem to be readily available here. 
(A derivation of GG identities in short-range systems in Ref.\ \cite{cms} is not suitable for our purposes because it is not done 
in the setting of metastates.) On the other hand, we have a set of identities involving each overlap $q_{[X]}$. We note that when
Gibbs states $\Gamma$ are $\kappa_J$-almost surely trivial, all AC and GG identities become vacuous.

We pause to record a particular identity \cite{gg} that results from the GG identities, for later use. Using $n=n_1$ and $f$ independent of $J$, 
and combining the cases $n=2$, $3$, one obtains \cite{gg} for any $X$ and $\nu$-almost every $J$ (we abbreviate $\Psi_l$ as $l$)
\bea
\lefteqn{\bbE_{\kappa_J w_\Gamma^\infty}q_{[X]}(1,2)q_{[X]}(3,4) =}&&\non\\ 
&=&\frac{1}{3}\bbE_{\kappa_J w_\Gamma^\infty}q_{[X]}(1,2)^2
+\frac{2}{3}\left(\bbE_{\kappa_J w_\Gamma^\infty}q_{[X]}(1,2)\right)^2.\qquad
\eea
This identity can be rearranged and expressed in the following illuminating form,
\be
{\rm Var}_{\kappa_J}\bbE_{w_\Gamma^\infty} q_{[X]}(1,2)=\frac{1}{2}\bbE_{\kappa_J}{\rm Var}_{w_\Gamma^\infty}q_{[X]}(1,2),
\label{ggvar}
\ee
which relates the variances due to $\kappa_J$ (metastate) and to $w_\Gamma$ (thermal) fluctuations. Either form is
satisfied in RSB, for example \cite{gg}.

The GG identities in the full form (i.e.\ for all powers $k$) have played a significant role in work on the infinite-range models; see 
Panchenko \cite{panchenko_book}. 
With the more restricted identities we have available here, we will only prove a result under stronger conditions on the overlap distribution. We 
recall that in so-called $1$-step RSB, or $1$RSB,  the pairwise overlap, say $q_{[X]}$, takes only two values, the larger of which is usually
the self-overlap; by our earlier results, for an indecomposable metastate, the two values must be constants for $\kappa_J$-almost every 
$\Gamma$, and for $\nu$-almost every $J$. In such a setting we can obtain the following result for the short-range case.
\newline
{\bf Corollary 4} (to Theorem 2): Assume (i) a short-range mixed $p$-spin model with translation-invariant $\nu$ 
such that ${\rm Var}\,J_X^{(2)}>0$ for all $X\in{\cal X}$, with an indecomposable metastate $\kappa_J$.
Suppose also (ii) that there is an $[X]$ such that, $\kappa^\dagger$-almost surely, the overlap $q_{[X]}(\alpha,\alpha')$ 
takes only two values (with nonzero probability of each) $q_0$, $q_1$, with $q_0< q_1$, where $q_1$ is the self-overlap of any 
pure state, and $q_0$ is the overlap of any two distinct pure states. Then the overlap $q_{[X]}(\Psi_\alpha,\Psi_{\alpha'})$ 
is $w_\Gamma\times w_\Gamma$-almost surely non-negative for $\kappa_J$ almost every $\Gamma$, $\nu$-almost every $J$, and if also 
(iii) all bonds $J_{X'}$ for $X'\in[X]$ include a centered Gaussian piece $J_{X'}^{(2)}$, 
then the distribution (induced from $\kappa_J$) of $(w_\Gamma(\alpha))_\alpha$, sorted into non-increasing order, 
is the Poisson-Dirichlet (PD) distribution (or ``process'') with parameter $x_1=1-\bbE_{\kappa^\dagger}\sum_\alpha w_\Gamma(\alpha)^2$,
for $\nu$-almost every $J$ [$x_1\in (0,1)$]. 
\newline
{\bf Proof}: The first statement (a special case of Talagrand's positivity principle \cite{panchenko_book}) follows directly from 
hypothesis (ii) because, after drawing $n$ pure states independently from $w_\Gamma$, the mixture of them with weights $1/n$ each must 
have non-negative $q_{[X]}$ self-overlap for all $n$. This self-overlap is
\be
q_{[X]}=[q_1+(n-1)q_0]/n
\ee
which is less than zero for sufficiently large $n$ if $q_0<0$, giving the result. Next, observe that $\kappa^\dagger$-almost every 
Gibbs state has a countably-infinite pure-state decomposition (labeled by $\alpha$) with weights $w_\Gamma(\alpha)$ of pure states 
$\Psi_\alpha$. This follows from hypothesis (ii) and indecomposability because (using Corollary 2 above), if $w_\Gamma$ were atomless, 
the $w_\Gamma\times w_\Gamma$ probability of drawing the same pure state twice, and so of obtaining overlap $q_{[X]}$ equal 
to the self-overlap, would be zero, while a trivial $\Gamma$ (atomic $w_\Gamma$) would give only a single value of the overlap.
Finally, assume hypotheses (i)--(iii), and consider the $q_{[X]}$ overlap of any pair of the pure states. Subtract $q_0$ from 
$q_{[X]}$, so that it is either zero or, when the pure states are the same, a positive constant $q_1-q_0$, 
which can be set to $1$ by rescaling. The resulting function is then the indicator function for the event that the two pure states are the 
same. Consider the $n=n_1$ GG identities obtained above for this function, with $\widetilde{f}$ a function of the same indicator function
(or overlaps) of the pure states. Then the analysis in Ref.\ \cite{panchenko_book}, pp.\ 46, 47 (with 
configurations $\sigma$ there replaced with pure states $\Psi_\alpha$), which utilizes Talagrand's identities, can be applied directly, and 
yields the final statement.   $\Box$

The result can be extended to the more general case in which the hypotheses (i), (ii) are the same except that in hypothesis (ii) the 
statements after $q_0<q_1$ are dropped. For definiteness, suppose that $q_1$ is {\em not} the $q_{[X]}$ self-overlap of a pure state, 
but is less than the latter; hence the $w_\Gamma\times w_\Gamma$ probability of drawing the same pure state twice must be zero, so 
$w_\Gamma$ is atomless. Then we can partition the pure states into distinct clusters (labeled $\alpha$), where those in the same cluster 
$w_\Gamma\times w_\Gamma$-almost surely have overlap $q_1$, while those in distinct clusters have overlap $q_0$. 
We lump together the pure states in each cluster $\alpha$ into a mixture, a single ``cluster state'', with total or lumped weights 
$W_\Gamma(\alpha)=\int_\alpha w_\Gamma(d\Psi)$ (the integral is over $\Psi$ in cluster $\alpha$); two distinct cluster states 
are at squared distance $d_{[X]}^2=2(q_1-q_0)$. (The situation discussed here is discussed further in Sec.\ \ref{subsec:gibbs} below.) 
In the remainder of the statement and proof, the weight $w_\Gamma(\alpha)$ must be replaced by $W_\Gamma(\alpha)$, and the 
pure states $\alpha$ by clusters $\alpha$. The overlap non-negativity follows by a similar argument as before, using the lumped (cluster) states. 
The GG identities hold for the overlaps of the cluster states and, using the analog of Corollary 2, which will be proved in Sec.\ 
\ref{subsec:gibbs}, there is a countable infinity of cluster states; the PD distribution of the lumped weights follows. 

Thus for these particular cases, corresponding to $1$-RSB, we find that the only allowed behavior is exactly as in the infinite-range models
($x=x_1$ is the position of the jump in $q_{[X]}(x)$ in terms of Parisi's $x$ variable). 
We note that the consistency in the $1$-RSB case of the PD distribution with certain identities (not unrelated to those obtained here) was 
established in Ref.\ \cite{ans15}, but of course that does not show that the PD distribution is necessary.

For the more general cases (including others in which the pure-state decomposition is uncountable), the proof of 
Panchenko's ultrametricity theorem \cite{panchenko_book} involves use of the full set of GG identities 
(i.e.\ with all $k\geq 1$ in the preceding discussion), which we do not have at present. We return to these cases in Sections 
\ref{subsec:pdultr} and \ref{subsubsec:furthgibbs} below.

\subsection{$\Sigma$ invariance of the metastate}
\label{sec:siginv}

The $\Theta$-covariance of the metastate and the stochastic-stability-type identities can be used to obtain an evolution, governed 
by a differential operator, that we introduce here, and term $\Sigma$ evolution ($\Sigma$ in honor of stochastic stability); it acts 
on functions of Gibbs states $\Gamma$, or alternatively 
on metastates, which not need be indecomposable (we use the same name for both of these, which are dual to one another). 
We show that the metastate is invariant under $\Sigma$ evolution. We also introduce corresponding stochastic, or $g$, evolutions 
on the states themselves, which relate to $\Sigma$ evolution in the same way Brownian motion relates to a diffusion or 
Fokker-Planck equation. We obtain results on partitions of pure states into a finite number of clusters, and prove Proposition 2. 

\subsubsection{$\Sigma$ invariance and evolution}
\label{subsubsec:siginv}

$\Sigma$ evolution corresponds to results in Refs.\ \cite{ac,ra,aa,ad,cg_book} that originated in infinite-range spin glass models in finite size, 
and were used to obtain results on PD distributions, Ruelle cascades, and ultrametricity in those systems (and in ``competing particle 
systems''). We aim to obtain the fullest possible 
statements in the metastate setting for infinite-size short-range systems, of which the most basic is the invariance (or stationarity, or stability) 
of the metastate itself under the evolution; by ``fullest possible'' we mean that, in terms of states, the results are not restricted to, 
for example, the derived distributions only on overlaps or other translation-invariant observables (compare Refs.\ \cite{ans15,ad}). 
The main results here concern metastates, but not exclusively indecomposable ones. 

First, we fix a choice of $X$, and consider a local transformation of a state $\Gamma$ in eq.\ (\ref{eq:covgam}). 
If $\Delta J_X=\lambda J_X^{(2)}$ (all other $\Delta J_{X'}^{(2)}=0$), 
where $\lambda$ is a real-valued parameter, then the first derivative with respect to $\lambda$ gives
\be
\frac{\partial}{\partial \lambda}\theta_{\Delta J_X^{(2)}}\Gamma(ds)=\beta J_X \left(s_X-\langle s_X\rangle_{\theta_{\Delta J_X^{(2)}}
\Gamma}\right)\theta_{\Delta J_X^{(2)}}\Gamma(ds),
\ee
where the multiplicative change (apart from $\beta$) is the combination appearing in the starting point for the stochastic-stability identities.
If now we generalize to 
\be
\Delta H=-\frac{\lambda}{W^d}\sum_{\bx:\bx\in\Lambda_W}J_{\theta_\bx X}^{(2)} s_{\theta_\bx X},
\ee 
then $\partial/\partial \lambda$ produces 
precisely the factor in the derivation of the identities, except for the $W\to\infty$ limit. We now specialize the derivation of the identities 
to the case of a single $\Gamma$, and use the notation $f$ for a function of $(J,\Gamma)$ ($\Gamma$, not spins), generally leaving the 
dependence on $J$ implicit. We should begin with a monomial in $J^{(2)}$ as in the previous argument, then later extend to more 
general functions as before. Application of 
$\partial/\partial \lambda$ to $\bbE_{\kappa_J} f(\theta_{\Delta J}\Gamma)$, and then setting $\lambda=0$, gives
\be
\beta\bbE_{\kappa_J}\left\langle\frac{\partial f}{\partial \Gamma(s)}\frac{1}{W^d}\sum_{\bx:\bx\in\Lambda_W}
J_{\theta_\bx X} ^{(2)}
(s_{\theta_\bx X}-\langle s_{\theta_\bx X}\rangle_{\Gamma})\right\rangle_{\Gamma}
\ee
[Note that in a partial (which might also be termed functional) derivative $\partial f/\partial \Gamma (s)$ of a function $f$ of a state 
$\Gamma$, we can ignore the fact that the $\Gamma(s)$ are constrained because $\Gamma$ is a probability distribution, because the factors 
$s_X-\langle s_X\rangle_{\Gamma}$ are 
present, due to the fact that the transformation $\theta_{\Delta J^{(2)}}$ preserves the normalization.] The expectation under $\nu^{(2)}$
of the $W\to\infty$ limit of this times a monomial in $J^{(2)}$ is now 
zero, because it is a modification of the translation average of the left hand side of eq.\ (\ref{ibp}), specialized to $n=n_1$, and with the 
fixed choice $l=1$ replaced by a sum over $l$, because all copies of $\Gamma$ are affected by $\theta_{\Delta J^{(2)}}$. 
Then by Gaussian integration by parts, and arguing as before to remove the $\nu^{(2)}$ expectation, we obtain the identities
\be
\bbE_{\kappa_J}\Sigma f =0,
\label{siginvbasic}
\ee
where the operator $\Sigma$ (or $\Sigma_{[X]}$) is defined by
\bea
\lefteqn{\Sigma f=}&&\non\\ 
&&{\rm Av}\left\langle \frac{\partial^2 f}{\partial \Gamma(s^{(1)})\partial \Gamma(s^{(2)})}\left(s_X^{(1)}
-\langle s_X\rangle_\Gamma\right)
\left(s_X^{(2)}-\langle s_X\rangle_\Gamma\right)\right\rangle_{\Gamma\times\Gamma}\non\\
&&{}-2{\rm Av}\left\langle \frac{\partial f}{\partial \Gamma(s)}\left(s_X-\langle s_X\rangle_\Gamma\right)
\right\rangle_\Gamma\langle s_X\rangle_\Gamma.
\label{sigdef}
\eea
Note that $\Sigma$ is a {\em linear} operator on functions $f$ of $(J,\Gamma)$.
The identity holds for $\nu$-almost every $J$, and makes sense when $f$ has bounded continuous 
second derivatives with respect to $\Gamma$, as well as obeying the conditions on its $J$ dependence stated earlier. 
As $[X]$ runs through all translation-equivalence classes, these identities form a more
general extended form of AC or stochastic-stability identities. Special cases of these identities 
were obtained in Ref.\ \cite{ans15}, by taking $\partial^2/\partial J_X^2$ of $\bbE_{\kappa_J}f$, followed by an average over 
translations of $X$, however the approach in that reference has to assume that $f$ is a translation-invariant function in order to obtain 
$\Sigma$ invariance, which is not required in the present approach, though a Gaussian $J^{(2)}$ piece is. 

We expect, but will not prove, that the functions with bounded continuous second derivatives with respect to $\Gamma$ 
are dense in the space of bounded continuous functions of $(J,\Gamma)$ (at least for the case of interest, in which 
$\Gamma$ is almost surely a Gibbs state); the expectations of such functions uniquely define a probability distribution \cite{billingsley_book2}. 
Then we can define the 
(topological vector space) transpose (or adjoint, or dual) of any $\Sigma$ (also denoted $\Sigma$, but acting to the left, just as a matrix 
acts to the left on row vectors, which is the same as the transpose of the matrix acting on the transpose of the row vectors 
to column vectors), 
which maps a signed measure $\kappa$ to a signed measure $\kappa\Sigma$, via the formula
\be
\int (\kappa\Sigma) (d\Gamma)f(\Gamma)=\int \kappa (d\Gamma)(\Sigma f)(\Gamma)
\ee
for all $f$ in the dense set.
(We can write down a formula for the transpose action on $\kappa$, but will not do so.)
Then assuming the identity above is valid on a dense set of $f$, we conclude the following.
\newline
{\bf Theorem 3}: Consider a short-range mixed $p$-spin model with translation-invariant $\nu$ such that ${\rm Var}\,J_X^{(2)}>0$ 
for all $X\in{\cal X}$, and a metastate $\kappa_J$. Then $\kappa_J$ is invariant under $\Sigma$, for all $[X]$: $\kappa_J\Sigma_{[X]} =0$ 
$\nu$-almost surely.
\newline
This is one of the main conclusions of this paper. 

\subsubsection{$g$ evolution}
\label{sec:gevol}

It is clear that $\Sigma$ is a kind of Laplacian; it might then be possible to define a diffusion-like process on the space of states for which 
$\Sigma$ is the generator. In the following, we will discuss the corresponding diffusion kernel for nonzero time $t$; on functions of states 
it would be $f\mapsto e^{t\Sigma}f$, $t\geq 0$, and then there would be a semigroup of kernels $\Sigma_t=e^{t\Sigma}$,
so $\Sigma$ (or more accurately, its closure) would be the generator of the semigroup. (As $\Sigma$ is an unbounded operator, 
the exponential should not be defined by the 
exponential series, but nonetheless can still be constructed under some conditions. This is discussed in e.g.\ Ref.\ \cite{reed_simon_book2}, 
Sec.\ X.8.) Extending the notion of the transpose, invariance of the metastate would be 
$\kappa_J\Sigma_t=\kappa_J$ ($\Sigma_t$ maps probability distributions to probability distributions, when it exists). 

We emphasize that, in general, we would not expect states to be invariant under $\Sigma$ evolution, however, so far we have not given 
a meaning to $\Sigma$-evolution of Gibbs states, as opposed to the effect $f\mapsto\Sigma_t f$ on a bounded continous function $f$ 
on Gibbs states. In order to do so, a reformulation of $\Sigma$-evolution will be useful. The results will be proved with less rigor than those 
earlier in this Section; there are some small gaps in the arguments, but we believe these are only technical and can be filled relatively easily.
Theoretical physicists will probably not be bothered by them. To at least some extent, the formalism here is only a convenient way to re-express 
the underlying invariance results, with which we begin, and which are rigorous.

First, we rewrite $\Sigma$ in terms of overlaps and pure states, using the discussion in Sec.\ \ref{subsec:stst}:
\bea
\lefteqn{\Sigma_{[X]} f=}\quad&&\non\\ 
&&\bbE_{w_\Gamma\times w_\Gamma}\left\langle \frac{\partial^2 f}{\partial \Gamma(s^{(1)})\partial \Gamma(s^{(2)})}
\right\rangle_{\Psi_1\times\Psi_2}\non\\
&&{}\qquad\times \left(q_{[X]}(\Psi_1,\Psi_2)-2q_{[X]}(\Psi_1,\Gamma)+q_{[X]}(\Gamma,\Gamma)\right)\non\\
&&{}-2\bbE_{w_\Gamma}\left\langle \frac{\partial f}{\partial \Gamma(s)}\right\rangle_\Psi\left(q_{[X]}(\Psi,\Gamma)
-q_{[X]}(\Gamma,\Gamma)\right).\qquad
\label{sigdefpure}
\eea
(We used the evident symmetry between $\Psi_1$ and $\Psi_2$ to simplify the expression.)

It is not difficult to produce processes involving Gaussian random variables, like Brownian motion, at given $J^{(1)}$, $J^{(2)}$, 
the distributions of which produce the same $\Sigma$ evolution in either form. For the first form, introduce for the given $X$ and finite $W$
\be
g_{XW}(s)=\frac{1}{W^{d/2}}\sum_{\bx:\bx\in\Lambda_W}g_{\theta_\bx X} s_{\theta_\bx X},
\ee
where $(g_{X'})_{X'\in{\cal X}}$ are centered Gaussians, independent of each other and of $J^{(1)}$, $J^{(2)}$, with variance $1$,
and define $\Delta J_W$ by $\Delta J_{X'}=-\lambda\beta^{-1} g_{X'}/W^{d/2}$ for $X'\in\{\theta_\bx X:\bx\in\Lambda_W\}$, all others zero. 
Then we write $\bbE_g$ for expectation over $g_{X'}$ (by abuse of notation). Taking the derivative of 
$\bbE_g\bbE_{\kappa_J} f(\theta_{\Delta J_W}\Gamma)$
with respect to $\lambda$, integrating by parts, applying translation average, dividing by $\lambda$, and then setting $\lambda=0$, 
produces an expression equal to $\bbE_{\kappa_J} \Sigma f$, because $g$ enters in the same way $J^{(2)}$ did. Thus, 
under a $J^{(2)}$ expectation, the effect of adding $-\lambda\beta^{-1}g_{XW}(s)$ to the Hamiltonian is equivalent to adding $\Delta H$ 
as in the preceding Subsection, to first order in $W^{-1}$ (because the variance of a sum of independent variables is the 
sum of the variances), up to an uninteresting $W$-independent rescaling of $\lambda$. This is close to the approach used originally 
\cite{ac,cg_book}; however, we still use our previous results to obtain the invariance equation $\bbE_{\kappa_J} \Sigma_{[X]} f=0$
for each $[X]$.

If we suppose that the $W\to\infty$ limit of $g_{XW}(s)$ makes sense, and denote it 
$g_{[X]}(s)$, then it is Gaussian with mean zero and covariance $\bbE_g g_{[X]}(s)g_{[X]}(s')=\widehat{q}_{[X]}(s,s')$ (so it 
only depends on the translation-equivalence class of $X$). The limit does indeed exist, in the sense that the random (and measurable) 
function $g_{XW}(s)$ of $s$ converges in distribution to $g_{[X]}(s)$ with covariance $\widehat{q}_{[X]}(s,s')$, where the 
latter exists for $\Gamma\times\Gamma$-almost every $(s,s')$, for $\kappa^\dagger$-almost every $(J,\Gamma)$, as discussed in 
Sec.\ \ref{models}. We usually wish only to discuss expectations over $g$, for which convergence in distribution is sufficient. In principle,
$\widehat{q}_{[X]}(s,s')$ depends on $(s,s')$, but we also know that it is $\Psi\times\Psi'$-almost surely constant
and equal to $q_{[X]}(\Psi,\Psi')$, for $w_\Gamma\times w_\Gamma$-almost every pair $(\Psi,\Psi')$ of pure states,
and also $\Psi\times\Psi$-almost surely constant for $w_\Gamma$-almost every $\Psi$, to obtain the self-overlaps. 
Thus we can also think of Gaussian random variables $g_{[X]}(\Psi)$ with mean zero and covariance 
\be
\bbE_g g_{[X]}(\Psi)g_{[X]}(\Psi')=q_{[X]}(\Psi,\Psi')
\ee 
for all $\Psi$, $\Psi'$ (the pure states exist for given $J$, and there is no need for $g_{[X]}(\Psi)$ to depend on a choice of $\Gamma$;
we assume that such a Gaussian process indexed by $\Psi$ actually exists).
Equivalently, we have
\be
\bbE_g \left|g_{[X]}(\Psi)-g_{[X]}(\Psi')\right|^2=d_{[X]}(\Psi,\Psi')^2
\label{eq:siginc}
\ee
the standard form for the so-called increments of a Gaussian process on a space with pseudometric $d_{[X]}$ \cite{vershynin_book}.
The $g_{[X]}(s)$ construction essentially reduces to this. Then $\bbE_g\bbE_{\kappa_J}f(\theta_{\lambda g_{[X]}}\Gamma)$ leads 
directly to $\bbE_{\kappa_J} \Sigma f$ with the second form (\ref{sigdefpure}) of $\Sigma$ (here we wrote $\theta_{\lambda g_{[X]}}$
in place of $\theta_{\Delta J_W}$). In connection with this, it is natural to rewrite eq.\ (\ref{sigdefpure}), viewing $f$ as a 
function of pure states $\Psi$ and of $w_\Gamma(\Psi)$, as 
\bea
\lefteqn{\Sigma_{[X]} f=}\quad&&\non\\ 
&&\bbE_{w_\Gamma\times w_\Gamma} \frac{\partial^2 f}{\partial w_\Gamma(\Psi_1)\partial w_\Gamma(\Psi_2)}\non\\
&&{}\qquad\times \left(q_{[X]}(\Psi_1,\Psi_2)-2q_{[X]}(\Psi_1,\Gamma)+q_{[X]}(\Gamma,\Gamma)\right)\non\\
&&{}-2\bbE_{w_\Gamma}\frac{\partial f}{\partial w_\Gamma(\Psi)}\left(q_{[X]}(\Psi,\Gamma)
-q_{[X]}(\Gamma,\Gamma)\right).\qquad
\label{sigdefpure'}
\eea
This will be the most useful form in what follows. Because pairwise overlaps are bilinear in the states, we can also write the 
combination of overlaps in the middle line more compactly as $q_{[X]}(\Psi_1-\Gamma,\Psi_2-\Gamma)$, and that in the last line as 
$q_{[X]}(\Psi-\Gamma,\Gamma)$; we will use such forms later.

We are not required to differentiate at $\lambda=0$, and we can instead tentatively identify the $\Sigma$ evolution for general $t$ 
as the mapping $f(\Gamma)\to(\Sigma_{t[X]} f)(\Gamma)=\bbE_g f(\theta_{\lambda g_{[X]}}\Gamma)$, with $t=\lambda^2/2$
(similar to a standard method for deriving the diffusion or Fokker-Planck equation from Brownian motion). 
From this it is clear that pure states are unaffected by $\Sigma_t$; the change in a pure state due to $g_{XW}$ cancels in the limit, 
again by the clustering 
property of pure states, as discussed in Sec.\ \ref{subsec:stst}, that is, they are invariant. The $\Sigma$ evolution changes only the weights 
$w_\Gamma$, if $\Gamma$ is a non-trivial mixed state. This makes contact with the ``competing particle systems'' set-up \cite{ra,aa}, 
where the particles correspond to a countable collection of pure states. For all $X$, we have the semigroup properties $\Sigma_{t+t',[X]}
=\Sigma_{t[X]}\Sigma_{t'[X]}=\Sigma_{t'[X]}\Sigma_{t[X]}$ and $\Sigma_{0[X]}={\rm id}$ (the identity). Notice also that $\Sigma_{t[X]}$, 
$\Sigma_{t[X']}$ for different $[X]$, $[X']$ commute, and that they commute with local transformations and translations.

More explicitly, $g$-evolution means that $\Gamma$ transforms to $\theta_{\lambda g_{[X]}}\Gamma$, where
\be
\theta_{\lambda g_{[X]}}\Gamma=\int w_\Gamma(d\Psi)^{\lambda g_{[X]}} \Psi
\ee
and $w_\Gamma^{\lambda g_{[X]}}$ is given by
\be
w_\Gamma(d\Psi)^{\lambda g_{[X]}}=\frac{e^{\lambda g_{[X]}(\Psi)-\frac{1}{2}\lambda^2q_{[X]}(\Psi,\Psi)}w_\Gamma(d\Psi)}
{\bbE_{w_\Gamma}e^{\lambda g_{[X]}(\Psi')-\frac{1}{2}\lambda^2q_{[X]}(\Psi',\Psi')}}.
\label{eq:gevolw}
\ee
The terms with $q_{[X]}(\Psi,\Psi)$ in the exponentials are required in general, in order that, if one formally expands in powers of 
$\lambda^2$, the $\Sigma_{t[X]}\equiv e^{t\Sigma_{[X]}}$ evolution constructed from $g_{[X]}(\Psi)$ reproduces that constructed 
from $g_{[X]}(s)$ at all orders (because the self-overlap of a pure state $\Psi$ is not equal to the self-overlap of a configuration $s$, 
which is $1$; they appeared also in Ref.\ \cite{argch,ad}), and the latter is what we obtain if we use $g_{XW}$ and take the limit as 
$W\to\infty$. In particular, they ensure that the correct action of $\Sigma_{[X]}$ is obtained at first order in $t$ (which is second order in 
$\lambda$). They cancel if $q_{[X]}(\Psi,\Psi)$ is the same for all $\Psi$ (at least for the given $\Gamma$), which is true in our case 
because of the result of NRS23 \cite{nrs23} or Proposition 1, but we will not make use of that here, because we will need the more general 
form shortly. The transformation of the measure $w_\Gamma$ makes sense if $e^{\lambda g(\Psi)}$ is a measurable function of $\Psi$. 

Then the action of the evolution operator $\Sigma_{t[X]}$ on either metastates or functions can be expressed in terms of 
a probability kernel $\Sigma_{t[X]}(\cdot\mid \cdot)$ mapping a state $\Gamma$ to a probability distribution on states, given by
\be
\Sigma_{t[X]} (\cdot\mid\Gamma) = \bbE_g \delta_{\theta_{\lambda g_{[X]}}\Gamma}(\cdot),
\ee
where $t=\lambda^2/2$. Then for measures on $\Gamma$, for any measurable set $A$ of states,
\be
\left(\kappa\Sigma_{t[X]}\right)(A)=\left(\kappa\cdot\Sigma_{t[X]}\right) (A),
\ee
while the action on functions is
\be
\left(\Sigma_{t[X]} f\right)(\Gamma)=\left(\Sigma_{t[X]}\cdot f\right)(\Gamma)
\ee
[the notations on the right-hand sides, involving a kernel, were defined in eqs.\ (\ref{eq:dotnot1}), (\ref{eq:dotnot2}) earlier].
We can also view the evolution in terms of Brownian fields $g_{t[X]}(\Psi)$, centered Gaussian functions of both $t$ and $\Psi$ 
for each $[X]$, with increments independent for distinct $t$, and 
\bea
\lefteqn{\bbE_g \left(g_{t+dt,[X]}(\Psi_1)-g_{t[X]}(\Psi_1\right)\left(g_{t+dt,[X]}(\Psi_2)-g_{t[X]}(\Psi_2\right)}
\qquad\qquad\qquad\qquad&&\non\\
\qquad\qquad&=&2q_{[X]}(\Psi_1,\Psi_2)dt;\qquad\qquad\qquad\qquad
\eea
the evolution of these fields produces the kernels $\Sigma_{t[X]} (\cdot\mid\cdot)$.

It may be useful to directly consider the effect of the perturbation by $g_{XW}$ on states and metastate. First, for bonds, the distribution
of $J_X+\lambda\beta^{-1}g_X/W^{d/2}$ tends to that of $J_X$ weakly as $W\to\infty$, so (because $\nu$ is determined by its 
finite-dimensional distributions), the distribution of the total bonds is unchanged in the limit, and any given $J_X$ is unchanged, almost surely
with respect to the distribution on $g$. Second, the specification $\gamma_{J\Lambda}$ 
for region $\Lambda$ is changed by a factor that involves only terms with at least one spin in $\Lambda$, and the total change in 
$H_\Lambda'$ goes to zero (in distribution) when $W\to\infty$. So the space ${\cal G}(\gamma)$ of Gibbs states is unchanged 
by the perturbation, and so is the set of pure states ${\rm ex}\,{\cal G}$, though that does not necessarily mean that an individual 
Gibbs state is invariant. Finally, for a pure state, we can consider the expectation of any $s_{X'}$, as the collection of these determines 
the state uniquely. To first order in $t$, the $\bbE_g$ expectation of the correlation is unchanged by the perturbation, due to the clustering 
properties of pure states. This confirms that, not only is ${\rm ex}\,{\cal G}(\gamma)$ 
invariant, but individual pure states are invariant under $\Sigma_t$ for all $t\geq0$, as inferred above.

\subsubsection{Application: proof of Proposition 2}
\label{subsubsec:prop2}

Now we will show how to use the result of this Section to extend some earlier results, and in particular prove Proposition 2. 
A general comment is that when we proved $\Sigma$ invariance of the metastate in the previous subsection, we had to use general functions
of a single $\Gamma$ drawn from $\kappa_J$, and the preceding identities were also that general. On the other hand, many desirable 
applications use only translation and locally covariant functions of $\Gamma$, or of $\Gamma$ and $\Psi$, and so on. For these, some earlier 
methods are available for the initial steps \cite{ans15}, and this has the advantage that it applies under the weaker condition that $\nu_X$ be 
n.i.p.\ for all $X$, without the need for a $J^{(2)}$ piece. Of course, these results can also be obtained using the method we have used up to 
now, with integration by parts on $J^{(2)}$. (Translation and local invariance will still be needed in later steps, as we will see.) On the other 
hand, as the methods adopted here are direct, the earlier results on stochastic stability and $\Sigma$ invariance are not even used, though
some of the notation (and $g$ evolution) will be, and the reasoning does not even trace back to Proposition 1. This also means that 
the possible difficulties encountered in the previous Sec.\ \ref{sec:gevol} do not arise here.
 
For Proposition 2, we wish to consider a set $A$ of $(J,\Gamma,\Psi)$ that is invariant, so in ${\cal I}_2$. For given $(J,\Gamma)$, this 
determines a set $A_{J\Gamma}$ that is covariant under local transformations. Then $(A_{J\Gamma})^c=(A^c)_{J\Gamma}$ is also 
covariant. Hence, for each $(J,\Gamma)$, these sets determine a covariant partition of the states $\Psi$ into two parts, labeled by 
$\alpha=1$, $2$, respectively (we use these labels as names of the sets, as well for indices for them). We can form a partial (or conditional) 
Gibbs state, or ``cluster state'', as the average of the pure states in that part (or ``cluster''),
\be
\Gamma_\alpha=\frac{1}{W_{\Gamma\alpha}}\int_\alpha w_\Gamma(d\Psi)\Psi,
\ee
with weights
\be
W_{\Gamma\alpha}=\int_\alpha w_{\Gamma}(d\Psi)
\ee
for each part,
so that 
\bea
\sum_{\alpha=1, 2}W_{\Gamma\alpha}&=&1,\\ 
\sum_{\alpha=1,2}W_{\Gamma\alpha}\Gamma_\alpha&=&\Gamma.
\eea 
If there is one $\alpha$ for which $W_{\Gamma\alpha}=0$, that $\Gamma_\alpha$ is indeterminate, but as usual this causes no difficulty.
If $W_{\Gamma1}\neq 0$ or $1$, then $\Gamma_1\neq \Gamma_2$.
Now we can begin.
\newline
{\bf Proof} of Proposition 2: We consider $\Sigma$ invariance of the distribution of $W_{\Gamma\alpha}$ ($W_{\Gamma1}$ alone 
would be sufficient because it determines $W_{\Gamma2}$ also, but we prefer the more symmetrical forms that follow, which will generalize). 
That is, for any continuous function $f$ of $W_{\Gamma1}$, $W_{\Gamma2}$ (for given $\Gamma$) with bounded continuous derivatives 
up to second order, for any $[X]$ we have the invariance equation
\be
\bbE_{\kappa_J}\Sigma^*_{[X]} f=0.
\label{invprop2}
\ee  
To obtain this under the hypotheses in Proposition 2, which are weaker than those used elsewhere in this Section, we use 
the approach of Ref.\ \cite{ans15}, mentioned earlier in this section, which did not involve integration by parts.
For completeness, we sketch the steps: Consider $\bbE_{\kappa_J} f$. By translation invariance of $f$ (which holds because $W_{\Gamma1}$, 
$W_{\Gamma2}$ are translation invariant) and ergodicity of $\nu$, this is $\nu$-almost surely independent of $J_X$ for every $X$.
Because of the n.i.p.\ property of $\nu_X$ for all $X$, we can take $\partial^2/\partial J_X^2$ of $\bbE_{\kappa_J}f$, followed 
by an average over translations of $X$. It is during this stage that 
the local covariance of $\kappa_J$ and of the sets $A_{J\Gamma}$ is used; the only change in $f$ with $J_X$ is through $w_\Gamma$ 
in the integral defining each $W_{\Gamma\alpha}$, and not through its domain of integration. In the result shown,
$\Sigma^*_{[X]}f$ takes the following form [which is just eq.\ (\ref{sigdefpure'}) specialized to these functions]:
\bea
\Sigma^*_{[X]} f&=&
\bbE_{W_\Gamma\times W_\Gamma} \frac{\partial^2 f}{\partial W_{\Gamma\alpha_1}\partial W_{\Gamma\alpha_2}}
q_{[X]}(\Gamma_{\alpha_1}-\Gamma,\Gamma_{\alpha_2}-\Gamma)\non\\
&&{}-2\bbE_{W_\Gamma}\frac{\partial f}{\partial W_{\Gamma\alpha}}q_{[X]}(\Gamma_\alpha-\Gamma,\Gamma)
.\qquad
\eea
Here $\bbE_{W_\Gamma}\cdots=\sum_{\alpha=1,2}W_{\Gamma\alpha}\cdots$, and similarly for 
$\bbE_{W_\Gamma\times W_\Gamma}$. This form assumes $W_{\Gamma1}\neq 0$ or $1$; otherwise $\Sigma_{[X]} f=0$ identically
(because such $W_{\Gamma1}$ is unchanged by any local change in $J_X$). Perhaps surprisingly, this has exactly the same form as eq.\ 
(\ref{sigdefpure'}), but with the two clusters $\Gamma_\alpha$, $\alpha=1$, $2$, in place of the pure states $\Psi$; this will allow 
the following argument to be made. The reason for the star on $\Sigma^*_{[X]}$ is that, unlike the case in eq.\ (\ref{sigdefpure'})
where the states are pure, the $\Gamma_\alpha$ do evolve under $\Sigma$ evolution, so the meaning of the operator is somewhat special,
and $\Sigma^*$ must not be confused with the general $\Sigma$.

First we separate the two distinct cases. If $W_{\Gamma\alpha}=1$ for $\alpha=0$ or $1$, then $\Gamma=\Gamma_\alpha$, 
and the $\kappa_J$ probability of this event is covariant, and so is that for $W_{\Gamma1}\in(0,1)$. Then the conditional distributions
for either event are also covariant, and can be treated as metastates in their own right. Thus we can assume until further notice that
$W_{\Gamma1}\in(0,1)$ $\kappa_J$-almost surely. Now $w_\Gamma$, which for example when it is purely atomic can be viewed 
as a distribution on a countable set of $\Psi$ together with their weights $w_\Gamma(\{\Psi\})$, reduces in the present case 
to the distribution with probabilities $W_{\Gamma1}$, $W_{\Gamma2}$ on the pairs $\Gamma_1$ $\Gamma_2$ (with no assumption that 
$w_\Gamma$ is atomic); this can be viewed as a distribution (on states) that happens to consist of two atoms. 
$\kappa_Jw_\Gamma$ reduces to a marginal probability distribution on $(\Gamma,\Gamma')$, where for given $\Gamma$ the conditional 
distribution can be described as $\Gamma'=\Gamma_\alpha$ with probability $W_{\Gamma\alpha}$, $\alpha=1$, $2$; $\kappa_J$ 
is recovered by integrating over $\Gamma'$. We will write this as $\kappa_Jw_\Gamma(\Gamma_1,\Gamma_2,W_{\Gamma1})$ on the triples 
$(\Gamma_1,\Gamma_2,W_{\Gamma1})$, where $W_{\Gamma\alpha}\in (0,1)$. From this we can obtain the marginal distribution 
$\kappa_Jw_\Gamma(W_{\Gamma1})$ 
[by integrating over $\Gamma$ (effectively over $\Gamma_1$, $\Gamma_2$) such that $W_{\Gamma1}$ takes a specified value], 
and the conditional distribution $\kappa_Jw_\Gamma(\Gamma_1,\Gamma_2|W_{\Gamma1})$, 
conditionally for given $W_{\Gamma1}$. We will now use these forms, with our usual notation 
$\bbE_{\kappa_Jw_\Gamma}\cdots$ for expectation under $\kappa_Jw_\Gamma(\Gamma_1,\Gamma_2,W_{\Gamma1})$, 
and $\bbE_{\kappa_Jw_\Gamma}(\cdots|W_{\Gamma1})$ for conditional expectation under 
$\kappa_Jw_\Gamma(\Gamma_1,\Gamma_2|W_{\Gamma1})$. 

In the invariance equation (\ref{invprop2}), we will now take combinations of overlaps, so that $q_{[X]}$ is replaced throughout 
by $q_{\rm tot}$
as defined in Sec.\ \ref{models}. We will also define $x=\ln (W_{\Gamma2}/W_{\Gamma1})$, so $x\in (-\infty,\infty)$;
we will write $f(x)$ in place of $f(W_{\Gamma1},W_{\Gamma2})$, and in fact in the present case we could have 
begun with the form $f(\ln [W_2/W_1])$ without loss of generality. Note that, while
\be
dW_{\Gamma1}=\frac{e^x}{(1+e^x)^2}dx,
\ee
in our set-theoretic notation for measures
\be
\kappa_Jw_\Gamma(dW_{\Gamma1})=\kappa_Jw_\Gamma(dx),
\ee
because the corresponding intervals, though described in different variables, are the same sets.
Then eq.\ (\ref{invprop2}) becomes
\be
\bbE_{\kappa_Jw_\Gamma}\,\bbE_{\kappa_Jw_\Gamma}(\Sigma^*_{\rm tot} f(x)|x)=0,
\ee 
by the usual rules for conditional probability. 

To derive the form of $\Sigma^*$ in the $x$ variable,
it is helpful to recall the $g$ evolution. We can introduce a reduced version of $g$-evolution, conditionally for given $\Gamma_1$, 
$\Gamma_2$. We use Gaussian fields $g_{\rm tot}(\alpha)$ with covariance matrix
\be
\bbE_g g_{\rm tot}(\alpha_1)g_{\rm tot}(\alpha_2)=q_{\rm tot}(\Gamma_{\alpha_1},\Gamma_{\alpha_2}).
\ee
Then $g$ evolution leaves $\Gamma_1$, $\Gamma_2$ invariant, by construction, and only affects $W_{\Gamma1}$. 
In terms of $x$, it simply changes additively by
\bea
x&\to& x+\lambda\left(g_{\rm tot}(2)-g_{\rm tot}(1)
\right)\non\\
&&{}+\frac{1}{2}\lambda^2[q_{\rm tot}(\Gamma_1,\Gamma_1)-q_{\rm tot}(\Gamma_2,\Gamma_2)].
\eea
At time $t=\lambda^2/2$ the variance of $\lambda\left(g_{\rm tot}(2)-g_{\rm tot}(1)\right)$ is
\be
\lambda^2\bbE_g  \left(g_{\rm tot}(1)-g_{\rm tot}(2)\right)^2=\lambda^2\,d_{\rm tot}(\Gamma_1,\Gamma_2)^2,
\ee
This $g$ evolution gives rise to a $\Sigma$ evolution that agrees with the given $\Sigma^*_{\rm tot}$ on functions $f(x)$, 
$\Sigma^*_{t\,\rm tot}f=e^{t\Sigma^*_{\rm tot}}f$, for given $\Gamma_1$, $\Gamma_2$. A standard derivation of the 
diffusion equation shows then that
\bea
\left.\frac{\partial}{\partial t}\Sigma^*_{t\,\rm tot}\,f\right|_{t=0}&=&\Sigma^*_{\rm tot}f\\
&=& d_{\rm tot}(\Gamma_1,\Gamma_2)^2\frac{\partial^2 f}{\partial  x^2}\\
&&{}+ [q_{\rm tot}(\Gamma_1,\Gamma_1)-q_{\rm tot}(\Gamma_2,\Gamma_2)]
\frac{\partial f}{\partial x}.\non
\eea
Only the conditional expectation of the last expression appears in the invariance equation. We can take that expectation and
arrive at
\bea
\Sigma^{**}_{\rm tot}f
&\equiv& \bbE_{\kappa_Jw_\Gamma}(\Sigma^*_{\rm tot}f(x)|x) \\
&=&D\frac{\partial ^2 f}{\partial  x^2}+V\frac{\partial f}{\partial x}
\eea
where
\bea
D&=& \bbE_{\kappa_Jw_\Gamma}\,d_{\rm tot}(\Gamma_1,\Gamma_2)^2, \\
V&=&\bbE_{\kappa_Jw_\Gamma}\,[q_{\rm tot}(\Gamma_1,\Gamma_1)-q_{\rm tot}(\Gamma_2,\Gamma_2)]
\eea
are the diffusion constant and drift velocity (these expectations are independent of $x$, so the conditioning has been 
dropped). We recall that $d_{\rm tot}$ is a metric, not a pseudometric, when applied to states drawn from translation-invariant distributions
(using Lemma 1), so because $\Gamma_1\neq \Gamma_2$
$\kappa_Jw_\Gamma$-almost surely, $d_{\rm tot}(\Gamma_1,\Gamma_2)^2>0$ $\kappa_Jw_\Gamma$-almost surely, and so $D>0$ 
$\kappa_Jw_\Gamma$-almost surely also. (Thus translation invariance of $A$ was used here.) The net drift velocity $V$ can have 
either sign or may vanish. 

The invariance equation (\ref{invprop2}) now reduces to 
\be
\int\kappa_Jw_\Gamma(dx)\Sigma^{**}_{\rm tot}f(x)=0,
\ee
for the same class of functions $f$.
Equivalently, $\kappa_Jw_\Gamma(x)$ is supposed to be annihilated by $\Sigma^{**}_{\rm tot}$ acting to the left, at least
for expectation of such a function. It is well known that, for diffusion on the real line 
(possibly with drift), there is no invariant probability distribution, and that will nearly conclude the proof. 

We will sketch some more self-contained argument for this last fact, formulated so as to avoid making additional assumptions 
(as may arise if we use integration by parts to obtain the action to the left). As in the case of $\Sigma^*$, 
it will be helpful to define a corresponding $\Sigma$ evolution, $\Sigma^{**}_{t\,\rm tot}=e^{t\Sigma^{**}_{\rm tot}}$. Explicitly, 
this is given by convolution with the function (we set $V=0$, but $V\neq 0$ is similar)
\be
P(x,x')=\frac{e^{-(x-x')^2/(4Dt)}}{\sqrt{4\pi Dt}}
\ee
for $t>0$. For $t>0$, $P$ obeys $\partial P/\partial t=D\partial^2 P/\partial x^2$, and tends to $\delta(x-x')$ as $t\to0$.
Then for $f$ a bounded continuous function of $x$, $\Sigma^{**}_{\rm tot}\Sigma^{**}_{t\,\rm tot}f=(d/dt)\Sigma^{**}_{t\,\rm tot}f$
for $t>0$. As the invariance equation holds for $f$ in the class of functions described, and $\Sigma^{**}_{t\,\rm tot}f$ is 
certainly belongs to that class, the invariance equation holds with $\Sigma^{**}_{t\,\rm tot}f$ in place of $f$. The (right-) derivative 
$(\partial/\partial t)\Sigma^{**}_{t\,\rm tot}f$ at $t=0$ exists if $f$ in that class (see 
Ref.\ \cite{reed_simon_book2}, p.\ 236). Then we integrate with respect to $t'$ from $0$ to $t$ to obtain
\be
\int\kappa_Jw_\Gamma(dx)\Sigma^{**}_{t\,\rm tot}f(x)=\int\kappa_Jw_\Gamma(dx)f(x)
\label{siginvnew}
\ee
for all $t\geq 0$ and any $f$ in the domain of $\Sigma^{**}_{\rm tot}$. Now choose $f$ to be
\be
f(x)=e^{-x^2/2},
\ee
so then
\be
\Sigma^{**}_{t\,\rm tot}f(x)=\frac{e^{-x^2/(2+4Dt)}}{\sqrt{1+2Dt}}
\ee
for all $t\geq 0$. Then the left-hand side of eq.\ (\ref{siginvnew}) is bounded:
\bea
\int\kappa_Jw_\Gamma(dx)\Sigma^{**}_{t\,\rm tot}f(x)&\leq&\sup_x \Sigma^{**}_{t\,\rm tot}f(x)\quad\quad\\
&=&\frac{1}{\sqrt{1+2Dt}},
\eea
which goes to zero as $t\to\infty$, whereas the right-hand side eq.\ (\ref{siginvnew}) is nonzero and independent of $t$.

This contradiction means that the metastate that we considered, for which $W_{\Gamma1}\in(0,1)$ $\kappa_Jw_\Gamma$-almost surely, 
cannot exist. Then the only remaining possibility is that $W_{\Gamma1}=0$ or $1$ $\kappa_Jw_\Gamma$-almost surely, 
which is the conclusion of Proposition 2. $\Box$

The conclusion holds similarly for any finite number $n>1$ of parts, in place of $n=2$ as here. This extended version 
of our Proposition 2 can also be easily obtained from Proposition 2 by expressing the choice of one of the $n$ parts as a sequence of 
binary choices. The NS09 result \cite{ns09}, that a finite number $n>1$ of {\em pure} states is not possible for Gibbs states 
drawn from a translation-invariant metastate, uses the corresponding invariance equation as here, but extended to $n$ parts, 
as a starting point. Arguin and Damron (AD) \cite{ad} pointed out how that result also follows from a stochastic stability 
approach involving random overlap structures (see their Corollary 4.6), somewhat like that used here more generally. 
Note, however, that the NS09 result is not a special case of the extended version of Proposition 2, because the latter assumes
there is a measurable correspondence between parts labeled $\alpha$ for distinct $(J,\Gamma)$ [not only for those $(J,\Gamma)$ 
related by a local transformation], while the former does not, but must treat all $n$ parts
on an equal footing (so permutation invariance of $f$ was used). 
Consequently, their argument takes a different form, and also assumes that the self-overlaps of the parts are almost-surely 
constant. We have seen that that assumption is not necessary in our case. See Ref.\ \cite{argch} for further discussion. 

We comment that our results show that the zero-one laws of Sec.\ \ref{zeroone} and the extended form of stochastic stability ($\Sigma$ 
invariance)
are essentially equivalent, in the following sense. From the weaker NRS23 zero-one law, Proposition 1, we were able to obtain the strong
version of stochastic stability (Theorem 2, Corollary 4, and Theorem 3), using only general arguments. Conversely, from a weak version 
of $\Sigma$ invariance (from Ref.\ \cite{ans15}) we were able to prove the stronger zero-one law, Proposition 2, again using only general 
arguments and Lemma 1. 

\subsubsection{Metastates with zero-dimensional support}
\label{subsubsec:zero}

Returning to the general discussion, by reformulating $\Sigma$ evolution as $g$ evolution, we can think of $\Sigma$ evolution 
as resulting from simultaneous (random) 
motion of each Gibbs state with random increments described by $g_{[X]}$. In general, a Gibbs state would not be invariant under 
$g$ evolution. However, in at least one case, states drawn from a metastate must be invariant.
Suppose that a metastate has an atom at $\Gamma=\Gamma_a$, so the decomposition of $\kappa_J$ into a countable number of 
atoms and an atomless part contains a term $\kappa_J(a)\delta_{\Gamma_a}$ where $\kappa_J(a)\leq 1$ is the weight of the atom. 
The atom gives nonzero probability $\kappa_J(a)$ to the set $\{\Gamma_a\}$, and that probability must be invariant, so the 
$\delta_{\Gamma_a}$ is invariant under $\Sigma$. In the dual point 
of view on states, we will now see that this implies that {\em $\Gamma_a$ must be $g_{[X]}$ invariant for all $[X]$}. 
(Of course, this is particularly clear for a trivial metastate, but is considerably more general.) In the next Section we will characterize
the $g$-invariant Gibbs states. 

First, consider the case of a metastate with a finite number of atoms. The atoms are at points $\Gamma_a$, $a=1$, 
\ldots, $m$, which are isolated. Any Gibbs state $\Gamma$ has a pure-state decomposition $w_\Gamma$, and so given the pure states 
involved, the Gibbs state can be viewed as parametrized by some number of 
parameters which, from NS09 \cite{ns09} or Corollary 2 above, in our case is either zero or (countably or uncountably) infinite. If the Gibbs state is not trivial, then 
under $g$ evolution these parameters change, in general, and so may fill out a space of Gibbs states. Recall that the support of a probability 
distribution is the smallest closed set that has probability $1$ (or equivalently, is the intersection of all closed sets that have probability $1$). 
So if the support of the metastate consists of isolated points, each a single Gibbs state, it is certainly a zero-dimensional set.
Because the metastate is $\Sigma$ invariant, each point (Gibbs state $\Gamma_a$) must be $g$ invariant, because if not, the 
$g$ evolution would generate a set of Gibbs states of dimension $>0$.

These observations can be formalized and extended by using the theory of topological dimensions of topological spaces;
we will not need the full theory, and instead only gve some definitions (those for $n>0$ could be skipped). 
First, a topological space $X$ (such as the support of the metastate in the relative or induced topology) is said to be 
zero dimensional at a point $p\in X$ if, for any open set $U\subseteq X$ to which $p$ belongs, there is an open set 
$V\subseteq U$, with $p\in V$, such that the boundary $\partial V\subseteq X$ is empty; 
$X$ is zero dimensional if it is zero dimensional at all $p\in X$ \cite{kechris_book,willard_book,hurwall_book}. 
A compact Hausdorff space $X$ is zero dimensional if and only if it is totally disconnected, where totally disconnected means that the 
connected components of $X$ are single points (see Ref.\ \cite{willard_book}, pp.\ 210--211). A zero-dimensional space can be infinite and 
need not have the discrete topology; for example, the Cantor set is an important example of a zero-dimensional space 
\cite{kechris_book,willard_book,hurwall_book}. More generally, for any topological space $X$, the inductive definition of the dimension 
at a given point $p\in X$ assigns a topologically-invariant number (the dimension), either $-1$ (if and only if $X=\emptyset$), 
a natural number, or infinity, to $p$ \cite{hurwall_book}; in the basic form, it does not distinguish countable from uncountable infinity. 
Inductively, we say \cite{hurwall_book} that the dimension at $p$ is $\leq n$ ($n\geq 0$ a natural number) if, for any open set 
$U\subseteq X$ to which 
$p$ belongs, there is an open set $V\subseteq U$, with $p\in V$, such that the boundary $\partial V\subseteq X$ has dimension $\leq n-1$, 
and $n=\infty$ if no such integer exists; here a set has dimension $\leq n-1$ if it has dimension $\leq n-1$ at every one of its points.
We say that $X$ has dimension $n$ at $p$ if it has dimension $\leq n$ at $p$ but not $\leq n-1$ at $p$, and $\infty$ at $p$ 
if it is not $\leq n$ for any finite $n$. Likewise, say that $X$ has dimension $n$ if it has dimension $\leq n$ at all $p\in X$ but there exists 
$p\in X$ at which the dimension is not $\leq n-1$, and $\infty$ if it is not $\leq n$ at all $p$ for any finite $n$. This defines the dimension 
for any $X$. There are other definitions of the dimension of $X$, which are 
equivalent to this one for separable metric spaces \cite{hurwall_book}. Another example of the theory 
is that Euclidean space $\bbR^n$ ($0< n<\infty$) indeed has topological dimension $n$. 

In terms of the topological dimension, we will say that the support of a metastate is $\kappa_J$-essentially $n$ dimensional
($n$ a natural number) if it is $\leq n$ dimensional at $\kappa_J$-almost every $\Gamma$, and $> n-1$ dimensional with nonzero 
$\kappa_J$ probability, while it is $\kappa_J$-essentially infinite dimensional if, for each $n$, there is nonzero $\kappa_J$ probability of 
$\Gamma$ at which the dimension is infinity (thus differing from the topological definition only in allowing for violations on null sets instead 
of ``every $\Gamma$'', but note that the topological definition still has to be used for the dimensions of the boundaries of the sets). This 
associates a number (the dimension of the support)  
to any metastate, and because the number is translation invariant, it is the same for $\nu$-almost every $J$, by ergodicity of $\nu$. 
We will say the support of a metastate is $\kappa_J$-almost surely $n$ dimensional, for $n$ a natural number or $\infty$, 
if the dimension at $\Gamma$ is $n$ for $\kappa_J$-almost every $\Gamma$. (For $n=0$, $\kappa_J$-essentially and $\kappa_J$-almost 
surely zero dimensional are identical.) Finally, say that the support is $\kappa_J$-almost surely finite dimensional if $n<\infty$ at $\kappa_J$-
almost every $\Gamma$; this is the negation of $\kappa_J$-essentially infinite-dimensional support. 
The support of a metastate is mapped by a homeomorphism under either a local transformation or 
a translation, so the dimension of the support at a Gibbs state $\Gamma$ is translation and locally invariant. Then the support of an 
indecomposable metastate must be $\kappa_J$-almost surely $n$ dimensional for some $n\geq 0$, possibly $n=\infty$. and takes the same 
value for $\nu$-almost every $J$. 

Because of all this, we can make the more general statement that, if $\Gamma$ is a Gibbs state in the support of a metastate and 
the support is zero dimensional at $\Gamma$, then $\Gamma$ must be $g$ invariant 
($\kappa_J$-almost surely, anyway), because again, if not, that would imply that the support of the metastate is not zero dimensional 
at such a Gibbs state. Clearly then, if a metastate is indecomposable, and its support is $\kappa_J$-almost surely zero 
dimensional, the conclusion holds $\kappa_J$-almost surely, and for $\nu$-almost every $J$. $\Gamma$ must also be $g$ invariant if it
is an atom of the metastate. We continue the discussion of $g$-invariant 
Gibbs states in the Section \ref{trivmet} below.

\subsubsection{$\Sigma$ invariance and ergodic decomposition}

Here we will discuss the overall picture of $\Sigma_t$ evolution and invariant distributions 
on $\Gamma$, assuming that it exists 
in this general form. We will now let $\Sigma$ (and $\Sigma_t$) stand for the indexed set $(\Sigma_{[X]})_{[X]}$ of $\Sigma_{[X]}$ 
for all $[X]$ [resp., $(\Sigma_{t[X]})_{t\geq0,[X]}$]. First, the theory of such commuting sets of probability kernels $\Sigma_t$ 
usually assumes that 
the kernels $\Sigma_{t[X]}(\cdot|\Gamma)$ exist for all $\Gamma$ in the relevant space (here ${\cal G}={\cal G}_J$, for the given $J$), 
however our analysis only showed that the overlaps $q_{[X]}(\Psi,\Psi')$ exist for almost every $\Gamma,\Psi,\Psi')$. We will assume 
that this causes no real difficulty. Second, if the system of commuting operators (kernels) exists then, for further purposes, suitable continuity 
properties are helpful. These continuity properties are defined in Ref.\ \cite{reed_simon_book2}, Sec.\ X.8,  Ref.\ \cite{liggett_book}, 
Chapter 1, and Ref.\ \cite{aliborder_book}, Ch.\ 19; correspondingly, the stochastic process ($g$ evolution) should be a ``Feller process''. 
Then because $\cal G$ is compact, it follows that invariant distributions $\kappa_J$ do exist (using the Markov-Kakutani fixed point theorem 
\cite{reed_simon_book}, p.\ 152). Moreover, any invariant distribution can be decomposed uniquely as an integral over ergodic components,
where the ergodic components are mutually singular invariant distributions (see e.g.\ Ref.\ \cite{phelps_book}, Sec.\ 12, where the 
arguments are for deterministic point transformations but can be easily adapted to a system of continuous probability operators, such 
as we are considering here).

Thus assuming the $\Sigma_t$ evolution exists and has suitable properties, it follows that, because an indecomposable metastate
is $\Sigma_t$ invariant, it can be decomposed further into ergodic invariant components. It might be that an indecomposable metastate
is already $\Sigma_t$ ergodic, but it is unclear at present in which nontrivial cases that may hold. If not, the ergodic components would 
not themselves be indecomposable, because there would be no way to form a correspondence between the components for distinct $J$ 
in general. (However, because the $\Sigma_t$ evolution commutes with local transformations and translations, the ergodic components 
would be $\Theta$ covariant near some given $J$.) Note that if there is nonzero $\kappa_J$ probability of $g$-invariant states $\Gamma$, 
then a $\delta$-function at one such $\Gamma$ is automatically $\Sigma_t$ ergodic, and this gives part of the ergodic decomposition 
in this case. Thus a nontrivial indecomposable metastate in which all Gibbs states are $g$ invariant is not $\Sigma_t$ ergodic,
and the metastate itself gives the $\Sigma_t$-ergodic decomposition; this includes the CS and CP cases, and an indecomposable metastate 
consisting of $m$ atoms of equal weight. On the other hand, if the Gibbs states drawn from an indecomposable metastate have a countably
infinite pure state decomposition then, by the results of the following Section, the orbit of one such Gibbs state under $g$ evolution would 
consist of different $w_\Gamma$ weights on the same pure states. If that orbit is a $\Sigma_t$-ergodic component, then the average of the 
Gibbs states on the orbit under that distribution should, by symmetry, presumably put equal probability on each of the pure states. But, as the 
pure states are countably infinite, no such distribution exists. Hence in this case an ergodic component cannot consist of a countable number 
of orbits, and must be an atomless distribution on an uncountable number of orbits, and also of Gibbs states.

\subsection{Poisson-Dirichlet and $k$-RSB case}
\label{subsec:pdultr}

In this subsection, we extend the result of Corollary 4 (in Subsec.\  \ref{subsec:stst} above) regarding 
the Poisson-Dirichlet (PD) distribution of weights and ultrametricity of the overlaps to more general cases, by making use of older results
\cite{ra,aa}. (The results will be extended a bit further in Subsec.\ \ref{subsubsec:furthgibbs} below.)

First, we extend the $\Sigma$-invariance as in eqs.\ (\ref{sigdefpure}) or (\ref{sigdefpure'}) to allow use of higher powers of an overlap
in place of the first power. 
This can be done when $f$ is a translation-invariant function of the state $\Gamma$, in particular, for $f$ a function of the weight $w_\Gamma$
and of the overlaps, say $q_{[X']}$, between pure states. 
To do so, we again use the approach of Ref.\ \cite{ans15}, which we recapitulate here. The expectation $\bbE_{\kappa_J}f$ is a 
translation-invariant function of $J$, and $\nu$ is ergodic, so the expectation is $\nu$-almost surely constant. Assuming $\nu$ is n.i.p.,
we can take the second derivative with respect to $J_X$ for $X=\{i_1,i_2,\ldots,i_r\}$ (of course, $i_m\neq i_n$ for $m\neq n$), and 
then the translation average over each of $i_m$ ($m=1$, \ldots, $r$) over the window $\Lambda_W$, followed by the limit $W\to\infty$; 
we write this averaging operation as ${\rm Av}_{\{i_1,\ldots,i_r\}}$). Each derivative acts on the weights $w_\Gamma$ and produces thermal 
expectations $\langle s_X\rangle_\Psi$ (for $X$ just specified). The terms with some $i_m=i_n$ for some $m\neq n$ should be dropped from 
the translation average; such terms constitute a negligible fraction of the terms in the $W\to\infty$ limit. Further, the thermal averages in a 
pure state $\Psi$ factorize when the sites are well separated, as discussed in Sec.\ \ref{models}, and then the second derivative produces 
in the limit (again up to contributions that clearly tend to zero) the $r$th power of the single site overlap $q_{[1]}^r$ in place of 
$q_{[1]}$ in the $\Sigma$ operator eq.\ (\ref{sigdefpure'}); it is natural to write such an operator as $\Sigma_{q_{[1]}^r}$, where 
$r=1$, $2$, \ldots, and we will use evident similar notations as well. We will assume that $f$ has bounded, continuous second derivatives
with respect to $w_\Gamma$.

This can be easily extended to a power of any overlap $q_{[X']}$. Replace $X$ by 
$X=\bigcup_{m=1}^r \theta_{{\bf x}_m}X'$, where the $r$ sets $\theta_{{\bf x}_m} X'$ are assumed pairwise disjoint, and for the 
translation average over the finite window $\Lambda_W$, fix $X'$ and sum over such $r$-tuples of $({\bf x}_m)$ such that all ${\bf x}_m$ 
lie in $\Lambda_W$. Then, by a similar argument, in the limit we obtain invariance under the operator $\Sigma_{q_{[X']}^r}$. This can be 
extended further by a slightly more involved construction to obtain powers such as $(a_{[X_1]}q_{[X_1]}+a_{[X_2]}q_{[X_2]})^r$, and so on. 
Ultimately, we can obtain all powers $r=1$, $2$, \ldots of any total overlap $q_{\rm tot}$ (it might be preferred to allow the number 
of terms in the sum to tend to infinity after the $W\to\infty$ limit has been taken). [The construction can also be used to obtain many other 
invariance equations, involving generalized overlaps \cite{nrs23} also; we do not consider these further.]

To make use of this result, we first borrow an idea from Arguin and Aizenman (AA) \cite{aa}. Suppose that the metastate is 
indecomposable, and that almost every Gibbs states has character (ii), that is, $w_\Gamma$ consists of a countable infinity of atoms;
here we will let $n$ labels these pure states $\Psi_n$. 
By indecomposability, for any given overlap type, such as $q_{[X]}$ or $\sum_{[X]}a_{[X]}q_{[X]}$, the self-overlap of 
$\mu_J$-almost every $\Psi$ takes a single value, say $q_{[X]}(\Psi,\Psi)$ in the first case, and that value is greater than or equal to the 
overlap of any other pair of pure states (for the same overlap type). If we normalize $q_{[X]}$ as 
$\widetilde{q}_{[X]}=q_{[X]}/q_{[X]}(\Psi,\Psi)$, then ${\widetilde{q}_{[X]}(n,n')}^r$ tends to $1$ or $0$ as $r\to\infty$; it is $1$ 
for the self-overlap, and possibly also for some pairs of distinct pure states, and otherwise $0$. We assume for now it is $1$ only for a 
self-overlap (so $({\widetilde{q}_{[X]}(n,n')}^\infty)_{n,n'}$ 
is the identity matrix; temporarily, we call this ``non-degeneracy'' of the original overlap, here $q_{[X]}$), and 
discuss the other case afterwards. Note that any total overlap is non-degenerate, by a version of Lemma 1. The invariance equation holds 
for all $r$, 
\be
\bbE_{\kappa_J}\Sigma_{{\widetilde{q}_{[X]}}^r}f=0,
\ee 
and also in the $r\to\infty$ limit (assuming $f$ has bounded continuous second derivatives, and using bounded convergence),
and note that we are considering a given fixed $J$ throughout. 

The idea now is that, in the $r=\infty$ limit, we can obtain a corresponding $g$ evolution in which the Gaussian variables $(g(\Psi_n))_n$  
are independent with variance $1$. In this case, which also corresponds to the case of $1$-RSB that we treated above using 
the GG identities in Corollary 4 (and which is a special case), one result of Ref.\ \cite{ra} is that the only invariant distribution on the weights 
$w_n\equiv w_\Gamma(\Psi_n)$ (which from now on we assume are arranged into non-increasing order) is a mixture of PD 
distributions, each with a parameter $x_1$, so the mixture is described by a distribution on $x_1\in(0,1)$ (a version of this statement 
goes back at least to Ref.\ \cite{mpv_book} in the context of the old cavity method). We will show that, by 
indecomposability, there is in fact a single fixed value of $x_1$. 

To address the value of $x_1$, we first recall that the PD distribution $PD(x_1)$ can be described by stating that
\be
w_n=\frac{u_n}{\sum_n u_n},
\ee
where $u_n$ are the points of a Poisson process with mean measure (or ``intensity'') $x_1 u^{-1-x_1}du$ \cite{panchenko_book,cg_book}.
The $u_n$ are distinct almost surely, and so can be assumed strictly decreasing.
For $0<x_1<1$, the sum $\sum_n u_n$ in the denominator converges almost surely. In the physics literature, the $u_n$ are described
by saying that $u_n=e^{-\beta f_n}$, where the $f_n$ are ``relative free energies'', and $\beta f_n$ are the points of a Poisson 
process with intensity an exponentially-increasing function of $\beta f$ (where $x_1$ describes the rate of exponential increase) 
\cite{mpv_book,cg_book}.

Given that the weights $w_n$ are drawn from a PD distribution, or a mixture thereof, we now show that the (possibly random) 
value of $x_1$ can be determined from a single sample $\Gamma$. Consider the sum $\sum_nw_n^x$ for real positive $x$ (it is a ``Dirichlet 
series''). The common denominator $(\sum_n u_n)^x$ is almost always finite, so we will consider only the sum $\sum_n u_n^x$, and ask 
whether or not it converges for given $x_1$. As $u_n$ are points of a Poisson process, for each $m=1$, $2$, \ldots, the number of points 
in the range $u\in[m^{-1/x_1},(m-1)^{-1/x_1})$ is a Poisson-distributed random variable $N_m$ with mean (conditioned on $x_1$)
$\int_{m^{-1/x_1}}^{(m-1)^{-1/x_1}} x_1 u^{-1-x_1}\,du=1$ for all $m$; $N_m$ for distinct $m$ are independent. 
Then the random series of interest can be bounded above and below,
\bea
\sum_{n: u_n<1} u_n&\leq& \sum_{m=2}^\infty N_m (m-1)^{-x/x_1},\\
\sum_{m=1}^\infty N_m m^{-x/x_1}&\leq &\sum_n u_n,
\eea
and the upper and lower bound are the same series in distribution (on the left-hand side of the first inequality, the number of terms omitted 
is almost-surely finite, so has a finite sum). Thus $\sum_n u_n$ converges if and only if $\sum_{m=1}^\infty N_m m^{-x/x_1}$ does.
Now the latter series is a function of the independent, identically-distributed random variables $N_m$, and the terms are non-negative. 
So by Kolmogorov's zero-one law, it either converges almost surely, or diverges almost surely. By examining its expectation, we see that
it converges almost surely if $x>x_1$, and diverges almost surely if $x\leq x_1$. Then the same statement holds for $\sum_n u_n^x$,
and for $\sum_n w_n^x$. Thus from a decreasing set of weights $(w_n)_n$ drawn from a PD distribution, we can determine the parameter 
value $x_1$ from which it arose almost surely.

In addition, the parameter value is unchanged under a local transformation. Under such a transformation [see eq.\ (\ref{loctransweight})],
say in a single $J_X$, $w_n$ changes by 
\bea
w_n\mapsto w_n'&=&\frac{r_n w_n}{\sum_n r_n w_n}\\
&=&\frac{r_n u_n}{\sum_n r_n u_n},
\eea
where $r_n=\langle e^{\beta\Delta J_X}\rangle_{\Psi_n}$. Then as $e^{-\beta|\Delta J_X|}\leq r_n\leq e^{\beta|\Delta J_X|}$,
another comparison-of-series argument shows that $\sum_n w_n'^x$ converges, or diverges, almost surely if and only if $\sum_nw_n^x$ 
does the same. Hence the parameter $x_1$ is a locally-invariant property of $\Gamma$, and by indecomposability of $\kappa_J$, $x_1$ 
must be $\kappa_J$-almost surely constant; the {\it a priori} possible mixture over values of $x_1$ is in fact trivial.

Thus we have proved the following:
\newline
{\bf Corollary $\bf 4'$}: Consider a short-range mixed $p$-spin model with the n.i.p.\ property for all $X'\in {\cal X}$ and an indecomposable 
(translation-invariant) metastate. Suppose that for $\kappa_J$-almost every Gibbs state
$\Gamma$, $w_\Gamma$ consists of a countable infinity of atoms with weights $w_n>0$ ($w_n$ non-increasing and $\sum_n w_n=1$),
and utilize an overlap, say some $q_{[X]}$, or any total overlap, that possesses the non-degeneracy property above. 
Then there is a parameter $x_1\in(0,1)$ such that the distribution on the weights $w_n$ is the PD $PD(x_1)$ distribution, 
$\nu$-almost surely.
\newline
We remark that, under the same conditions, $1-x_1=\bbE_{\kappa_J}\sum_n w_n^2$, and that the $PD(x_1)$ distribution is a feature of 
RSB also, under a corresponding condition \cite{mpv_book}. There $x_1$ is the value of Parisi's $x$ such that there is a plateau in $q(x)$ 
extending from $x_1$ to $1^-$.

Now we turn to a discussion of the non-degeneracy condition. Suppose that some overlap type,
say $q_{[X]}$, is degenerate in the sense that there is nonzero $w_\Gamma\times w_\Gamma$ probability for $\Psi\neq \Psi'$ with 
$d_{[X]}(\Psi,\Psi')=0$ (i.e.\ the two pure states have overlap equal to the self-overlap of either one); thus the pseudometric $d_{[X]}$ 
cannot be said to be almost surely a metric. It is a standard fact from metric space theory (see Ref.\ \cite{willard_book}, p.\ 20) that a true 
metric can be obtained from any pseudometric: using the triangle inequality, the relation $d_{[X]}(\Psi,\Psi')=0$ is an equivalence relation, 
and there is a metric on the space of equivalence classes (clusters of pure states), defined as the nonzero $d_{[X]}$ distance between 
any two pure states, one from one cluster, one from the other. Then the argument leading to Corollary $4'$ now shows that the total weights 
of these clusters are PD distributed, extending the conclusion of Corollary $4'$. The parameter $x_1'$, say, in that distribution must be less 
than that, say $x_1$, obtained from any non-degenerate overlap. This picture is perfectly consistent if the Ruelle cascades describe the full 
(countable) set of weights, because the cascades exhibit just this property under lumping together states into clusters using the 
ultrametric. However, as we are referring to different overlaps, this already involves some form of overlap equivalence (see Sec.\ 
\ref{subsec:ultra}). 

We now extend the result using the full conclusion from AA \cite{aa}. We assume the same conditions, including some choice of given overlap 
type, say $q_{[X]}$ or a total overlap, except that the non-degeneracy property will now be dropped in light of the preceding discussion. 
In addition, we will assume that the overlap takes only $k+1<\infty$ distinct values (and see the discussion in Sec.\ \ref{subsubsec:overl}). 
Without loss of generality we can pass to the normalized overlap $\widetilde{q}_{[X]}$ so that the self-overlaps are $1$ for 
$\mu_J$-almost every $\Psi$. Following AA, we will refer to the pair $((w_n)_n,(\widetilde{q}_{n,n'})_{n,n'})$ as a random overlap 
structure (or ROSt). Then from indecomposability of the metastate and the result of AA \cite{aa} we immediately obtain the following:
\newline
{\bf Corollary $\bf 4''$}: Assume the same conditions as in Corollary $4'$, except for the non-degeneracy, and also that the overlap takes 
only $k+1$ distinct values $q_{(0)}< \cdots <q_{(k)}$. Then the ROSt is distributed as the Ruelle cascade, with a fixed set of $k$ parameters 
$0=x_{(0)}<x_{(1)}< \cdots <x_{(k)} < x_{(k+1)}=1$, $\nu$-almost surely. In particular ultrametricity, eq.\ (\ref{eq:ultramet}), holds, 
and $q_{(0)}\geq 0$, both $\nu$-almost surely.
\newline
[In the Ruelle cascade, the parameters describe, for each $l=1$, \ldots, $k$, PD distributions $PD(x_{(l)})$ on the total 
weight of clusters of 
states such that the overlap of any pair in the same cluster is at least $q_{(l)}$, and these parameters can be shown to be constants by 
a similar argument as in Corollary $4'$.] Thus this situation is described exactly by $k$-RSB \cite{mpv_book,cg_book}. 
The differences $x_{(l+1)}-x_{(l)}$, $l=0$, \ldots, $k$, are the weights of the $k+1$ $\delta$-functions in $P_{J[X]}$, and note that $x_{(k)}$
corresponds to the previous $x_1$.  

In addition to assuming that the self-overlap is the same for every pure state, the analysis of AA refers to ``overlap indecomposability'',
the property that for each pure state $n$, the overlaps $q_{[X]}(n,n')$ take all of the $k+1$ values. For an indecomposable metastate, this is a 
consequence of Corollary 1 above. 

Now we discuss overlap equivalence.
If we consider two overlap types, say $q_{[X_1]}$, $q_{[X_2]}$, both of which take only a finite number of values, then we can first apply 
Corollary $4''$ to each. It might be that, as functions of pairs of pure states, one overlap is a function of the other ($k$ takes the same value 
for both), and then the clusters defined by the ultrametric must be the same for both. But it might also be (as in the special case of degeneracy 
with the self-overlap above), that one overlap takes the same value on distinct pairs that have different values for the other overlap, and 
it is possible that the situation is reversed for some other values of the overlaps (then $k$ may take different values for the two overlaps). 
Then some $\delta$-functions in $P_{J[X_1]}(q)$ may split when we compare with $P_{J[X_2]}(q)$, while others merge. If we consider 
an overlap type $a_{[X_1]}q_{[X_1]}+a_{[X_2]}q_{[X_2]}$ (with positive coefficients), then the degeneracy of both overlap types would 
be split (it could be split further if we consider another type $[X_3]$, and so on, until we reach a total overlap). Use of this 
combination, still with a finite number of values, would again lead to a finite Ruelle cascade. Then we may say, as a consequence of these 
observations, that the original two overlaps (or their corresponding pseudometrics) possess a common refinement. This is the general 
form of overlap equivalence, in the context of overlaps taking a finite number of values. It is compatible with the appearance of Ruelle 
cascades with the various parameter sets. We conclude that overlap equivalence does hold in the short-range systems, at least in this 
restricted context.

It is now very natural to expect that Corollary $4''$ can be extended so as to drop at least the assumption of a finite number of values of the 
overlap. It might be possible to achieve this by approximating a general $q_{[X]}(x)$ function by a step function with a finite number of steps 
(as in the above), but this is not clear at present. 

\section{Further analysis of Gibbs states drawn from a metastate}
\label{trivmet}

In this Section, we give some further analysis of Gibbs states drawn from a metastate. Some such analysis has already been 
given in Secs.\ \ref{indecomp}, \ref{sec:stst}, but here until Subsec.\ \ref{subsubsec:earl} most of the results make no use of the 
constructions and results from Secs.\ \ref{indecomp} and \ref{sec:stst}, in particular, of indecomposable metastates, 
to which we will make only passing reference until near the end, though $g$ evolution will be used in Subsec.\ \ref{subsec:ginv}. 
After reviewing older results of NS, the first results of this section are for the structure of Gibbs states that are atoms in a metastate,
or otherwise $g$ invariant; in particular, a trivial metastate is a single atom, and indecomposable.
Then we introduce the Dovbysh-Sudakov (DS) representation, and use it to discuss the structure of Gibbs states drawn from metastate,
leading to a classification of Gibbs states into types I, II, and III, each with a subclassification as a or b. 

\subsection{More on Gibbs states}
\label{sec:moregibbs}

First we prove a version of the result from NS07 \cite{ns07}, which asserts that, within the metastate set-up, 
if there is nonzero $\kappa_J$ probability of nontrivial Gibbs states (for given $J$), then the pure-state
decomposition of the metastate average state (MAS) must be uncountable. In the form 
in which we state the result, it concerns atoms of the pure-state decomposition of the MAS. 
The measure in the latter is $\mu_J(\Psi)=\int \kappa_J(d\Gamma)dw_\Gamma(\Psi)$, where the integral
is over $\Gamma$s. In order for $\mu_J$ to possess an atom, say at $\Psi=\Gamma_\alpha$, it is necessary
that some $\Gamma$s in the support of $\kappa_J$ have such an atom in $w_\Gamma$, so that $w_\alpha\equiv 
w_\Gamma(\{\Gamma_\alpha\})>0$, and those $\Gamma$s form a set that must have nonzero $\kappa_J$ probability. 
The weight of the atom in $\mu_J$ is then $\mu_J(\alpha)\equiv
\mu_J(\{\Gamma_\alpha\})=\int \kappa_J(d\Gamma)w_\alpha>0$. {\it A priori} it is not obvious that $\kappa_J$ 
necessarily has an atom at such $\Gamma$; 
an atomless component of $\kappa_J$ could also produce an atom. However, the Theorem of NS07 below rules that out.
Before stating the result, we note that, for given $J$, the atoms of $\mu_J$ form an, at most countable, set of $\Gamma_\alpha$, 
each with nonzero weight $\mu_J(\alpha)$, and that $\sum_\alpha \mu_J(\alpha)\leq 1$. In order to work with functions 
of $J$ that are defined for ($\nu$-almost every) $J$, for given $J$ we will sort the $\mu_J(\alpha)$ into descending order, and note
that there may be ties or degeneracies, that is some weights may be equal, in which case the rank-ordered weights can be
described by distinct weights and their (finite) multiplicities. As the weights are translation invariant, by the ergodicity of $\nu$
under translations they are $\nu$-almost surely constant, and so the multiplicities of the distinct weights are also $\nu$-almost 
surely constant. The statement and proof of the following differ somewhat from, and extend, those in NS07.
\newline
{\bf Theorem 4} (NS07): Assume the same hypotheses as in Proposition 1. If $\mu_J$ has any atoms, then they arise from atoms 
of $\kappa_J$ at Gibbs states $\Gamma$ that are trivial, and exist for $\nu$-almost every $J$.
\newline
{\bf Proof}: For given $J$, under a change $\Delta J_X$, 
to first order the change in a weight $\mu_J(\alpha)$ is
\be
\beta\Delta J_X \int \kappa_J(d\Gamma) w_\alpha(\langle s_X\rangle_\alpha-\langle s_X\rangle_\Gamma),
\ee
using covariance of the metastate and of $\Gamma_\alpha$, and eq.\ (\ref{loctransweight}) for the change in $w_\alpha$.
$J'_X=J_X+\Delta J_X$ with arbitrarily small $\Delta J_X$ exists with nonzero probability by the n.i.p.\ property,
and so by translation ergodicity this quantity (with $\beta\Delta J_X$ removed) must be zero for all $X$, $\nu$-almost surely.
(Here we use the remarks that precede the Theorem.) This can be restated as
\be
\int \kappa_J(d\Gamma) w_{\alpha}\left[(1-w_{\alpha})\langle s_X\rangle_{\alpha}
-\int_{\{\Gamma_\alpha\}^c} w_\Gamma(d\Psi)
\langle s_X\rangle_{\Psi}\right]=0
\ee
for all $X$. Here the domain of integration $\{\Gamma_\alpha\}^c$ includes all pure states other than $\Gamma_\alpha$.
If there is nonzero $\kappa_J$ probability of $\Gamma$ such that $w_\alpha\neq 0$ or $1$, then this implies
\be
\langle s_X\rangle_{\alpha}= \frac{\int \kappa_J(d\Gamma) w_{\alpha}\int_{\{\Gamma_\alpha\}^c} w_\Gamma(d\Psi)
\langle s_X\rangle_{\Psi}}{\int \kappa_J(d\Gamma) w_{\alpha}(1-w_{\alpha})}
\ee
for all $X$. But this implies that $\Gamma_{\alpha}$ is a convex combination of pure states $\Psi\neq \Gamma_\alpha$, 
which is impossible because $\Gamma_\alpha$ is a pure state. Hence $w_{\alpha}=0$ or $1$ for $\kappa_J$-almost 
every $\Gamma$. Thus the Gibbs states $\Gamma$ that contribute to the weight $\mu_J(\alpha)$ are trivial, 
$\Gamma=\Gamma_\alpha$, and so $\kappa_J$ itself has an atom of weight $\mu_J(\alpha)>0$ at  
$\Gamma=\Gamma_{\alpha}$. The triviality for $\nu$-almost every $J$ follows from the remarks before the proof. $\Box$

We note that it follows from Theorem 4 and the earlier Proposition 4 that a $\mu_J$ with atoms may arise from a decomposable
$\kappa_J$. A side remark is that the proof of Theorem 4 can be viewed as a special case of the proof of the zero-one law, in which 
the set $\{\Gamma_\alpha\}$ plays the role of a covariant set of pure states, even though, in the case that $\mu_J(\alpha)$ is 
degenerate with other weights, the set may not be well-defined (i.e.\ by a choice of one of the pure states) for $\nu$-almost every~$J$. 

A particular case of Theorem 4 arises for a trivial metastate, which will be one of our concerns in this Section. If $\kappa_J$
is trivial, then there is a unique Gibbs state $\Gamma$ for each $J$, which we can view as a {\em function} of $J$ that we call 
$\Gamma_J$; by covariance, $\Gamma_J$ is covariant under local transformations. In this case $\mu_J=w_{\Gamma_J}$ and 
$\mu^\dagger=\nu w_{\Gamma_J}$. Such a metastate is of course
indecomposable, and then by Corollary 2 above $\Gamma_J$ has one of the three characters (i)--(iii) identified in Sec.\ \ref{subsec:prop}
for $\nu$-almost every $J$; in fact in the present case, as noted already, the use of Theorem 1 in Corollary 2 can be eliminated by 
making direct use of translation ergodicity of $\nu=\kappa^\dagger$ to obtain the same result. But now, by Theorem 4 (in which 
translation ergodicity was again used), we find that if $w_{\Gamma_J}$ has nonzero $\nu$-probability of having any atoms, then 
$\Gamma_J$ is trivial $\nu$-almost surely. So either $\Gamma_J$ is trivial $\nu$-almost surely, or its pure-state decomposition 
is atomless $\nu$-almost surely; character (ii) has been eliminated. 

More generally we can consider the atomic part of the metastate. If an atom of the metastate is a Gibbs state with atoms in its pure-state
decomposition, with or without an atomless part, then the atoms give rise to atoms of $\mu_J$. Then Theorem 4
implies that the atoms in the metastate are Gibbs states that are either trivial ($\kappa^\dagger$-almost surely) or atomless 
($\kappa^\dagger$-almost surely), 
as for the special case of trivial metastate. In particular, this is the case for an indecomposable metastate consisting of a finite number $m$ 
of atoms of equal weight. Then for such an indecomposable metastate, either all of the $m$ Gibbs states for given $J$ are trivial,
$\kappa^\dagger$-almost surely, or else all are atomless $\kappa^\dagger$-almost surely. 

The contrapositive of Theorem 4 says, in particular, that if there is nonzero $\kappa_J$-probability that the pure-state decomposition 
of $\Gamma$ has more than one atom then $\Gamma$ belongs to the atomless part of the 
metastate (which in particular would be nontrivial). (This extends the conclusion of NS07 to allow for an atomless part of $\Gamma$,
however separate arguments above already eliminated the possibility of an atomless part in this case; see the discussion before 
Corollary 2.) 
Then the pure-state decomposition $\mu_J$ of the corresponding part of the MAS would also be atomless, meaning that the atoms 
of the Gibbs states change with $\Gamma$, due to the continuous metastate. This result can now be combined with the result of NS09 
\cite{ns09},
so that the number of atoms in such $\Gamma$ must be countably infinite, and for an indecomposable metastate the character is the
same for $\kappa^\dagger$-almost every $(J,\Gamma)$, and $\mu_J$ is atomless $\nu$-almost surely.

\subsection{$g$-invariant and dynamically-frozen Gibbs states}
\label{subsec:ginv}

We recall from Sec.\ \ref{subsubsec:zero} that, under some conditions on the disorder and for given $J$, if a $\Theta$-covariant metastate 
has any atoms, then each atom $\Gamma_a$ is a Gibbs state that is $g_{[X]}$ invariant
for every $X$ for which ${\rm Var}\,J_X^{(2)}>0$. The same holds for $\kappa_J$-almost every Gibbs state in any 
metastate with zero-dimensional support (for $\nu$-almost every $J$). Here we determine the consequences when such invariance
holds for all $[X]$ for a set of $\Gamma$ with positive $\kappa_J$ probability, for given $J$, and hence for $\nu$-almost every $J$. 
Then if also the metastate is indecomposable, invariance for all $[X]$ holds in fact $\kappa_J$-almost surely. Once again, the arguments 
may possibly be slightly less rigorous than those earlier in this Section.

We consider $g$ evolution, assuming that such a process exists; in case it does not another proof, using a process that may be 
better defined, will be given in Sec.\ \ref{subsubsec:gevo} below.
We first consider a single $[X]$, so to simplify writing let $g_{[X]}(\Psi)=g(\Psi)$, and also write $\Gamma$ for $\Gamma_a$ 
until further notice. Then we recall that $\bbE_g g(\Psi_1)g(\Psi_2)=q_{[X]}(\Psi_1,\Psi_2)$, $w_\Gamma\times w_\Gamma$-almost surely. 
$g$-evolution means that $w_\Gamma$ transforms to $w_\Gamma^{\lambda g_{[X]}}$, eq.\ (\ref{eq:gevolw}).
As the pure states $\Psi$ do not change under $g$ evolution, we then require that $w_\Gamma'=w_\Gamma$ as a distribution.
Taking $d/d\lambda$ at $\lambda=0$, we require
\be
[g(\Psi)-\bbE_{w_\Gamma}g(\Psi)]w_\Gamma(d\Psi)=0
\ee
as a signed measure, and so $g(\Psi)=\bbE_{w_\Gamma}g(\Psi)$ as centered Gaussian random fields ($w_\Gamma$-almost surely), 
implying that the covariances are equal, which gives
\bea
q_{[X]}(\Psi_1,\Psi_2)&=&q_{[X]}(\Psi_1,\Gamma)\\
&=&q_{[X]}(\Gamma,\Gamma),
\eea
that is, the overlaps in the first line are constant, $w_\Gamma\times w_\Gamma$- and $w_\Gamma$-almost surely, respectively.
All this holds for all $[X]$. 
[Then the overlap distribution $P_{J[X]\Gamma}(q)$ is trivial, that is, a single $\delta$-function, and this is true for all $[X]$.] 
If $\Psi_1$ is an atom of $w_\Gamma$, then there is nonzero $w_\Gamma\times w_\Gamma$ probability that $\Psi_2=\Psi_1$, 
and then it follows that $d_{[X]}(\Psi_1,\Gamma)^2=0$ for the pair $(\Gamma,\Psi_1)$ and for all $[X]$. But then, using the distribution 
$\kappa^\dagger w_\Gamma$ on triples $(J,\Gamma,\Psi)$ and applying a version of Lemma 1, such $\Gamma$ can occur with positive 
$\kappa_J$ probability only if $\Psi_1=\Gamma$, that is, such $\Gamma$s are trivial. The only alternative remaining is that $w_\Gamma$ 
is atomless, with overlaps of pairs of pure states having the form above. For atoms in the metastate, these alternatives of trivial or atomless 
$w_\Gamma$ reproduce the consequence of NS09, Theorem 4 in the preceding section. 

In the particular case of a trivial metastate, we can also obtain the result that pairwise overlaps
are constant from eq.\ (\ref{ggvar}) (assuming there is nonzero $J_X^{(2)}$ for all $X$), because in this case the left-hand side is zero. 
We also point out that, given our conclusion about overlaps, the $\Sigma$ evolution becomes trivial,
and $\Sigma$ invariance gives no further information about the distribution $w_\Gamma$. Thus each Gibbs state at which the support
is zero-dimensional, and also each Gibbs state that is an atom in the metastate, is either a trivial or an atomless mixture of pure states, 
and is $g$ invariant. The converse statements also hold for $g$-invariant Gibbs states. We summarize with a Theorem.
\newline
{\bf Theorem 5}: Assume the hypotheses of Theorem 3, and that there is positive $\kappa_J$ probability for Gibbs states
$\Gamma$ that are $g_{[X]}$ invariant for all $[X]$. Then for such $\Gamma$ either (i) it is trivial (i.e.\ pure), or (ii) its 
pure-state decomposition $w_\Gamma$ is atomless and, for each $[X]$, the overlap $q_{[X]}(\Psi_1,\Psi_2)$ is constant for 
$w_\Gamma\times w_\Gamma$-almost every pair $(\Psi_1,\Psi_2)$. For $\kappa_J$ indecomposable, one of (i), (ii) holds for
$\kappa_J$-almost every $\Gamma$.
Conversely, if (i) or (ii) holds for a Gibbs state $\Gamma$ drawn from a metastate, then $\Gamma$ is $g$-invariant.

The special case of a trivial metastate (i.e.\ one that consists of a single atom), with the Gibbs state atomless,
corresponds to a mean-field solution for infinite-range Potts and $p$-spin models found in Refs.\ \cite{kirk}, after related behavior was found in 
a dynamical treatment below a temperature at which the dynamics froze (i.e.\ ceased being ergodic), and so the phase was said to be 
``dynamically frozen'' (DF). It will be useful to extend this term here. 

In general, if a Gibbs state $\Gamma$ (not necessarily one drawn from a metastate) has non-trivial $w_\Gamma$ 
(not a single atom) and also, for some particular $[X]$, $q_{[X]}(\Psi_1,\Psi_2)$ is constant for $w_\Gamma\times 
w_\Gamma$-almost every pair $(\Psi_1,\Psi_2)$, then we say that $\Gamma$ is {\it DF with respect to $q_{[X]}$}; 
the latter condition implies that $P_{J\Gamma[X]}(q)$ is a single $\delta$-function. 
We call a Gibbs state $\Gamma$ DF (without qualification) when it is DF with respect to $q_{[X]}$ for all $[X]$. 
We emphasize that the definition says nothing about how $\Gamma$ may depend on $J$. (In the infinite-range case, 
the property holds whenever a state is DF with respect to $q_{[X]}$ for $[X]=[1]$, 
as found in the references given.)
We emphasize again that, in our setting in which $\Gamma$ is drawn from a translation-covariant metastate, a $\Gamma$ 
that is DF necessarily has atomless $w_\Gamma$ $\kappa_J$-almost surely, as we have just proved.

In the infinite-range case, in the DF phase there appeared to be an extensive entropy of ordered (frozen) states (each ergodic for the 
dynamics). In Refs.\ \cite{ktw,bb}, it was assumed that in a short-range system, this still holds for some sort of corresponding 
so-called states, which would then be ``metastable'', and would form a ``mosaic'' of regions, within each of which the 
distribution on the spins is dominated by a few of the metastable states [we will not discuss the remaining random first-order 
transition (RFOT) into another phase at a lower temperature]. It seems that the mosaic was intended to be a pure state. 
It was tacitly assumed that there results a {\em single} such mosaic or pure state, which implies that in fact there would be no DF phase 
and no freezing transition into it, but it is not clear to us why that should be so \cite{hr,read22}. The complexity of pure states in any 
Gibbs state must be sub-extensive in any case \cite{vEvH}, but it is not clear that if the DF phase occurs in a mean-field solution 
for a short-range system, then it must imply an extensive complexity of pure states. Thus, in the short-range case, a DF phase with 
subextensive complexity and an extensive contribution to the entropy that goes to zero at the lower-$T$ RFOT, where a jump in the 
specific heat occurs \cite{ktw,bb}, does not seem to be ruled out.

\subsection{Dovbysh-Sudakov representation and classification of Gibbs states}
\label{subsec:gibbs}

We cannot say as much about Gibbs states drawn from an indecomposable metastate that are not $g$ invariant (in which case the 
metastate must be atomless) as we could for $g$-invariant ones. In general, the Dovbysh-Sudakov (DS) representation of the weight 
$w_\Gamma$ is a very useful tool, and we discuss it here (it will also be useful in Sec.\ \ref{sec:mas}); its use in SGs goes back to 
Refs.\ \cite{ad,argch}. 
We give a direct proof of a reformulation of the DS representation as it relates to our problem, which reveals 
more of the structure than the original DS result usually used for SGs. Then we turn to $\Sigma$ evolution and the classification of Gibbs states.

\subsubsection{Dovbysh-Sudakov representation}

First we discuss the Dovbysh-Sudakov (DS) representation for random positive-semidefinite infinite exchangeable matrices \cite{ds}.
In our applications in the present Section, such a matrix arises by drawing (independently) a sequence of pure states 
$(\Psi_l)_{l\in \bbN}$ from the weight $w_\Gamma$ 
of a Gibbs state $\Gamma$, and forming the matrix of overlaps $(q_{[X]}(\Psi_l,\Psi_{l'}))_{l,l'}$ for some overlap type $[X]$,
as in Sec.\ \ref{models}
(again, this could be a total overlap $q_{\rm tot}$). This matrix is random because the $\Psi_l$ are drawn from a distribution $w_\Gamma$, 
which is itself random [it depends on $(J,\Gamma)$], but the matrix is easily seen to be positive semidefinite because of the relation of overlaps 
to inner products in a real vector space (it is a Gram matrix). ``Exchangeable'' (or ``weakly exchangeable'') means that the distribution 
of the matrix is invariant under simultaneous permutations of rows and columns, both by the same permutation, where the permutation must  
leave all but finitely many indices fixed. (In the following Section, we will also 
apply the DS representation to pure states drawn from a MAS, and to Gibbs states drawn 
from a metastate.) Several earlier works have applied the DS representation in SG theory, but to the matrix of overlaps of spin configurations
\cite{aa,ad,argch,panchenko_book}; except for Ref.\ \cite{ad}, all of these use global rather than window overlaps. 

The DS representation \cite{ds} says that, given a weakly-exchangeable random positive-semidefinite infinite matrix, there is a random 
distribution on pairs $({\bf v},h)$ that consist of a vector $\bf v$ in a {\em separable} real Hilbert space $\cal H$, with inner product of 
vectors $\bf v$, ${\bf v}'$ denoted ${\bf v}\cdot{\bf v}'$, and a non-negative 
real number $h$, such that, drawing a sequence of the pairs $({\bf v},h)$, the distribution of the matrix formed from their pairwise inner 
products, with $h$ on the diagonal,
\be
{\bf v}_l\cdot{\bf v}_{l'} +h_l\delta_{l,l'},
\ee 
is the same as that of the given random matrix, conditionally on the randomness (Panchenko \cite{panchenko_book} gives an exposition).
The representation is unique up to isometries (i.e.\ orthogonal linear maps) of the Hilbert space; we write $O({\cal H})$ for the group
of orthogonal linear maps of $\cal H$. In the case of spin configurations (rather 
than pure states), the diagonal overlaps are always $1$, and then only the off-diagonal $l\neq l'$ entries are of 
interest (but the diagonal entries are still relevant to positive semidefiniteness). We will often be interested only in the distribution on $\bf v$,
the marginal distribution ignoring $h$, and call that the DS distribution (on ${\bf v}\in\cal H$), with notation such as $w_{\Gamma{\rm DS}[X]}$
(note that it depends on the choice of overlap $[X]$); we call the distribution on $({\bf v},h)$ the full DS representation. The DS distribution on 
$\cal H$ is determined by only the off-diagonal entries of the matrix of overlaps. In our situation, the DS distribution for spin configurations
is the same as that for pure states.

\subsubsection{Reformulation and proof}

One may wish to think of the DS distribution as simply the image of $w_\Gamma$ under a map (determined by $w_\Gamma$) 
of pure states into $\cal H$, and (as we will see) it would be useful if this map were measurable with respect to the natural (Borel)
$\sigma$-algebra determined by the norm in $\cal H$; we will refer to this map as the DS map. In the form in which 
we have so far discussed the DS distribution, it is not evident that such a map even exists. For this reason, we here prove the existence 
and properties of such a map directly, thus reformulating the meaning of the DS distribution, and then discuss the resulting properties. 
[We note, though, that the DS representation is more general because (apart from including the self-overlaps) it applies to any exchangeable 
positive semidefinite matrix, whereas we begin with the distribution $w_\Gamma$, from which such a random matrix can be obtained
by drawing a sequence and computing the overlaps. The result we obtain here may be well known in the general setting of such a, 
possibly random, probability space equipped with a bounded positive semidefinite bilinear form, but we are unaware of a reference.]
These results are the key to the remainder of this Section, and to most of Sec.\ \ref{sec:mas} also.

Throughout the discussion, we assume a choice of overlap type $q$, such as $q=q_{[X]}$ or $q_{\rm tot}$, and that the pairwise 
(and self-overlaps) $q(\Psi,\Psi')$ and [resp., $q(\Psi)$] exist for $w_\Gamma\times w_\Gamma$-
(resp., $w_\Gamma$) almost every pair $(\Psi,\Psi')$ (resp., single $\Psi$); we have seen that in our setting these hold for 
$\kappa^\dagger$-almost every $(J,\Gamma)$, and for any choice of overlap type. We view these overlaps as defining a single object 
$Q=(q(\Psi,\Psi'))_{\Psi,\Psi'}$, which may be viewed as a matrix with possibly continuous index set. It defines a symmetric bilinear 
form (also denoted $Q$) by using the weight $w_\Gamma$: for $f$, $g$ integrable functions [belonging to $L^1(w_\Gamma)$], let
\be
fQg\equiv\int\int f(\Psi)q(\Psi,\Psi')g(\Psi')w_\Gamma(d\Psi)w_\Gamma(d\Psi'),
\ee
which is bounded,
\bea
|fQg|&\leq&\sup_{\Psi,\Psi'}|q(\Psi,\Psi')|\,||f||_1||g||_1\\
&\leq&||f||_1||g||_1,
\eea
and positive semidefinite. (These statements are in addition to the similar ones for $Q$ as a bilinear form on finite formal linear 
combinations of $\Psi$ without use of $w_\Gamma$, and also for a combination of such finite linear combinations with integrals under 
$w_\Gamma$; these points will help during the proof.) Also define the $Q$-norm $||f||_Q$ or $||f||$ by $||f||^2=fQf$. Our goal is to 
use $Q$ to obtain a distribution on a Hilbert space that retains as much information as possible about the distribution $w_\Gamma$ 
on pure states. 

We note that if $Q$ is a finite-dimensional positive-semidefinite symmetric matrix, which we view as a bilinear form on a finite-dimensional 
vector space $V$, written as a matrix using an arbitrary basis for $V$, then it can be expressed 
as $Q=R^TR$ ($^T$ denotes transpose), where $R$ is a matrix that may be rectangular; the number of rows in $R$ can be chosen to equal 
the rank of $Q$, and $R$ is determined only up to left multiplication by an orthogonal matrix. This says that $Q$ is the matrix of inner 
products (the Gram matrix) of a set of finite-dimensional column vectors (the columns of $R$) in the space of column vectors equipped 
with its standard 
positive-definite inner product. $Q$ may not be positive definite, but here that occurs only because some of the vectors may be linearly 
dependent. A sketch of the proof of the result goes as follows; for $f$, $g\in V$ write the form as $fQg$. Let $V_0\subseteq V$ be the subset 
$V_0=\{f\in V:fQf=0\}$, which we can call the kernel of $Q$. Using the Cauchy-Schwarz inequality, one can check 
that $V_0$ is a (closed) vector subspace of $V$, and that for $f\in V_0$, $fQg=0$ for all $g\in V$. The quotient space $V/V_0$ inherits
a positive definite bilinear form, and is a Hilbert space $\cal H$ (here, finite dimensional). 
This result can be extended even to an uncountable number of rows and columns (where $f$, $g$ are nonzero at only a finite number of 
entries); when $\cal H$ is infinite-dimensional, completion 
of $V/V_0$ in the norm topology may be required to obtain the Hilbert space $\cal H$. 

This gives some idea of what we need to do, 
but is not exactly what we need, because it did not make use of the weight $w_\Gamma$ in the definition of the form $Q$
we gave above; instead it used finite sums.
For our discussion, we define pure states $\Psi$, $\Psi'$ to be 
{\em congruent} (with respect to the given overlap type $q$), written $\Psi\sim\Psi'$, if $q(\Psi,\Psi'')=q(\Psi',\Psi'')$ for 
$w_\Gamma$-almost every $\Psi''$ (this definition follows Ref.\ \cite{ad}, although they define it only for
a $w_\Gamma$ that is purely atomic, with $\Psi$, $\Psi'$, $\Psi''$ referring to the atoms only). Clearly, congruence is an 
equivalence relation; we refer to the equivalence classes as congruence classes. We define ${\cal L}^1={\cal L}^1(w_\Gamma)$,
the space of all Borel measurable functions (see Appendix \ref{app:back}) that are normalizable in the 
$w_\Gamma$ $1$-norm, that is the $L^1$ norm $||f||_1<\infty$, and the Banach space 
$L^1$ is ${\cal L}^1$ modulo functions whose $L^1$ norm is zero. We define $L^p$ and ${\cal L}^p$ similarly.
We assume given $J$, $\Gamma$, and we will omit mention of $\kappa^\dagger$-almost every such $J$, $\Gamma$. 
To avoid tedious repetition, we suppress mention of sets of $w_\Gamma$ probability zero that arise because the overlap 
may only be defined almost everywhere, and treat both $q(\Psi,\Psi)$ and $q(\Psi,\Psi')$
as well defined for all $\Psi$ or $(\Psi,\Psi')$; we do not suppress almost-everywhere statements that may be required even when the overlap 
is defined everywhere. 

We have the following
\newline
{\bf Proposition 5}: For a Gibbs state $\Gamma$ with pure state decomposition $w_\Gamma$ and an overlap $q$ as discussed, 
there is a measurable map $\phi$ from ${\rm ex}\,{\cal G}$ (defined for $w_\Gamma$-almost every $\Psi$) into a separable real 
Hilbert space $\cal H$, such that (i) for the inner product in $\cal H$, 
\be
\phi(\Psi)\cdot\phi(\Psi')=q(\Psi,\Psi')
\ee
for $w_\Gamma\times w_\Gamma$-almost every pair $(\Psi,\Psi')$. In addition, $\phi$ obeys (ii) $\phi(\Psi)=\phi(\Psi')$
if $\Psi\sim\Psi'$, and (iii) $||\phi(\Psi)||^2\leq q(\Psi,\Psi)$, everywhere that $\phi$ is defined. Any function $f\in{\cal L}^1$ maps to
a vector in $\cal H$ given by a so-called Bochner integral $\int f(\Psi)\phi(\Psi)w_\Gamma(d\Psi)$, and the distribution $w_\Gamma$ 
can be pushed forward to a distribution on $\cal H$. If $\phi'$ is another measurable map into a separable real Hilbert space ${\cal H}'$ 
with property (i), then there is a linear map from $\cal H$ to ${\cal H}'$ taking the range of $\phi(\Psi)$ isometrically into the range 
of $\phi'(\Psi)$ $w_\Gamma$-almost everywhere, and a similar inverse map; thus if ${\cal H}'=\cal H$, there is an 
orthogonal linear map $U\in O({\cal H})$ such that $\phi'(\Psi)=U\phi(\Psi)$ at $w_\Gamma$-almost every $\Psi$.
In particular, given property (i), $||\phi(\Psi)||$ is determined $w_\Gamma$-almost everywhere.
\newline
{\bf Proof}: We begin by constructing a Hilbert space $\cal H$ (we call this construction ``canonical''). We have the space 
$V={\cal L}^1$ of functions $f(\Psi)$ on $\Psi$, together with the bilinear form $Q$; let $X$ temporarily denote the space of states $\Psi$
(or the support of $w_\Gamma$). We follow the argument given before the statement of the 
Proposition to obtain the quotient space $V/V_0$ and complete it (if necessary) to obtain a Hilbert space $\cal H$.
Generic vectors in $\cal H$ will be denoted by $\bf v$, $\bf w$, \ldots, the inner product by ${\bf v}\cdot{\bf w}$,
and the norm-square again by $||{\bf v}||^2={\bf v}\cdot{\bf v}=||{\bf v}||^2_Q$. 
The $L^1$ norm defines a topology on $V$ that is separable. The $Q$ norm, $||f||_Q$, on functions 
(or equivalence classes of functions) $f\in V$ defines the $Q$-norm topology on $V$, which may not be a Hausdorff topology. Using the above 
bound, with $g=f$, we can show that a sequence that is convergent in the $L^1$ norm is also convergent in the $Q$ norm. Hence the 
$Q$-norm topology is weaker than the $L^1$-norm topology, and as $V$ with the former topology is separable, so are $V/V_0$ 
and its completion, the Hilbert space ${\cal H}$. If $V/V_0=\{0\}$, that is, $q(\Psi,\Psi')=0$ for almost every pair $(\Psi,\Psi')$, 
then ${\cal H}=\{0\}$ (the zero-dimensional Hilbert space) and the result is trivial, so assume $V/V_0\neq \{0\}$.

First consider the function 
\be
Qg(\Psi)\equiv \int q(\Psi,\Psi')g(\Psi')w_\Gamma(d\Psi')
\ee 
for any $g\in L^1$, which is defined for $w_\Gamma$-almost every $\Psi$, and is bounded by $||g||_1$. It can be viewed as a linear 
functional (or operator) $Q\cdot(\Psi)$ into $\bbR$, defined by $g\mapsto Qg(\Psi)$ for $g\in V$ or $V/V_0$. As such, it has 
(operator) norm defined as usual by 
\be
||Q\cdot(\Psi)||_{\rm op}=\sup_{g\in V/V_0: g\neq 0}\frac{|Qg(\Psi)|}{||g||_Q}.
\ee
In principle the supremum could be infinite, however, $Qg(\Psi)$ can be viewed as the bilinear form 
applied to the single point $\Psi$ and the integral weighted by $gw_\Gamma(d\Psi')$, as mentioned before the Proposition. Then we have a 
version of the Cauchy-Schwarz inequality:
\be
|Qg(\Psi)|\leq q(\Psi,\Psi)^{1/2}(gQg)^{1/2}.
\ee
This implies that the norm $||Q\cdot(\Psi)||_{\rm op}\leq q(\Psi,\Psi)^{1/2}$, and so $Q\cdot(\Psi)$ is bounded. Hence the linear functional 
can be extended from $V/V_0$ to a bounded (equivalently, continuous) real-valued linear functional defined on all ${\bf w}\in{\cal H}$ 
(in place of $g$) that has the same norm. Any continuous linear functional on a Hilbert space is given by the inner product with some 
particular vector in $\cal H$, which we denote by $\phi(\Psi)$ or $\bf v$, and the norm of the vector is the same as the norm of the linear 
functional, so $||\phi(\Psi)||=||Q\cdot(\Psi)||_{\rm op}$ (see Ref.\ \cite{reed_simon_book}, pp.\ 9 and 43). Hence we have 
established the existence ($w_\Gamma$-almost everywhere) of a map $\phi$ from $X$ into $\cal H$, and property (iii).

As $\cal H$ is separable, it has a countable orthonormal basis. Let $e_\alpha(\Psi)$, $\alpha=1$, $2$, \ldots, be such a basis of  
(equivalence classes of) functions, so $e_\alpha Qe_{\alpha'}=\delta_{\alpha\alpha'}$; we should keep in mind that the $e_\alpha(\Psi)$s 
are only defined modulo addition of functions from $V_0$, including those that are zero $w_\Gamma$-almost everywhere. 
[Actually, $Q$ can be viewed as a linear map on ${\cal L}^1$ or on ${\cal L}^2\subset {\cal L}^1$, given by $g\mapsto Qg$,
and it is a Hilbert-Schmidt operator, and hence compact, with only non-negative eigenvalues, and the positive eigenvalues tend to $0$. 
Then representatives of the $e_\alpha$s can be chosen to be the set of eigenfunctions of strictly positive eigenvalue. These eigenfunctions 
are elements of ${\cal L}^2$, and can be chosen orthogonal in ${\cal L}^2$ as well as with respect to $Q$, because $Q$ is symmetric.
This choice of representatives is not essential in the following.]
The inner products of a fixed vector (function) ${\bf v}\in{\cal H}$ with the $e_\alpha$s define the components of $\bf v$ in the basis,
and serve as coordinates. Then the components of $\phi(\Psi)$ must be $\phi(\Psi)^\alpha=e^\alpha(\Psi)$, where 
\be
e^\alpha(\Psi)\equiv\int q(\Psi,\Psi')e_\alpha(\Psi')w_\Gamma(d\Psi')
\ee 
for each $\alpha$. Each $e^\alpha$ is a real-valued Borel measurable function (not an equivalence class of functions) of $\Psi$.
From the expression for $e^\alpha(\Psi)$, we see that the map $\phi$ of $\Psi$ to a vector ${\bf v}=\phi(\Psi)$ in $\cal H$ 
is constant on congruence classes, which is property (ii); members of a congruence class cannot be separated using $Q$.
A general function $f$ in $V/V_0$ can be represented by the sum $f=\sum_\alpha f^\alpha e_\alpha$, with $Q$-norm-square 
$\sum_\alpha|f^\alpha|^2<||f||_1^2$. Here the components are explicitly 
\bea
f^\alpha&=&\int\int f(\Psi) q(\Psi,\Psi')e_\alpha(\Psi')w_\Gamma(d\Psi)w_\Gamma(d\Psi')\non\\
&=&\int f(\Psi)e^\alpha(\Psi)w_\Gamma(d\Psi),
\eea
and this also holds for the components of the image in $V/V_0$ of an arbitrary $f\in{\cal L}^1$.  
Thus the image of $f$ as a vector in $\cal H$ is some sort of integral over $\phi(\Psi)$, which we will investigate below. 

For $w_\Gamma$-almost every given $\Psi$, the components $\phi(\Psi)^\alpha=e^\alpha(\Psi)$ (as above) are genuine functions 
of $\Psi$ (i.e.\ they are in ${\cal L}^1$) for each $\alpha$. These components define a vector ${\bf v}=\phi(\Psi)$ in ${\cal H}\cong \ell^2$ 
(the isomorphism using the orthonormal basis), provided that $\sum_\alpha \phi(\Psi)^\alpha\phi(\Psi)^\alpha<\infty$, which we already 
proved. 
Hence $\phi(\Psi)=\sum_\alpha\phi(\Psi)^\alpha e_\alpha$ does converge (in $Q$-norm). 
Then the inner product $\phi(\Psi)\cdot {\bf w}$ is a measurable function of $\Psi$ for any fixed ${\bf w}\in{\cal H}$ (it is the limit of 
partial sums over components, the partial sums are measurable, and the limit exists for almost every $\Psi$); such a map 
$\Psi\mapsto\phi(\Psi)$ is said to 
be weakly measurable. A function from a probability space into a separable Hilbert space is weakly measurable if and only if it is Borel 
measurable, that is, measurable with respect to the Borel $\sigma$-algebra induced from the $Q$-norm topology on $\cal H$ 
(Ref.\ \cite{reed_simon_book}, p.\ 116), so we have now proved Borel measurability of the map $\Psi\mapsto \phi(\Psi)$. By a similar 
argument, the $Q$-norm $||\phi(\Psi)||$ is also measurable.

For $fQg$, using the components $\phi(\Psi)^\alpha$, which are measurable functions of $\Psi$, and $f^\alpha$, $g^\alpha$, we have 
\bea
fQg&=&\sum_\alpha f^\alpha g^\alpha\non\\
&=&\sum_\alpha\int f(\Psi)\phi(\Psi)^\alpha\phi(\Psi')^\alpha g(\Psi') w_\Gamma(d\Psi) w_\Gamma(d\Psi')\non\\
&=&\int f(\Psi)\phi(\Psi)\cdot\phi(\Psi') g(\Psi')w_\Gamma(d\Psi) w_\Gamma(d\Psi')
\eea 
for all $f$, $g\in L^1$, which immediately gives property (i),
\be
\phi(\Psi)\cdot\phi(\Psi')=q(\Psi,\Psi')
\ee 
for $w_\Gamma\times w_\Gamma$-almost every pair $(\Psi,\Psi')$. 

A function $f\in {\cal L}^1$ maps to a vector in $\cal H$, 
which we can now write as $\int f(\Psi)\phi(\Psi)w_\Gamma (d\Psi)$, provided we make some remarks about integration of 
vector-valued functions. We know that $f(\Psi)\phi(\Psi)$ is a measurable function of $\Psi$, and so is its norm, while $||\phi(\Psi)||
\leq q(\Psi,\Psi)^{1/2}\leq 1$. It follows that $\int ||f(\Psi)\phi(\Psi)||w_\Gamma (d\Psi)<\infty$, and our vector-valued integral with 
values in $\cal H$ is well defined as a Bochner integral, the strongest sense of integrability in this context (see e.g.\ Ref.\ \cite{aliborder_book}, 
Section 11.8, and especially Theorem 11.44). We can of course define the push-forward of $w_\Gamma$, written $\phi_\ast w_\Gamma$, to 
a probability distribution on $\cal H$ (the DS distribution), by $\phi_\ast w_\Gamma(A)=w_\Gamma(\phi^{-1}(A))$ for any measurable set 
$A\subset \cal H$. 

Suppose that $\phi'$ is a measurable map into ${\cal H}'$ that satisfies property (i) (we denote the inner product and norm on ${\cal H}'$ by 
the same symbols as for $\cal H$; the meaning should be clear in context). Then the vector $\phi'(\Psi)$ defines a linear functional
on ${\cal H}'$ with norm $||\phi'(\Psi)||_{\rm op}=||\phi'(\Psi)||$. The Bochner integral can still be used, so any $g\in {\cal L}^1$ maps 
to a vector in ${\cal H}'$. Then, by using (i) and the earlier part of the proof, the operator norm must equal 
$||\phi'(\Psi)||_{\rm op}=||\phi(\Psi)||$ for $w_\Gamma$-almost every $\Psi$ (it must agree on $V$, and the rest follows). Then the map 
defined as taking $\phi(\Psi)$ to $\phi'(\Psi)$ is an isometry for $w_\Gamma$-almost every $\Psi$. Note that $\cal H$ as constructed above is 
the ``smallest'' one that satisfies the conditions and that, for that $\cal H$, ${\cal H}'$ would be isometrically isomorphic to 
${\cal H}\oplus{\cal H}''$ for some real Hilbert space ${\cal H}''$. This completes the proof. $\Box$

We emphasize that equality does not necessarily hold in (iii), in particular it does not necessarily follow from (i), because the 
set of pairs $(\Psi,\Psi')$ with $\Psi'=\Psi$ may have zero $w_\Gamma\times w_\Gamma$ probability. Note that we will prove some further 
statements in the same general setting as the Proposition after discussing a couple of examples.

We can obtain the corresponding reformulation of the full DS representation with little additional work. For any $\Psi$, define
$h=q(\Psi,\Psi)-||\phi(\Psi)||^2$. Then we obtain the push-forward of $w_\Gamma$ to a distribution on $({\bf v},h)\in{\cal H}\times \bbR_+$,
where $\bbR_+$ denotes the non-negative real numbers, with the DS distribution as the marginal on only $\bf v$.

Finally, we can show that what we have called the DS distribution actually agrees with the DS definition. Suppose we draw a sequence 
$({\bf v}_l)_{l\in \bbN}$, with ${\bf v}_l\in{\cal H}$, from $(\phi_\ast w_\Gamma)^\infty$. Then the matrix of off-diagonal overlaps 
is equal in distribution with that of a sequence $(\Psi_l)_l$ drawn from $w_\Gamma^\infty$ because of property (i). Similarly, 
the full versions agree for the matrices with diagonal elements included. Accordingly, we will now use the notation $w_{\Gamma{\rm DS}[X]}$
(or similar) for $\phi_\ast w_\Gamma$ constructed as above for an overlap $q_{[X]}$ (or so on).

\subsubsection{Examples for the DS distribution}

Next we consider in some examples how the DS map, for some overlap type, maps the pure-state decomposition $w_\Gamma$ 
of a given Gibbs state $\Gamma$ onto the DS distribution $w_{\Gamma{\rm DS}[X]}$ on Hilbert space. First, if $w_\Gamma$ consists solely
of atoms at pure states, then each pure state maps to a distinct vector in $\cal H$. In particular, for an atom, say at $\Psi_0$, 
there is nonzero $w_\Gamma\times w_\Gamma$ probability for a pair $(\Psi,\Psi')$ to both be $\Psi_0$, which implies by (i) that $\Psi_0$
maps to a vector $\bf v$ with $||{\bf v}||^2=q(\Psi_0,\Psi_0)$. Thus if $w_\Gamma$ is purely atomic, the DS map $\phi$ is one to one
almost everywhere.

Now suppose that $\Gamma$ is a DF state with respect to $q$, and also for the sake of argument that the pairwise $q$ overlaps differ from 
the $q$ self-overlap. Then because $w_\Gamma$ is atomless, there is zero probability 
of drawing the same pure state twice. The pairwise overlaps are almost surely equal, so (i) can be satisfied if $w_\Gamma$-almost 
every $\Psi$ maps to the same vector $\bf v$, with $||{\bf v}||^2$ equal to the pairwise overlap, which is strictly less 
than the self-overlap of the pure states. By uniqueness up to an isometry, this is the only form the DS distribution can have for a DF state
with respect to $q$ ($\cal H$ is one dimensional). The DS map is definitely not injective in this case. 
[In this case the topology induced on the uncountable set of pure states by the metric $d_{\rm tot}$ is discrete, so not separable.
An injective DS representation for this case would require a nonseparable Hilbert space,
with infinitesimal weight at each of an uncountable number of isolated linearly-independent vectors, which is impossible.]
Again, more generally, if there is a cluster of pure states (rather than all pure states) such that the pairwise overlaps $q$ of members
of the cluster are almost surely equal, and the total weight of the cluster is nonzero, then the cluster maps to an atom of 
$w_{\Gamma{\rm DS}}$.

Another example is a Gibbs state $\Gamma$, which is a distribution on spin configurations. By results in Sec.\ \ref{models}, the DS 
distribution of $\Gamma$ is the same as that of $w_\Gamma$. In the literature on SK models, the DS distribution has been used as a 
construction of an ``asymptotic Gibbs'' distribution \cite{ad,panchenko_book}. From the examples involving DF states, we see that 
in the case of Gibbs (i.e.\ DLR) states in short-range systems, such a distribution may not faithfully describe the weight $w_\Gamma$ 
of the Gibbs state, that is, if states like DF states can occur.

\subsubsection{Conditional distributions and local cluster states}

In the following, a central role is played by certain cluster states, which we call local cluster states (local in $\bf v$, and also for given $J$, 
$\Gamma$). These are based on a general construct, the conditional distributions that exist because we have a measurable map $\phi$ 
from pure states to vectors ${\bf v}\in{\cal H}$ (the existence of these conditionals is one advantage of the reformulation of the DS
distribution). This means we can consider the conditional 
distribution $w_{\Gamma{\bf v}}$ for pure states, conditioned on the vector $\bf v$ to which the pure states map (see Ref.\ 
\cite{einsward_book}, Ch.\ 5; in principle the conditional distribution depends on a choice of overlap type, such as $[X]$, but we will not 
show this in the notation). Then $w_\Gamma$ can be represented by an integral over the DS distribution, by the rules for 
conditional probability:
\be
w_\Gamma=\int w_{\Gamma{\rm DS}[X]}(d{\bf v})w_{\Gamma{\bf v}}.
\ee  
This is completely general, and holds in the setting of Proposition 5. In our situation, from the conditional distribution, we obtain 
the local cluster states in the usual way,
\be
\Gamma_{\bf v}=\int w_{\Gamma{\bf v}}(d\Psi)\Psi,
\ee
and so we also obtain a decomposition of $\Gamma$ as an integral over $\bf v$ of $\Gamma_{\bf v}$.
The decomposition of $w_\Gamma$ shown is unique, given $w_{\Gamma{\rm DS}[X]}$ on $\cal H$
and the DS map from pure states into $\cal H$. It is invariant under the orthogonal transformations of $\cal H$, which cancel
when the integral over $\bf v$ is taken. The local cluster states will play a role analogous to that of pure states when we work 
with the DS distribution. We note that the local cluster states for distinct $\bf v$ are mutually singular when viewed either as distributions 
on pure states or on spin configurations. 

With the conditional probability, we can discuss a converse to the result of Sec.\ \ref{subsec:ginv} above, which again is very general. 
Suppose there is an atom of $w_{\Gamma{\rm DS}[X]}$ (or for any other overlap type), say at ${\bf v}_0$. 
Because it is an atom, there is nonzero $w_\Gamma\times w_\Gamma$ probability of drawing a pair
$(\Psi,\Psi')$ of pure states that both map to ${\bf v}_0$ under $\phi$. By property (i), the norm-square $||{\bf v}_0||^2=q_{[X]}(\Psi,\Psi')$ 
for $w_{\Gamma{\bf v}_0}\times w_{\Gamma{\bf v}_0}$-almost every such pair. This means the local cluster state at ${\bf v}_0$ 
is either trivial or a DF state for $[X]$. 

Another easy deduction is the following. For this, $w_{\Gamma{\rm DS}[X]}$ does not need to be atomic. Suppose (for some overlap type, say 
$[X]$) that $\Psi$, $\Psi'$ are both mapped to the same $\bf v$ by $\phi$, ${\bf v}=\phi(\Psi)=\phi(\Psi')$. Then, for any $g\in{\cal L}^1$, 
$Qg(\Psi)=Qg(\Psi')$ almost surely (using the notation of the proof of Proposition 5). So conditionally on $\bf v$, $\Psi\sim\Psi'$ almost surely. 
That is, for $w_{\Gamma{\rm DS}[X]}$-almost every $\bf v$, $\Psi$ and $\Psi'$ belong to the same congruence class, 
$w_{\Gamma{\bf v}}\times w_{\Gamma{\bf v}}$-almost surely; this is a strong converse to property (ii). So for 
$w_{\Gamma{\rm DS}[X]}$-almost every $\bf v$, the conditional $w_{\Gamma{\bf v}}$ at $\bf v$ is dominated by a single congruence 
class, in the sense that one class has $w_{\Gamma{\bf v}}$ probability one.

For a strong converse to the examples above, consider the following. Draw $\bf v$ from $w_{\Gamma{\rm DS}[X]}$ for any overlap type,
such as $[X]$, and again write $q$ for the overlap. In $\cal H$, form the ball $A_{{\bf v}\varepsilon}=\{{\bf v}':||{\bf v}'-{\bf v}||
\leq\varepsilon\}$. On the space of $\Psi$, form the conditional distribution $w_{\Gamma{\bf v}\varepsilon}$ of $w_\Gamma$ conditioned 
on $\phi^{-1}(A_{{\bf v}\varepsilon})$, noting that the probability $w_\Gamma(\phi^{-1}(A_{{\bf v}\varepsilon}))>0$ for 
$w_{\Gamma{\rm DS}[X]}$-almost every $\bf v$. It is obvious that as $\varepsilon\to 0$, $w_{\Gamma{\bf v}\varepsilon}$ converges
in the weak* sense to $w_{\Gamma{\bf v}}$ (note the latter 
is essentially unique \cite{einsward_book}). Drawing $\Psi_1$, $\Psi_2$ from $w_{\Gamma{\bf v}\varepsilon}\times w_{\Gamma{\bf v}
\varepsilon}$, using the definition of $A_{{\bf v}\varepsilon}$ and property (i), but here for $w_{\Gamma{\bf v}\varepsilon}$ in place of 
$w_\Gamma$ (which is a consequence), we have, for any $\varepsilon$,
\bea
-\varepsilon^2-2\varepsilon||{\bf v}||&\leq&q(\Psi_1,\Psi_2)-||{\bf v}||^2\non\\
&\leq&{}\varepsilon^2+2\varepsilon||{\bf v}||,
\eea
for $w_{\Gamma{\bf v}\varepsilon}\times w_{\Gamma{\bf v}\varepsilon}$-almost every $(\Psi_1,\Psi_2)$. Hence it is clear that the 
probability distribution on $q$ induced from $w_{\Gamma{\bf v}\varepsilon}\times w_{\Gamma{\bf v}\varepsilon}$ on $(\Psi_1,\Psi_2)$
tends to a $\delta$-function as $\varepsilon\to0$. Thus in the limit we can conclude that, for $w_{\Gamma{\rm DS}[X]}$-almost every 
$\bf v$, $q(\Psi_1,\Psi_2)=||{\bf v}||^2$ for $w_{\Gamma{\bf v}}\times w_{\Gamma{\bf v}}$-almost every $(\Psi_1,\Psi_2)$. 
This includes, but is not limited to, the special case of atoms in $w_{\Gamma{\rm DS}[X]}$.
(Again, this result is very general.) An alternative slick way to obtain the same result is to consider the DS distribution of 
$w_{\Gamma{\bf v}}$ itself. Clearly it must be a single atom, with the same conditional with which we began, which implies the result.

In order to make statements for all $[X]$ at once, we turn to the use of a total overlap, and discuss how the DS distribution 
depends on the choice. Recall that we defined the more general overlaps $\sum_{[X]}a_{[X]}q_{[X]}(\Psi,\Psi')$, which we 
here denote as $q_a$, writing the indexed set of coefficients as $a=(a_{[X]})_{[X]}$. We say $a\geq0$ ($a>0$) if $a_{[X]}\geq0$ 
for all $[X]$ (resp., $a_{[X]}>0$ for all $[X]$). A total overlap was defined as such a linear combination with $a>0$ and 
$\sum_{[X]}a_{[X]}<\infty$, so $a$ is in the sequence space $l^1$, and the set of such $a$ is a cone, which is the interior of the cone 
$\{a:a\in l^1, a\geq0\}$. For each $q_a$, $a\geq 0$, we have a corresponding pseudometric $d_a$. We also define the corresponding
positive-semidefinite bilinear forms $Q_a$ for $a\geq 0$ as before, and use the corresponding notation $\phi_a$, ${\cal H}_a$, $||{\bf v}||_a$,
and $w_{\Gamma{\rm DS}a}$ to show the dependence on $a$. 

Now consider the effect of a change from $a$ to $a+\delta a$, 
where $\delta a\geq 0$, $\delta a\in l^1$ ensures this is still in the cone in $l^1$ (also assume $\delta a$ is not identically zero). 
Then $Q_{a+\delta a}=Q_a+Q_{\delta a}$ and, referring to the proof of Proposition 5, we note that if $f\in{\cal L}^1$ is in the kernel
of $Q_a$, it may or may not be in the kernel of $Q_{\delta a}$. Then the kernel $\ker Q_{a+\delta a}$ may be a strict subspace of 
$\ker Q_a$, in which case we have a proper inclusion ${\cal H}_{a+\delta}\subset{\cal H}_a$ as a closed subspace.
This means that some points in the range of $\phi_a$ may split on passing to the range of $\phi_{a+\delta }$, and if so the weight
in the DS distribution at that point is distributed among the resulting points; the reverse effect that points merge on adding $\delta a \geq 0$ 
cannot occur. If $\ker Q_{a+\delta }=\ker Q_a$, then we can identify ${\cal H}_a$ with ${\cal H}_{a+\delta}$
and again simply call them $\cal H$. Then there is an invertible bounded linear map of $\cal H$ into itself, such that $\phi_a(\Psi)$ maps to 
$\phi_{a+\delta a}(\Psi)$ for (almost every) $\Psi$. Now because $a$, $\delta a\in l^1$, their entries are bounded, and if $a>0$ is in the 
interior of the cone, then there exists a finite $\lambda>1$ and a $\delta a'\geq0$ such that $a+\delta a+\delta a'=\lambda a$ 
(where $\lambda a$ is defined as multiplication by a scalar on $l^1$), which clearly brings $\phi$ back to $\lambda^{1/2}\phi$. Hence for 
$a>0$, passing to $a+\delta a$ can always be described by an invertible linear map on $\cal H$.
Put another way, for $a>0$,
\be
\ker Q_a=\bigcap_{[X]} \ker Q_{[X]}
\ee 
and is independent of $a$, as long as $a>0$. So all $a>0$ give DS distributions that are the same up to an invertible linear map
on $\cal H$. This means the local cluster states $\Gamma_{\bf v}$ and their weights $w_{\Gamma{\bf v}}$ are covariant under 
a change in $a$; only the label $\bf v$ changes, and the DS distribution $w_{\Gamma{\rm DS}a}$ changes only by a Jacobian factor.

Combining these results, we see that if $q=q_a$ then, for the conditional distributions $w_{\Gamma{\bf v}}$ 
for $q_a$, $q_a(\Psi_1,\Psi_2)$ is the same for $w_{\Gamma{\bf v}}\times w_{\Gamma{\bf v}}$-almost every $(\Psi_1,\Psi_2)$
and for all $a>0$. Then varying $a$ by $\delta a\geq 0$ (and using covariance in $\bf v$), we see that $q_{[X]}$ is constant for 
$w_{\Gamma{\bf v}}\times w_{\Gamma{\bf v}}$-almost every pair and all $[X]$. It now follows, exactly as in the argument 
given earlier in this Section, that the local cluster state $\Gamma_{\bf v}$ is either trivial 
or a DF state, and this holds for $w_{\Gamma{\rm DS}[X]}$-almost every $\bf v$. [Note that in the case of a trivial state, the self-overlap
$q_a(\Psi,\Psi)$ is the same as $||\phi_a(\Psi)||^2$, while for a DF state it is strictly larger (Lemma 1 again).] Thus this can be viewed as a 
generalization of the earlier result from a $g$-invariant Gibbs state to any local cluster state. 
From now on we assume $a>0$ whenever $q_a$ is used.

For local cluster states, we have the pseudometric
\be
d_a(\Gamma_{\bf v},\Gamma_{{\bf v}'})=||{\bf v}-{\bf v}'||_a.
\ee
This is in fact a metric on local cluster states when they are drawn from $w_{\Gamma{\rm DS}a}$ (as well as being a metric 
on vectors in $\cal H$): if $d_a=0$, then $d_{[X]}=0$ for all $[X]$, and then a version of Lemma 1 implies that the
$\kappa^\dagger (w_{\Gamma{\rm DS}a}\times w_{\Gamma{\rm DS}a})$ probability that this holds and that
$\Gamma_{\bf v}\neq\Gamma_{{\bf v}'}$ is zero. 

\subsubsection{Extension of earlier results}
\label{subsubsec:earl}

We can consider whether results obtained earlier that were phrased in terms involving pure states also hold if $\Psi$s
(which almost surely will be pure states) are replaced by vectors $\bf v$, and $w_\Gamma$ by $w_{\Gamma{\rm DS}[X]}$ for some $[X]$ 
or $\rm tot$. First, for the covariance properties of the DS distribution, translation covariance is unaffected. 
Local transformations should be considered as applying to $w_{\Gamma{\rm DS}a}$ with the local cluster states
in the role formerly played by pure states. We pointed out in Sec.\ \ref{models} that similar forms hold for any partition
of $\Gamma$ into disjoint sets of pure states, provided the partition itself transforms covariantly. Here a ``partition'' really means
a sub-$\sigma$-algebra of the Borel $\sigma$-algebra of the space of pure states, and in the present case the sub-$\sigma$-algebra is 
determined by the measurable map $\phi$ \cite{einsward_book}. The map is determined by use of the overlap $q_a$, which is invariant
under a local transformation, when the values for corresponding pairs of $\Psi$s are compared. Thus eq.\ (\ref{loctransweight}) 
applies here with $w_{\Gamma{\rm DS}a}$ in place of $w_\Gamma$, $\bf v$ in place of $\Psi$, and expectation in $\Gamma_{\bf v}$ 
in place of expectation in $\Psi$. The DS distribution is only defined up to an $O({\cal H})$ transformation, but we can choose 
an $O({\cal H})$ transformation for any $\theta_{\Delta J}$ so that it defines a partition into $\Gamma_v$s that transforms covariantly.
(Actually, if we use the canonical space $\cal H$ from the proof of Proposition 5, it is automatically covariant.) 

Each result obtained previously for pure-state decompositions now also applies here, provided the proof did not use aspects of $\Psi$s
being pure states, other than the uniqueness of the decomposition of $\Gamma$ into pure states. We can use the same $\sigma$-algebras 
of invariant sets ${\cal I}_n$ of, for example, for $n=2$, $(J,\Gamma,\Psi)$, but now in practice the dummy label $\Psi$ 
will take values in the local cluster states $\Gamma_{\bf v}$ for $\Gamma$, which may depend on which $\Gamma$ they came from 
(unlike pure states, it is not clear if they can be defined without first considering a $\Gamma$). (In making these extensions of results, 
it may be useful to realize that the map $\phi_*\circ\widetilde{\phi}_*$ that maps $\Gamma$ to $w_{\Gamma{\rm DS}a}$ 
is measurable; here $\widetilde{\phi}_*$, which maps $\Gamma$ to $w_\Gamma$ is measurable and was defined in App.\ \ref{app:furth}.)

Thus for example, Proposition 2, the extended zero-one law, can be rephrased for covariant sets (or properties) 
as a zero-one law for $(\Gamma,\Gamma_{\bf v})$. The proof used a simple version of $\Sigma_a$ invariance for a choice of $a>0$, 
which goes through for the DS distribution as we have discussed. In fact, in that proof, two cluster states for disjoint clusters were formed.
When we do the same for two clusters of local cluster states, the resulting clusters are simply particular cases of those in the 
earlier proof, meaning it is not necessary to work through the entire proof again. The only difference now is that the zero-one law applies 
to $w_{\Gamma{\rm DS}a}$-almost every local cluster state, for $\kappa_J$-almost every $\Gamma$ and $\nu$-almost every $J$. 
If the metastate is indecomposable, it holds for $\kappa_Jw_{\Gamma{\rm DS}a}$-almost every $(\Gamma,\Gamma_{\bf v})$. 
An example application is that the norm-square $||{\bf v}||_a^2$ is the same for $w_{\Gamma{\rm DS}a}$-almost every local cluster 
state $\Gamma_{\bf v}$ for given $\Gamma$, and if $\kappa_J$ is indecomposable it is the same for $\kappa^\dagger 
w_{\Gamma{\rm DS}a}$-almost every $(J,\Gamma,\Gamma_{\bf v})$. (A corresponding statement was obtained for infinite-range models 
based on the GG identities; see Ref.\ \cite{panchenko_book}, Thm.\ 2.15.) This norm-square is also the supremum $\sup q_a$ of the overlap 
distribution. This result simplifies the relation between overlaps and $d_a$ pseudometrics on pairs of local cluster states, in the same way 
as the relation between overlaps and the $d_{[X]}$ pseudometrics on pairs of pure states was simplified when the self-overlaps of 
pure states are all equal.  

Another extension of an earlier result to the DS distribution for a total overlap, $q_a$, and to the decomposition of $\Gamma$ 
into local cluster states, concerns the character of the Gibbs states. First, we recall the NS09 result \cite{ns09}, which said that 
a Gibbs state cannot have a pure-state decomposition in which the number of atoms is strictly between 1 and infinity. The exact 
same proof applies for the atoms in the DS distribution: if the number of atoms is finite and larger than $1$, there is a contradiction 
(this uses the fact that the norm squares of $\bf v$ of the atoms are all equal, as just discussed). If the atoms are trivial states, 
this is nothing new, so it is strictly an extension only if the local cluster states of the atoms are DF states. (A similar result as this 
was obtained in Ref.\ \cite{ad}, though they only considered equivalence classes of finite numbers of pure states whereas 
we have DF states.)

Next, above we also proved Corollary 2, that the pure-state decomposition cannot be a mixture of both atoms and an atomless part. 
The proof used Proposition 2. So now we obtain precisely the same statement for the DS distribution: it is either purely atomic or atomless,
and in the case of atoms the number is either $1$ or countably infinite. Again, this is only an extension when the local cluster states 
are DF states. For an indecomposable metastate, this character of the DS distribution is the same for $\kappa^\dagger$-almost 
every $(J,\Gamma)$.

\subsubsection{Classification of Gibbs states by form of DS representation}

These various observations now lead us to propose a classification of Gibbs states drawn from a metastate. While one could 
propose classifications based on different properties according to taste, we propose to classify them according to the form 
of the DS representation. The rationale is based particularly on the fact that $\Sigma$ invariance (stochastic stability) works within 
the DS distribution, as we will discuss after this, and is likely to play a major role in future work on the structure of the Gibbs states. 

For the classification, we consider a choice of total overlap $q_a$, and find the DS distribution for given $\Gamma$. 
According to what was just proved, the DS distribution has one of three forms: it is either a single atom, or a countable infinity of
atoms, or atomless. We call these types I, II, and III, respectively. We can subdivide each type into one of two further types,
using the remaining information in the full DS representation of $w_\Gamma$:
recall that the norm-square of $w_{\Gamma{\rm DS}a}$-almost every vector $\bf v$ is the same, and the self-overlap of
$w_\Gamma$-almost every pure state is the same. Then either these two numbers are equal, or the second is larger than the first;
in the first case, the local cluster state is trivial, while in the second case it is a DF state.
We call these types a and b, respectively. This distinction can apply to any of types I, II, and III, so in all we have six types
(or {\em may} have; we do not prove that each type actually occurs in practice); see Table \ref{tab:gibbs}. In addition to abbreviations 
defined earlier, note that in the Table CP and CS refer to ``chaotic pairs'' (CP) and ``chaotic singles'' (CS), in which the metastate 
is nontrivial, but the Gibbs states are trivial; see Refs.\ \cite{ns96a,ns96b,read14,nrs_rev}. (CP and CS can arise, or be viewed, as 
another degenerate limit of RSB 
\cite{read14,hr}.) For an indecomposable metastate, $\kappa_J$-almost every Gibbs state is of the same one of these six types. In terms of 
the earlier characters of Gibbs states, characters (i) and (ii) are types Ia and IIa, respectively, while character (iii) has been divided into 
the four remaining types.

By what we proved earlier, this classification into types does not depend on the choice of a total overlap, 
and the I--II--III classification is independent of the choice of $a$. The a--b distinction is clearly unaffected by a 
change $\delta a>0$ within the cone $a>0$, because each (i.e.\ almost every) local cluster, described by $w_{\Gamma{\bf v}}$, 
remains either a single atom or atomless. So the full type classification is completely robust under a change in the choice of total overlap, 
as long as only $a>0$ is considered. 

\begin{table}
\begin{tabular}{lc|c|c|}
&&a: trivial&b: DF
\\ \hline
I:&trivial&RS, SD, CS, CP&DF
\\ \hline
II:&countable $\infty$&RSB&RSB?
\\ \hline
III:&atomless&RSB?&RSB?
\\ \hline
\end{tabular}
\caption{Table of forms of the DS representation for Gibbs states drawn from an indecomposable metastate, for some total overlap type 
$q_a$ ($a>0$) and $0<T<\infty$. Rows are labeled by the form of the DS distribution in Hilbert space, where 
``countable $\infty$'' refers to the number of atoms. Columns are labeled according to whether the local cluster states are trivial (pure) 
or DF. Names for some scenarios
in which a case occurs are indicated, with a ? when there is uncertainty; abbreviations are defined in the main text.}
\label{tab:gibbs}
\end{table}

Note that, in type II, $P_{J\Gamma a}(q)$ has a $\delta$-function 
at the supremum of $q_a$. Conversely, suppose there is a $\delta$-function at $\sup q_a$; note that this property is covariant, and for 
an indecomposable metastate is either true for almost every Gibbs state, or false for almost every one. As the supremum is attained
only by the norm-square of a vector in $\cal H$, the $\delta$-function means there is nonzero probability of drawing the same
cluster (i.e.\ vector $\bf v$) twice, and so $\Gamma$ must be type II. Type IIa is the form 
found in the interpretation of most cases of non-trivial standard RSB mean-field solutions. Type III does not seem to have arisen in that context 
as far as we are aware. Our six distinct types refine earlier descriptions of the allowed possibilities (e.g.\ Ref.\ \cite{nrs_rev}).

\subsubsection{DS distribution and $g$ evolution}
\label{subsubsec:gevo}

The DS distribution is useful (perhaps essential) for constructing stochastic stability or $\Sigma$ evolution arguments also \cite{aa,ad,argch}. 
When we attempted to define the $g_{[X]}$ evolution in Sec.\ \ref{sec:gevol}, we wanted to have centered Gaussian random fields with 
$\bbE_gg_{[X]}(\Psi_1)g_{[X]}(\Psi_2)=q_{[X]}(\Psi_1,\Psi_2)$ for all $\Psi_1$, $\Psi_2$  including for $\Psi_1=\Psi_2$. 
But this is more than is required in order to reproduce the desired $\Sigma$ evolution. $\lambda g_{[X]}(\Psi)$ always appears 
inside an expectation under $w_\Gamma$, and as we noted there, the self-overlaps $q_{[X]}(\Psi,\Psi)$ of a $\Psi$ generated 
by expanding and doing the $g$ expectation order by order under an integral cancel. On the other hand, if one considers 
$\int w_\Gamma(d\Psi)q_{[X]}(\Psi,\Psi')^m$, $m$ an integer, where $\Psi'$ is a spectator also drawn from $w_\Gamma$, 
or other functions of the overlap under the integral, then these are reproduced by the DS distribution and integration over $\bf v$ 
in place of $\Psi$ [but  $\int w_\Gamma(d\Psi)q_{[X]}(\Psi,\Psi)$ may not be].
Only integrals of this form, for which the DS distribution reproduces the result
of using $w_\Gamma$, actually arise. (This is an informal account, but should give the idea.) 

Then for $g$ itself, we can use simply
a Gaussian field $g_{[X]}({\bf v})$, a centered Gaussian random field on $\cal H$, with covariance ${\bf v}\cdot{\bf v}'$. Such a 
random field definitely exists, and can be constructed by using an independent standard Gaussian variable for each vector in an 
orthonormal basis, such as $e_\alpha$ above; then $g_{[X]}({\bf v})$ is a linear combination of these random variables, with coefficients 
the components of $\bf v$ in that basis, and is almost surely a measurable function of $\bf v$ \cite{ad,argch}. Then we can view $g_{[X]}$ 
evolution as acting on $w_{\Gamma{\rm DS}[X]}$; 
\be
w_{\Gamma{\rm DS}[X]}(d{\bf v})^{\lambda g_{[X]}}=\frac{e^{\lambda g_{[X]}({\bf v})-\frac{1}{2}\lambda^2
||{\bf v}||^2}w_{\Gamma{\rm DS}[X]}(d{\bf v})}
{\bbE_{w_{\Gamma{\rm DS}[X]}}e^{\lambda g_{[X]}({\bf v}')-\frac{1}{2}\lambda^2||{\bf v}'||^2}}
\label{eq:ginvDS}
\ee
is then a genuine measure (and mutually absolutely continuous with $w_{\Gamma{\rm DS}[X]}$), in which 
$||{\bf v}||^2$ now appears in place of $q_{[X]}(\Psi,\Psi)$.
This probably makes the construction of $g$ or $\Sigma$ evolution much better defined than in our earlier description, 
and gives rise to desirable continuity properties of $\Sigma_t$ \cite{argch}. We can pull 
the construction back to make better sense of $g$ evolution in terms of $w_\Gamma(\Psi)$ itself, by defining
$g_{[X]}(\Psi)=g_{[X]}({\bf v})|_{{\bf v}=\phi(\Psi)}$. Note, however, that
earlier we wanted $g_{[X]}(\Psi)$ to be a random function of $\Psi$, independent of $\Gamma$, whereas here the construction
involves $\Gamma$. It is also worthwhile to notice that, for the increments of the original $g_{[X]}(\Psi)$ we had eq.\ (\ref{eq:siginc}), 
which involved the pseudometric $d_{[X]}$. But now if we use $g_{[X]}({\bf v})$, the increments have the same form, but involve 
$d_{[X]}(\Gamma_{\bf v},\Gamma_{{\bf v}'})=||{\bf v}-{\bf v}'||_{[X]}$ (i.e.\ the $Q$ norm based on $q_{[X]}$) instead. 
This observation does not affect what we said above.

As an example of use of the $\Sigma$ evolution in terms of the DS distribution, we consider again the problem of finding the Gibbs states
$\Gamma$ that are $g$ invariant for all $[X]$. Take the DS distribution of $\Gamma$ for some $[X]$. We want 
$w_{\Gamma{\rm DS}[X]}^{\lambda g_{[X]}}=w_{\Gamma{\rm DS}[X]}$ for all $\lambda>0$ and almost every $g_{[X]}$.
Clearly this does not happen unless $w_{\Gamma{\rm DS}[X]}$ is a single atom. If this holds for all $[X]$
then (from the definition of a DF state) it immediately implies that $\Gamma$ is either trivial or a DF state. 

All of this is helpful when we wish to consider $\Sigma$ evolution for all $\Sigma_{[X]}$ simultaneously. 
In terms of distributions on pure states, it makes sense to pick one $a>0$ and use the corresponding DS distribution
to express $w_\Gamma$ as above, and use the transformation above to express any desired $\Sigma_{a+\delta a}$ evolution
on $w_\Gamma$ in terms of this one choice. We use this point of the view in the following. The local cluster states are 
invariant under $g_{a}$ evolution or $g_{\delta a}$ evolution, $\delta a\geq0$,
or in particular $g_{[X]}$ invariant for every $[X]$ (note that when referred to the DS Hilbert space for $q_a$, the covariance
of $g_{[X]}({\bf v})$ will not be ${\bf v}\cdot{\bf v}'$, but some other bilinear form that depends on $[X]$). 
In fact, by what we proved above, they would be invariant even if we set up $g$-evolution for $w_{\Gamma{\bf v}}$
instead of for $w_{\Gamma{\rm DS}a}$ (but note we have no basis for asserting that they must be invariant for this version, 
and care must be taken to distinguish these different statements).
The $g$ invariant Gibbs states are a special case.

Thus type I Gibbs states are precisely the $g$-invariant Gibbs states, while types II and III are not $g$-invariant (though their local cluster 
states are), and $\Sigma$ invariance of the metastate becomes an important resource when characterizing metastates. The difference between 
types a and b is completely invisible to $\Sigma$ evolution, so arguments based on the latter cannot help us rule out one of a and b 
(e.g.\ for a given type I, II, or III) if it does not rule out the other. 

\subsubsection{Further structure of Gibbs states}
\label{subsubsec:furthgibbs}

We have seen that if, for $\Gamma$ drawn from a metastate $\kappa_J$, the overlap distribution $P_{J\Gamma a}(q)$ is a single 
$\delta$-function for any $a>0$, then the Gibbs state is either trivial or a DF state, in other words it is type I. For an indecomposable 
metastate this then holds for $\kappa_J$-almost every Gibbs state $\Gamma$. If instead the $\kappa_J$ expectation $P_{J a}(q)$ 
consists of a finite number, greater than $1$, of $\delta$-functions, then the DS distribution must consist only of atoms at isolated points, 
and the number of these must be countable, so such Gibbs states are type II (either IIa or IIb). If there are exactly two 
$\delta$-functions (with nonzero weight in each), then 
we already saw in Corollary 5, together with the following discussion, that (under some assumptions on the disorder) the distribution 
of what we may now identify as the (decreasing) cluster weights $w_{\Gamma{\rm DS}a}$ (the ``lumped'' weights) must be the PD 
distribution; thus these would be examples of both types IIa and IIb. Note that the pure state and overlap structure of these Gibbs states 
are ultrametric, and for the type IIb case the ultrametricity is somewhat nontrivial, in that there are two levels of the ultrametric tree, 
one for overlaps of distinct pure states within each cluster, the other for those between the clusters.

More generally, we can extend the results of Sec.\ \ref{subsec:pdultr} to the present setting. For type II Gibbs states, whether type a or b, 
there is a countable infinity of atoms in the DS dstribution; the earlier results were cases of type IIa only. Now we need to use normalized 
overlaps such as $\widetilde{q}_{[X]}$
as before, except that here they are normalized by the common value of $||{\bf v}||^2$ in the DS Hilbert space corresponding to
the same choice of overlap, not necessarily by the self-overlap of a pure state as used in Sec.\ \ref{subsec:pdultr}. Then Corollary 4$'$ 
extends immediately to say that, for an indecomposable 
metastate with Gibbs states that are almost-surely type II, the distribution of the weights of the DS atoms is almost surely $PD(x_1)$, 
with a fixed parameter $x_1\in(0,1)$. Note that a PD distribution cannot occur for Gibbs states of types I (for which the DS distribution
is trivial) or III (where there are no atoms in the DS distribution). 

Similarly, if there are $k+1\geq 2$ atoms in $P_{J a}(q)$, so the $q_a$ overlap takes only $k+1$ values, then the 
results of Sec.\  \ref{subsec:pdultr} apply to these type II Gibbs states, and Corollary 4$''$ extends to say that ultrametricity 
holds, the weights of the DS atoms together with the overlaps (the ROSt) follow a Ruelle cascade, and the overlaps are non-negative. 
In other words, the only possible form is exactly as in $k$-RSB. The comments about overlap equivalence extend to the present case also.
Note that again, the additional cases covered here, which are type IIb, are ultrametric in terms of overlaps of pure states, not only of 
local cluster states.

This is as far as we can go using the methods presently available. We have not proved ultrametricity or the Ruelle cascade form
for the remaining cases of type II or any cases of type III. But because of the results we do have, it is reasonable to expect
that these properties hold in the remaining cases, and that it may be possible to prove this based again on stochastic stability.
Some other results should also be mentioned in this connection \cite{bk,bs}.

As a final remark for this Section, we point out that, for Gibbs states with non-trivial DS distribution (those of types II and III),
there are infinitely many local cluster states involved, and so there is an infinity of distinct ways for the state to change under $g$ evolution.
In terms of the discussion of the topological dimension of the support of the metastate in Sec.\ \ref{subsubsec:zero}, this implies that 
the support
of the metastate at such a Gibbs state must be infinite dimensional. The contrapositive statement is that, if at some $\Gamma$ 
the support of the metastate is finite dimensional, then that Gibbs state must be type I; this strengthens the earlier statement, where 
the hypothesis was zero-dimensional support at some point rather than finite dimensional. The stronger statement only becomes possible by 
using the more detailed type classification, which was enabled by the analysis of this Section.

\section{Further analysis of metastates and the MAS}
\label{sec:mas}

The results of this section are largely independent of those for Gibbs states in Sec.\ \ref{trivmet}, but we make use of the DS distribution that 
was described in Subsec.\ \ref{subsec:gibbs}.

\subsection{Relative weak mixing}

To round out the discussion, we turn finally to the MAS $\rho_J$ of an indecomposable metastate $\kappa_J$, making use of results from earlier 
sections. For the MAS $\rho_J$ of an indecomposable metastate $\kappa_J$, the questions 
that arise at this stage mainly concern what happens when two pure states are drawn from it independently, that is, when 
a pair $(\Psi_1,\Psi_2)$ is drawn from $\mu_J\times\mu_J$ at given $J$ (of course, one can also consider drawing more than two;
recall that $\mu_J$ is the pure-state decomposition of the MAS $\rho_J$). 
In particular, we can ask about the overlap distributions $P_{J\rho[X]}(q)$ induced from $\mu_J\times\mu_J$. We recall that
these distributions are always self-averaged \cite{ns96a}. 

One of the main results of Ref.\ \cite{read14} was that RSB theory predicts that $P_{J\rho[X]}(q)$ is trivial, a single $\delta$-function at 
$q=q_{[X]}(x=0)$, at least for $[X]=[1]$; this was mentioned as a possibility by NS, as far back as Ref.\ \cite{ns96a}. (In fact, for overlaps 
$q_{[1]\Lambda}$ defined for finite regions $\Lambda$, it was shown \cite{read14} by rescaling the deviation 
of $q$ from its expectation that the distribution of the scaled variable is a Gaussian when $d>6$, a central-limit theorem type of result, but that
will not enter the present discussion.) That is, the $[X]$ overlap is $\mu_J\times\mu_J$-almost surely a constant. RSB also predicts that the 
constant $q_{[X]}(0)$ must be less than or equal to $\inf q_{[X]}$, the infimum of the support of $P_{J[X]}(q)$. (The inequality could be strict, 
but we will have little to say about that.) The constancy of the overlap
in no way contradicts the possibly non-trivial distribution of overlaps within a single $\Gamma$, provided it occurs when the metastate 
$\kappa_J$ is atomless, so that there is zero probability of drawing the same $\Gamma$ twice. In the contrary case in which both the 
metastate and Gibbs states drawn from it are trivial, it is also trivial that each $q_{[X]}$ is almost surely constant; in the following, we
will assume that either the metastate or Gibbs states drawn from it (or both) are nontrivial. Note that trivial $P_{J\rho[X]}(q)$
must occur for all $[X]$ in the case of a trivial metastate with the single Gibbs state non-trivial (a DF state), by results in the preceding 
Section.

We will discuss what happens when the $P_{J\rho[X]}(q)$ is trivial for {\em all} $[X]$ ($\nu$-almost surely). First, note that 
if $P_{J\rho[X]}(q)$ is trivial for all $[X]$, then the MAS $\rho_J$ is in fact a DF state \cite{hr}, as we defined the term in 
Sec.\ \ref{subsec:ginv}, 
and as in the special case of trivial metastate with a nontrivial Gibbs state.  

So far, we discussed the implications of $P_{J\rho[X]}(q)$ being trivial for all $[X]$. Next we will discuss conditions under which it may hold
(when not already excluded).
We discussed in Sec.\ \ref{indecomp} how a $\Pi$-invariant $\kappa^\dagger$ is extremal if and only if it is trivial on the $\sigma$-algebra
of invariant sets ${\cal I}_1$, or alternatively ${\cal I}_1(\kappa^\dagger)$ of sets of pairs $(J,\Gamma)$.
We have also seen in Corollary 3$'$ that, for an indecomposable metastate, $\mu^\dagger=\nu \mu_J$ is trivial on 
the sub-$\sigma$-algebra ${\cal I}_1(\mu^\dagger)$, here of sets of pairs $(J,\Psi)$: that is, any almost-surely invariant set of 
pure states has $\mu_J$ probability $0$ or $1$, $\nu$-almost surely. 
We further mentioned earlier that, in terms of ergodic theory, $\kappa^\dagger$ is analogous to an ergodic invariant extension of 
an ergodic invariant $\nu$, and so also is $\mu^\dagger$. A very natural, and much discussed, stronger property is that an extension may be 
{\em relatively weak mixing}, also known as a weakly mixing extension \cite{einsward_book,glasner_book}; the terms 
``relative'' and ``extension'' refer to $\nu$ (this stronger property for an extension is the analog 
of weak mixing as a strengthening of ordinary ergodicity \cite{einsward_book}; strong mixing will not enter the present discussion). 
This motivates the following definition: for an indecomposable metastate $\kappa_J$, we say that 
$\mu^\dagger$ is {\em relatively weak mixing}, or is a {\em weakly mixing extension}, if $\nu(\mu_J\times \mu_J)$ is trivial 
on ${\cal I}_2(\nu(\mu_J\times \mu_J))$. It will also sometimes be convenient to apply the same term to any of
$\kappa_J$ or $\mu_J$, as well as $\mu^\dagger$, that is, to a metastate or to (the pure-state decomposition of) a MAS.

If $\kappa_J$ is relatively weak mixing, then it follows immediately that, for each $[X]$, the pairwise overlap $q_{[X]}(\Psi_1,\Psi_2)$
is $\nu(\mu_J\times\mu_J)$- (and in view of ergodicity of $\nu$, $\mu_J\times\mu_J$-) almost surely constant. This implies that
$P_{J\rho[X]}(q)$ is trivial for all $[X]$. Thus we see that the latter property would be a direct consequence of a property that plays
a prominent role in modern ergodic theory. We will only prove statements about the overlaps, and we will use the same term ``relative weak 
mixing'' when the pairwise overlaps are constant, without considering the more strict definition discussed here.
Relative weak mixing seems like a reasonable property to hypothesize, even though it is not
possible in, for example, the case of an indecomposable metastate with $m>1$ atoms. Of course, relative weak mixing, in the strict sense, may 
be stronger than required; it could be that under some conditions $P_{J\rho[X]}(q)$ is trivial for all $[X]$, but that $\kappa_J$ is not relatively 
weak mixing in the full (or strict) sense. 

In the general ergodic-theory setting, relative weak mixing is strictly stronger than an extension being ergodic. Relative weak mixing for either
$\kappa_J$ or $\rho_J$, or even simply triviality of $P_{J\rho[X]}(q)$, require examination of properties involving two pure states drawn 
from two Gibbs states, and such results do not generally follow from the results on indecomposability so far. Consequently, further work is 
required in order to prove results in this direction. We will be content to discuss only the question of whether overlaps
of pure states drawn from $\mu_J$ are almost-surely equal [i.e.\ triviality of $P_{J\rho[X]}(q)$] (to which we refer as relative weak mixing). 
This leads to the somewhat surprising appearance of possible hidden symmetry in the metastate itself.

\subsection{Overlaps and singularity of Gibbs states}
\label{subsec:ovsing}

First, we consider overlaps of pure states drawn from two Gibbs states, where the latter are drawn independently from a metastate.
We will do this by applying $\Sigma$ invariance of the metastate, and use it to show that (as in the discussion in Sec.\ \ref{subsec:ginv} 
of $g$-invariant Gibbs states, which is a special case) 
the overlap between a pair of pure states $(\Psi_1,\Psi_2)$ is the same $w_{\Gamma_1}\times w_{\Gamma_2}$-almost surely, for 
$\kappa_J\times\kappa_J$-almost every pair of Gibbs states 
$(\Gamma_1,\Gamma_2)$. This result, which holds for any metastate, may be considered an extension of the zero-one law, or 
single-replica equivalence, of Proposition 1 or NRS23 \cite{nrs23} to the case of pairs of pure states drawn independently from $\mu_J$, 
and is a step towards the relative weak mixing question (technically, we prove it under the slightly stronger conditions 
that there is a Gaussian $J^{(2)}$ piece in the disorder). The proof is modeled on that of Proposition 2. The result implies that the overlap
of the two Gibbs states themselves is equal to the common value for the pure states. In addition, the result 
and its proof hold without change if we use two metastates, rather than two copies of the same one, when drawing two Gibbs states, 
and also even if the two metastates are for different Hamiltonians, temperature, or disorder 
distributions. For simplicity, we delay describing these generalizations until afterwards.

We will prove
\newline{\bf Proposition 6}: Under the same hypotheses as in Theorem 3, for $\kappa_J\times\kappa_J$-almost every given pair 
$(\Gamma_1,\Gamma_2)$
and for every $[X]$ the overlap $q_{[X]}(\Psi_1,\Psi_2)$ has the same value for $w_{\Gamma_1}\times w_{\Gamma_2}$-almost every pair 
$(\Psi_1,\Psi_2)$. [We emphasize that the value of the overlap $q_{[X]}(\Psi_1,\Psi_2)$ can depend on  $(\Gamma_1,\Gamma_2)$.]
\newline
{\bf Proof}: We will use the indicator function ${\bf 1}_{A_J}$ for the covariant set $A_J$ of pairs $(\Psi_1,\Psi_2)$ based on their overlap 
$q_{[X]}(\Psi_1,\Psi_2)$; if the limit involved in the definition of the overlap does not exist, we will here define 
$q_{[X]}(\Psi_1,\Psi_2)=-\infty$. Then let $A=\{(J,\Psi_1,\Psi_2):q_{[X]}(\Psi_1,\Psi_2)\geq q_0\}$ ($q_0$ a real number). 
$A$ is a member of ${\cal I}_2$, the invariant sets of $(J,\Psi_1,\Psi_2)$, and in practice $\Psi_{1,2}$ will be pure states 
almost surely. 
We start with one pure state $\Psi_1$ drawn from $w_{\Gamma_1}$, 
for $\Gamma_1$ drawn from $\kappa_J$, and for now keep $\Gamma_1$, $\Psi_1$ fixed. 
Then we are interested in the set $A_{J\Psi_1}$, together with its weight $w_{\Gamma_2}(A_{J\Psi_1})$
for $\Gamma_2$ drawn independently from $\kappa_J$, and in the expectation over $\Gamma_2$ of a bounded continuous function $f$ of 
this weight, which can be written as 
\be
\bbE_{\kappa_J} f(\bbE_{w_{\Gamma_2}}{\bf 1}_{A_J}(\Psi_1,\Psi_2)).
\ee
Now we recall the $\Sigma$ invariance statement, eq.\ (\ref{siginvbasic}), in which $f$ is a function of a state, $\Gamma_2$ here,
and here consider $\Sigma_af$ for some $a>0$ (i.e.\ for a choice of total overlap $q_a$; see Secs.\ \ref{models} and \ref{trivmet}).
For given $\Psi_1$, we want to construct $f$ so that $f(\Gamma_2)=f(\bbE_{w_{\Gamma_1}}{\bf 1}_{A_J}(\Psi_1,\Psi_2))$.
Such a function can be approximated uniformly by polynomials in the weight $w_{\Gamma_2}(A_{J\Psi_1})$, which lies in $[0,1]$.
${\bf 1}_{A_J}(\Psi_1,\Psi_2)=\Theta(q_{[X]}(\Psi_1,\Psi_2)-q_0)$, where here $\Theta$ denotes the Heaviside step function, $\Theta(x)=1$
if $x\geq 0$, $0$ otherwise (with $x$ allowed to be infinite). As the overlap takes values in $[-1,1]$, $\Theta$ can be approximated
by polynomials in its argument, which are dense (with respect to the supremum norm) in the space of continuous functions 
on the interval, and hence also, when applied to $q_{[X]}(\Psi_1,\Psi_2)-q_0$, dense in $L^1(w_{\Gamma_2})$. 
The expectation of a polynomial, such as $\bbE_{w_{\Gamma_2}}q_{[X]}(\Psi_1,\Psi_2)^m$ ($m=1$, $2$, \ldots), can 
be approximated, using the finite $W$ definition of the overlap $\widehat{q}_{X\Lambda}(\Psi_1,s^{(2)})$, as 
$\langle \widehat{q}_{[X]\Lambda}(\Psi_1,s^{(2)})^m\rangle_{\Gamma_2}$. This is because, as $\Lambda\to\infty$, 
$\widehat{q}_{X\Lambda}(\Psi_1,s^{(2)})\to q_{[X]}(\Psi_1,\Psi_2)$ $\Psi_2$-almost surely, so the same is true for the $m$th power.
Now in 
\be
\bbE_{\kappa_J}f(\langle \widehat{q}_{[X]\Lambda}(\Psi_1,s^{(2)})^m\rangle_{\Gamma_2})
\ee
(for $f$ now a polynomial), the expectations $\langle s_{\theta_\bx X}\rangle_{\Psi_1}$ appear merely as coefficients in a sum
and are viewed as fixed 
(all that will matter at the following step is that for the $\Psi_1$ and $\Psi_2$ of interest, the limit defining the overlap exists, 
$w_{\Gamma_1}\times w_{\Gamma_2}$-almost surely). Hence eq.\ (\ref{siginvbasic}) does apply to these polynomials, 
where $\Sigma_a$ acts only on each copy of $\Gamma_2$ and its pure-state 
decomposition. Then the $W\to\infty$ limit can be taken through the expectation (by e.g.\ bounded convergence); note that then, 
for the overlap, $\Sigma_a q_{[X]}(\Psi_1,\Psi_2)=0$. Using a polynomial approximation to $\Theta$, that limit can be taken also. 
With a final limit, we arrive at the general continuous $f$, however to ensure that $\Sigma_af$ is well defined and that the limit can be 
taken, we assume not only that $f$ is bounded and continuous on $[0,1]$, but also that its second derivatives are continuous and 
bounded on $(0,1)$; as in the proof of Proposition 2, this class of functions will suffice for the proof. For such $f$, we now have 
\be
\bbE_{\kappa_J}\Sigma_af(w_{\Gamma_2}(A_{J\Psi_1}))=0,
\ee
$\mu_J$-almost every $\Psi_1$, and $\nu$-almost every $J$.
The remainder of the proof is similar to that of Proposition 2. For given $\Psi_1$, $A_{J\Psi_1}$ is a covariant set of $\Psi_2$, analogous to 
$A_{J\Gamma}$ in the proof of Proposition 2. Then $\Sigma_a$ reduces to $\Sigma_a^*$ on functions of 
$W_1\equiv w_{\Gamma_2}(A_{J\Psi_1})$, $W_2\equiv w_{\Gamma_2}(A_{J\Psi_1}^c)$, and $\Gamma_2$ can be 
viewed as a mixture of cluster states $\Gamma_{21}$, $\Gamma_{22}$, with weights $W_1$, $W_2$. Then a very similar argument shows 
that $w_{\Gamma_2}(A_{J\Psi_1})=0$ or $1$, for $\kappa_J$-almost every $\Gamma_2$, and this holds 
for $\kappa^\dagger w_{\Gamma_1}$-almost every $(J,\Gamma_1,\Psi_1)$. 
Now by varying $q_0$, this means that for given $\Psi_1$ and $\Gamma_2$, the overlap $q_{[X]}(\Psi_1,\Psi_2)$ does not depend on 
$\Psi_2$. As $(\Gamma_1,\Gamma_2)$ can be drawn simultaneously from $\kappa_J\times\kappa_J$ at the beginning, 
and then $(\Psi_1,\Psi_2)$ can be drawn from $w_{\Gamma_1}\times w_{\Gamma_2}$,
this means by symmetry that, for $\kappa_J\times\kappa_J$-almost every given pair $(\Gamma_1,\Gamma_2)$ and every $[X]$, 
the overlap $q_{[X]}(\Psi_1,\Psi_2)$ takes the same value for $w_{\Gamma_1}\times w_{\Gamma_2}$-almost every pair $(\Psi_1,\Psi_2)$.
$\Box$

By examining the proof, one can see that there is no reason why the same metastate must be used to choose both $\Gamma_1$, 
$\Gamma_2$; two metastates $\kappa_J^{(1)}$, $\kappa_J^{(2)}$ can be used instead. As symmetry between $\Psi_1$, $\Psi_2$ is then lost, 
the argument must be repeated with $1$ and $2$ switched to arrive at the conclusion, which is now for 
$\kappa_J^{(1)}\times\kappa_J^{(2)}$-almost every given pair $(\Gamma_1,\Gamma_2)$. Here the metastates were for the same 
disorder, so implicitly for the same Hamiltonian and temperature, but all of these can also be taken to be different. In general, the two systems 
can experience disorder $J_{(1)}$, $J_{(2)}$, respectively (not to be confused with the pieces $J^{(1)}$, $J^{(2)}$ in a single system), 
with a joint distribution $\nu(J_{(1)},J_{(2)})$ which may not be a product, so $J_{(1)X}$, $J_{(2)X}$ may not independent, though we will 
continue to assume independence between different $X$. Then the joint distribution for the two systems would be 
$\nu(J_{(1)},J_{(2)})(\kappa^{(1)}_{J_{(1)}}\times\kappa^{(2)}_{J_{(2)}})$, but the argument would go through with no other change
(note that the overlaps of pure states are still well defined, even when the distributions on $J$ are different, 
as the only requirement for that is translation invariance of the distributions). 

Reverting to the simpler version in the statement of Proposition 6, 
we point out that Proposition 6 implies that, for $\kappa_J\times\kappa_J$-almost every $(\Gamma_1,\Gamma_2)$, the overlap
between the two Gibbs states, $q_{[X]}(\Gamma_1,\Gamma_2)$, is equal to the common value $q_{[X]}(\Psi_1,\Psi_2)$ for 
$w_{\Gamma_1}\times w_{\Gamma_2}$-almost every pair of pure states. 
This implies that, for each overlap type, say $[X]$ or $a$, 
the DS distributions on the DS Hilbert space $\cal H$ for $\kappa_J$ and for $\mu_J$ are equal, 
\be
\kappa_{J{\rm DS}a}=\mu_{J{\rm DS}a},
\ee 
for any metastate. Note that the equality of the DS distributions implies 
that $\mu_{J{\rm DS}[X]}$ is locally covariant (and obviously translation covariant), just like a metastate (with vectors $\bf v$ in place of 
states $\Gamma$). Finally, we remark that if $\kappa_J$-almost every Gibbs states is trivial (as in the chaotic singles scenario) then we have 
$\kappa_J=\mu_J$ and the above result is obvious. The case in which almost every Gibbs state is a DF state is similar.
We will discuss the local cluster properties later in this Section. 

In the following, we will use the Lebesgue decomposition of two measures on the same space, so we briefly recall it here.
We have already defined absolute continuity of one measure $\mu_1$ with respect to another, $\mu_2$, written $\mu_1\ll \mu_2$. 
(Here these are not intended to be related to MASs, but are simply two measures.)
We also say that two measures $\mu_1$, $\mu_2$ are singular with respect to one another, or mutually singular, 
written $\mu_1\perp\mu_2$, if there are measurable sets $A$, $B$, such that $A\cap B=\emptyset$, $A\cup B$ is the whole space, 
$\mu_1(A)=0$, and $\mu_2(B)=0$ \cite{royden_book}; note that singularity is always a symmetric relation. By the Lebesgue decomposition 
\cite{royden_book,aliborder_book,garling_book}, given two measures $\mu^{(1)}$, $\mu^{(2)}$ , there are measures 
$\mu^{(1)}_1$, $\mu^{(1)}_2$ such that $\mu^{(1)}=\mu^{(1)}_1+\mu^{(1)}_2$, where 
$\mu^{(1)}_1\ll \mu^{(2)}$, and $\mu^{(1)}_2\perp \mu^{(2)}$ (hence $\mu^{(1)}_1\perp\mu^{(1)}_2$), and the decomposition is unique. 
Then we can also form the Lebesgue decomposition of $\mu^{(2)}$ with respect to $\mu^{(1)}_1$ (it is already singular relative 
to $\mu^{(1)}_2$), so $\mu^{(2)}_1\ll \mu^{(1)}_1$, and $\mu^{(2)}_2\perp\mu^{(1)}_1$. Thus $\mu^{(1)}_1$, $\mu^{(2)}_1$ 
are mutually absolutely continuous, or ``equivalent'', meaning that any set that has zero measure for one has zero measure 
for the other. We may imagine that  $\mu^{(1)}_1$, $\mu^{(2)}_1$ share a set on which they put full measure, while the other measures 
are disjoint (like a Venn diagram for two sets). 

We now sketch an argument, similar to that above, to show that, for $\kappa_J\times\kappa_J$-almost every pair 
$(\Gamma_1,\Gamma_2)$ drawn independently from a metastate $\kappa_J$, either the weights $w_{\Gamma_1}$, $w_{\Gamma_2}$ are 
mutually singular or
else $w_{\Gamma_1}=w_{\Gamma_2}$, that is, $\Gamma_1=\Gamma_2$. 
We will use some cluster states of these Gibbs states, with notation as defined in Sec.\ \ref{sec:siginv} (again, two parts will be sufficient).
Suppose $w_{\Gamma_1}$ and $w_{\Gamma_2}$ are not mutually singular. Using the Lebesgue decomposition of $w_{\Gamma_1}$, 
$w_{\Gamma_2}$, let $W_{\Gamma_i1}>0$ be the total weights of the set of pure states in $\Gamma_i$ ($i=1$, $2$) that are common to 
both $w_{\Gamma_1}$ and $w_{\Gamma_2}$ (i.e.\ the weights under the mutually absolutely continuous parts of $w_{\Gamma_i}$), and 
$W_{\Gamma_i2}$ the total weights of the two remaining sets of pure states, so $\sum_\alpha 
W_{\Gamma_i\alpha}=1$ for $i=1$, $2$. Define cluster states $\Gamma_{i\alpha}$ for those parts for which $W_{i\alpha}\neq0$.

Now we keep $\Gamma_1$ fixed, and consider the probability distribution for the weight $W_{21}$. 
Under $g_a$-evolution, the weights $w_{\Gamma_2}^{\lambda g_a}$ and $w_{\Gamma_2}$ are mutually absolutely continuous,
so the part of $w_{\Gamma_2}$ that is mutually absolutely continuous with part of $w_{\Gamma_1}$ remains so. This allows us to apply 
$\Sigma$ evolution to this probability distribution, similar to the proof of Proposition 6. (Because we do not have an explicit construction 
of the function as a limit of functions built on spins, as we did in proving Proposition 6, we do not insist that the present proof is rigorous.) 
The conclusion is again that $W_{21}=0$ or $1$ for $\kappa_J$-almost every $\Gamma_2$, and $\kappa_J$-almost every given 
$\Gamma_1$. By symmetry, the same holds for $\Gamma_1$, $\Gamma_2$ exchanged. Then $w_{\Gamma_i}$ are either mutually 
singular or equivalent, $\kappa_J\times\kappa_J$-almost surely. If they are equivalent then, by using Proposition 6, we have 
$d_a(\Gamma_1,\Gamma_2)^2=0$. To prove this claim, we separate cases. By Corollary 2, we can assume each $w_{\Gamma_i}$
is either purely atomic or else atomless. If one is atomic and the other atomless, they are mutually singular. If both are purely atomic,
then the overlap between any two pure states, one drawn from each $w_{\Gamma_i}$, must equal the self-overlap of any of the pure states 
involved by Proposition 6, and the claim follows. If both are atomless, then again almost every pairwise overlap of pure states is the 
same value, and again the claim follows. So in both cases, using a version of Lemma 1, $\Gamma_1=\Gamma_2$ (so $w_{\Gamma_1}
=w_{\Gamma_2}$). 

While the last argument here may have flaws, it strongly suggests that the conclusion is valid.
We should note that the possibility $\Gamma_1=\Gamma_2$ has $\kappa_J\times\kappa_J$ probability zero unless $\kappa_J$ 
has an atom at $\Gamma_1=\Gamma_2$, in which case it follows again from the proof (see the two cases near the end, and use Lemma 1 for 
the first) that $\Gamma_1$ is either trivial or DF, consistent with results in Sec.\ \ref{subsec:ginv}. Finally, as for Proposition 6, the argument 
also extends without essential change to two metastates, with the same
conclusion for $\kappa_J^{(1)}\times\kappa_J^{(2)}$-almost every given pair $(\Gamma_1,\Gamma_2)$ (note that two distinct 
indecomposable metastates are mutually singular on Gibbs states, in which case $\Gamma_1=\Gamma_2$ is not possible).  

\subsection{Hidden symmetry in metastates and MASs}

Using Proposition 6 in the preceding Subsection, for an indecomposable metastate we can now analyze the structure of the DS distribution of 
either the metastate or its MAS. This will reveal a hidden symmetry structure common to both of these; when the hidden symmetry is trivial
the relative weak mixing property holds. A heuristic argument then suggests that relative weak mixing will occur generically in models of the 
mixed $p$-spin type discussed in this paper. The arguments in the 
present Subsection are mostly independent of those in the preceding Sections, except for the basic results on indecomposable metastates
in Sec.\ \ref{indecomp}, and the models discussed in App.\ \ref{app:infanis} will also make an appearance.

We use the DS distribution of $\kappa_J$ or $\mu_J$, which encodes all the 
relations that employ the single overlap type $q$. Consider an overlap type, 
which we will write as $q_{[X]}$, though before we could use a total overlap $q_a$), and also the corresponding (pseudo-) 
metric $d_{[X]}$ or $d_a$. For given $\Gamma$, consider the $\kappa_J$ probability of the set of $\Gamma_1$ such that 
$q_{[X]}(\Gamma,\Gamma_1)\geq q_0$, $q_0$ a constant. This probability is a translationally and locally covariant function of 
$\Gamma_1$ and so, by indecomposability of $\kappa_J$, it is constant for $\kappa^\dagger$-almost every $(J,\Gamma)$. 
The infimum of the values of $q_0$ at which this probability is zero could depend on $\Gamma$, but for
$\kappa_J$ indecomposable it must be the same for $\kappa_J$-almost every $\Gamma$. This value is the norm-square 
$||{\bf v}||^2$ of the vector representing $\Gamma$ or $\Psi$ in the DS space. 
Similarly, for a given $\Gamma$, the $\kappa_J\times\kappa_J\times\cdots$ probability that, 
additional states $\Gamma_1$, $\Gamma_2$, \ldots, are in some (measurable) geometric relation with one another and $\Gamma$, 
as defined using overlaps, must be the same for $\kappa_J$-almost every $\Gamma$. 

We will consider only the support of the DS distribution of $\kappa_J$ or of $\mu_J$ in the DS Hilbert space; doing so may lose some 
fine structure of 
the distribution, but will nonetheless give us interesting results. Denote the support by $M$. Then $M$ is a closed subset of a sphere, 
$||{\bf v}||^2=$ constant, in $\cal H$, and it again inherits the metric $||{\bf v}-{\bf v}'||$ from the 
norm on $\cal H$. Any point in $M$ can be mapped to any other point by an orthogonal linear 
map of $\cal H$; let $O({\cal H})$ denote by the group of all orthogonal linear mappings on $\cal H$. From the preceding observations, 
we now see that we have a distribution on $M$ that is invariant under a group of linear maps in $O({\cal H})$ which act transitively on 
$M$, and which are isometries of $M$ because they preserve the metric. That is, $M$ is a homogeneous space for the group 
of isometries, and we have an invariant probability distribution $\mu_J$ on $M$. 

Denote the group of isometries of $M$ by $Iso(M,d_{[X]}')$ or 
by $G$. An isometry of $M$ can be described as a homeomorphism $f$ of $M$ onto itself, and we can define a uniform topology on the 
space of isometries, induced in a natural way from the uniform topology on $M$-valued functions on $M$; it is defined 
using the metric $d_u(f,f')=\sup_{{\bf v}\in M}||f({\bf v})-
f'({\bf v})||$ on pairs of homeomorphisms $f$, $f'$. For this topology, $G$ is a topological group (i.e.\ the group operations are 
continuous), and indeed it is a Polish group (a topological group that is a separable completely metrizable, or Polish, space; see 
\cite{kechris_book}, p. 60). Note that this topology on a subset of $O({\cal H})$ is not in general the same as the operator norm topology 
defined on $O({\cal H})$ viewed as a subset of the bounded operators on $\cal H$; in fact, the former topology is weaker than the latter 
\cite{reed_simon_book,willard_book}. Because $M$ is a homogeneous space, for any given 
point (vector) $\bf v$ in $M$ there is a closed subgroup $H$ (sometimes written $G_{\bf v}$) of $G$ that leaves $\bf v$ fixed, 
and then $M\cong G/H$. (The $H$ or $G_{\bf v}$ for different $\bf v$ are isomorphic, in fact, they are conjugate in $G$.) 
As examples, note that if $\kappa_J$ consists 
of $m$ atoms, then $M$ consists of $m$ points, while if $\kappa_J$ is atomless then $M$ is uncountable. 

We now state the general result, or structure theorem, that applies in this setting. The proof will explain the notation used.
\newline
{\bf Theorem 6}: Let $\cal H$ be a separable real Hilbert space, and $M$ a closed subset in $\cal H$ contained in a sphere centered 
at the origin in $\cal H$, equipped with the metric induced from the norm on $\cal H$, and suppose that $M\cong G/H$ is a homogeneous 
space for a group $G$ of orthogonal transformations of $\cal H$, that there is a $G$-invariant probability distribution $\mu$ on $M$, and 
let $G$ have the uniform topology defined just above. Then $G$ is a topological group, and $M$ and $G$ are compact. There is an infinite 
sequence of compact Lie groups $G_k$ ($k=1$, $2$, \ldots; it might be that there is a $k_0\geq 1$ such that 
all $G_k$ are the same for $k\geq k_0$) with a family of continuous surjective homomorphisms $G_{k+1}\to G_k$ for all $k$, and 
$G\cong\underleftarrow{\lim}_k\,G_k$ is the ``inverse'' (or ``projective'') limit of $G_k$ as $k\to\infty$, determined uniquely (up to 
isomorphism) by the homomorphisms.
There is a corresponding sequence of closed Lie subgroups $H_k\subseteq G_k$, to which the homomorphisms can be restricted, and  
$H\cong \underleftarrow{\lim}_k\,H_k$. Finally, $M\cong \underleftarrow{\lim}_k\,M_k$ where, for each $k$, $M_k\cong G_k/H_k$ is 
a finite- (possibly zero-) 
differentiable manifold with a finite number of connected components, and there are smooth surjective maps $M_{k+1}\to M_k$ for all $k$.  
There is a $G_k$-invariant probability distribution $\mu_k$ on $M_k$, and $\mu\cong \underleftarrow{\lim}_k\,\mu_k$ as measures.

{\bf Proof}: $M$ is by hypothesis a closed subset of $\cal H$, and because $\cal H$ is complete, so is $M$, and thus $M$ is a Polish space,
with the preferred metric inherited from $\cal H$. In a metric space, a subset is compact if and only if it is complete 
and totally bounded. Here, ``totally bounded'' means that for any $\varepsilon>0$, $M$ can be covered by finitely-many closed balls of radius 
$<\varepsilon$. In our case, $M$ is totally bounded because, for any $\varepsilon$, (i) any closed ball in $M$ of radius $\varepsilon/2$ has 
the same $\mu$ measure, (ii) only a finite number of pairwise non-intersecting such balls in $M$ can pack $M$, because otherwise the total 
measure 
of the balls would be infinite, while the total measure of $M$ is $\mu(M)=1$, and (iii) the smallest cardinality of a covering by closed balls 
of radius $\varepsilon$ is less than or equal to the maximum cardinality of a packing by non-intersecting closed balls of radius $\varepsilon/2$ 
(see Ref.\ \cite{vershynin_book}, p.\ 76; the two cardinalities are respectively the covering and packing numbers of $M$). Hence $M$ is 
compact, and also separable, because a totally-bounded metric space is separable. Thus $M$ is a compact Polish space.

It is a fact that the group $G$ of isometries (with the uniform topology defined before the Theorem) of a compact Polish space $M$ is compact 
(see Ref.\ \cite{kechris_book}, p. 60, or Ref.\ \cite{garling_book}, p.\ 63). [The proof can be obtained using the equicontinuity 
of any family of isometries of a fixed space $M$ together with Ascoli's theorem, for which see pp. 29, 30 of Ref.\ \cite{reed_simon_book}; the 
statements and proofs there easily extend to the situation here, or see Ref.\ \cite{garling_book}, p.\ 63.] Hence $G$ is a compact Polish group. 

The remaining statements are essentially standard ones about compact groups, general references for which are Ref.\ \cite{brotd_book}, 
Chapter 4, Ref.\ \cite{bump_book}, Chapters 1, 4, and Ref.\ \cite{price_book}, Chapter 6; we sketch some of the arguments.
Because $G$ is compact, there is a probability distribution on $G$, that is invariant under the action of $G$ from the left; it is the (left)
Haar measure, which we denote by $\mu_G$. The Hilbert space of functions (usually complex functions are used)
that are square-integrable with respect to $\mu_G$ will be denoted simply by $L^2(G)$; it is separable because $G$ is compact and the 
topology on $G$ is metric. $G$ acts on the functions in $L^2(G)$, and this action is unitary, so we have a representation of $G$ in $L^2(G)$, 
called the left regular representation. We want the map $G\times L^2(G)\to L^2(G)$, taking a group element and a 
function to a function, to be jointly continuous. In fact, because $G$ and $L^2(G)$ are Polish, it is sufficient that the 
action be continuous separately, that is both as a function $G\to L^2(G)$ for fixed $f\in L^2(G)$, and as a function 
$L^2(G) \to L^2(G)$ for fixed $g\in G$ (see Ref.\ \cite{kechris_book}, p. 62), and these are not difficult to verify. 

Given a jointly continuous unitary action of a compact group $G$ on a separable Hilbert space, the Peter-Weyl theorem 
\cite{brotd_book,bump_book} says (in part) 
that the Hilbert space decomposes as an orthogonal direct sum of finite-dimensional irreducible representations (there may be multiplicities 
larger than $1$ for each of the latter). Any finite-dimensional subspace invariant under the $G$ action gives a continuous quotient of $G$ 
onto a subgroup of the unitary group of the subspace (with its standard topology), and the subgroup is closed because $G$ is compact 
and the unitary group is Hausdorff; by Cartan's closed-subgroup theorem (see Ref.\ 
\cite{brotd_book}, p.\ 28), this closed subgroup is a compact 
Lie group [note that the term ``Lie'' implies that the connected component of the identity is finite (possibly zero) dimensional]. Thus we have 
obtained compact Lie groups $G_k$, which will increase or stay the same on passing to an invariant subspace that contains the previous one, 
with a natural continuous quotient map $G \to G_k$, and hence a family of quotients $G_{k+1}\to G_k$. Restricting the former map to $H$, 
we also obtain maps $H\to H_k\subseteq G_k$, and $H_{k+1}\to H_k$ for all $k$. As $H$ is closed in $G$, it is compact, and so $H_k$ is also 
compact and closed in $G_k$ for each $k$. Letting $k\to\infty$, the invariant subspaces in $L^2(G)$ asymptotically exhaust 
$L^2(G)$, and so $G\cong \underleftarrow{\lim}_k\,G_k$ ($\cong$ meaning a topological isomorphism of topological groups) and 
$H\cong \underleftarrow{\lim}_k\,H_k$. For each $k$, the closed-subgroup theorem implies that $G_k$ has a finite number of connected 
components. The remaining statements follow; see Ref.\ \cite{kechris_book}, p.\ 109, for the inverse limit of measurable spaces and of 
measures.  $\Box$

The Theorem is largely topological in nature, with little use of the geometry of the Hilbert space $\cal H$,
once it is known that $M$ is a homogeneous space.
We remark that for each $k$, given $M_k$ and $G_k$, the topology and the invariant probability measure on both spaces, 
which are finite-dimensional manifolds with finitely-many connected components, are determined uniquely, and the sequence of these
determines those on $M$ and $G$ uniquely also. On the other hand, the metric on $M$ (and hence that on $G$) is not unique, 
as we will see in an example in a moment.
We note that particular cases of the structures in the theorem arise when $M$ (resp., $G$) is a finite or countably infinite
direct product of compact finite-dimensional homogeneous spaces (resp., compact Lie groups); the inverse limits in the theorem can be viewed 
as a generalization of such products. Indeed, for $G$ there is a sense in which it is close to a direct product. (Some readers may wish to 
skip the remainder of this paragraph.) Let $G_{(0)}$ [resp., $G_{k(0)}$]
be the connected component of the identity in $G$ (resp., $G_k$); $G_{(0)}$ [resp., $G_{k(0)}$] is a closed normal subgroup in $G$ 
(resp., $G_k$). A connected compact Lie group, such as $G_{k(0)}$, is the quotient of a finite direct product of finitely-many copies of $SO(2)$ 
and of simple, simply-connected, compact Lie groups by a finite subgroup of the center of the product \cite{brotd_book}. 
[The distinct simple, simply-connected, compact Lie groups are $SU(n)$ ($n\geq 2$), $Spin(n)$ (the simply-connected cover of $SO(n)$; 
$n\geq 7$), $Sp(2n)$ ($n\geq 2$), $E_6$, $E_7$, $E_8$, $F_4$, and $G_2$ (excuse the clash of notation!) \cite{brotd_book,bump_book}.] 
In the inverse limit, the statement for $G_{(0)}$ (see Ref.\ \cite{price_book}, pp.\ 145--146, 158) is the same except that the finite product is 
replaced by a countable one (because our $G$ is separable; Ref.\ \cite{price_book} is more general), and the finite subgroup of the center 
is replaced by a ``profinite'' one (i.e.\ the inverse limit of finite groups; such a group is necessarily compact and totally disconnected, and more 
generally one can speak of profinite sets in a similar way). Finally, each quotient $G_k/G_{k(0)}$ is a finite group, as we have seen, so 
$G/G_{(0)}$ is profinite (Ref.\ \cite{willard_book}, p.\ 215); the set of connected components of $M$, with the quotient topology 
obtained from identifying points within each connected component, can be viewed as a homogeneous space for $G/G_{(0)}$. 

Next we want to analyze the geometry of $M$ as a subset of $\cal H$. We will do this in terms of the action of $G$ on the Hilbert space 
$\cal H$, but {\it a priori} we have only its action on the subset $M$. We can consider the vector space 
spanned by linear combinations of vectors in $M$. To do this in most generality, we need to consider integrals of vector-valued functions
of the form 
\be
{\bf v}=\int_M g({\bf v}'){\bf v}'\mu(d{\bf v}'),
\ee
where $g({\bf v}')$ is a real-valued function on ${\bf v}'\in M$. We define this as a Bochner integral of vector-valued functions on 
$M$ taking values in $\cal H$, so that convergence occurs in the norm topology on $\cal H$ (see e.g.\ Ref.\ 
\cite{aliborder_book}, Section 11.8). In our setting, the above integral exists and $\bf v$ is in $\cal H$ (so normalizable) if and only if 
$g({\bf v}')$ is integrable with respect to $\mu$ as a function on $M$, that is $g$ is measurable and $\int_M |g({\bf v}')|\mu(d{\bf v}')<\infty$ 
(the logical equivalence holds because $||{\bf v}'||$ is constant on $M$). We take the closure of the space of vectors 
that results from using all $g\in L^1(M,\mu)$. This might not be all of $\cal H$, however in the DS representation of $\mu_J$ we were not 
obligated to use any Hilbert space larger than this closed subspace, so without loss of generality we can think of it simply as $\cal H$.
Of course, if $M$ is a finite set of $m$ vectors (with equal weight $1/m$ each), $\cal H$ will be $m$ dimensional, and might in principle 
be finite dimensional in other cases also.

Now $G$ acts on $\cal H$ in a simple way, induced from its action on $M$ by using the Bochner integral above. This action is a jointly 
continuous map from $G\times {\cal H}\to{\cal H}$ as before. Then the Peter-Weyl theorem again implies that $\cal H$ can be decomposed 
as an orthogonal direct sum of irreducible representations of $G$ (note again that these are finite dimensional). From the preceding discussion, 
each of these is an irreducible faithful representation of a compact Lie group. 

At this stage, an example will illuminate the possible structures. Suppose that $M$ (as a metric space) is the sphere $S^2$, with the group of 
isometries $G=O(3)$, which is a non-Abelian group. These are what arise in the WF infinite anisotropy models, which we will discuss shortly,
and we will use (almost) the same notation. When the sphere is viewed as the
unit (without loss of generality) sphere in Euclidean space $\bbR^3$, we will use unit vectors $\bf m$ to denote vectors in $S^2$. 
There are many invariant metrics on the sphere, for example, the chord distance, $|{\bf m}-{\bf m}'|
=\sqrt{2(1-{\bf m}\cdot{\bf m}')}$,
and the arc length along the shortest geodesic connecting the two points, $\arccos ({\bf m}\cdot{\bf m}')$. (Note that the same group $G$ 
can arise for different metrics on $M$, with the same topology though not necessarily the same metric on $G$.)

Clearly, if our overlap $q_{[X]}({\bf m},{\bf m}')$ of two points (pure states, corresponding to $\bf m$, ${\bf m}'$) is 
$q_{[X]}({\bf m},{\bf m}')={\bf m}\cdot{\bf m}'$
(up to a possible normalization), then the DS Hilbert space will be simply $\bbR^3$. But in general the overlap is an 
$O(3)$-invariant function of $\bf m$, ${\bf m}'$, and instead we will have $q_{[X]}({\bf m},{\bf m}')=
\sum_{l=1}^\infty b_{[X]l}({\bf m}\cdot{\bf m}')^l$, because the terms shown in the series span the space of $O(3)$ invariants
that can be constructed from two vectors. The tensors with components the products of components of ${\bf m}=(m_1,m_2,m_3)^T$, 
namely $m_{i_1}m_{i_2}\cdots m_{i_l}$, which are symmetric, span the irreducible representation of $SO(3)$ of angular momentum 
$l=0$, $1$, \ldots, provided we subtract traces on pairs of indices so that in fact we have traceless symmetric tensors of degree $l$.
Then, apart from similar subtractions of terms, the powers $({\bf m}\cdot{\bf m}')^l$ correspond to the $O(3)$-invariant inner product
on tensors in the $l$th irreducible representation. When $q_{[X]}$ is used as the inner product to construct the DS Hilbert space, 
this discussion essentially shows that for a general overlap in this example the DS Hilbert space is necessarily a direct sum over all $l\geq 1$ of 
$O(3)$-invariant $2l+1$-dimensional subspaces of multiplicity $1$ for each $l\geq 1$, and thus is infinite dimensional even 
when $M\cong S^2$ [and similarly for $S^{m-1}$ for all $m\geq 2$, where again only the irreducible representations of $O(m)$ 
corresponding to traceless symmetric tensors occur]. This remains true in general for overlaps that are linear combinations 
of $q_{[X]}$. Further, under sufficiently small changes in the coefficients in those linear combinations, the space $M$ will change by a 
$G$-equivariant homeomorphism, remaining topologically an $S^2$. 

It is interesting that in the WF infinite-anisotropy models \cite{wf} (see Appendix \ref{app:infanis} below), which in their spin-glass phases 
feature nontrivial indecomposable 
metastates but $\kappa_J$-almost surely trivial Gibbs states, the Gibbs (or pure) states can be described by points on a sphere. In the 
original models, $XY$ spins were used, so the sphere is a circle, but the models can be immediately generalized to models of spins that are 
$m$-component unit vectors, and then a sphere $\cong S^{m-1}$ results. (Their models used infinite anisotropy, so that an Ising degree of 
freedom was left at each site, and there is an Ising spin-flip symmetry. Similar models without the spin-flip symmetry can be obtained 
by adding weak isotropic random fields at each site, and then the Gibbs states are trivial, rather than flip-related pairs. When the Gibbs states 
are flip-related pairs, one should use only overlaps with even $|X|$, and in the series expansions $b_{[X]l}=0$ 
for $l$ odd. In these cases 
$M\cong \bbR \mathbb{P}^{m-1}$, real projective space, instead of $S^{m-1}$; $\bbR \mathbb{P}^{m-1}\cong S^{m-1}/\bbZ_2$, 
where the group $\bbZ_2$ acts on $S^{m-1}$ by mapping ${\bf m}\to-{\bf m}$ for all $\bf m$.) The models can be 
generalized further by using spins 
that each take values in the same homogeneous space $M$ that has the same form as one of those above, with uniform anisotropy. 
(The compactness of $M$ makes both the disorder and the Gibbs states on these models straightforward to construct.)
Then the Gibbs states will be described by points on almost the same space, though at a given $T>0$ it may be that only a finite-dimensional 
quotient space $M_k$ shows up, because of thermal disordering, though a larger space will be revealed as the temperature is lowered. 
These constructions may give (up to homeomorphism) all of the spaces $M$ discussed above, at least as $T\to0$. 

Returning to the general discussion, we can characterize the geometry of $M\subset{\cal H}$ a little further. If we take an increasing sequence
of finite-dimensional $G$-invariant subspaces ${\cal H}_k\subseteq{\cal H}$, so ${\cal H}\cong \underleftarrow{\lim}_k {\cal H}_k$, then $
\cal H$ can be projected onto ${\cal H}_k$ by orthogonal projection, and we can identify the image of $M$ as $M_k\subseteq {\cal H}_k$, 
with properties as above ($G$ has a quotient $G_k$ that acts faithfully on ${\cal H}_k$, and $M_k\cong G_k/H_k$). For ${\bf v}\in M$, 
define the continuous map $f_k:M\to M_k$ by this projection, so $f_k({\bf v})\in{\cal H}_k$, and also $f_k$ is $G$-equivariant (i.e.\ it 
commutes with the action of $G$). For any 
given ${\bf v}\in M$, $f_k({\bf v})\to {\bf v}$ in norm convergence as $k\to\infty$. Because $M$ is compact and $f_k$ are continuous, 
this is in fact uniform convergence of the sequence of maps $f_k\to f$, where $f$ is the injective inclusion map of $M$ into $\cal H$, 
$f({\bf v})={\bf v}$. 
This is a general fact (we did not use the property that ${\cal H}_k$ are invariant subspaces): a compact set $M\in {\cal H}$, or rather 
its inclusion map $f$ (as just defined) can be uniformly approximated by maps $f_k:M\to M_k\subseteq {\cal H}_k$ where $M_k$ 
lies in a finite-dimensional subspace ${\cal H}_k$, and in our case $M_k$ consists of finitely-many, finite-dimensional connected manifolds. 
Thus, as a metric space but without direct reference to the embedding into either $\cal H$ or ${\cal H}_k$, our $M\cong G/H$ may consist 
of a finite number of finite-dimensional connected components $M_k\cong G_k/H_k$ (as in the example and the WF models) or, 
more generally, it can be uniformly and $G$-equivariantly approximated by metric spaces of that type; in either case we can then say 
that $M$ is approximately finite dimensional with finitely-many connected components (AFF).

It is worthwhile to consider what all this means for self overlaps of Gibbs states. The self overlap of a Gibbs state $\Gamma$ drawn from 
$\kappa_J$ is not necessarily locally invariant, because it depends on the weights of the pure states. If the metastate is atomless, as
it must be for an indecomposable metastate for which Gibbs states are not $g$ invariant, the self overlap appears in the full DS representation, 
which involves not only a vector in $\cal H$, but also a distribution on positive real numbers, so that the distribution of the self-overlaps which 
appear on the diagonal in the array of overlaps of a countable number of samples can be reproduced. Then for the atomless 
indecomposable case, this distribution on self overlaps (minus the norm-square of the DS vector) may be nontrivial, but this does not affect the 
validity of the result above. 

The full DS representation also implies that there is a distribution on $q_{[X]}(\Gamma,\Gamma)-||{\bf v}||^2\geq 0$ where $\bf v$ is the 
DS vector to which $\Gamma$ maps. Thus the infimum of the support of the distribution on $q_{[X]}(\Gamma,\Gamma)$ must be 
greater than or equal to $||{\bf v}||^2\geq0$, which is the supremum of the $q_{[X]}$ overlaps between pure states drawn from two Gibbs 
states drawn independently from $\kappa_J$; this supremum is certainly non-negative. In connection with RSB, it was argued 
\cite{read14} that there is a unique value of the 
pairwise overlap of pure states drawn from $\mu_J$, which is the value of $q_{[X]}(0)$ in RSB in terms of Parisi's $q(x)$ function (extended 
here to all types of overlaps), and there $q_{[X]}(0)\leq q_{[X]}(x)$ for all $x\in (0,1]$ is expected. In the present case, we find that 
$0\leq \sup q_{[X]}(\Gamma_1,\Gamma_2)\leq\inf q_{[X]}(\Gamma,\Gamma)$. There is a question about whether all overlaps are 
non-negative, as they are in RSB theory. In rigorous work on the SK model, this is called Talagrand's positivity principle (see 
Ref.\ \cite{panchenko_book}, p.\ 65). Clearly 
$\inf q_{[X]}(\Gamma,\Gamma)$ must be greater than or equal to $\inf q_{[X]}(\Psi_1,\Psi_2)$. If $\inf q_{[X]}
(\Gamma,\Gamma)$ is equal to $\inf q_{[X]}(\Psi_1,\Psi_2)$ for pure states in a single $\Gamma$, this would prove non-negativity of overlaps 
in our case. It is not currently clear whether this is true, that is that the probability distribution of $q_{[X]}(\Gamma,\Gamma)$ is nonzero down 
to the lower bound. To approach the lower bound, a Gibbs state $\Gamma$ must approximate a DF state, in that it must have, for example,  
most of the $w_\Gamma$ weight on a large number of pure states all of whose pairwise overlaps are close to the minimum value, and the 
weight of each pure state must be small. It may be that stochastic stability or $g$ evolution implies this, but we will not attempt to prove 
this here.

We can also make similar statements about conditional distributions for metastates as we did for Gibbs states in the previous Section. 
That is, we have
\be
\kappa_J=\int_{\cal H} \kappa_{J{\rm DS}a}(d{\bf v}) \kappa_{J{\bf v}}.
\ee
We can call the conditional distributions $\kappa_{J{\bf v}}$ ``local cluster metastates''. For an indecomposable metastate, the 
expression on the right-hand side (in which the integral can be restricted to $M$) can clearly be viewed as extending the description of an 
indecomposable metastate consisting of $m$ atoms of equal weight to the general case. Again, as in that case, although it resembles 
a decomposition of a metastate as a mixture, it does not mean the metastate is decomposable, because there may be no (measurable) way 
to identify corresponding vectors $\bf v$ for different $J$ that are very different (i.e.\ not related by a local transformation). If $M$ is a 
single point (i.e.\ when relative weak mixing holds), then this decomposition is trivial. In complete generality, the local cluster metastates 
obey relations like those for local cluster states, which were general results about the DS distribution. That is, in terms of the corresponding 
average states $\rho_{J{\bf v}}=\int\kappa_{J{\bf v}}(d\Gamma) \Gamma$, which naturally we can call ``local cluster MASs'', 
$q_a(\rho_{J{\bf v}},\rho_{J{\bf v}})=||{\bf v}||^2_a$, and $\kappa_{J{\bf v}}$ has the property, like a DF state, that 
for $\kappa_{J{\rm DS}a}$-almost every $\bf v$, $\kappa_{J{\rm DS}{\bf v}}\times \kappa_{J{\rm DS}{\bf v}}$-almost every pair 
$(\Gamma_1,\Gamma_2)$ has the same overlap $q_a(\Gamma_1,\Gamma_2)=||{\bf v}||^2_a$ [and hence the same 
$q_{[X]}(\Gamma_1,\Gamma_2)$ for each $[X]$], and for an indecomposable metastate this value is (resp., these values are) 
the same for $\kappa_{J{\rm DS}a}$-almost every $\bf v$. We discuss the analogous results for the MAS and its pure-state 
decomposition $\mu_J$ in the following Subsection.

\subsection{Uniqueness and mutual singularity of MASs}

We already saw for two indecomposable metastates for the same bonds $J$ that they are either mutually singular or identical as distributions
on Gibbs states, and that the same alternative holds for $\nu$-almost every $J$. In this subsection we address analogous questions 
that involve comparing the pure-state decompositions of two MASs, 
say $\mu_J^{(1)}$, $\mu_J^{(2)}$ arising from two metastates  $\kappa_J^{(1)}$, $\kappa_J^{(2)}$ for the same disorder, and also 
the same questions for distributions on pure states that are defined for all $J$, having similar properties as a MAS.

First, we point out that while a metastate $\kappa_J$ is locally covariant, that is, by eq.\ (\ref{eq:loccov})
\be
\kappa_{J+\Delta J}(d\Gamma)=
\kappa_J(d[\theta_{\Delta J}^{-1}\Gamma])
\ee 
(for $J$ and $J+\Delta J$ such that both sides are defined), this is not in general expected for $\mu_J$ of a MAS. If $\kappa_J$-almost 
every Gibbs state $\Gamma$ is trivial, then $\mu_{J+\Delta J}(d\Psi)=\mu_J(d[\theta_{\Delta J}^{-1}\Psi])$ follows from local 
covariance of the metastate. But in general,
\bea
\lefteqn{\kappa_{J+\Delta J}(d\Gamma)w_\Gamma(d\Psi)=}&&\\
&&\qquad{}\Delta J_X \kappa_J(d[\theta_{\Delta J}^{-1}\Gamma])w_{\theta_{\Delta J}^{-1}\Gamma}
(d[\theta_{\Delta J}^{-1}\Psi])(\langle s_X\rangle_\Psi-\langle s_X\rangle_\Gamma)\non
\eea
at first order in $\Delta J=\Delta J_X$, and so local covariance of $\mu_J$ does not hold. However, the first-order transformation 
shows that, apart from the change in $\Gamma$, $\Psi$, the joint distribution $\kappa_J w_\Gamma$ after the change in $J$ is 
absolutely continuous with the respect to that 
before. In particular, it follows that $\mu_{J+\Delta J}(\Psi)$ is absolutely continuous with respect to $\mu_J(\theta_{\Delta J}^{-1}\Psi)$,
and by considering the inverse transformation as well, $\mu_{J+\Delta J}(\Psi)$ and $\mu_J(\theta_{\Delta J}^{-1}\Psi)$
are (measure-theoretically) equivalent, that is they have the same null sets, even when they are not equal.

The different transformation behaviors imply that even if the $\mu_J^{(1)}$, $\mu_J^{(2)}$ for two metastates appear to be the 
same for one given $J$, they may not be the 
same if $J$ changes by a local transformation. For example, two extreme cases of an indecomposable metastate are that in which all Gibbs 
states are trivial, but the metastate $\kappa_J^{(1)}$ is not (chaotic singles; we will also assume its $\mu_J^{(1)}$ is relative weak mixing), 
and that in which the metastate $\kappa_J^{(2)}$ is trivial but the single Gibbs state for each $J$ is a DF state. Both have a MAS that at any 
fixed $J$ is a DF state \cite{hr}, so one might think they could be the same. But in the first case we have $\mu^{(1)}_J$ locally covariant 
as just explained, while in the second case $\mu^{(2)}_J$ transforms as $w_{J\Gamma}(\Psi)$ for a single Gibbs state, and these are 
different. Consequently, it is impossible for $\mu_J^{(1)}$ and $\mu_J^{(2)}$ to be equal for all, or $\nu$-almost every, $J$. In the 
intermediate cases when both the Gibbs states and the metastate are nontrivial, things are less clear, but nonetheless we see that the 
pure-state decomposition $\mu_J$ of a MAS does contain some information about the type of metastate from which it came.  

The next arguments are for the following more general set-up. We consider states $\rho_J$ that are well defined, and are Gibbs states, 
for $\nu$-almost every $J$, and also consider their pure state decompositions $\mu_J$ (distributions on pure states). Either of these uniquely 
determines the other, and we will pass between notations for corresponding objects without comment. Both of them are random measures,
meaning that, for $\nu$-almost every $J$ we have a probability distribution $\rho_J$ on spin configurations $s$ and, for any given 
measurable set $A$ of spin configurations, $\rho_J(A)$ is a measurable function of $J$ (and similarly for $\mu_J$ as a distribution on 
states $\Psi$). In addition, we assume both are covariant under translations, but in general we assume nothing at present about 
covariance under local transformations. From either $\rho_J$ or $\mu_J$, we can obtain the DS distribution (which is the same 
for both) on the DS Hilbert space $\cal H$. 

From the DS distribution on $\cal H$, we again obtain conditional distributions and a representation of $\mu_J$ as distributions
on pure states:
\be
\mu_J=\int_{\cal H} \mu_{J{\rm DS}{\rm tot}}(d{\bf v}) \mu_{J{\bf v}},
\ee
where $\mu_{J{\bf v}}$ is a conditional probability distribution on $\Psi$, defined for $\mu_{J{\rm DS}{\rm tot}}$-almost every $\bf v$,
and supported on the $\Psi$s that map to $\bf v$. [In the case when $\rho_J$ is a MAS, we already called the corresponding 
$\rho_{J{\bf v}}$ a ``local cluster MAS'', and the definitions agree because
\be
\mu_{J{\bf v}}=\int \kappa_{J{\bf v}}(d\Gamma) w_\Gamma
\ee
and $\kappa_{J{\rm DS}a}=\mu_{J{\rm DS}a}$.] The DS distribution of $\mu_{J{\bf v}}$ is a $\delta$-function at $\bf v$, 
meaning that $\mu_{J{\bf v}}\times \mu_{J{\bf v}}$-almost every pair $\Psi_1$, $\Psi_2$ have the same pairwise overlap. Then for 
each $\bf v$ at which $\mu_{J{\bf v}}$ is defined, either $\rho_{J{\bf v}}$ is a trivial Gibbs state ($\mu_{J{\bf v}}$ is a $\delta$-function), 
or $\rho_{J{\bf v}}$ is a DF state. We can see that this decomposition of $\mu_J$ generalizes that which occurred for the MAS of an 
indecomposable metastate consisting of $m$ atoms of equal weight $1/m$, where the atoms correspond to distinct $\bf v$; in that case 
also, the same two alternatives were found (in that case, they had to be the same type for every Gibbs state, by indecomposability).

In the particular case that $\rho_J$ is a MAS, then we know that the DS distribution of $\mu_J$ is locally covariant, and coincides with that of 
$\kappa_J$. Hence any non-covariance of $\mu_J$ is now confined within each $\mu_{J{\bf v}}$. While each 
$\mu_{J{\bf v}}$ moves covariantly under a local transformation starting from given $J$, this does not mean that the above decomposition
gives a decomposition of $\kappa_J$, because there may be no measurable way to say that $\mu_{J{\bf v}}$s correspond at widely 
different $J$. That is what occurs if $\kappa_J$ is indecomposable, and in that case the domain of integration can be restricted to $M$. 
(Again, an example is the case of an indecomposable metastate with $m$ atoms of equal weight.) 

We will now show that the map from the space of $\mu_J$ (or that of $\rho_J$) that arise from indecomposable metastates to that 
of distributions $\mu_{J{\rm DS}[X]}$ is one-to-one for all $X$ or for a total overlap. 
Because the construction of the DS map $\phi$ uses $\mu_J$ we must be careful with the definitions here. If we had defined a map from 
the space of all pure states into $\cal H$ and used it to discuss this problem, we would have a fixed map, however, our map depends on
$\mu_J$, and we need to resolve a possible relative ambiguity by an orthogonal linear map in $O({\cal H})$. We can do this
for the case in which we wish to compare two distributions $\mu_J^{(1)}$, $\mu_J^{(2)}$, as follows. We consider a convex combination 
$\mu_{J\lambda}=\lambda\mu_J^{(1)}+(1-\lambda)\mu_J^{(2)}$ with $0<\lambda<1$, and find its DS distribution for $q_a$ or $q_{[X]}$. 
This effectively maps the corresponding DS distributions of $\mu_J^{(1)}$, $\mu_J^{(2)}$ into the same space, with any pure states 
$\Psi$ that appear in both distributions mapped to the same point: $\mu_{J\lambda{\rm DS}[X]}=\lambda\mu_{J{\rm DS}[X]}^{(1)}
+(1-\lambda)\mu_{J{\rm DS}[X]}^{(2)}$. (This construction can of course be generalized to convex combinations of more than two $\mu_J$s.)
We can now ask how $\mu_{J{\rm DS}[X]}^{(1)}$ and $\mu_{J{\rm DS}[X]}^{(2)}$ are related, 
as distributions on the same space, and in particular, what happens if they are equal for all $q_{[X]}$, or for $q_{[X]}$ replaced by 
$q_a$, and for $\nu$-almost every $J$. 

A key property of the $q_{[X]}$ DS distribution of $\mu_J^{(1)}$, $\mu_J^{(2)}$ (or of other distributions on pure states, as considered 
earlier) is that it is sufficient for computing both the self-overlap of either of $\rho_J^{(i)}$ ($i=1$, $2$), and also the mutual overlap 
$q_{[X]}(\rho_J^{(1)}$, $\rho_J^{(2)})$; in particular, the part of the full DS representation that describes self-overlaps of each pure state 
is not needed. These self- or pairwise overlaps can be expressed as double integrals (i.e.\ over $\bf v$ and ${\bf v}'$) of overlaps of 
corresponding local cluster states, and the overlaps of those are simply ${\bf v}\cdot {\bf v}'$,
from the general results in the previous Section (Proposition 5 and following). 
Now suppose that, on the common $\cal H$, $\mu_{J{\rm DS}[X]}^{(1)}=\mu_{J{\rm DS}[X]}^{(2)}$. 
Then it follows from the preceding remarks that the two self-overlaps and the mutual overlap are all equal, and so that 
$d_{[X]}(\rho_J^{(1)}, \rho_J^{(2)})^2=0$. 
If these are true for $\nu$-almost every $J$ and for all $[X]$ (or equivalently for $d_a^2$), then it follows by a version 
of Lemma $1$ that $\rho_J^{(1)}=\rho_J^{(2)}$, or equivalently $\mu_J^{(1)}=\mu_J^{(2)}$, for $\nu$-almost every $J$. 
This is the desired result showing that, in the sense we defined, the map from the space of $\mu_J$s to that of $\mu_{J{\rm DS}a}$s 
is one to one. Note that we do not have a corresponding statement for the map from the space of $\kappa_J$s to that of 
$\kappa_{J{\rm DS}a}$s. 

We now turn to the question of mutual singularity versus equality of MASs $\mu_J^{(1)}$, $\mu_J^{(2)}$ as distributions on $\Psi$,
when these are obtained from two indecomposable metastates $\kappa_J^{(1)}$, $\kappa_J^{(2)}$. 
Consider their DS distributions on a common Hilbert space, by the construction above, and using a total overlap $q_a$. 
Then each $\mu_{J{\rm DS}a}^{(i)}$ is a distribution with support in a space $M^{(i)}$ which is a homogeneous space for a 
group $G^{(i)}$ ($G^{(i)}$ could be trivial). Now apply the Lebesgue decomposition (we drop the suffix $a$ as it remains fixed for now). 
Suppose that $\mu_{J{\rm DS}}^{(1)}$ decomposes as the sum of two mutually singular parts, $\mu_{J{\rm DS}1}^{(1)}$ that is 
absolutely continuous with respect to $\mu_{J{\rm DS}}^{(2)}$, and $\mu_{J{\rm DS}2}^{(1)}$ that is mutually singular with 
$\mu_{J{\rm DS}}^{(2)}$. Both parts are translationally and locally covariant (as mentioned earlier) so the probability of either of these 
two parts is translation (and locally) covariant, and so the same for $\nu$-almost every $J$. But because of Proposition 6, 
these parts can be viewed as distributions on Gibbs states $\Gamma$, and they can be pulled back to give mutually-singular distributions 
$\kappa_{J1}^{(1)}$, $\kappa_{J2}^{(1)}$ on $\Gamma$, 
which are translation and locally covariant, giving a decomposition of $\kappa_J^{(1)}$ into two normalized parts for 
$\nu$-almost every $J$ with $J$-independent coefficients. But $\kappa_J^{(1)}$ is indecomposable, so this means that one or other 
part has zero weight. That is, $\mu_{J{\rm DS}}^{(1)}$ is either absolutely continuous or singular with respect to $\mu_{J{\rm DS}}^{(2)}$. 
The same argument with the two metastates interchanged implies that $\mu_{J{\rm DS}}^{(1)}$, 
$\mu_{J{\rm DS}}^{(2)}$ are either mutually singular $\nu$-almost surely, or equivalent $\nu$-almost surely.
If they are mutually singular, then $\mu_J^{(1)}$ and $\mu_J^{(2)}$ are mutually singular also, and so are $\rho_J^{(1)}$ and 
$\rho_J^{(2)}$. In this case, $d_a(\rho_J^{(1)},\rho_J^{(2)})>0$ for $\nu$-almost every $J$.

If $\mu_{J{\rm DS}}^{(1)}$ and $\mu_{J{\rm DS}}^{(2)}$ are equivalent, then their supports $M^{(1)}$, $M^{(2)}$ are equal 
(in the common Hilbert space $\cal H$, and for any $J$). Then the two probability distributions $\mu_{J{\rm DS}}^{(i)}$ must be 
the same, because they are determined by symmetry of $M^{(i)}$ as a homogeneous space. Now consider the MAS states 
$\rho_J^{(i)}$ themselves. By the remarks of this Subsection, the distance $d_a(\rho_J^{(1)},\rho_J^{(2)})=0$, and by Lemma 1 
$\mu_J^{(1)}=\mu_J^{(2)}$ (or $\rho_J^{(1)}=\rho_J^{(2)}$), $\nu$-almost surely. 

Moreover, the local cluster MASs of the two must be the same as well; we can show this directly. Recall the convex combination 
$\mu_{J\lambda}$ of the two MASs,
with parameter $\lambda$, where now $\mu_{J{\rm DS}}^{(1)}$ and $\mu_{J{\rm DS}}^{(2)}$ coincide. Then the conditional 
$\mu_{J\lambda{\bf v}}$ takes the usual form, and is the convex combination of $\mu_{J{\bf v}}^{(i)}$ with the same parameter $\lambda$.
It follows that, for all $[X]$, for $\mu_{J{\rm DS}}^{(1)}=\mu_{J{\rm DS}}^{(2)}$-almost every $\bf v$, and $\nu$-almost every $J$, 
the pairwise $q_{[X]}$ overlap of $\Psi_1$, $\Psi_2$ takes the same value ($=||{\bf v}||^2$) for 
$\mu_{J{\bf v}}^{(1)}\times\mu_{J{\bf v}}^{(2)}$-almost every $(\Psi_1,\Psi_2)$. Then 
$d_{[X]}(\rho_{J{\bf v}}^{(1)},\rho_{J{\bf v}}^{(2)})=0$ for all $[X]$, almost every $\bf v$,  and $\nu$-almost every $J$, so by a 
version of Lemma 1, $\rho_{J{\bf v}}^{(1)}=\rho_{J{\bf v}}^{(2)}$ for $\mu_{J{\rm DS}}^{(1)}=\mu_{J{\rm DS}}^{(2)}$-almost 
every $\bf v$, $\nu$-almost surely. 

We have arrived at a striking conclusion: for two indecomposable metastates, their MASs $\rho_J^{(i)}$, or their pure-state decompositions 
$\mu_J^{(i)}$, are either mutually singular or identical, and if identical this holds for their local cluster MASs also. For a decomposable 
metastate, we obtain a unique decomposition of its MAS into mutually singular parts, just as for a Gibbs state at a single $J$, or for 
a metastate. 
Readers are cautioned that, when two indecomposable metastates $\kappa_J^{(i)}$, $i=1$, $2$, have the same MASs 
$\rho_J^{(1)}=\rho_J^{(2)}$, we cannot necessarily conclude that then $\kappa_J^{(1)}=\kappa_J^{(2)}$, because we do not have a unique 
decomposition of $\rho_J$ into Gibbs states $\Gamma$ as an expectation under a unique $\kappa_J$ as we do for pure state decomposition 
(we can decompose a given $\kappa_J$ into a mixture of indecomposable metastates, but that is not sufficient). Hence it is not ruled out that
a decomposable metastate might give a $\mu_J$ that is indecomposable in the current sense (i.e.\ not decomposable into a mixture of mutually 
singular MASs). 

In some particular cases, we can show that distinct (mutually singular) metastates have distinct (mutually singular) MASs. For example, 
suppose there are two indecomposable metastates, $\kappa_J^{(1)}$, which is trivial, and $\kappa_J^{(2)}$, which is atomless and 
has the relative weak mixing property. Then the $q_{\rm tot}$ DS distribution of either is an atom at a single vector in $\cal H$; 
suppose it is the same for both, so their $\kappa_{J{\rm DS}{\rm tot}}^{(i)}$ are equivalent. Then their MASs, which each involve 
a single local cluster MAS, must be the same. But if in addition the unique Gibbs state drawn from $\kappa_J^{(1)}$ is trivial, 
their MASs (local cluster states) are clearly distinct, because that of $\kappa_J^{(2)}$ is atomless. Similarly, if $\kappa_J^{(1)}$ and 
$\kappa_J^{(2)}$ are both trivial, but the Gibbs state of the first is trivial and that of the second is a DF state, then again the 
MASs are distinct. A slightly different argument for these cases extends to any pair of indecomposable metastates that have the same DS 
distribution on $\cal H$, but the self-overlaps of pure states in the first take one value, and those in the second a different value. 
Then their MASs cannot be the same. In all these cases, the contradiction means that in fact the DS distributions
of the two metastates must be mutually singular. It is possible that this type of argument can be extended to all cases, but doing 
so might require more information about the Gibbs states, derived perhaps from stochastic stability.
 
\subsection{Application to maturation MAS}

A extension of the results of the last Subsection is of some interest, and will be used here. The statement we will prove is very general. 
Suppose we have two states $\rho_J^{(i)}$, $i=1$, $2$, that is, probability distributions on Ising spins, and we require that they exist 
for all $J$, and be covariant under translations of $J$; they do not have to be Gibbs states. Suppose that $\rho_J^{(1)}$ has the property 
that $\rho_J^{(1)}\times\rho_J^{(1)}$-almost every pair $(s,s')$ of spin configurations has the same value of the overlap 
$\widehat{q}_{[X]}(s,s')$, for all $[X]$ and $\nu$-almost every 
$J$. Thus, for each $[X]$, the distribution of the overlap $\widehat{q}_{[X]}$ is a single $\delta$-function. Suppose that $\rho_J^{(2)}$ 
is absolutely continuous with respect to $\rho_J^{(1)}$, $\rho_J^{(2)}\ll\rho_J^{(1)}$, for $\nu$-almost every $J$, that is, no set of spin 
configurations has nonzero probability in $\rho_J^{(2)}$ if it has zero probability in $\rho_J^{(1)}$. Absolute continuity implies that, 
for all $[X]$ and $\nu$-almost every $J$, $\rho_J^{(1)}\times\rho_J^{(2)}$-almost every, and also $\rho_J^{(2)}\times\rho_J^{(2)}$-almost 
every, pair $(s,s')$ has the same $\widehat{q}_{[X]}$ overlap as for $\rho_J^{(1)}\times\rho_J^{(1)}$-almost every pair. 
But then $d_{[X]}(\rho_J^{(1)},\rho_J^{(2)})^2=0$ for all $[X]$, $\nu$-almost surely. Hence, by a version of Lemma 1, 
$\rho_J^{(2)}=\rho_J^{(1)}$, $\nu$-almost surely. 

Under the same conditions, if in addition $\rho_J^{(1)}$ 
is a Gibbs state, then the $\widehat{q}$ overlap and the corresponding overlap $q_{X]}$ of pure states are the same, conditionally 
on a pair of pure states, and so the latter also has trivial distribution, because the former reduces to the latter (as discussed in Sec.\ 
\ref{models}); thus relative weak mixing holds. Then we can conclude that $\rho_J^{(2)}$ is also a Gibbs state, and so the pure-state 
decompositions of both states are the same, $\mu_J^{(2)}=\mu_J^{(1)}$. (If we assume both states are Gibbs states, we can use a similar
argument based instead on the hypothesis that $\mu_J^{(2)}\ll\mu_J^{(1)}$, with the same result.)

We apply this to the so-called maturation MAS. This arises in a context distinct from the equilibrium situation we have considered throughout the 
paper so far. We again suppose that we have an infinite Ising model on the same $d$-dimensional lattice $\bbZ^d$ as before, the same 
Hamiltonian $H$, and some choice of temperature $T$. But now we begin at time $t=0$ (this $t$ has nothing to do with the more fictitious 
one we used in connection with $\Sigma$ and $g$ evolution) with a uniform distribution on all the spin configurations; this can be viewed 
as the Gibbs state for infinite temperature. Then we introduce on the Ising spin system some choice of stochastic dynamics that obeys 
detailed balance (see e.g.\ Ref.\ \cite{liggett_book}); for example, it could be that used in a Monte Carlo simulation. Given the initial uniform 
state at $t=0$, one can then study, for example, the correlations among the spins at one later time $t>0$, and these will evolve with time. 
Physically, this set-up represents a very rapid quench of an infinite system from infinite temperature to some finite temperature $T>0$, and 
it would be expected that as $t$ increases from $0$ the state begins to equilibrate locally, but changes as effects of more distant regions 
propagate into the local region. 

A plausible hypothesis is that, if the state that evolves from the uniform state at $t=0$ could be followed to asymptotically large times, 
for $\nu$-almost every $J$ it would eventually approach a Gibbs state (in the sense of weak* convergence, that all correlation 
functions converge) for the same $H$ and $T$. This can be shown to hold for non-disordered ferromagnetic systems under similar 
conditions (see e.g.\ Ref.\ \cite{liggett_book}, Chapter IV, supplemented by a compactness argument to ensure existence of a limit for 
the uniform initial state, at least along some sequence of times $t_n$ that tend to infinity), and it is tempting to assume 
it holds. Technically, convergence should occur along a sequence of times (tending to infinity) independent of $J$, 
and can be better described as a limit for the joint distribution of bonds $J$ and a state; as with the equilibrium metastate, there is again an 
unsolved question of uniqueness of the limit. We call this Gibbs state the maturation MAS, 
or MMAS, in terminology inspired by WF \cite{wf}, and denote it $\rho_J^{\rm M}$. 
We will also consider an equilibrium metastate $\kappa_J$ for the system at the same temperature $T$ and its MAS $\rho_J$, as discussed 
elsewhere in this paper. (The possibility that one or both of these is not unique will be ignored here.) It has been suggested that these two 
states could be the same: 
\be
\rho_J^{\rm M}=\rho_J
\ee 
for $\nu$-almost every $J$. This was a brief remark in NS99 \cite{ns99}, and was discussed in the context of some models by WF 
\cite{wf}. For models like EA models, including their long-range analogs, it was proposed and studied numerically in Refs.\ \cite{mhy,wy,jry}, 
and called a statics-dynamics relation. Of course, it is of most interest in the low-temperature region. 

Here we will assume that the equilibrium metastate has the relative weak mixing property, and note that this 
allows scenarios such as SD, RSB, and CS. For the MMAS, we note that the translation-invariant initial state and 
translation-covariant dynamics imply that the MMAS is translation covariant. Heuristically, one expects that the system tries to equilibrate into 
some of the same pure states as in equilibrium and, because of the uniform initial state, that no pure state obtains much higher weight 
in the pure-state decomposition $\mu_J^{\rm M}$ than in equilibrium. We can formulate these ideas mathematically by saying that
$\mu_J^{\rm M}$ should be absolutely continuous with respect to $\mu_J$, $\mu_J^{\rm M}\ll\mu_J$ (in particular, there are no atoms 
in $\rho_J^{\rm M}$ unless corresponding ones occur in $\mu_J$). In fact, we do not have to assume that $\rho_J^{\rm M}$ is a Gibbs 
state; as above, it is sufficient to assume only that $\rho_J^{\rm M}\ll\rho_J$ (as distributions on spin configurations). Then from the argument 
above, we find that indeed the MMAS and MAS must be equal, $\rho_J^{\rm M}=\rho_J$ for $\nu$-almost every $J$, and
that $\rho_J^{\rm M}$ is a Gibbs state. Consequently, any correlation function (not only ones that are squared or translation averaged) 
must be the same in both states. 

We should comment on spin-flip symmetry, which is present in many of the discussions and simulations mentioned here. 
Our discussion so far has assumed spin-flip symmetry is absent. As we commented earlier, if instead the Hamiltonian (and the
equilbrium metastate and dynamics) are spin-flip invariant, we can 
apply all the same results if we consider only $X$ with $|X|$ even, and replace pure states by the flip-invariant mixture of spin-flip 
related pairs of pure states. In this case, the single-site overlap of two such mixtures is zero by symmetry, while a two-point correlation 
such as $\langle s_i s_j\rangle$ can be considered; the latter will usually be nonzero, and when its square is translation averaged it 
becomes a $q_{[X]}$-self-overlap for $X=\{i,j\}$. The latter are commonly used in simulations.
The single-site overlap (or the overlap for any other $X$ with $|X|$ odd) is nonzero between pure states, and takes at least two values 
$\pm|q_{[1]}|$ by symmetry. This does not violate our assumption that each overlap is the same for almost every pair of flip-invariant 
mixtures of two pure states, not pairs of pure states; only overlaps with even $|X|$ of these mixtures are nonzero. 
Then it again follows from the previous argument, and use of symmetry, that $\rho_J^{\rm M}=\rho_J$ for $\nu$-almost every $J$. 
Correlations $\langle s_X\rangle_{\rho_J}$ with $|X|$ odd again vanish by symmetry.

\section{Summary and discussion}
\label{sec:disc}

\begin{table}
\begin{tabular}{c|c|c|c|c|}
&I&II&III
\\ \hline
trivial&RS, SD, DF&$\emptyset$&$\emptyset$
\\ \hline
atomless&CP, CS&RSB&RSB?
\\ \hline
\end{tabular}
\caption{Table of results for $0<T<\infty$, for type of local cluster metastate (rows) versus type of Gibbs state drawn from it (columns);
types a, b are subsumed into a single column in each case. 
The notation is the same as in Table I.}
\label{tab:metgibbs}
\end{table}

We already gave an informal, less technical, summary of this work in the Introduction, which the reader may wish to review. Here we 
will summarize and comment on the results in the light of the full paper. We again assume throughout that a metastate is translation covariant.

A succinct summary of our results goes as follows. We describe an indecomposable metastate, with the understanding that doing so 
should include a description of typical Gibbs states drawn from it. Combining results from the culminating Sections \ref{trivmet} 
and \ref{sec:mas}, part of what we have found is the characterization of the structure of any indecomposable metastate by specifying: 
i) a compact homogeneous space $M$, which is a subspace of a sphere in real Hilbert space $\cal H$, and so a symmetry group 
$G\subset O({\cal H})$ (possibly trivial) such that $M\cong G/H$ for some subgroup $H\subset G$ ($M$ is equipped with 
a $G$-invariant probability distribution, the marginal of the metastate); ii) a choice between 
trivial and atomless structure, the same for every ``local cluster metastate'', the conditional distribution of the metastate when conditioned
on a point in $M$; iii) a choice of a type of Gibbs state out of a set of six (types I, II, and III, with further choice of a or b in each case), 
the same for almost every Gibbs state drawn from almost every local cluster metastate. In iii), the choice of type of Gibbs state 
is independent of the type of local cluster metastate, except that it must be type I when the latter is trivial.

The definitions of the six types of Gibbs states (drawn from a metastate) were already presented in the form of Table \ref{tab:gibbs} 
in Sec.\ \ref{trivmet}. We show in Table \ref{tab:metgibbs} the allowed types of Gibbs state for either of the two types of local cluster 
metastate; there is no need to separate type a from type b Gibbs states. The pure-state decomposition $\mu_J$ of the local cluster MAS 
is trivial for the first row, and atomless for every case in the second row. In the case of a trivial DS distribution for the metastate 
(i.e.\ relative weak mixing), the local cluster metastate becomes the metastate itself (and similarly for the MAS).

We reiterate that our results limit the allowed possibilities for a metastate and for Gibbs states drawn it, and do not establish existence
in any case (though existence of metastates follows from Refs.\ \cite{aw,ns96b}, and another proof can be found in App.\ 
\ref{app:furth}). Hence, further work might eliminate 
some possibilities viewed here as allowed. The local covariance property of any metastate was essential in a number of results that we proved.
It led in particular (in conjunction with translation covariance) to stochastic stability, or $\Sigma$ invariance, of any metastate. In practice, the 
strong form of this result was not used, and instead we used somewhat weaker versions, and the related $g$ evolution.

An obvious gap in our results is that we dd not prove in every case that non-trivial Gibbs states have properties such as ultrametricity 
and the Ruelle cascades that are closely associated with RSB, but only for type II Gibbs states with a finite number of values of the overlap. 
We suspect that these properties will emerge for all non-trivial Gibbs states in short-range systems. Then we can make a conjecture: 
RSB (and RS) describe relatively weak mixing indecomposable metastates, or more generally, all possible local 
cluster metastates, and the Gibbs states drawn from these. In relation to RSB as a mean-field theory, the following analogy is interesting. 
Usually, for example in an Ising ferromagnet, mean-field theory proceeds by assuming some order parameter and deriving an equation for it,
with a nonzero solution in the ordered phase (at low temperature and zero magnetic field, in the example). Each nonzero solution of the 
mean-field equations corresponds to an ordered state, and when spontaneous breaking of a symmetry (under which the Hamiltonian is 
invariant) is involved, the different solutions map to one another under a symmetry operation. Thus mean-field theory, applied to a short-range 
system, gives some account of the distinct pure states of the model. In contrast, the RSB mean-field theory in the SG phase describes a 
distribution on nontrivial Gibbs states, not a single pure state. But now, if we view RSB as describing the local cluster metastates, 
the analogy with conventional mean-field theories is restored, with local cluster metastates in analogy with pure states. 

We comment that stochastic stability and $\Sigma$ evolution are connected with the fluctuations of the 
weights of pure states (or of local cluster states) in a given Gibbs state $\Gamma$ (drawn from a metastate) due to the randomness 
of the bonds in a subregion of linear size $W$, other bonds remaining fixed; see Secs.\ \ref{subsubsec:siginv}, \ref{sec:gevol}, and 
\ref{subsubsec:earl}. There, we fixed the initial values of those bonds, and scaled the strength of the {\em additional} disorder 
in the subregion by $W^{-d/2}$, such that a finite limit of the change in the relative free energies between local cluster states as 
$W\to\infty$ was obtained (it was nonzero when the distance $d_a$ between the cluster states was nonzero), which led to the 
conclusion of $\Sigma$-invariance of the metastate. This also suggests that the fluctuations in the relative free energies between 
local cluster states due to the {\em total} disorder (in the original scaling) would scale with the square root $W^{d/2}$ of the volume. 
Indeed, very recent work \cite{ns24} proved {\em lower bounds} of order 
$W^{d/2}$ on the size of those fluctuations with the total disorder in the subregion under a related condition (``incongruence'' \cite{hf}); 
the result agrees with an expectation from Ref.\ \cite{fha} (and with the preceding remarks). 

In Ref.\ \cite{fha}, the authors tried to argue that such fluctuations destroy any RSB scenario in a short-range SG. A logical 
endpoint of this line of thought could be instead that $\Sigma$-invariance of the metastate must hold, as we have shown. That does not by any 
means immediately rule out RSB in general; work on this (in the SK model) found in fact that ultrametricity must hold 
\cite{aa,panchenko_book} (see also Ref.\ \cite{ans15}, which found that, for any $k$,  $k$-RSB is consistent in the short-range case). 
On the other hand, metastates had not been 
discovered in 1987, and when the metastate is (perhaps implicitly) treated as trivial, $\Sigma$-invariance reduces to $g$-invariance of
the Gibbs state, and implies that it is either trivial or a DF state, as we showed, which vindicates Ref.\ \cite{fha} to some extent. 
(Thus, other types of Gibbs states or of RSB require a non-trivial local cluster metastate; see Table \ref{tab:metgibbs}. 
The use of local cluster states because of our close analysis of $\Sigma$ 
invariance implies that a generalization of the notion of incongruence is in use, which then allows the DF local cluster states.) 

In these results, stochastic stability was used, but not the full $\Pi$ invariance of $\kappa^\dagger$. It is logically conceivable that use
of the latter could limit the allowed possibilities for Gibbs states (drawn from a metastate) further than we have found here (or than was 
found in the past). For example, for a trivial metastate, or more generally for any atom in a metastate, it might be that use of the 
full $\Pi$ invariance could eliminate the 
possibility of a DF state, which are immune to arguments based on stochastic stability (and this might extend to eliminating DF local cluster 
states in general, that is, all Gibbs states of type b), but this is not clear 
at present. Such an argument might be viewed as accounting for the fluctuations of the relative weights of pure states in a mixed Gibbs state.
[DF states, in the form seemingly implied by RSB mean-field theory, have been discussed in the past \cite{ktw,bb}; it was argued 
that they do not arise in the short-range case, however, we already discussed that critically in Sec.\ \ref{subsec:ginv}.] 
Similarly, (non-)existence of Gibbs states of types II or III could be tackled likewise. We have not proved an existence result for these, 
but neither are we aware of an argument that nontrivial metastates with type II or III Gibbs states do not exist, and it may be that 
investigation will lead to additional necessary conditions for their existence. Various works aimed at proving ultrametricity include results 
that say that distributions with the structure of the Ruelle cascades are stable under stochastic stability or possess Bolthausen-Sznitman 
invariance \cite{aa,bs,panchenko_book}, which suggests that (in the short-range context) they may also satisfy the constraints arising from 
$\Upsilon$ invariance.  

Even to those readers prepared to accept that RSB is a possibility, some of the structures found here may seem surprising or unlikely to occur 
in practice. These include (in-)decomposable metastates, DF states, and non-trivial metasymmetry. We will discuss these issues next.
We already commented on DF local cluster states. DF states also arise in a wider sense and
probably will not be ruled out in general, because these include MASs in some of the examples in Sec.\ \ref{sec:examples}, 
which seem truly to exist. Overall, at present we see no reason to expect DF (including local cluster) states to be ruled out in general.

For decomposable metastates, we can point out that if there is more than one indecomposable metastate (for given $J$ and temperature $T$),
then it is possible to construct a decomposable metastate as a mixture. Hence the latter cannot be ruled out in our infinite-size point of view 
unless one proves that there is a unique indecomposable metastate (under the same conditions). However, we comment on finite size below.
For metasymmetry, our arguments used indecomposability of the $\Pi$-invariant metastate and the DS distribution. The map of a metastate 
to its DS distribution maps each Gibbs state to a single point, and then the DS distribution is $\Theta$ covariant,
as is the DS distribution of the MAS. So in the infinite-size point of view used here, it is not clear if any further argument at the level of 
Gibbs states could rule out, or further restrict, either decomposable metastates or nontrivial metasymmetry.

From the discussions so far, it is clear that, because our starting point was a definition of metastates in the abstract for an infinite-size system, 
we may not have accounted for all effects that may appear when the constructions of a metastate as a limit of finite-size systems is studied. 
These may involve the fluctuations of the relative free energies of different states occurring at different sizes or as (approximately defined) 
parts of a state at one size, and could imply additional stability requirements on the distributions (i.e.\ on the metastate $\kappa_J$ or on the 
pure state decompositions $w_\Gamma$) that we did not consider in this paper. At present we can only speculate about what these might be.

For a decomposable metastate, there is nonzero probability that two Gibbs states drawn independently could differ macroscopically,
or that the pure states in their respective decompositions could so differ; the latter possibility was already discussed a little in NRS23 
\cite{nrs23}. In the latter, we mentioned what occurs in a first-order phase transition; the basic NRS23 zero-one law, Proposition 1 above,
rules out such behavior in a single Gibbs state almost surely, so that pure states with distinct macroscopic behavior (invariant observable
properties) can only appear in the decompositions of distinct Gibbs states, and so for a decomposable metastate. The underlying idea here
(or interpretation of the result)
is that a perturbation of the Hamiltonian can distinguish such pure states, splitting their free energies, so that they would not both occur in a 
single Gibbs state. Our results in this paper do not rule out decomposable metastates in the abstract (infinite-size) setting. However, if we 
consider taking the limit from finite size, as in the AW and NS constructions of metastates, then decomposable metastates appear likely to be 
non-generic---they would only occur by tuning some parameters in the model. A basic example for this is the RFFM, discussed in Sec.\ 
\ref{sec:examples}. In this case, there is equal probability for each of two Gibbs states, both of them pure states, which have different 
(in fact, opposite) magnetization per site. If a uniform magnetic field is now applied to the whole system at finite size, then (in so far
as we can speak heuristically of the two distinct ordered states in finite size), their free energy densities will be split,
and only the pure state favored by (and somewhat modified by) the field will be present when the limit is taken, because it arises in a larger 
fraction of sample, that tends to $1$ in the limit. Then the construction of a metastate will give a trivial, so indecomposable, metastate. 
[In this system, the decomposable metastate apparently only occurs at all when the distribution of the independent random fields 
at each site is symmetric under reversing its sign; more generally, we could require
such symmetry for all $\nu_X$ with $|X|$ odd.] We imagine that similar effects can occur in other cases, some of which
may involve random rather than deterministic perturbations. Thus we expect decomposable metastates to be non-generic when a 
metastate is constructed from finite size systems with a generic disorder distribution. 

There may also be reasons for misgivings about a nontrivial metasymmetry of an indecomposable metastate, as we discussed it in the previous 
Section; such phenomena have not been anticipated in the past, even in RSB. One point is that the homogeneous space $M\subset{\cal H}$ 
has the AFF property. In this approximate sense, points in the DS representation of the metastate or MAS are parametrized by a few (i.e.\ 
finitely many) parameters. This contrasts with the spaces that are either a single point or are infinite-dimensional to which we are 
accustomed in non-trivial scenarios. If $M$ is not a single point, there is a compact group $G$ of nontrivial symmetries of the DS distribution 
of the metastate. (Note that this discussion applies in particular in the case of an indecomposable metastate of $m>1$ atoms.) 
All this suggests that such behavior can arise only if it is a result of some simpler underlying structure as it was in the WF models \cite{wf} 
and several of the other examples given earlier in Sec.\ \ref{sec:examples}, in all of which there was some symmetry 
of the model (either a symmetry of the disorder distribution, or a global symmetry acting on the spins, which was then disrupted by a 
random boundary condition); that symmetry given by a group $G$ as above was then recovered in the DS representation of the MAS. 
In other words, we suspect that this behavior cannot arise spontaneously when no particular structure or symmetry was present in the 
microscopic model, and then relative weak mixing (that is, the space $M$ is a single point) would be the behavior in generic 
models. Put another way, we suspect that if one begins with a model with the hidden symmetry, it will be possible to perturb it by adding
terms or disorder in finite size, consistent with any desired generic symmetries, but so as to break the hidden symmetry. 
Thus use of finite-size systems may be essential to prove results in this direction.
[Of course, spontaneous breaking of a global symmetry of $\nu$-almost every random Hamiltonian $H$ can lead to the 
presence of symmetries of the metastate or MAS, but that is not the topic of discussion here.]

Clearly, there is much left to be done, but we hope that the results given here serve as a sketch of the landscape that will be explored 
further in future work on disordered classical spin systems.

{\it Note added:} When this paper was near completion, D.L. Stein pointed out that given a translation-invariant $\nu$ and a 
$\Pi_0$-invariant $\kappa^\dagger$, a $\Pi$-invariant $\kappa^\dagger$ can be obtained by translation averaging \cite{ns01} (as mentioned in 
Sec.\ \ref{models}). This implies that results obtained here for $\Theta$-invariant (or ${\cal I}_n$-measurable, $n\geq 1$) observables 
or sets and a $\Pi$-invariant $\kappa^\dagger$ also apply for a $\Pi_0$-invariant $\kappa^\dagger$, that is, for metastates in general, 
which do not have to be translation covariant, provided $\nu$ and the observables are translation invariant. We will not list all these results 
here.

\acknowledgements

We thank C.M. Newman, D.L. Stein, L.-P. Arguin, and M.A. Moore for stimulating discussions and correspondence.
The author is grateful for support from NSF grant no.\ DMR-1724923. 

\begin{appendix}
\section{Background and further results}
\subsection{Background material}
\label{app:back}

Here we briefly recall only some very basic background information used in the main text. 
For full explanation of the background material here, see some of Refs.\ 
\cite{chung_book,breiman_book,royden_book,billingsley_book2,kechris_book,einsward_book,
aliborder_book,phelps_book,garling_book}.

First, the space $\widehat{J}$ of all $J=(J_X)_X$ is a product space of a countable infinity of factors; 
each $J_X$ takes values in $\bbR$ (equipped with its usual topology). We use the product topology on this space, in which 
a basis of open sets consist of the product of an open set in each of a finite number of the factors, times $\bbR$ 
in the others. In this topology, the space is metrizable, complete, and separable (such a space is termed ``Polish''). 

For the space $S$ of spin configurations, which we assume is a product over $i$ of copies of the 
same compact metric space for each $i$, we likewise use the product topology. The product space is uncountable, Polish, 
and also compact; for Ising spins it is the Cantor space. 
We also use the space $C(S)$ of bounded continuous real-valued functions on $S$. As $S$ is compact, by the 
Stone-Weierstrass theorem any algebra of bounded continuous functions on $S$ that separates points is dense in $C(S)$ 
[we always use the supremum norm, or uniform, topology on $C(S)$]. We can use the algebra of polynomials constructed as linear 
combinations of products of the components $s_{ie}$ of the spins, where each $s_i$ is an $m$-component 
unit vector in general. Then it is sufficient to check properties involving $C(S)$ on these functions only. 

On $S$ (and on other spaces), we use the Borel $\sigma$-algebra, which in this case we denote $\cal S$, 
that is, the $\sigma$-algebra generated by the open sets, in order to do measure theory (as our spaces are concrete, 
we usually refer to distributions \cite{breiman_book} rather than to measures in the abstract). (We use the Borel 
$\sigma$-algebra ${\cal J}$ on $\widehat{J}$ to define probability distributions there, for example $\nu$; we can also restrict
it to the support of $\nu$.) The space of (Borel) 
probability distributions, or states, $\Gamma$ on $\cal S$ can be characterized as the 
subspace ${\cal P}(S)$ of the dual space of linear functionals (integrals) on $C(S)$ such that the functional 
of a non-negative function is non-negative, and the functional of the function $1$ is $1$. 

${\cal P}(S)$ can itself be given the weak* topology, the weakest topology such that the evaluation maps are continuous 
for every function in $C(S)$. With this topology, ${\cal P}(S)$ is itself compact and Polish; see Ref.\  \cite{aliborder_book}, Ch.\ 15. 
It can be given the Borel $\sigma$-algebra induced from the Borel $\sigma$-algebra determined by the weak* topology. 
Then the product space $\widehat{J}\times{\cal P}(S)$ of pairs $(J,\Gamma)$ and other similar spaces 
on which we will consider probability distributions (using the product of $\sigma$-algebras on the product space) are 
also Polish, making them very nice spaces with which to work.  
The space of Gibbs states ${\cal G}(\gamma_J)$ [or ${\cal G}_J$] for a given specification $\gamma_J$ is a subspace
${\cal G}(\gamma_J)\subset {\cal P}(S)$ which can be endowed with the (relative) topology and $\sigma$-algebra
induced from those of ${\cal P}(S)$, so it is also separable and metrizable; the same holds for the space of extremal
Gibbs states ${\rm ex}\,{\cal G}(\gamma_J)$.  

We impose the additional condition on the specification $\gamma_J$ that, for each $\Lambda$, 
$\gamma_{J\Lambda}$ is $\nu$-almost surely a continuous function $S\to{\cal P}(S)$, or equivalently the map 
\be
f(s)\to\int f(s')\gamma_{J\Lambda}(ds'\mid s),
\ee
for $f\in C(S)$, maps $C(S)$ into itself (see also the following Subsection). There is a corresponding (dual) map on states $\Gamma$ of 
${\cal P}(S)$ into itself, which is weak*-weak* continuous (see Ref.\ \cite{aliborder_book}, Chapter 19; these 
properties are known as the Feller property for Markov transition probabilities). The DLR equations characterize 
Gibbs states $\Gamma_J$ as precisely the fixed points of all the maps $\gamma_J$. The continuity of these 
maps then implies that the Gibbs states form a closed, hence compact, convex subset of ${\cal P}(S)$, and 
the Choquet theory \cite{phelps_book} then asserts the existence of a {\em unique} \cite{simon_book} 
decomposition into extremal or pure Gibbs states, provided ${\cal G}(\gamma_J)$ is not empty. 
The latter follows, because the existence of DLR states for given $J$ holds because, for any $s$, the net (under 
inclusion $\Lambda\subset \Lambda'$) $(\gamma_\Lambda(\cdot|s))_\Lambda$ has a weak* cluster point by compactness 
of ${\cal P}(S)$ (Ref.\ \cite{georgii_book}, Chapter 4). See also the following Subsection.

The maps $\theta_\bx$ and $\theta_{\Delta J}$ of Gibbs or pure states are weak*-weak* continuous homeomorphisms of 
${\cal P}(S)$ and of ${\cal G}(\gamma_J)$. 

We denote the sub-$\sigma$-algebra of $\Theta$-invariant Borel sets in ${\cal J}$ by $\cal I$. 
The sets in $\cal J$ invariant only under all $\theta_\bx$ (translations) form the $\sigma$-algebra $\widehat{\cal I}$ 
of translation-invariant Borel sets, which frequently arises in ergodic theory \cite{breiman_book}. The sets in 
$\cal J$ invariant under $\theta_{\Delta J}$ for all $\Delta J$ such that $\Delta J_X$ is non-zero for only 
a finite number of $X$, form the tail $\sigma$-algebra $\cal T$ of sets that are independent of the values of 
any finite number of $J_X$. Thus ${\cal I}=\widehat{\cal I}\wedge{\cal T}$, the intersection of the 
$\sigma$-algebras. $\widehat{\cal I}\subseteq {\cal T}$ up to symmetric differences that are $\nu$-null 
sets \cite{georgii_book}, so modulo $\nu$-null sets $\widehat{\cal I}$ and $\cal I$ are also the same. 

Here and elsewhere, we have departed slightly from Georgii's treatment \cite{georgii_book} (see Chapters 2, 4, 7) 
in that we use the bounded continuous functions $C(S)$ on $S$, the weak* topology on ${\cal P}(S)$, and corresponding Borel 
$\sigma$-algebras, rather than quasilocal measurable functions and the associated evaluation $\sigma$-algebra. 
Our choice means that we can utilize the methods from Refs.\
\cite{chung_book,breiman_book,royden_book,billingsley_book2,einsward_book,kechris_book,aliborder_book}
throughout; this may have some cost in terms of generality. 
However, the proofs of the Theorems appear very robust; we leave the study of such technicalities for another occasion
(but see also the following Subsection).

\subsection{Additional proofs}
\label{app:furth}

Here we give some results and proofs that were postponed from Sections \ref{models} and \ \ref{indecomp}.
First, we discuss measurability questions for the pure-state decomposition with weight $w_\Gamma$ of a given Gibbs (DLR) state $\Gamma$.
In the main text (and earlier in Ref.\ \cite{nrs23}), we use $\kappa_J$ expectations 
of functions of $w_\Gamma$ and of functions of the pure states. Then we need to check that these expectations are measurable functions 
of $J$ (they will usually be translation invariant, and the plan will be to apply the ergodicity of $\nu$). Hence we should check that 
$w_\Gamma$ is a jointly measurable function of $(J,\Gamma)$. 
Here we sketch a proof.

The first elementary point we wish to make (extending one in Subsec.\ \ref{app:back}) is that, for Ising spins and under 
the short-range (or absolutely summable) condition on $J$ [i.e.\ that the (``seminorms'') $\sum_{X:i\in X}|J_X|$ be $<\infty$ for all $i$; 
cf.\ (\ref{eq:shortrange}) and following discussion], $\gamma_{J\Lambda}(s|_\Lambda\mid s|_{\Lambda^c})$ is a 
{\em jointly}-continuous function of $J$ and $s=(s|_\Lambda,s|_{\Lambda^c})$, for all finite $\Lambda$. At a less elementary level, 
it will be useful to know that the topology determined by the family of seminorms on the space 
of $J$ that obey the short-range condition can be metrized and is then complete, so the space is a Fr\'echet
space (see Georgii \cite{georgii_book}, p.\ 29), which we denote by $\mathbb{F}$. Viewed as topological spaces, 
$\mathbb{F}\subset \widehat{J}$ with the relative topology (the same sequences in $\mathbb{F}$ converge in both topologies, namely 
$J^{(n)}\to J$ if and only $J_X^{(n)}\to J_X$ for all $X$ \cite{willard_book}). Hence $\mathbb{F}$ is also separable (or this can be seen
directly; cf. Ref.\ \cite{reed_simon_book}, Ch.\ 3), and so is Polish. $\gamma_{J\Lambda}(s|_\Lambda\mid s|_{\Lambda^c})$ 
is continuous with respect to the product topology on $\mathbb{F}\times{\cal S}$. The DLR conditions then produce a correspondence
(a set-valued map \cite{aliborder_book}) $J\mapsto {\cal G}_J$ from $\mathbb{F}$ into ${\cal P}(S)$, where ${\cal G}_J$ is closed 
and compact for each $J\in \mathbb{F}$. This gives rise to the ``graph'' of the correspondence, that is, the set of pairs 
$(J,\Gamma)\in\mathbb{F}\times{\cal P}(S)$ with $\Gamma\in{\cal G}_J$, which (by joint continuity of the maps on states 
with respect to $J$ as well as $\Gamma$) is a closed subset of $\mathbb{F}\times{\cal P}(S)$, and hence the correspondence 
is ``upper hemicontinuous'' by the closed-graph theorem for correspondences \cite{aliborder_book}. These results extend to spins
taking values in a compact metric space and a continuous specification, with a similar separable Fr\'echet space $\mathbb{F}$, on making 
suitable changes. (These results are also in Georgii, Ch.\ 4, though he uses conditions and a topology that differ from ours in some cases.)

Next, given a Gibbs state for a (short-range) specification $\gamma_J$, we can recover $\gamma_J$ or 
$J$ as a measurable function of the given $\Gamma$. Suppose first that $J$ is strictly short range (see Sec.\ \ref{models}). For a given finite 
subregion $\Lambda$, form the conditional distribution of $\Gamma$ on a set $i\in\Lambda'-\Lambda$ of the spins $s_i$ outside $\Lambda$, 
but in $\Lambda'$, for $\Lambda'\supset\Lambda$, for $\Lambda'$ sufficiently large that for all interaction terms $-J_Xs_X$ with 
$X\cap\Lambda\neq\emptyset$, $X\subseteq\Lambda'$ (this $\Lambda'$ exists because $J$ is strictly short range). Then using 
eqs.\ (\ref{eq:specdef}) and following, we can uniquely recover all $J_X$ for all $X$ (i.e.\ $X\subseteq\Lambda'$) that involve 
some $i\in\Lambda$ (i.e.\ $X\cap\Lambda\neq\emptyset$) by Fourier-Walsh analysis. We let $\Lambda\to\infty$ though a (strictly) increasing 
sequence, say $(\Lambda_n)_n$, of finite sets, that is $\Lambda_n\subset \Lambda_{n'}$ if $n<n'$, such that any finite $\Lambda$ 
is contained in $\Lambda_n$ for all sufficiently large $n$ [such a sequence $(\Lambda_n)_n$ is termed cofinal]. Then in the limit, 
we recover $J_X$ for all $X\in{\cal X}$, by operations that are (weak*) measurable in $\Gamma$. 

For $J$ that is short range, but not 
strictly short range, we can recover the specification $\gamma_J(s)=(\gamma_{J\Lambda}(s))_\Lambda$, or $(H_\Lambda'(s))_\Lambda$ 
(see Sec.\ \ref{models}), as follows. Using the same set-up of finite $\Lambda$, $\Lambda'$, $\Lambda\subset\Lambda'$, 
from general principles of conditional probability we have 
$\Gamma(s|_\Lambda\mid s|_{\Lambda'-\Lambda})=\bbE_\Gamma(\Gamma(s|_\Lambda\mid s|_{\Lambda^c})\mid s|_{\Lambda'-\Lambda})$, 
and from eq.\ (\ref{eq:specdef}) this is
\be
\Gamma(s|_\Lambda\mid s|_{\Lambda'-\Lambda})=\int_{s|_{\Lambda'^c}} \Gamma(ds)\gamma_{J\Lambda}(s|_\Lambda\mid 
s|_{\Lambda^c}),
\ee
for all $s|_\Lambda$, $s|_{\Lambda'-\Lambda}$ (because we consider only Ising spins, the conditional on the left exists in the 
sense of ordinary discrete probability theory; note that there is nonzero probability for any configuration $s|_{\Lambda'}$ for any finite 
$\Lambda'$). Using the formula for $\gamma_{J\Lambda}$, this can be bounded above and below in the form 
\bea
\beta^{-1}\ln \Gamma(s|_\Lambda\mid s|_{\Lambda'-\Lambda})&\leq&\sum_{X:X\cap\Lambda\neq\emptyset,X\subset \Lambda'}
J_Xs_X\non\\
&&{}-\beta^{-1}\ln\sum_{s|_\Lambda} e^{\beta\sum_{X:X\cap\Lambda\neq\emptyset,X\subseteq\Lambda'}J_Xs_X}\non\\
&&{}+2\sum_{X:X\cap\Lambda\neq\emptyset,X\cap\Lambda'^c\neq\emptyset}|J_X|,
\eea
along with a similar lower bound in which the final sum has the opposite sign.
Then by the short-range condition, 
as $\Lambda'\to\infty$ (along a cofinal sequence) the final sum tends to zero, and the rest converges; hence we have shown that
\be
\Gamma(s|_\Lambda\mid s|_{\Lambda'-\Lambda})\to\gamma_{J\Lambda}(s|_\Lambda\mid s|_{\Lambda^c})
\ee 
as $\Lambda'\to\infty$, for all $\Lambda$, $s|_\Lambda$, and $s|_{\Lambda^c}$, whenever $J$ is short range (and $T>0$).
As pointwise limits of measurable functions are measurable, this shows that $\gamma_{J\Lambda}$ is a measurable function of $\Gamma$ 
for all $\Lambda$ when $J$ is short range. From that we recover $J$ by Fourier-Walsh analysis of $\ln\gamma_{J\Lambda}$, which 
again is measurable in $\Gamma$ as long as $J$ is short range (see also the following discussion). 

These observations imply that rather than working with the fibered product space $\mathbb{F}\times{\cal G}_\cdot$, 
equipped with its natural $\sigma$-algebra, we could work instead with the space of all DLR states for all short-range specification, say 
$\bigcup_{J\in\mathbb{F}}{\cal G}_J$, or the same with $J$ restricted to the support of $\nu$ (or with $J_X$ restricted to the support 
of $\nu_X$ for all $X$); either of the latter is a subset of ${\cal P}(S)$. Then after we introduce $\kappa^\dagger$,
$J$ is a random variable, measurable with respect to the same $\sigma$-algebra consisting of sets of $\Gamma$ (it is unusual in that 
we stipulate the marginal distribution $\nu$ on $J$, with the short-range property). Then there is no difficulty working with the $\kappa_J$ 
expectations mentioned above, provided only expectations of measurable functions of $\Gamma$ are considered; those expectations 
are simply conditional expectations under $\kappa^\dagger$, conditioned on $J$. (For simplicity, we leave Sec.\ \ref{indecomp}, and the 
material to follow, written in terms of $J$, $\Gamma$ as if those are independent variables; this is harmless.)

Next we need to consider the pure-state decomposition of a short-range $\Gamma$. First we consider this for $J$ 
fixed, and establish measurability of $w_\Gamma$ with respect to $\Gamma$; we will explain what exactly this means. 
To obtain the pure-state decomposition of the given $\Gamma$, we consider the approach of Georgii \cite{georgii_book}, Sec.\ 7.3
(which is based on work of Dynkin and F\"ollmer).
In this Section, he uses the product Borel $\sigma$-algebra on the space of spin configurations, but evaluation $\sigma$-algebras 
on spaces of probability distributions, defined by evaluating a measure on a measurable set of the underlying space. 
These may be more generally useful but in fact, for the space of probability distributions ${\cal P}(X)$ on any separable 
metrizable space $X$ [and also for any subset of ${\cal P}(X)$, such as ${\rm ex}\,{\cal G}_J\subset {\cal G}_J$ when ${\cal G}_J$ is 
separable metrizable], the Borel $\sigma$-algebra 
${\cal B}({\cal P}(X))$ of the weak* topology on ${\cal P}(X)$ is generated by the evaluation maps $e_A:\mu\mapsto \mu(A)$ 
for all Borel measurable sets $A\subseteq X$, or alternatively by the evaluation maps $e_f:f\mapsto\int fd\mu$ for all bounded measurable 
functions $f$ on $X$. For a proof, see Ref.\ \cite{kechris_book}, Thm.\ 17.24 on page 112. Then if we consider models 
of spins where each spin takes values in the same compact metric space (for example Ising spins or $m$-component unit-vector spins), 
then Georgii's evaluation $\sigma$-algebra on ${\cal P}(S)$ is the same as the one we use. 
Further, if $X$ is a ``standard Borel space'', then ${\cal P}(X)$ is also standard Borel, and the constructions of the Borel $\sigma$-algebra 
and the space of probability distributions can be iterated \cite{aliborder_book,einsward_book,kechris_book}. (A measurable space 
is a standard Borel space if it is isomorphic to a Borel subset of a Polish space with the induced Borel $\sigma$-algebra 
\cite{kechris_book,einsward_book,aliborder_book,georgii_book}; a countable product of standard Borel spaces is again standard Borel.)

If we assume that the specification is short-range and continuous in $s$, as discussed above for Ising spins, and that each spin takes values
in a compact metric space, then the space of pure states ${\rm ex}\,{\cal G}_J$ is a $G_\delta$ subset of ${\cal G}_J$ in the weak* topology 
(see e.g.\ Phelps \cite{phelps_book}, p.\ 5), and so a separable metric space, to which the preceding remarks apply as well as to ${\cal G}_J$. 
Hence we can restrict the Borel $\sigma$-algebra ${\cal B}({\cal G}_J)$ on ${\cal G}_J$ to ${\cal B}({\rm ex}\,{\cal G}_J)$ on 
${\rm ex}\,{\cal G}_J$, then consider ${\cal B}({\cal P}({\rm ex}\,{\cal G}_J))$ on ${\cal P}({\rm ex}\,{\cal G})$, and then 
measurability of $w_\Gamma$ means Borel measurable with respect to the latter and to the Borel $\sigma$-algebra ${\cal B}({\cal G}_J)$
on ${\cal G}_J$, that is, on sets of Gibbs states $\Gamma$s. (If the conditions on the specification and the space of spin configurations 
are not met, ${\rm ex}\,{\cal G}_J$ may not be a Borel set, and Georgii's evaluation $\sigma$-algebras may be necessary. In the following, 
we will continue to assume those conditions hold.)

Next, the proof of Georgii \cite{georgii_book}, Prop.\ 7.22, applies to show [in step 3)] that the map $\widetilde{\phi}:S\to {\rm ex}\,{\cal G}_J$ 
defined from the limit 
$\lim_{\Lambda\to\infty}\gamma_{J\Lambda}(\cdot\mid s)\equiv\widehat{\gamma}^s(\cdot)$ (properties of which can be established 
as in Prop.\ 7.25) by $\widetilde{\phi}(s)=\widehat{\gamma}^s(\cdot)$, which is a pure state $\Gamma$-almost surely 
for any $\Gamma\in{\cal G}_J$, is Borel measurable. It then follows that we can define the push-forward map 
$\Gamma\mapsto \widetilde{\phi}_*\Gamma\equiv w_\Gamma$ from ${\cal G}_J$ to 
${\cal P}({\rm ex}\,{\cal G}_J)$, and this map is also Borel measurable; see step 3) in Georgii \cite{georgii_book}, 
Ref.\ \cite{kechris_book}, p.\ 116, or Ref.\ \cite{einsward_book}, pp.\ 145--147. (In Georgii's set-up, measurability both of $\phi$ and of 
his map $\Gamma\mapsto w_\Gamma$ with respect to his evaluation $\sigma$-algebras follow in one line from his statements, though 
he does not appear to say so anywhere.) Then $w_\Gamma$ is essentially like a marginal distribution of $\Gamma$ on the 
sub-$\sigma$-algebra $\widetilde{\phi}^{-1}({\cal B}({\rm ex}\,{\cal G}_J))\subset\cal S$, and ``represents'' $\Gamma$ 
\cite{georgii_book,phelps_book} (i.e., gives its pure-state decomposition), 
\be
\Gamma=\int w_\Gamma(d\Psi)\, \Psi,
\ee
exactly as desired (the pure states play the role of conditional probability distributions; see Ref.\ \cite{einsward_book}, Sec.\ 5.3). 
(We went into a little detail here because the point of view seems illuminating.) Finally, uniqueness of $w_\Gamma$ for given 
$\Gamma$ (i.e.\ that ${\rm ex}\,{\cal G}_J$ is a simplex) can be proved as in step 4) of the proof of Prop.\ 7.22. 
In addition, $\widetilde{\phi}_*$ is surjective, and so is a Borel-measurable affine bijection from ${\cal G}_J$ to 
${\cal P}({\rm ex}\,{\cal G}_J)$, as in \cite{georgii_book}, Thm.\ 7.26.
The inverse map defined by $\widetilde{\phi}_*^{-1}:w\in{\cal P}({\rm ex}\,{\cal G}_J)\mapsto \int w(d\Psi)\,\Psi \in {\cal G}_J$ is actually 
continuous (see Ref.\ \cite{phelps_book}, Prop.\ 1.1 on p.\ 3), and hence measurable. Thus $\widetilde{\phi}_*$ is in fact an affine 
Borel isomorphism from ${\cal G}_J$ to 
${\cal P}({\rm ex}\,{\cal G}_J)$ (i.e.\ a measurable bijection with measurable inverse). We emphasize that this proof is fairly general, 
and goes through if the spins take values in a compact metric space and the specification is continuous. 
In that setting, the result can also be obtained using Choquet theory, and measurability of  $w_\Gamma$ in $\Gamma$ is discussed in that
context in Ref.\ \cite{phelps_book}, Sec.\ 11. (Georgii's more general treatment will be useful in the following, however.)

Even though $\gamma_J$ can be recovered measurably from $\Gamma$, it is still necessary to check that the pure state decomposition
map $\widetilde{\phi}_*:\Gamma\mapsto w_\Gamma$ is jointly measurable with respect to $(J,\Gamma)$ or $(\gamma_J,\Gamma)$, 
not only measurable at fixed $\gamma_J$. The main point here is that the limit 
$\lim_{\Lambda\to\infty} \gamma_{J\Lambda}(A\mid s)=\widehat{\gamma}_J(A\mid s)$ ($A\in{\cal S}$) is jointly measurable in $(J,s)$, wherever the limit 
exists, because it is a pointwise limit of continuous functions, and we can extend Georgii's proof of Prop.\ 7.25, step 1), to obtain this.  

Next, some further comments on existence of metastates are in order. While we pointed out that the AW and NS constructions
from finite size imply existence of metastates under the general infinite-size definition that we use, it may also be useful to 
establish this directly, without resorting to finite size. This will be based on the formulation that a metastate $\kappa_J$ is the conditional
of an $\Upsilon$-invariant joint distribution $\kappa^\dagger$ on pairs $(J,\Gamma)$, such that the set ${\cal G}_J$
of Gibbs states has $\kappa_J$ probability $1$ for $\nu$-almost every $J$; it amounts to another construction of metastates. 
We use a cofinal sequence $(Y_n)_n$ of sets of $X$; for example, we could use 
$Y_n=\{X:X\subseteq \Lambda_n\}$, where $\Lambda_n$ is a cube of (odd) side $n$ centered at the origin of $\mathbb{Z}^d$. 
Assume we are given a pair $(J,\Gamma)$, where we can assume that $J$ is drawn from a translation-invariant distribution $\nu$ 
and that $\Gamma$ is a Gibbs state for $J$; such a Gibbs state exists, as we pointed out above in App.\ \ref{app:back}.
We consider the sequence of probability distributions $\pi_{Y_n}(\cdot|(J,\Gamma))$. Each of these 
is invariant under $\pi_X$ for all $X\in Y_n$, and any pair $(J',\Gamma')$ drawn from one of them consists of $J'=J+\Delta J$, 
where $\Delta J_X=0$ for $X\not\in Y_n$, and $\Gamma'=\theta_{\Delta J}\Gamma$ is a local transformation of $\Gamma$ 
by $\theta_{\Delta J}$; thus $(J,\Gamma)$ plays the role of a boundary condition (this is somewhat similar to the AW construction, 
except that that uses the unique Gibbs state in each finite size $\Lambda_n$, possibly with some boundary condition).  
By arguments similar to those for the AW metastate \cite{aw,ns97,newman_book}, 
we can obtain a weak* subsequence limit as $n\to\infty$. This distribution $\kappa^\dagger$ is a probability distribution 
under which, for given $J$, ${\cal G}_J$ has probability $1$, and it is $\Pi_0$ invariant. In particular, metastates exist; the space 
${\cal K}^\dagger$ of $\Pi_0$-invariant $\kappa^\dagger$ (with the given $\nu$) is not empty. This argument resembles that for the 
existence of Gibbs states for given $J$ as in Ref.\ \cite{georgii_book}, Ch.\ 4. A $\Pi$-invariant $\kappa^\dagger$ 
can then be obtained by translation averaging \cite{ns01}. 

Note that a resulting metastate of either type may depend on the subsequence 
used, and may be decomposable. On the other hand, if we draw $(J,\Gamma)$ from an extremal $\Pi_0$-invariant $\kappa^\dagger$, 
then in the same construction the (unique) weak* limit exists and recovers the same extremal $\kappa^\dagger$, $\lim_{n\to\infty}
\pi_{Y_n}(\cdot|(J,\Gamma))=\kappa^\dagger(\cdot)$, for $\kappa^\dagger$-almost every $(J,\Gamma)$; for the proof, compare Georgii, 
pp.\ 122--124 \cite{georgii_book}, especially Theorem 7.12(b), and note that Remark 7.13, and the discussion preceding it, point out that 
the result holds under broad conditions, which in fact include ours. Unfortunately, neither of the two preceding results guarantees existence of 
indecomposable metastates if additional properties of them are prescribed. 

Next, we prove that the $\sigma$-algebras ${\cal I}_{1\Upsilon}(\kappa^\dagger)$ of $\kappa^\dagger$-almost surely 
$\Upsilon$ invariant sets, and ${\cal I}_{1\Upsilon}$ of strictly invariant sets [which may also be written respectively as
${\cal I}_{1\Phi}(\kappa^\dagger)$ and ${\cal I}_{1\Phi}$], defined in the text, are related: 
the first is the $\kappa^\dagger$-completion of the second. 
By this we mean that, for any set $A\in{\cal I}_{1\Upsilon}(\kappa^\dagger)$, there is a set
$B\in{\cal I}_{1\Upsilon}$ such that $\kappa^\dagger(A\bigtriangleup B)=0$. Here we will write 
$\omega$ for a pair $(J,\Gamma)$. 
First consider $\Upsilon=\Pi_0$. Similar to Georgii \cite{georgii_book}, p.\ 118, construct $B$ as
\be
B=\bigcap_{Y:Y\in(Y_n)_n}\bigcup_{Y':Y\subseteq Y'}\{\omega:\pi_{Y'}(A|\omega)=1\},
\ee
where $Y$ runs through a cofinal sequence $(Y_n)_n$. 
Thus $B$ is defined as the $\limsup$ of the sets shown. Each set $\{\omega:\pi_{Y'}(A|\omega)=1\}$ is strictly invariant 
under $\pi_X$ for all $X\in Y'$, and because any $X$ is a member of $Y_n$ for all sufficiently large $n$, $B$ is strictly invariant 
under all $\pi_X$. The $Y'$ and $Y$ are countable, so $B$ is measurable, and hence belongs to ${\cal I}_{1\Pi_0}$. 
$\limsup$ of sets corresponds to $\limsup$ of indicator functions, so
\bea
{\bf 1}_B&=&\limsup {\bf 1}_{\{\pi_{Y'}(A|\cdot)=1\}}\\
&=&\limsup {\bf 1}_A \hbox{, $\kappa^\dagger$-almost surely}\\
&=&{\bf 1}_A\hbox{, $\kappa^\dagger$-almost surely},
\eea
where the second equality holds because $A$ belongs to ${\cal I}_{1\Pi_0}(\kappa^\dagger)$, and the third because $A$ 
does not depend on $n$. Hence $\kappa^\dagger(A\bigtriangleup B)=0$ as required. This completes the proof for $\Pi_0$. 

For translation invariant $\kappa^\dagger$, we must do the same for ${\cal I}_{1\Pi}(\kappa^\dagger)$, ${\cal I}_{1\Pi}$. If 
$A\in{\cal I}_{1\Pi}(\kappa^\dagger)$, we first form $B$ as we did for $\Pi_0$, then define 
\be
C=\bigcup_{{\bf x}\in\mathbb{Z}^d}\theta_{\bf x} B,
\ee 
so that $C$ is strictly invariant under translations. For each $\bf x$, 
\bea
\theta_{\bf x}B&=&\bigcap_{Y}\bigcup_{Y':Y\subseteq Y'}\{\theta_\bx \omega:\pi_{Y'}(A|\omega)=1\}
\non\\
&=&\bigcap_{Y}\bigcup_{Y':Y\subseteq Y'}\{\omega:\pi_{Y'}(A|\theta_\bx^{-1}\omega)=1\}\non\\
&=&\bigcap_{Y}\bigcup_{Y':Y\subseteq Y'}\{\omega:\pi_{\theta_\bx Y'}(\theta_\bx A|\omega)=1\},
\eea
where we made use of obvious translation covariance properties of $\pi_X$.  For given $\bf x$ and $Y$, if $Y\subseteq Y'$ 
and $Y'$ is sufficiently large, then $Y\subseteq \theta_{\bf x} Y'$ also, so the last line can be expressed as the $\limsup$ 
(in the same form 
as before) of $\{\omega:\pi_{Y'}(\theta_{\bf x} A|\omega)=1\}$. Then by the same argument as for $B$, $\theta_{\bf x} B$ 
is strictly invariant under all $\pi_X$. Hence $C$ is $\Pi$-invariant, and belongs to ${\cal I}_{1\Pi}$ because only a countable number 
of intersections and unions were used. Then
\bea
\kappa^\dagger(A\bigtriangleup C)&\leq&\kappa^\dagger\left(\bigcup_{{\bf x}\in\mathbb{Z}^d}( A\bigtriangleup 
\theta_{\bf x}B)\right) \\
&\leq&\sum_{{\bf x}}
\kappa^\dagger(A\bigtriangleup \theta_{\bf x} B).
\eea
$A$ belongs to ${\cal I}_{1\Pi}(\kappa^\dagger)$ by hypothesis, which implies that the 
$\kappa^\dagger$-probability of the symmetric difference of $\{\omega:\pi_{Y'}(\theta_{\bf x} A|\omega)=1\}$ 
and $A$ is zero. Then ${\bf 1}_{\theta_{\bf x}B}={\bf 1}_A$ $\kappa^\dagger$-almost surely by an argument similar to that above, 
so $\kappa^\dagger(A\bigtriangleup \theta_{\bf x} B)=0$, and finally $\kappa^\dagger(A\bigtriangleup C)=0$, which completes the 
proof for $\Pi$.  Some consequences of these results are discussed in the main text.

Finally, we discuss the decomposition of an $\Upsilon$-invariant $\kappa^\dagger$ as a mixture of  
extremal $\kappa^\dagger$, using some distribution on the latter. The underlying idea is that
it should be possible to recover the full measure $\kappa^\dagger$ from something like a ($\kappa^\dagger$-dependent) 
marginal distribution on ${\cal I}_{1\Upsilon}$ (i.e.\ the restriction of $\kappa^\dagger$ to ${\cal I}_{1\Upsilon}$), 
together with the conditional probability distribution $\kappa^\dagger(\cdot|{\cal I}_{1\Upsilon})$; 
further, we may hope that there is a {\em fixed} ($\kappa^\dagger$-independent) probability kernel $\pi_\Upsilon$ from 
${\cal I}_{1\Upsilon}$ to the full $\sigma$-algebra, sich that for all $\Upsilon$-invariant $\kappa^\dagger$ and all $A$, 
the conditional obeys $\kappa^\dagger(A|{\cal I}_{1\Upsilon})(\omega)=\pi_\Upsilon(A|\omega)$ for $\kappa^\dagger$-almost 
every $\omega$. Then the kernel is essentially the collection of extremal $\kappa^\dagger$s (for $\Upsilon$), which can then 
be characterized by the statement that an extremal $\kappa^\dagger$ obeys, for all measurable $A$,
\be
\kappa^\dagger(A)=\pi_\Upsilon(A|\omega)
\ee
$\kappa^\dagger$-almost surely. Given $\pi_\Upsilon$, the 
existence of a unique decomposition of a $\kappa^\dagger$ as a mixture of (that is, as an integral under some distribution 
on the space of) extremal $\kappa^\dagger$s follows again by Proposition 7.22 in Ref.\ \cite{georgii_book}. The existence of conditional 
measures $\kappa^\dagger_\omega$ can be proved as in Theorem 5.14 
on p.\ 135 of Ref.\ \cite{einsward_book} (``measure disintegration''; see also Ref.\ \cite{kechris_book}), but that does not establish 
their independence from $\kappa^\dagger$. To obtain the stronger result we will follow the constructive approach given in Georgii's book 
\cite{georgii_book}. 

For $\Pi_0$, the kernel $\pi_{\Pi_0}\equiv \pi_0$ is obtained by taking the limit (in a suitable sense) of 
eq.\ (\ref{kdagker}), that is $\pi_0$ should be the limit of $\pi_Y$ as $Y\to\infty$ through a cofinal sequence. 
The situation is analogous to the pure-state decomposition for Gibbs states, with the probability kernels
$(\pi_Y)_Y$ corresponding to the specification $\gamma=(\gamma_\Lambda)_\Lambda$. The proof of the existence of 
the kernel $\pi_0$ is given in Georgii \cite{georgii_book}, Proposition 7.25, and on pp.\ 124, 136 he points out that the proof 
carries over to a more general case in the form we need, provided the space of $\omega$ is a standard Borel space 
\cite{georgii_book,kechris_book}. 
The last property holds because, if each spin takes values in a compact metric space and the specification is continuous, 
the spaces $\mathbb{F}$, ${\cal G}_J$, and 
$\mathbb{F} \times {\cal P}(S)$ are Polish, and the closed subset of all $(J,\Gamma\in{\cal G}_J)$ in the latter (as discussed above) 
is then also Polish and hence standard Borel. To show the existence of the desired probability 
kernel $\pi_0$, first define the $\Theta_0$-invariant set
\be
\Omega_0=\left\{\omega:\lim_{n\to\infty}\pi_{Y_n}(A|\omega)\hbox{ exists for all $A\in {\cal C}$}\right\},
\ee
where $Y_n$ is a cofinal sequence of $Y$, as before, and $\cal C$ is a countable ``core'' for the standard Borel space $\Omega$ of 
all $\omega$; such a core always exists (Ref.\ \cite{georgii_book}, Appendix 4A). Now we can define $\pi_0(A|\omega)$ for 
$\omega\in\Omega_0$ and $A\in{\cal C}$ as 
$\pi_0(A|\omega)=\lim_{n\to\infty}\pi_{Y_n}(A|\omega)$, and then it can be extended to a kernel from ${\cal I}_{1\Pi_0}$ 
to the full $\sigma$-algebra, and the remaining required properties can be proved \cite{georgii_book}. 

The case of $\Pi$ is analogous to that of translation-invariant Gibbs distributions in Ref.\ \cite{georgii_book}, Chapter 14, which however
has simplifying features that do not appear to be available here. Instead, using notation
already defined above in this Section, first define
\bea
\lefteqn{\Omega_1=\left\{\omega:\theta_\bx\omega\in\Omega_0\;\forall \bx \in \mathbb{Z}^d,
\vphantom{\sum_{\bx\in\Lambda_n}}\right.}\;\;&&\non\\
&&\left.\lim_{n\to\infty}\frac{1}{|\Lambda_n|}\sum_{\bx\in\Lambda_n}\pi_0(A|\theta_\bx \omega)
\hbox{ exists for all $A\in {\cal C}$}\right\}.\;\;\;\;
\eea
$\Omega_1$ is well defined, translation invariant (if the limit exists for $\omega$, it is easy to see that it also exists,
and is the same, for $\theta_{\bx'}\omega$), and further $\Omega_1 \in {\cal I}_{1\Pi_0}$ by invariance of $\pi_Y(A|\omega)$ 
under any local transformation of $\omega$; hence
$\Omega_1\in {\cal I}_{1\Pi}$. Then define $\pi_1(A|\omega)$ for $\omega\in\Omega_1$ and $A\in{\cal C}$ as 
\be
\pi_1(A|\omega)=\lim_{n\to\infty}\frac{1}{|\Lambda_n|}\sum_{\bx\in\Lambda_n}\pi_0(A|\theta_\bx \omega).
\ee
The rest of the proof is similar to those of Georgii \cite{georgii_book}, Theorem 14.10 and Proposition 7.25,
using the ergodic theorem for translations and results for the extended $\pi_0$ also. 
[A perhaps easier alternative to the preceding approach for $\Pi$ is that the {\em extended}, or final,
$\pi_0(A|\cdot)$ can replace ${\bf 1}_A(\cdot)$ in the proof of Georgii's Theorem 14.10, and then the set called $\Omega_0$ there 
becomes $\Omega_1'$, defined in the same way as $\Omega_1$ above, except that the condition 
$\theta_\bx\omega\in\Omega_0\;\forall \bx \in \mathbb{Z}^d$ is dropped; $\Omega_1'$ belongs to ${\cal I}_{1\Pi}$ 
by invariance of $\pi_0(A|\omega)$ under any local transformation of $\omega$. $\pi_1(A|\omega)$ above is replaced by 
$\pi'_1(A|\omega)$ for $\omega\in\Omega_1'$ and $A\in {\cal C}$, defined by the same formula as for $\pi_1$ above.
The remainder of the proof is similar to before.]

\section{A class of infinite-anisotropy models}
\label{app:infanis}

Another model considered by WF \cite{wf} will be analyzed here; again, we sketch the arguments. 
The model consists of classical XY, not Ising, 
spins with short-range constant ferromagnetic bonds and independently random, isotropically-distributed, easy-axis anisotropy 
of infinite strength at each site; infinite easy-axis anisotropy means that for each site $i$ there is a unit vector ${\bf n}_i$,
and the XY spin is a unit vector ${\bf s}_i$ that is only allowed to be ${\bf s}_i=\pm{\bf n}_i$. (For each $i$, ${\bf n}_i$ and 
${}-{\bf n}_i$ define the same model, so they are equivalent.) There is a global symmetry of the Hamiltonian under reversing 
the sign of all XY spins. There are similar models, with similar properties, for $m$-component unit vector spins for all $m>1$, 
in which the anisotropy vectors 
${\bf n}_i$ are isotropically distributed on the sphere $S^{m-1}$. It was argued \cite{wf} that, in $d>4$ dimensions and under 
some conditions, at low temperatures there is a continuum of pure states with ferromagnetic order of the vector spins;
let the translation-average of the $m$-vector magnetizations $\langle {\bf s}_i\rangle_\Gamma$ be ${\bf m} = (m_a)_{a=1}^m$. 
It is not difficult to (heuristically) exhibit mean-field solutions of this form for an infinite-range version of the model. In the 
infinite-range model, the ordered states correspond to nonzero solutions to
\be
{\bf m}=\frac{1}{\sigma_m}\int d^{m-1}n\,{\bf n}\tanh(\beta J_0 {\bf n}\cdot{\bf m}),
\ee
where the integral is over $\bf n$ on the unit sphere $S^{m-1}$ in $\mathbb{R}^m$ with its standard measure, $\sigma_m$ 
is the ``surface area'' of $S^{m-1}$, so $\int d^{m-1}n\,1=\sigma_m=2\pi^{m/2}/\Gamma(m/2)$ (here $\Gamma$ 
is the Euler Gamma function), and the ferromagnetic bonds are, for a finite number $N$ of sites, $J_0/N$ for each distinct pair $\{i,j\}$ 
of sites ($J_0>0$). Ordered states appear below the critical temperature $J_0/m$. In the short-range models of WF, the claim \cite{wf}
is that, by an Imry-Ma--type argument (though at strong disorder rather than weak), 
the ordered phase survives in the short-range models for $d>4$, at least if interactions are not too short ranged, but not for $d<4$. 
All pure states have the same value of $|{\bf m}|$, by symmetry of the distribution.

In these models, we can view a local transformation as changing the direction of a finite number of the anisotropies, under which $\bf m$ 
for a pure state will not change. One expects that the metastate $\kappa_J$ is nontrivial, for reasons similar to those in the RFFM, though 
in this case each Gibbs state is a symmetric mixture of two pure states that are related by the global symmetry. [The role of 
$J\equiv(J_{ij})_{ij}$ is played in these models by $({\bf n}_i)_i$ or, with the sign redundancy removed, for each $i$ by a projection 
operator ${\bf P}^{({\bf n}_i)}$, in other words a second-rank tensor with components $P^{({\bf n}_i)}_{ab}\equiv n_{ia} n_{ib}$, 
onto a one-dimensional subspace in $\mathbb{R}^m$ (at the moment it makes no difference which description is used); we will 
continue to write this as $J$.] Each of these Gibbs states determines a one-dimensional subspace of $\mathbb{R}^m$, so they can 
be distinguished from one another by using the projection operator $P^{({\bf m})}_{ab}\equiv m_a m_b/|{\bf m}|^2$, 
or equivalently by a ``director'', the direction of 
$\bf m$ modulo ${\bf m}\to{}-{\bf m}$ (so it lies in real projective space $\mathbb{RP}^{m-1}$, not in $S^{m-1}$, just as 
the anisotropy for each $i$ does, when viewed modulo ${\bf n}_i\to{}-{\bf n}_i$). If $O(m)$-invariance is preserved in the construction
of the metastate, then $\kappa_J$ can be viewed as an $O(m)$-invariant distribution on $\mathbb{RP}^{m-1}$. 
But by Theorem $1'$ 
(generalized here to $m$-vector spins), more than one of the Gibbs states cannot occur in a single indecomposable metastate, 
so the metastate must be decomposable, and will decompose into an $O(m)$-invariant atomless mixture of trivial metastates. 

With the disorder given by the product of uniform distributions on projections ${\bf P}^{({\bf n}_i)}$, we can also view the system, 
as WF suggest, as consisting of Ising spins $s_i$ which, given a choice of sign for ${\bf n}_i$ for each $i$, 
can be defined as ${\bf s}_i=s_i{\bf n}_i$. The interaction term between sites $i$, $j$ that have ferromagnetic bond $J_{ij}>0$ 
becomes $-J_{ij}{\bf n}_i\cdot{\bf n}_j s_is_j$, so the ``Ising'' bonds $J_{ij}{\bf n}_i\cdot{\bf n}_j$ between Ising spins are random 
because of the anisotropy, but correlated. If we are given only the model described in terms of these Ising spins and bonds, 
with no information about, say, the sign choice for choosing the ${\bf n}_i$s,
it may appear that the underlying structure has been hidden, and that the metastate, which is still nontrivial, with each Gibbs state
a symmetric mixture of flip-related pure states, could now be indecomposable;
we now analyze this. First, there are many ways of choosing the signs of ${\bf n}_i$ for all $i$. One way is
to select a hemisphere in $S^{m-1}$, the same for all $i$, and define the sign of ${\bf n}_i$ so that it lies in that hemisphere
(some choice must be made on the boundary of the hemisphere to fully define this, but as ${\bf n}_i$ is isotropically distributed, 
this affects only a vanishingly small fraction of the sites, so we can ignore this issue). For example, we could define the hemisphere 
as that in which the first component of ${\bf n}_i$ is positive. If one of these choices was in fact used, then 
${\bf n}_i\cdot{\bf n}_j\geq0$ for all $i$, $j$, so the Ising bonds are ferromagnetic. The pure states
at low $T$ will have a magnetization of the Ising spins, given by $\bf m$ projected onto the unit vector that defines the choice 
of positive hemisphere. The values of these form a continuum in the range between $|{\bf m}|$ and ${}-|{\bf m}|$, and partially 
distinguish the ordered states, so the metastate of the model in this case is definitely decomposable, though not necessarily into trivial 
metastates. However, the original question still stands for other sign choices; another particular choice would be
to introduce an additional random sign for each $i$, which in fact is already present if the model is defined as the product over $i$ 
of isotropic distributions on unit vectors ${\bf n}_i$ in $S^{m-1}$ for each $i$. That is the obvious case to consider next, 
and appears to be what WF had in mind.

As a next step, we ask whether it is possible to reconstruct the original $m$-vector model from the Ising model, that is to 
recover the ordered set $({\bf n}_i)_i$, modulo a global $O(m)$ transformation, from the ordered set of 
$(J_{ij}{\bf n}_i\cdot{\bf n}_j)_{(i,j)}$. We will consider only the case in which it is given that
all nonzero $J_{ij}$ are equal and positive; the common value is the supremum of the set of all $J_{ij}{\bf n}_i\cdot{\bf n}_j$,
and without loss of generality we set it to $1$. Then the bond values determine the angle between ${\bf n}_i$ and ${\bf n}_j$, 
and must satisfy some constraints from the geometry. 
It is not difficult to see, especially in the $m=2$ (XY) case, and assuming that $J_{ij}=1$ at least for all nearest neighbors in 
$\mathbb{Z}^d$, that with $\nu$-probability one there is a unique set of ${\bf n}_i$ that give the set of bonds, modulo 
a global rotation. Hence the probability distribution on the Ising bonds is equivalent to the product distribution $\nu$ on 
$J=({\bf n}_i)_i$ modulo the action of global rotation on that distribution, so is certainly not a product distribution,
and not strictly speaking included in even our most general formulations so far. We will consider the metastate $\kappa_b$ 
conditioned on the set of Ising bonds $J_{ij}{\bf n}_i\cdot{\bf n}_j$. Covariance and decomposability continue to make sense. 
We point out that the set of bonds $J_{ij}{\bf n}_i\cdot{\bf n}_j$ are analogous to the bonds $\xi_i\xi_j$
for $i$, $j$ adjacent sites in the Mattis-type models, and the latter are exactly the $m=1$ version of the WF models;
of course, the latter are much more interesting for $m>1$. 

It is natural to consider what happens if we begin with the analysis above for the model with isotropically-distributed ${\bf n}_i$
for each $i$, and then project it down to the Ising model. First, we note that under global $O(m)$ transformation of $({\bf s}_i)_i$ and 
$({\bf n}_i)_i$, the relation ${\bf s}_i=s_i{\bf n}_i$ is preserved, that is the Ising spins are invariant. If $J=({\bf n}_i)_i$,
then a pair $(J,\Gamma)$, where $\Gamma$ is a state (a probability distribution) on $({\bf s}_i)_i$, can be viewed as a similar
pair in which (using the given $J$), $\Gamma$ is now viewed as a state of the Ising spins, and then a global transformation only 
affects $J$. Next, in the analysis above, it was natural to consider the $m$-vector magnetization per site $\bf m$ as an order 
parameter, and then trivial metastates (extremal $\kappa^\dagger$s) can be identified for all $J$ by using ${\bf P}^{({\bf m})}$, 
independent of $J$, to distinguish them. The group $O(m)$ acts diagonally 
on the pairs $(J,{\bf P}^{({\bf m})})$ (i.e.\ simultaneously on each). The quotient of the space of $J$ by $O(m)$ [the space 
of $O(m)$ orbits] can be viewed as the space of Ising bonds, as we have seen. The quotient of the space of pairs 
$(J,{\bf P}^{({\bf m})})$ also makes sense. As we have seen, if the Gibbs state for each ${\bf P}^{({\bf m})}$ for given $J$ 
is viewed as a state of Ising spins, then it is $O(m)$-invariant. So the space of orbits of $(J,{\bf P}^{({\bf m})})$ is a space of Ising 
bonds and certain Ising Gibbs states.

In the original $m$-vector models, the extremal $\kappa^\dagger$ in the decomposition are trivial on the invariant 
$\sigma$-algebra ${\cal I}_{1\Upsilon}$ or ${\cal I}_{1\Phi}$. In order to 
project down to the Ising description, we first note that $\Upsilon$ is $O(m)$-covariant (cf.\ Ref.\ \cite{georgii_book}),
and so the probability kernels $\pi_Y$ effectively project to corresponding probability kernels on the Ising description,
acting on the space of pairs, each consisting of an Ising bond configuration and an Ising state.
Then we can define a corresponding $\sigma$-algebra of invariant sets of such pairs. By the preceding paragraph, 
the inverse image of an invariant set in this $\sigma$-algebra can be described in the original variables as a set 
of pairs [i.e.\ $({\bf n}_i)_i$ and a Gibbs state for ${\bf s}$] that are $\Upsilon$- or $\Phi$- invariant, and in addition invariant 
under the diagonal action of $O(m)$ on the pair. Now, each extremal $\kappa^\dagger$ in the latter variables is concentrated 
on a set that contains essentially all $J$ (by ergodicity of the product $\nu$), and for each $J$ a corresponding Gibbs state 
characterized by a ${\bf P}^{({\bf m})}$ that is independent of $J$; this set belongs to the
$\sigma$-algebra ${\cal I}_\Upsilon$, but is not $O(m)$-invariant: it includes almost all rotations of $J$,
but no rotations of the Gibbs state characterized by ${\bf P}^{({\bf m})}$, independent of $J$. To obtain an $O(m)$-invariant set,
we must act with the diagonal action of $O(m)$ on the pairs, and then all $O(m)$ transformations of ${\bf P}^{({\bf m})}$
are included as well as all $J$. That is, the $O(m)$-invariant invariant sets are larger, and the $\sigma$-algebra that they form 
corresponding smaller (it is a sub-$\sigma$-algebra of ${\cal I}_\Upsilon$), and unable to resolve distinctions that were 
formerly visible. Hence the extremal $\kappa^\dagger$s 
in the Ising description subsume all those of the $m$-vector description, meaning that there is a single extremal $\kappa^\dagger$, 
that is, the metastate $\kappa_b$ in these WF infinite-anisotropy models is indecomposable. By Theorem $1'$, there is no invariant 
observable that can distinguish the ordered Gibbs states in the Ising language.

It is interesting to consider also the pairwise overlaps in these models. In the $m$-vector language, a class of overlaps can be defined
by taking the inner product of each $\langle {\bf s}_i\rangle_\Gamma$ in $\Gamma$ with that in another Gibbs state $\Gamma'$, 
followed by taking the product of these over $i$ in some set $X$ of sites, and then translation average. Because each ${\bf s}_i=
\pm {\bf n}_i$, in Ising language these overlaps simply give the usual Ising overlaps we have used elsewhere in the paper. First 
consider pure states in the $m$-vector language. In the infinite-range model mentioned earlier, for two ordered states 
described by $\bf m$, ${\bf m}'$, the single-site overlap ($[X]=[1]$) is given by
\be
q_{[1]}({\bf m},{\bf m'}) =\frac{1}{\sigma_m}\int d^{m-1}n\,\tanh(\beta J_0 {\bf n}\cdot{\bf m})\tanh(\beta J_0 {\bf n}
\cdot{\bf m}').
\ee
For general $X$, the overlap is a power of this, $q_{[X]}({\bf m},{\bf m'}) = q_{[1]}({\bf m},{\bf m'})^{|X|}$, similar to the case 
in the SK model. Clearly for $m>1$ these overlaps take a continuous range of values in the ordered phase. The picture should be 
similar for the overlaps of pure states in the ordered phase of the short-range model in $d>4$ dimensions, though now the different 
overlaps will not be powers of one another. 
The picture does not change for overlaps of pure states in the Ising language, even though 
$\bf m$ for a pure state is not defined (as an invariant observable, that is). Hence the overlap distributions $P_{[X]\rho}(q)$ 
in the MAS $\rho$ of the Ising system are 
nontrivial, not simply one or two $\delta$-functions. We see that in these models $\mu^\dagger$ is not 
relatively weak mixing (see Sec.\ \ref{sec:mas}). If instead of pure states we consider overlaps of Gibbs states $\Gamma$ 
drawn from $\kappa_b$, 
the tensors $\langle\bigotimes_{i\in X} {\bf s}_i\rangle_\Gamma$ or their Ising counterparts $\langle s_X\rangle_\Gamma$ 
vanish because of spin-inversion (or -flip) symmetry for $|X|$ odd, but not for $|X|$ even, and for the latter cases similar results 
will hold for the overlaps. Hence also $\kappa^\dagger$ is not relatively weak mixing.

We can say more about the overlap distributions in the short-range models by using $O(m)$
invariance. First, in the $m$-vector language, in the ordered phase the pure state characterized by a choice of ${\bf m}/|{\bf m}|$ 
can be obtained for given disorder by imposing a corresponding boundary condition before taking the thermodynamic limit.
Thus the pure states can be obtained without use of a metastate. The single-site overlap, defined above, for pure states $\bf m$, 
${\bf m}'$ can then be calculated, and due to the ergodic theorem and the $O(m)$ invariance of $\nu$, its value is invariant 
under $O(m)$ transformation acting diagonally on $\bf m$ and ${\bf m}'$ with the disorder $({\bf n}_i)_i$ held fixed; 
the self-overlap of each pure state is also $O(m)$ invariant. All of this is true also in the infinite-range model, as can be
seen from the formula above. The overlap and self-overlap together define a (possibly only pseudo-) 
metric on the sphere $S^{m-1}$ (points in which correspond to ${\bf m}/|{\bf m}|$), similarly to the Ising case discussed in 
the main text. The overlap is not in general a multiple of ${\bf m}\cdot{\bf m}'$, as can be seen from the formula above already
in the infinite-range model. However, this does hold in the infinite-range model when $\beta J_0|{\bf m}|$ is very small 
(so that $\tanh$ can be linearized, giving $q_{[1]}({\bf m},{\bf m'})=(\beta J_0)^2 {\bf m}\cdot{\bf m}'/m$), in particular 
close to the phase transition. Then if in the short-range model, for each $m\geq 2$, we suppose that the overlap is proportional 
to ${\bf m}\cdot{\bf m}'$ (perhaps under a similar condition), then using the pure-state decomposition of the MAS $\rho$, with 
distribution $\mu_J$ that is uniform on $S^{m-1}$, the distribution of the single-site overlaps in the MAS will be given exactly 
by the distribution of ${\bf m}\cdot{\bf m}'$ up to some rescaling, that is by
\be
P_{[1]\rho}(q)=\frac{\Gamma(m/2)}{q_{[1]\rm EA}\Gamma((m-1)/2)\sqrt{\pi}}\left[1-\left(\frac{q}{q_{[1]\rm EA}}
\right)^2\right]^{\frac{m-3}{2}}
\ee
for $q\in({}-q_{[1]\rm EA},q_{[1]\rm EA})$,
where again $\Gamma$ is the Euler Gamma function and $q_{[1]\rm EA}>0$ is the single-site self-overlap or largest overlap.
This serves to illustrate the appearance of a nontrivial overlap distribution. Further, even when the overlap is not proportional 
to ${\bf m}\cdot{\bf m}'$ over the full range of values of the latter, it will still hold to leading order as $q_{[1]}$ approaches 
${}\pm q_{[1]\rm EA}$ that generically (perhaps not at $T=0$)
\be
1\mp q_{[1]}/q_{[1]\rm EA}\sim 1\mp{\bf m}\cdot{\bf m}'/|{\bf m}|^2,
\ee
so, for each $m$, the power-law as $q$ approaches ${}\pm q_{[1]\rm EA}$ should be universal.
The overlap distribution in the Ising language is the same as in $m$-vector language. More general overlaps $q_{[X]}$ 
(defined as above) are again $O(m)$ invariant; each will be some function of ${\bf m}\cdot{\bf m}'$, but it will likely 
depend on $[X]$ due to both fluctuations and correlations between the sites.

\end{appendix}

\end{document}